%% file: thesis.tex
\documentclass[12pt]{carletoncsthesis}
\pdfoutput=1
\nosignaturepage
\nopermissionpage
\nolistoftables

\usepackage{amsmath}
\usepackage{graphicx}
\usepackage{latexsym}
\usepackage{stmaryrd}
\usepackage{setspace}
\usepackage{color}

\usepackage{amsthm}
\newtheorem{thm}{Theorem}[section]
\newtheorem{lemma}[thm]{Lemma}
\newtheorem{corollary}[thm]{Corollary}
\newtheorem{definition}[thm]{Definition}

\usepackage[ruled,vlined,nofillcomment,algosection,linesnumbered]{algorithm2e}
\Setnlsty{textsl}{\color[gray]{0.5}}{}
\Setnlskip{1em}
\SetArgSty{}
\SetCommentSty{textit}
\SetKwComment{tcc}{--- }{ ---}
\SetKwComment{tcp}{ --- }{}
\SetInd{0.25em}{1.5em}
\dontprintsemicolon

\begin{document}

\mcs
\title{
	Practical Parallel External Memory Algorithms\\
	via Simulation of Parallel Algorithms
}
\author{David E. Robillard}
\defenceday{7}
\defencemonth{December}
\defenceyear{2009}
\convocation{February}{2010}
\supervisor{Anil Maheshwari}
\supervisor{David A. Hutchinson}
\reader{F. Dehne}
\reader{A. Naik}

\titlepagematter

\begin{abstract}
This thesis introduces PEMS2, an improvement to PEMS (Parallel External
Memory System).  PEMS executes Bulk-Synchronous Parallel (BSP) algorithms in
an External Memory (EM) context, enabling computation with very large data
sets which exceed the size of main memory.  Many parallel algorithms have been
designed and implemented for Bulk-Synchronous Parallel models of computation.
Such algorithms generally assume that the entire data set is stored in
main memory at once.  PEMS overcomes this limitation without requiring any
modification to the algorithm by using disk space as memory for additional
``virtual processors''.  Previous work has shown this to be a promising
approach which scales well as computational resources (i.e.\ processors and
disks) are added.  However, the technique incurs significant overhead when
compared with purpose-built EM algorithms.  PEMS2 introduces refinements to
the simulation process intended to reduce this overhead as well as the amount
of disk space required to run the simulation.  New functionality is also
introduced, including asynchronous I/O and support for multi-core processors.
Experimental results show that these changes significantly improve the runtime
of the simulation.  PEMS2 narrows the performance gap between simulated BSP
algorithms and their hand-crafted EM counterparts, providing a practical
system for using BSP algorithms with data sets which exceed the size of RAM.
\end{abstract}

\begin{acknowledgements}
I owe my deepest gratitude to my supervisors, Anil Maheshwari and David
Hutchinson, for their guidance and inspiration during the course of this work.

Anil Maheshwari's assistance with theoretical matters, guiding advice, and
handling of my ``implement first and ask questions later'' tendency have
helped me immensely in becoming a more effective student and researcher.
I would also like to thank him for financial support, and taking on the
silent unknown undergraduate from the back of the room as his student.

David Hutchinson's intuitive grasp of performance issues, vision for
this project, and help with writing have had a great impact on this thesis.
My abilities as a writer have improved immeasurably as a result of his
suggestions.

PEMS2 could not exist without Mohammad Nikseresht, whom I thank for the
PEMS1 implementation and setting up a development site for the project.

The HPCVL administrators, Ryan Taylor and later Mohammad Nikseresht, have
been most helpful during experiments.  Both have responded patiently to my
configuration requests, and dealt swiftly with any issues encountered during
the development and experimentation processes.

Finally I would like to thank my family, who, despite not having the
slightest clue what Computer Science is, have always supported me in my
academic endeavors.
\end{acknowledgements}


\frontmatter
\listofalgorithms

\mainmatter

\chapter{Introduction}
\thispagestyle{empty}

\section{Background and Motivation}

External Memory (EM) algorithms are designed to work with data sets much
larger than main memory.  Though strictly defined in more general terms,
EM models typically consider a 2-tier memory hierarchy: ``main memory''
(RAM) and ``external memory'' (disk).  Algorithms designed for these models
explicitly transfer blocks of data between these levels of memory, attempting
to minimize the number of transfers between them.  In addition to minimizing
data transfer, EM algorithms may also be designed to access external memory
(e.g.\ disk) in an efficient pattern to minimize expensive disk seeking.

Unfortunately, most algorithms are designed for Random Access Memory (RAM)
models rather than EM.  RAM algorithms work with a single level of memory,
and assume a read or write at any location has a fixed constant cost.
Because RAM algorithms do not consider locality of reference a performance
factor, translating a RAM algorithm into an EM algorithm with acceptable
performance is not simple or automatable in the general case.  There are,
however, certain classes of algorithms which can work well in an EM context
despite not being designed with EM specifically in mind.

The goal of this thesis is to enable the practical use of such algorithms
on problems that exceed the size of RAM, allowing an algorithm to scale
beyond the limits of main memory without requiring a complete rewrite.
Parallel algorithms are particularly desirable in this context since very
large problems may exceed the resources of a single machine and sequential
computation with data of this magnitude in reasonable time is generally
not feasible.

\section{Computational Models}

\subsection{Parallel Disk Model (PDM)}

The multiple disk model originally proposed by Vitter and Shriver
\cite{pdm1}\cite{pdm2}, usually referred to as the PDM model, is commonly
used for designing disk-based algorithms.  In PDM, an algorithm has access
to an internal random access memory of size $M$, and $D$ disks which transfer
in blocks of size $B$.  I/O is fully parallel and blocked, i.e.\ a transfer of
size $BD$ (to $D$ disks) is considered a single I/O operation.  The complexity
of an algorithm is measured exclusively in terms of the number of such I/O
operations, ignoring other factors such as computation time.  This reflects
the reality that disk access is orders of magnitude more expensive than RAM
access.  Computation time of an algorithm may also be given for algorithms
with especially high computational requirements, though typically I/O time
dwarfs computation time by a large enough margin that computation does not
significantly contribute to the total run time.

\subsection{Bulk Synchronous Parallel (BSP) and Related Models}

The BSP model \cite{bsp} was proposed as an abstract ``bridging model for
parallel computation''.  BSP serves as a common model for both system/hardware
and algorithm/software designers which allows for accurate performance analysis
on a wide range of parallel computers.  BSP considers a set of processors
each with independent local memory that communicate by sending messages
between each other.  Computation proceeds in a series of synchronised
``supersteps'', each of which consists of a ``computation superstep'' followed
by a ``communication superstep''.  The total runtime of an algorithm is thus
the sum of the computation time, communication time, and synchronisation time.

A superstep where each processor sends and receives $O(h)$ data is called an
``h-relation''.  BSP* \cite{bsp*} and Coarse Grained Multicomputer (CGM)
\cite{cgm}, other common models of parallel computation, are special cases
of BSP with restrictions on $h$ to ensure a more coarse grained computation.
In practical terms BSP* and CGM algorithms proceed in an identical fashion
to BSP algorithms, i.e.\ in a series of supersteps.  Accordingly, the three
are considered equivalent for much of this thesis and collectively referred
to as ``BSP-like''.

Though all BSP-like models function similarly, the performance characteristics
of restricted models have important implications when used with EM.
CGM requires that each processor works with $O(\frac{N}{v})$ local data
(i.e.\ $h = O(\frac{N}{v})$).  This ensures balanced computation and
communication with coarse granularity.  Synchronisation overhead is thus
minimized, while processor and disk parallelism is exploited efficiently.
Since communication and synchronisation in PEMS is relatively expensive due
to disk I/O, these characteristics are especially desirable in this context.
The applications presented in Chapter~\ref{experiments} are CGM algorithms.

BSP-like algorithms are useful on a wide variety of configurations,
particularly the common and inexpensive ``cluster'' style of parallel
computer composed of several commodity machines connected by a switched
Ethernet network.

\begin{figure}[h]
\begin{center}
	\resizebox{!}{0.25\textheight}{
	\includegraphics{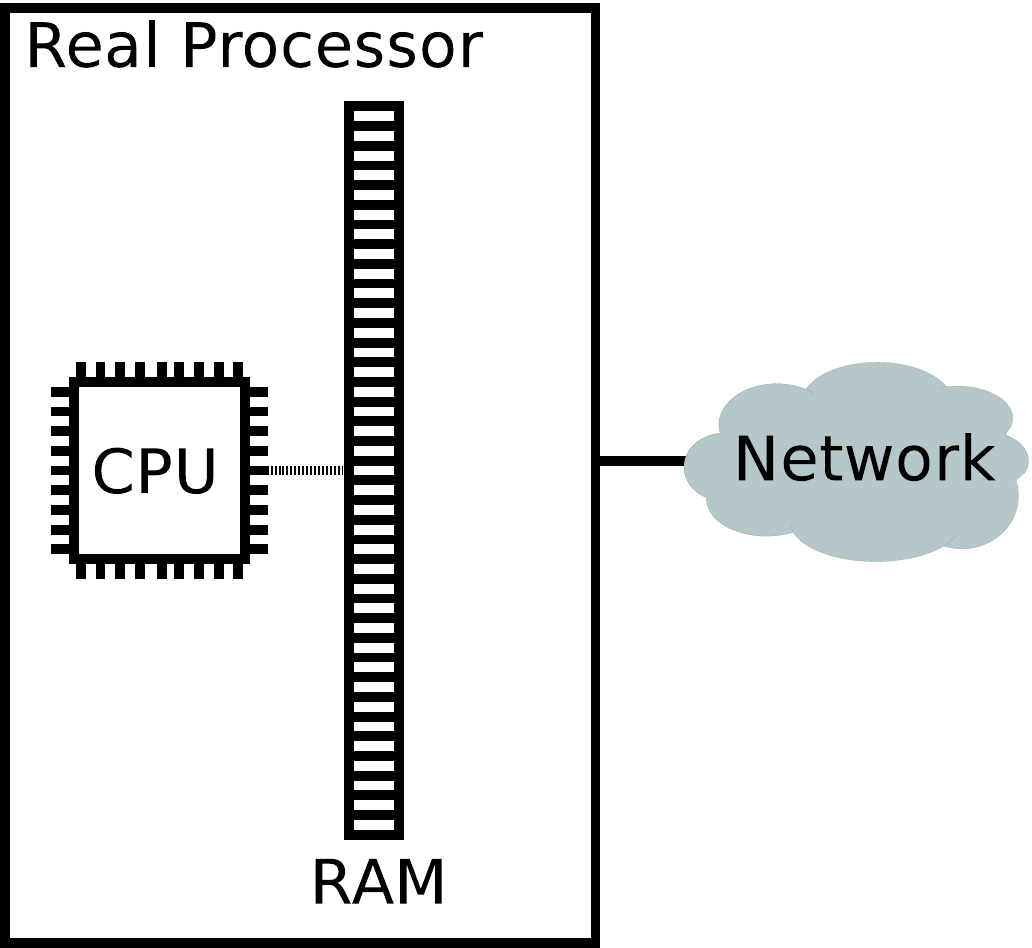}
	} \\
\end{center}
\caption{BSP-like Model}
\label{bsp_model}
\end{figure}

\subsection{EM-BSP Models}

\begin{figure}[h]
\begin{center}
	\resizebox{!}{0.25\textheight}{
	\includegraphics{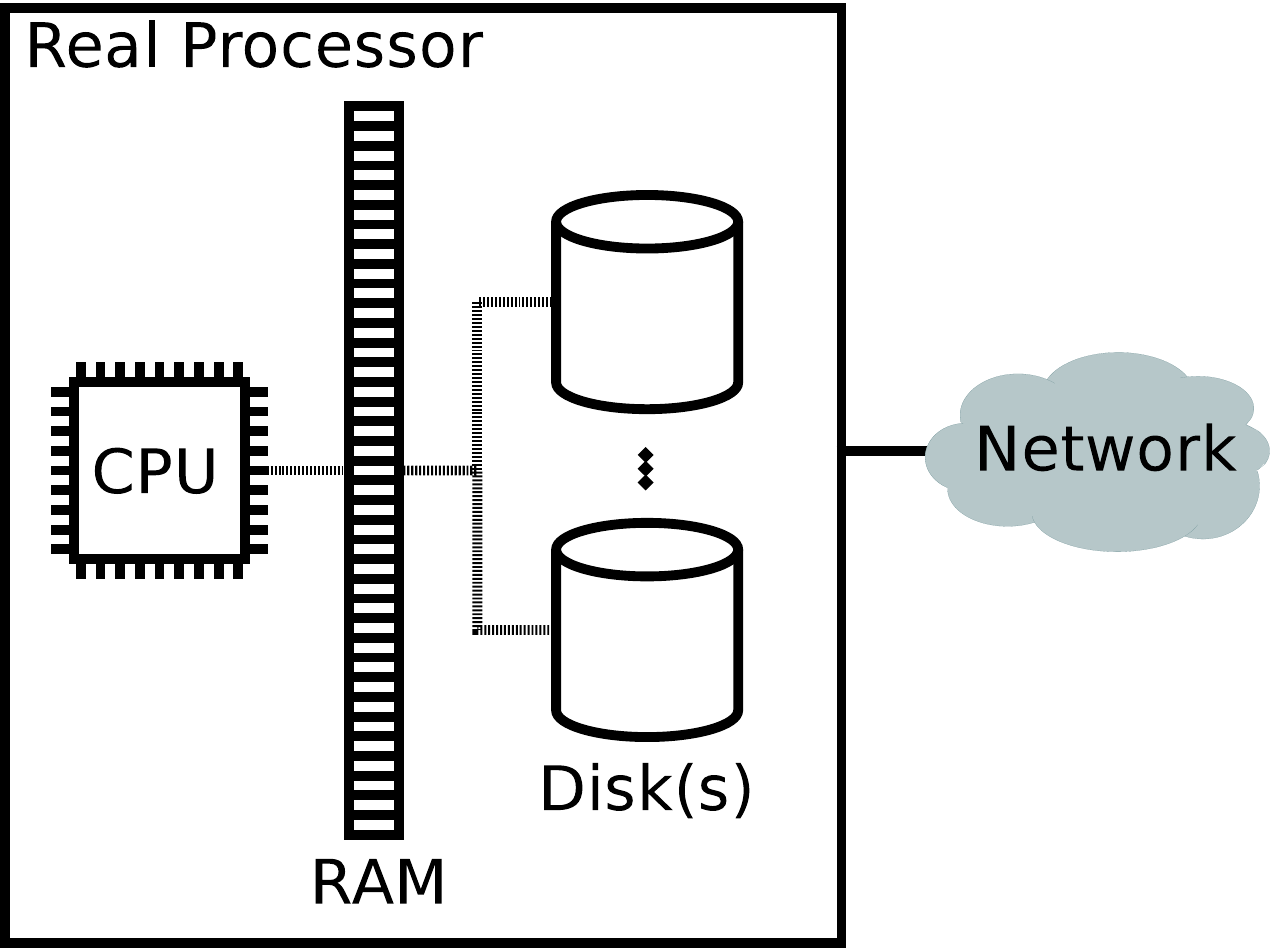}
	} \\
$D=2$~Disks
\end{center}
\caption{EM-BSP Model}
\label{pems1_model}
\end{figure}

The parallel and distributed memory nature of BSP-like algorithms is
advantageous from an EM perspective since, as in PDM, a collection of parallel
disks can perform I/O much faster than a single disk.  The EM-BSP, EM-BSP*,
and EM-CGM models \cite{dhthesis}\cite{emsimulation}\cite{bspem} augment
the corresponding BSP-like model by adding local disk(s) to each machine.
Such a configuration is shown in Fig.~\ref{pems1_model}.

Computation proceeds in supersteps as in BSP, except each processor may
access local disk as necessary during the computation superstep.  Thus,
the EM-BSP models can be considered a hybrid of the BSP-like models and
the PDM model: synchronisation and communication is inherited from BSP,
and I/O from PDM.

\section{Previous Work}

\subsection{STXXL}
\label{stxxl-sec}

\begin{figure}[ht]
\begin{center}
	\resizebox{0.8\textwidth}{!}{
	\includegraphics{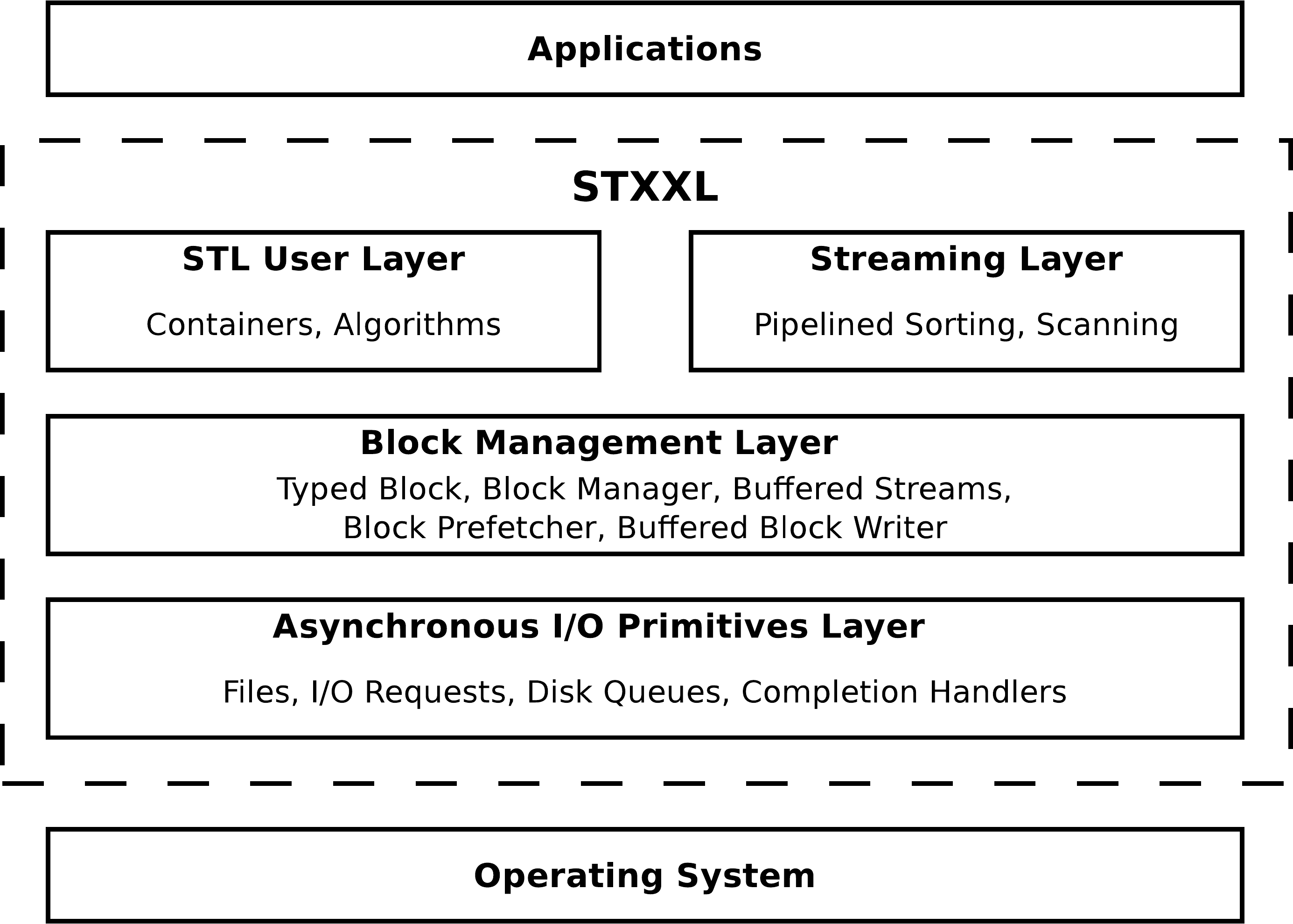}
	}
\\
(Reproduced from \cite{stxxl})
\end{center}
\caption{STXXL Design}
\label{stxxl_layers}
\end{figure}

STXXL is a C++ library for EM algorithms.  STXXL is composed of many layers, as
shown in Fig.~\ref{stxxl_layers} (reproduced from \cite{stxxl}).  Higher level
layers in STXXL make use of the lower level layers, though user applications
may directly use any layer, bypassing higher level functionality if desired.

The lower level Block Management and Asynchronous I/O Primitive layers provide
generic functionality useful to EM algorithms, such as asynchronous
I/O and transparent parallel disk access.

The STL User Layer provides an implementation of the C++ Standard Template
Library (STL), the algorithms and data structures component of the C++
standard library.  This layer can be used to write C++ code in the standard
style that functions as an EM algorithm, or simplify the porting of existing
C++ RAM algorithms to EM.

The Streaming Layer provides additional functionality that does not fit
within the confines of the STL API.  EM-specific techniques such as pipelining
and I/O optimal scanning are implemented in this layer.

STXXL provides a rich suite of EM code, making it simple to write advanced
EM algorithms at a relatively high level.  Notably, all layers above and
including the Block Management Layer transparently support parallel disks.
Thus, applications built with STXXL can take advantage of parallel disk
performance without any specific effort required on behalf of the
application developer.  The Asynchronous I/O Primitives layer provides a
simple, low-level, and portable interface to asynchronous I/O.  Since the
asynchronous I/O interface of operating systems is typically more complex
and varies between systems, this layer is useful to applications that require
asynchronous I/O but not the higher level functionality of STXXL.

\subsection{Cache-Oblivious Algorithms}

Cache-Oblivious algorithms \cite{cacheoblivious} are designed with I/O
efficiency in mind (unlike RAM algorithms), but without any explicit
block size parameters (unlike EM algorithms).  For example, traditional
EM algorithms explicitly transfer blocks of some size $B$ between disk
and main memory.  The algorithm implementation must know the value of $B$
at run time.  In contrast, a cache-oblivious algorithm is unaware of (or
oblivious to) any such parameter, and may transfer data with arbitrary size
and alignment much like a RAM algorithm.  However, unlike most RAM algorithms,
cache-oblivious algorithms are analysed in terms of memory transfers of an
{\em arbitrary} size $B$, and aim to minimize the number of transfers much
like an EM algorithm\footnote{Cache-oblivious literature typically uses
$L$ (for ``line''), rather than $B$.  This thesis consistently uses $B$ and
``block'' regardless of whether cache or disk is being discussed.}. Thus,
an efficient cache-oblivious algorithm is efficient for any $B$ and does
not require modification to perform well on various systems.

This approach is particularly useful in the presence of cache hierarchies,
where many levels of cache are in use at one time, each with a different
block size.  I/O efficiency is an increasingly important performance
factor, even for algorithms that work only with internal memory (RAM).
On modern systems, a cache miss can be several hundred times slower than a
cache hit \cite{intelperf}.  Cache-oblivious literature often presents this
problem in the context of a modern processor's cache and memory hierarchy,
though the block size independent nature naturally applies where the lowest
level of the memory hierarchy is disk.  This suggests cache-oblivious
algorithms are a promising strategy for the design of algorithms that show
good performance across a very wide range of problems sizes.

\subsection{MPI}

MPI (Message Passing Interface) \cite{mpidocs} is an Application Programming
Interface (API) for distributed memory parallel programming.  MPI provides
communication and synchronisation functions useful for many types of parallel
program.  Most relevant to this thesis are the ``collective communication''
MPI functions, since these can be used to implement BSP-like algorithms.

Collective communication functions in MPI synchronise all processors, then
perform communication.  There are many different styles of communication
available, such as {\tt MPI\_Gather} (each processor sends a message
to a single processor) or {\tt MPI\_Alltoall} (each processor sends a
message to every other processor)\footnote{A more detailed description
of the collective communication functions described here can be found in
Chapter~\ref{comm-algs}}.

In a BSP-like program implemented with MPI, a call to a collective
communication represents a communication superstep and subsequent superstep
barrier.  In this way, a BSP-like algorithm can be implemented as a series
of MPI collective communication calls interleaved with computation code.

MPI is a widely used interface for distributed memory parallel programming
with many implementations for a variety of systems.

\subsection{EM-BSP Simulation}

Many parallel algorithms intended to work with large data sets have been
designed for BSP-like models.  Though these algorithms scale to larger data
sets than single processor RAM algorithms by exploiting the memory available
to several machines, unfortunately they do not generally make use of disk
and are thus limited to problems that fit entirely within main memory.

Fortunately, it is possible to use these existing algorithms with data larger
than main memory via simulation in the EM-BSP models\footnote{This idea was
introduced with the original presentation of the EM-BSP models
\cite{dhthesis}\cite{bspem}\cite{emsimulation}}.
The basic idea is to simulate a number of ``virtual processors'', each with
memory small enough to fit into ``real processor'' main memory.  A subset of
these virtual processors is executed at once, while the (virtual) memories of
others are swapped out to disk.  Thus it is possible to run a bulk-synchronous
algorithm with total memory size exceeding that of real main memory, limited
only by the amount of available disk space.

To illustrate, consider a BSP-like algorithm that requires 128 processors,
each with 1 GiB of RAM.  If these resources are available, the algorithm may
be executed directly.  However, this is not the case if only 32 processors are
available.  Nevertheless, the algorithm may be executed using these limited
resources via simulation as follows: for each superstep, rather than run 128
processes in parallel, run 32 processes in parallel, storing any generated
messages on disk.  Then, another round of 32 processes is executed in a similar
fashion, and so on until all 128 processes have been executed.  At the end
of this process, all computation for the superstep has been completed and
all communication is stored on disk, so the next superstep may begin.

In practice, this strategy can be implemented as a library which provides
communication functions for use by BSP-like applications.  In particular, no
special operating system level support is required.  All details pertaining
to external memory can be managed by this library; the application code
need not be changed.

\subsection{PEMS}

\begin{figure}[ht]
\begin{center}
	\resizebox{!}{0.5\textheight}{
	\includegraphics{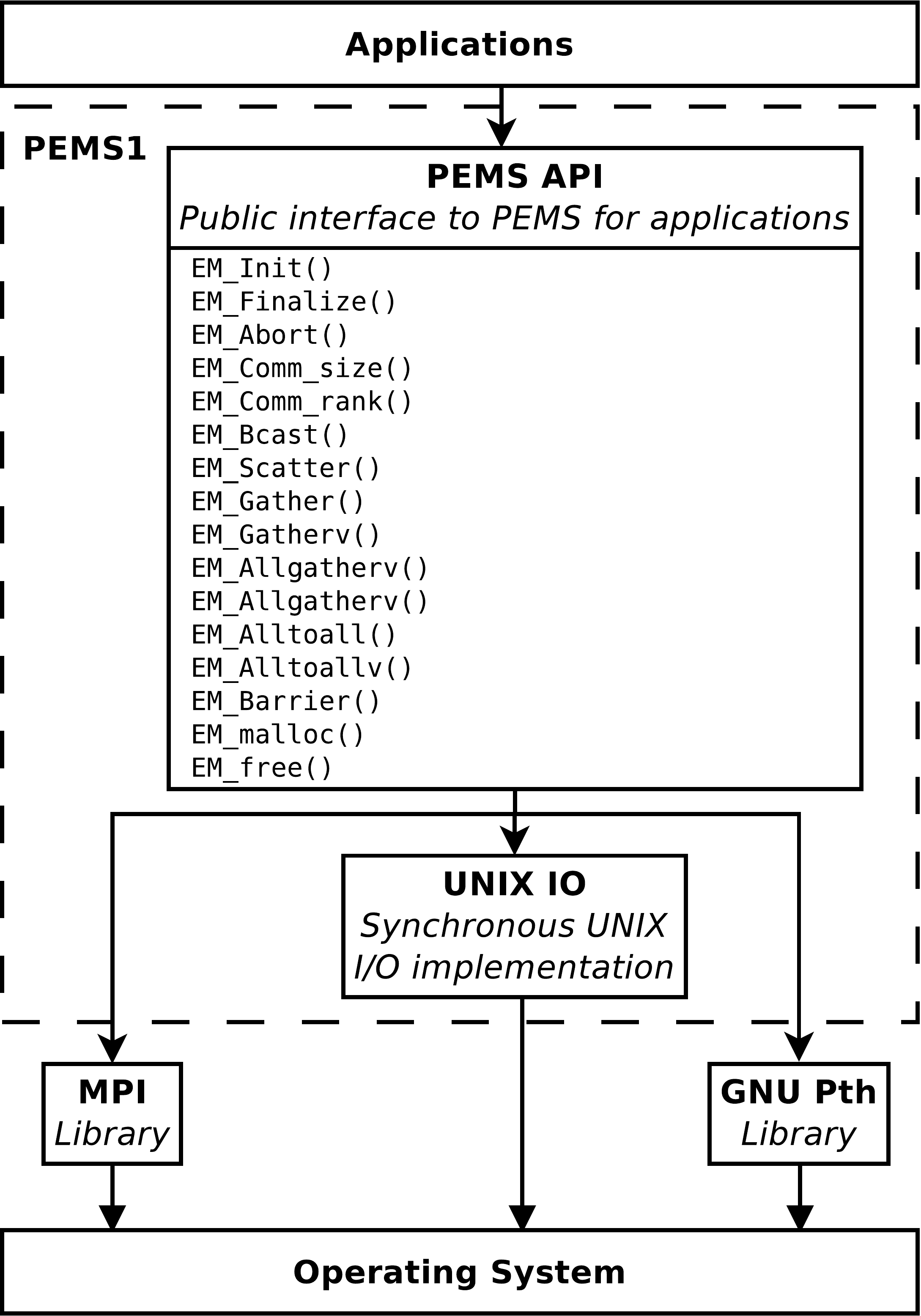}
	}
\caption{PEMS1 Design}
\label{pems1_design}
\end{center}
\end{figure}

PEMS1 \cite{mnthesis} (Parallel External Memory System) is an implementation of
the EM-BSP simulation technique which provides an API similar to that of MPI.
Fig.~\ref{pems1_design} shows an overview of the PEMS1 design.

Significant modifications to PEMS1 have been made as a part of this thesis.
Where the distinction is necessary the previous implementation is referred to
as ``PEMS1'', and this improved version as ``PEMS2''.  Both are collectively
referred to as ``PEMS'' where appropriate.

PEMS1 is implemented as a library which transparently handles virtual
processor swapping, synchronisation, memory allocation, and communication.
The user program is an MPI-like program, but communication may be deferred
to disk to allow the simulation of more processors than are actually available.

The interface to PEMS1 is, with a few exceptions, semantically identical
to a subset of MPI, though functions names have a different prefix to avoid
conflicts\footnote{This has been resolved in PEMS2, see \S\ref{contributions}}.
When the applications calls collective communication functions, PEMS1
internally performs the necessary network or I/O operations, swapping virtual
processors in and out as required.  Though much occurs ``behind the scenes'',
from the application's point of view the collective communication operation
has been completed exactly as if it had been performed directly by MPI.

Internally, the system's MPI library is used to perform communication between
virtual processors on separate real processors.  I/O is performed using the
operating system's I/O interface; specifically that of POSIX, the standard
common to all UNIX-like systems such as GNU/Linux, Solaris, or Mac OS X.

Thread support in PEMS1 is handled via the GNU Pth library, which implements
user-space threads.  This is advantageous for single-core processors since
thread switching does not incur the overhead of a kernel-level context switch.
However, user-space threads do not allow for true thread concurrency on
multi-core machines.

PEMS1 has been shown to scale well in practice on sorting and list
ranking problems significantly larger than the total amount of available
RAM \cite{experimentswith}.  The distinguishing characteristic of PEMS is
that existing algorithms not explicitly designed as EM algorithms may be
efficiently used with external memory.  In addition to the large number of
suitable (BSP-like) existing algorithms, a considerable advantage of this
approach is the ability to exploit distributed memory parallel computers.
While it is possible to implement distributed memory algorithms using STXXL
or a cache-oblivious approach, the algorithm must be deliberately designed
to have this ability -- a significantly more difficult task than designing
a sequential EM algorithm.  Algorithms designed to support both distributed
memory parallelism and external memory are relatively rare.  In contrast,
{\em all} algorithms that work with PEMS are inherently capable of executing
on a distributed memory parallel computer.

Because of this ability, PEMS can easily scale to extremely large problem
sizes without requiring any modification to the algorithm.  If, for example,
one wanted to use a straightforward STXXL application on a problem too
large to feasibly handle with a single computer, the algorithm may require
a significant redesign in order to scale further.  A PEMS application,
however, can easily scale to very large problem sizes by adding disk and/or
processor resources as necessary.

Similarly, if a given problem takes an unacceptable amount of time, processor
and/or disk resources may be added to improve the run time.  Experiments in
Chapter~\ref{experiments} and previous work on PEMS1 \cite{mnthesis} show that,
though the STXXL sort is faster than PEMS given equivalent computational
resources, computation resources can be added until PEMS out-performs the
STXXL sort.

\section{Summary of Contributions}
\label{contributions}

This thesis presents PEMS2, an enhanced version of PEMS1 with new functionality
and improved performance and usability.  These enhancements include fundamental
changes to the simulation process, such as new I/O drivers and multi-core
support; as well as new communication primitives with improved performance
characteristics.

While PEMS1 supported parallelism across several machines in a cluster
configuration, SMP (or ``multi-core'') on each of these machines was not
explicitly supported.  Though at the time, most commodity machines were
single core, recently multi-core has become ubiquitous.  PEMS2 introduces
support for multi-core machines, allowing the simulation to take advantage
of multiple cores with less overhead than simply running several local
MPI processes.  The computation performed by the simulated algorithm is
executed in parallel across many cores, allowing for speedup in computation
heavy algorithms.  However, even for algorithms that are I/O bound, many of
the improved communication primitives achieve an I/O reduction proportional
to the number of local cores available.

PEMS1 used explicit, blocking, aligned I/O operations exclusively.
While the improvements presented here can also work in the same fashion,
the implementation has been redesigned to allow simple switching between
various I/O ``drivers''.  In conjunction with other changes, this allows
for the use of asynchronous I/O, or memory-mapped I/O, both of which have
significantly different performance characteristics to traditional blocking
I/O.  In particular, memory-mapped I/O is interesting because a superstep
does not necessarily incur a swap of the entire context as with explicit I/O.
With memory mapped I/O the memory access characteristics of the simulated
algorithm dictate the I/O performed during simulation, allowing algorithms to
take advantage of desirable memory access characteristics when used with PEMS.
Experiments show that these I/O strategies are beneficial in some cases, but
not always an improvement depending on the nature of the simulated algorithm.

The most powerful communication primitive in PEMS1, {\sc Alltoallv}, has been
redesigned to use a new message delivery strategy which avoids the need for
an area on disk reserved for delivery.  The new algorithm thus requires less
I/O and disk space to perform the same task.  In practice this also eases
configuration since the user no longer needs to calculate the message volume
of a given algorithm in order to allocate disk space to virtual processors.

Several common collective communication primitives are merely restricted cases
of Alltoallv, including all of the primitives implemented in PEMS1.  However,
these can often be implemented much more efficiently than the equivalent
call to Alltoallv.  These primitives, as well as Alltoallv itself, have been
optimised to eliminate any unnecessary swapping.  In particular, with the
introduction of multi-core support, ``rooted'' communication primitives can be
implemented more efficiently using appropriate synchronisation techniques.
``Rooted'' communication primitives are those which send to or receive
from some root virtual processor, as opposed to Alltoallv in which all
processors communicate as equals.  This thesis introduces a small set of thread
synchronisation primitives that handle swapping in such cases, to ensure a
minimal amount of I/O is performed.

Also introduced in PEMS2 is a new type of collective communication function
that performs communication as well as computation, unlike those implemented
in PEMS1 which perform communication alone.  {\sc Reduce}, and similar methods,
are beneficial to certain algorithms since the system can perform the combined
communication and computation more efficiently than a user program could by
using communication primitives alone.  These operations are defined by MPI and
used in many BSP algorithms, expanding the useful scope of the implementation.

In addition to these fundamental changes, the implementation has been
thoroughly rewritten with the intention of being straightforward to use with
existing or new MPI programs on any appropriate system.  MPI programs can be
compiled against PEMS2 without modification\footnote{Assuming, of course,
that the program is restricted to the set of calls implemented by PEMS2},
making it straightforward to simulate any existing MPI algorithm for problems
sizes vastly exceeding the amount of available main memory.  All parameters of
PEMS2 can be passed at run-time to the program through command line arguments,
simplifying automated or manual experimentation.  An integrated benchmarking
system can record the overall run time of a simulation or a fine-grained
breakdown of run-time at each superstep.  Benchmark results are written to
a gnuplot compatible file which can be used to generate plots like those in
this thesis.  A comprehensive test suite adapted and augmented from several
existing MPI test suites ensures correctness at the application level for
any configuration.  PEMS2 is freely available on the web \cite{pems2site}
under an Open Source license and is straightforward to compile and use on
any UNIX system.

\section{Thesis Outline}

The notation, terminology, and variables used throughout are described in
Appendix~\ref{conventions}.

To more thoroughly introduce the reader to the context of this thesis,
Chapter~\ref{pems1-ch} describes in detail the approach to EM-BSP
simulation taken in PEMS1, and limitations which PEMS2 aims to improve.
Subsequent chapters describe how these limitations are addressed in PEMS2:
Chapter~\ref{overview-ch} gives a brief overview of the architecture of
PEMS2.  Chapter~\ref{multicore} describes the modifications necessary to
allow the simulation to take advantage of multi-core processors, including
the synchronisation primitives referenced in later chapters.  The various
styles of I/O available in PEMS2 are described in Chapter~\ref{io-drivers}.
The choice of I/O style does not affect the implementation of communication
algorithms, but may affect analysis; the consequences of this choice are also
discussed in Chapter~\ref{io-drivers}.  Chapter~\ref{simulation} describes
a new message delivery strategy which differs significantly from that used
in PEMS1, using the {\sc Alltoallv} operation as an example.

The communication algorithms presented in Chapter~\ref{comm-algs} make use
of the material in preceding chapters to implement several new communication
methods with improved performance characteristics.  Chapter~\ref{experiments}
then presents several applications built using these methods along with
theoretical and experimental performance analysis.  This data shows the
claims of improvement in previous chapters translate into ``real-world''
application scenarios.

Finally, Chapter~\ref{conclusions-ch} discusses conclusions that can be
drawn from PEMS2 results, and suggests potential directions for future work.

\chapter{Overview of PEMS1}
\thispagestyle{empty}
\label{pems1-ch}

\section{Overview}
\label{overview}

For clarity this overview describes the case where a single real processor
is used; the strategy for simulation with multiple real processors is
similar, and addressed in detail in later sections.

PEMS1 implements EM-BSP simulation by assigning a thread to each simulated
virtual processor.  Since there are $v$ virtual processors, $v$ threads
exist simultaneously.  However, only a single thread executes at a given time.

The application, being a BSP-like algorithm, is a series of computation
supersteps separated by calls to PEMS communication functions.  These functions
serve as both communication supersteps and superstep barriers.  When one is
called, PEMS performs the necessary communication, then swaps the calling
thread out out memory.  At this point there is a superstep barrier, so the
thread yields and another thread is swapped in, which will eventually call
the same communication function and reach the same barrier\footnote{Recall
that all virtual processors run identical programs}.  Thus, all threads will
eventually synchronise at this barrier and the next superstep can begin.
Several barriers may actually be used to implement a collective communication
function in PEMS, but there is always at least one as required by the BSP
model and its derivatives.

\begin{figure}[h]
\begin{center}
	\resizebox{0.4\textwidth}{!}{
	\includegraphics{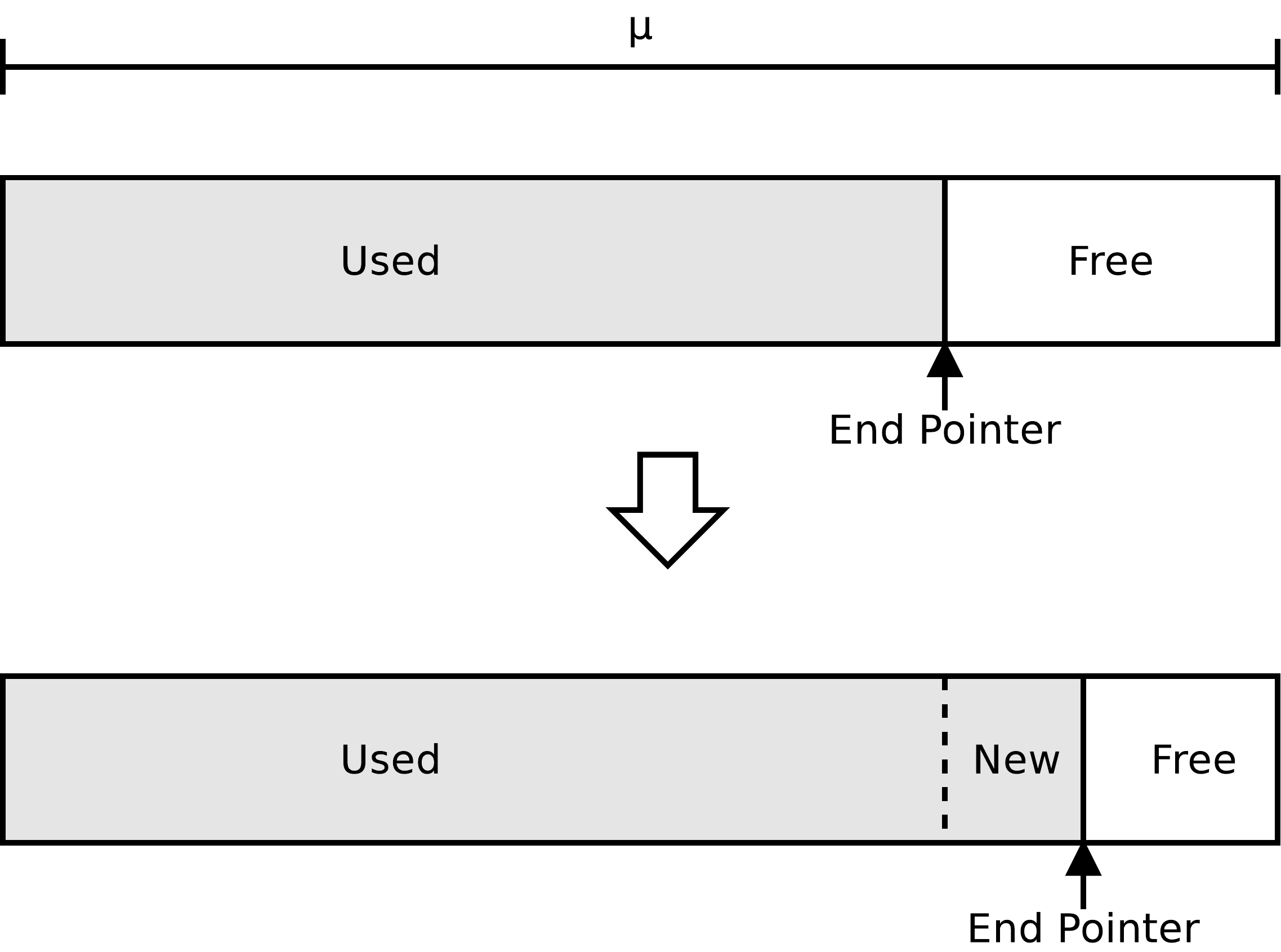}
	}
\end{center}
\caption{Memory Allocation in PEMS1}
\label{allocation-fig}
\end{figure}

This process of swapping and synchronisation is relatively straightforward
to implement.  PEMS1 implements allocation using a basic ``bump pointer''
allocator, which simply allocates memory in a contiguous range by appending
new allocations and ``bumping'' (increasing) the end pointer.
 Fig.~\ref{allocation-fig} shows an example of such an allocation.
To swap, this area of memory is read from / written to disk in a single read
/ write operation, respectively.  Whenever a virtual processor is executing,
its context is swapped in to the same area of RAM.  This ensures the address
to a given memory location remains constant so pointers in the application
remain valid.

Implementing communication is more complex.  Following the literature
associated with PEMS1 \cite{mnthesis}\cite{experimentswith}, the communication
strategy used is described here using the {\sc Alltoallv} call as an example.
{\sc Alltoallv} is the most powerful collective communication method:
all others implemented in PEMS1 can be considered simple cases of
{\sc Alltoallv}.

\section{Alltoallv}

PEMS1 performs message delivery using a special disk area separate from the
virtual processor contexts, called the ``indirect area''.  The indirect area
is statically partitioned such that each virtual processor has a dedicated
region of some fixed size for message delivery.  {\tt Alltoallv} is performed
in two internal supersteps: messages are first written by the sender to the
indirect area in a block-aligned and parallel fashion, then read from the
indirect area by the receiver and delivered to the receiver's context on disk.

Alg.~\ref{simple-alltoall-seq} shows a straightforward implementation of this
approach for a single processor.

Note the algorithm style used here differs slightly from
that used in previous work on PEMS \cite{mnthesis} and EM-BSP
\cite{dhthesis}\cite{bspem}\cite{emsimulation}.  The style used here is more
implementation directed, omitting lines such as ``for i in $0{\dots}v-1$ do in
parallel'' which do not actually occur in this type of multi-threaded code.
When reading or analysing algorithms in this style it is best to think from
the perspective of a single thread executing the code.  For example, all
algorithms in this thesis are written from a perspective such as ``first,
deliver {\em my} messages''; not ``first, each virtual processor delivers
their messages''.  The reader must keep in mind that several threads perform
these actions simultaneously.

The notation $m_{i \rightarrow j}$ denotes the message sent from virtual
processor $i$ to virtual processor $j$.

\begin{algorithm}[h]
\KwData{$\mathcal{S}$ : Array of pointers to $v$ messages to send}
\KwData{$\mathcal{R}$ : Array of pointers to $v$ messages to receive}
\BlankLine
	\tcc{Send Messages}
	\ForEach{message $m_{\rho{\rightarrow}i}$ in $\mathcal{S}$}{
		Write $m_{\rho \rightarrow i}$ to $i$'s indirect area on disk\;
	}
	Swap out\;
\BlankLine\tcc{Finished Internal Superstep 1}
\tcc{Begin Internal Superstep 2}\BlankLine
	\tcc{Receive messages}
	Swap in\;
	\ForEach{message $m_{i{\rightarrow}\rho}$ in $\mathcal{R}$}{
		Read $m_{i \rightarrow \rho}$ from indirect area on disk
		to the $i^{th}$ location in $\mathcal{R}$\;
	}
	Swap out\;
\BlankLine\tcc{Finished Virtual Superstep}\BlankLine
\caption{{\sc Simple-Alltoallv-Seq}}
\label{simple-alltoall-seq}
\end{algorithm}

In the analysis of Alg.~\ref{simple-alltoall-seq} the following variables are used:
\begin{itemize}
\item[$\omega$ ] An arbitrary bound on the simulated algorithm's message size
(i.e.\ the size of a message sent from one {\em virtual} processor to another).
\item[$\mu$ ] The (maximum) size of a single virtual processor's context
(i.e.\ the maximum amount of memory allocated by any virtual processor)
\item[$B$ ] The size of a disk block
\end{itemize}

These variables remain free in the stated run times for various communication
methods presented in this thesis.  Their actual value depends on the
characteristics of a particular application or system configuration.
All three are assumed to have the same unit, such as bytes.

To convert I/O volume to run time, the following coefficients are used:
\begin{itemize}
\item[$G$ ] The time required to read / write a single block from / to disk
for message delivery
\item[$S$ ] The time required to read / write a single block from / to disk
for swapping
\end{itemize}

$G$ and $S$ are identical\footnote{Except with memory-mapped I/O,
see \S\ref{mmap}}, but different coefficients are used to keep terms
related to swapping and terms related to message delivery separate.
Chapter~\ref{multicore} describes the reasons for this in further detail.

The notation $\llceil{\omega}\rrceil$ means ``$\omega$ rounded up to the
next multiple of $B$''.

The terms ``I/O volume'' or ``amount of I/O'' are used to refer to an amount of
I/O in the same unit as $\mu$ and $\omega$, e.g.\ bytes.  Specifically, it does
not refer to number of I/O operations in blocks, often referred to as ``I/Os''
(note plurality) in EM literature.  I/O volume is stated separately because
later sections in this thesis investigate non-blocked I/O, and comparison of
these approaches is best described in terms of volume.  This is done only to
facilitate discussion and simplify analysis, the total run time of algorithms
is given in terms of I/O operations (``I/Os''), as is typical in EM literature.

The notation ``$I_{a}$'' is used to refer to the amount of I/O performed by
line $a$ of the algorithm.  The amount of I/O performed by a range of lines
(inclusive) is denoted ``$I_{a{\dots}b}$''.  The same notation used with $T$
rather than $I$ refers to the time taken, rather than the amount of I/O.
For example, line $4$ in
Alg.~\ref{simple-alltoall-seq} refers to ``Swap in'', therefore:
\begin{align*}
I_{4} &= \mu \\
T_{4} &= S\frac{\mu}{B}
\end{align*}

Recall that all $v$ virtual processors execute the same code.  Thus,
if a given line in the algorithm performs $x$ I/O, in total the line is
responsible for $vx$ I/O, unless the line is conditionally executed by only
some virtual processors.  To avoid excessive repetition, phrases such as ``for
each virtual processor'' are omitted from proof explanations where it is clear
that all virtual processors perform the same actions (as is the case here).

The variables and notation used here are used consistently throughout this
thesis, and documented (with others) in Appendix~\ref{conventions}.

\begin{lemma}
\label{simple-alltoall-seq-io}
Alg.~\ref{simple-alltoall-seq} (PEMS1 single processor {\sc Alltoallv})
performs $4v\mu + 2v^2\omega$ total I/O.
\end{lemma}
\begin{proof}
For each of the $v$ virtual processors:
\\
The loop at line 1 first writes all $v$ outgoing messages, each of size $\omega$:
\begin{align*}
I_{1{\dots}2} &= v^2\omega
\intertext{Line 3 swaps out the partition of size $\mu$:}
I_{3} &= v\mu
\intertext{Line 4 swaps in the partition of size $\mu$:}
I_{4} &= v\mu
\intertext{The loop at line 5 reads all $v$ incoming messages, each of size $\omega$:}
I_{5{\dots}6} &= v^2\omega
\intertext{Line 7 swaps out the partition of size $\mu$:}
I_{7} &= v\mu
\intertext{Finally, line 8 swaps in the partition of size $\mu$:}
I_{8} &= v\mu
\intertext{The total I/O performed by the algorithm, in the unit of $\mu$ and $\omega$ (e.g.\ bytes), is therefore:}
I_{\text{simple-alltoall-seq}} &= I_{1{\dots}2} + I_{3} + I_{4} + I_{5{\dots}6} + I_{7} + I_{8} \\
&=	  \left( v^2\omega \right)
	+ \left( v\mu \right)
	+ \left( v\mu \right)
	+ \left( v^2\omega \right)
	+ \left( v\mu \right)
	+ \left( v\mu \right) \\
&=	4v\mu + 2v^2\omega \\
\end{align*}
\end{proof}

\begin{thm}
\label{simple-alltoall-seq-time}
Alg.~\ref{simple-alltoall-seq} (PEMS1 single processor {\sc Alltoallv})
takes $S\frac{4\mu}{B} + G2v^2\frac{\llceil{\omega}\rrceil}{B} + 2L$ time.
\end{thm}
\begin{proof}
Follows directly from Lem.~\ref{simple-alltoall-seq-io}, since
Alg.~\ref{simple-alltoall-seq} performs no network communication and no
significant computation.

Since messages are delivered one at a time (i.e.\ each message
delivery is a separate I/O operation), message deliveries are each of
size $\frac{\llceil{\omega}\rrceil}{B}$ ($\llceil{\omega}\rrceil$ meaning
``$\omega$ rounded up to the next multiple of $B$'').  Thus, if messages are
smaller than a single block then overhead is accumulated for every message.
However, this is not a performance problem since a single block of I/O is
the minimal amount of time possible for an I/O operation.

There are two internal superstep barriers, contributing $2L$ to the total
run time.
\end{proof}

\begin{thm}
\label{simple-alltoall-seq-mem}
Alg.~\ref{simple-alltoall-seq} (PEMS1 single processor {\sc Alltoallv})
requires $v\mu + v^2\omega$ disk space.
\end{thm}
\begin{proof}
Each virtual processor requires $\mu$ disk space for its context regardless
of message delivery.  $v^2$ messages are delivered in total, each of size
$\omega$, therefore an additional $v^2\omega$ space is required for the
indirect area.
\end{proof}

\section{Potential for Improvement}
\label{potential}

While experiments with PEMS1 have shown desirable scalability characteristics,
the system has significant overhead which requires the use of considerably more
computational resources to match the performance of comparable EM algorithms.
Though the ability to take advantage of several machines with parallel disks
is a considerable advantage, reducing this overhead will make the system
more competitive on a wider range of systems and problem sizes.

Additionally, the use of a separate area for message delivery introduces
scalability problems with large contexts (see \S\ref{s-disk-space})
and makes tuning difficult in practice since the user must know in advance
the bounds on a given algorithm's communication volume in order to allocate
disk.

This thesis introduces several new strategies and capabilities for PEMS
intended to address these issues.

As is usually the case with EM algorithms, the most significant source
of overhead in PEMS1 is unnecessary I/O.  There are two cases where PEMS
must perform I/O: swapping and message delivery.

\subsection{Swapping}
\label{improvements-swap}

Each internal superstep barrier in Alg.~\ref{simple-alltoall-seq} implies a
swap out and a subsequent swap in of each virtual processor.  However, many
such swaps can be avoided by making more extensive use of the ``direct delivery
to context'' technique described in the PEMS1 literature \cite{mnthesis}.
This technique is based on the observation that a subsequent swap-in in
the second internal superstep is not required, since messages can be written
directly to the context on disk.  That is, instead of swapping in the context,
modifying it in memory, then swapping the context back out; the message can
simply be written directly to the appropriate location on disk.

Thus, if the second loop delivers messages directly this way, it is not
necessary to swap in at the first barrier.  The final swap-out is also avoided
because the context on disk is already known to be consistent\footnote{In
fact a swap out {\em can't} occur here because the context is not swapped
in, so a swap out would write garbage data to disk}, hence this avoids
$2\mu$ I/O per virtual processor.

Swapping with finer granularity can avoid slightly more unnecessary I/O: the
swap out at the first internal superstep barrier swaps out the entire context,
however this is not necessary.  Virtual processors receive messages to some
area within their context, so when the {\tt Alltoallv} call is completed and
control is returned to user code this region will have been overwritten with
the received messages.  Therefore, it is not necessary to swap out this region
(the ``receive buffer'') at the initial superstep barrier.

Some swapping can also be avoided at superstep barriers: in a straightforward
implementation, all threads swap at superstep barriers.  However, for the
last thread to execute in the superstep this is not necessary.  The order
of execution within a superstep is undefined, so it is wasteful to swap out
this thread's context and allow a different thread to swap in and execute
first in the next superstep.  Instead, the last thread can simply remain
swapped in through the barrier and be the first thread to run in the following
superstep, thus avoiding one swap per superstep.  More generally, in the case
of multi-core, threads execute in parallel rounds of $k$ threads at a time
therefore this technique avoids $k$ swaps per superstep.

\subsection{Message Delivery}
\label{pems1_delivery}

Alg.~\ref{simple-alltoall-seq} writes all messages to be delivered to the
indirect area on disk.  There is potential for improvement here based on two
observations:
\begin{enumerate} 
\item Each message that must be written in the first loop is a part of the
sending virtual processor's context, and therefore will be written to disk
regardless at the first barrier (when the sender's context is swapped out).
Thus, the previous algorithm results in each message being written to disk
{\em twice}.
\item If the receiving virtual processor of a message is local and has already
executed this superstep, then the final destination of the message is known
and the message can be delivered directly to the destination context on disk.
This avoids reading the message from disk again in order to deliver it to
the receiver.
\end{enumerate}

There is an additional downside to delivering messages via a separate disk
area: because the indirect area is large and separate from the area on disk
where contexts are stored, delivery of messages (and subsequently swapping
out) involves seeking across a very large area of disk.  In the worst case
this results in constantly seeking back and forth between the contexts area
and the indirect area.  This is potentially a serious performance issue,
particularly for large $\mu$ or $v$.  Since disk seeking is extremely
expensive, the performance impact of this behaviour could be as significant
as the actual amount of I/O performed -- or even more so.  The addition
of multi-core support compounds the problem due to several threads seeking
simultaneously.  Reducing or eliminating this effect is therefore a promising
path to improving the performance of PEMS in practice.

Of course, indirect message delivery is not done without reason: the messages
are aligned and distributed among disks in a way designed to achieve fully
parallel disk I/O, and support ``direct'' I/O which requires all operations to
be block aligned.  \S\ref{delivery} describes new methods of retaining these
desirable characteristics without writing messages to a separate area on disk.

\subsection{Communication Balancing}
\label{pems1-balancing}

The original EM-BSP simulation algorithms (and PEMS1) require an upper
bound on communication volume so disk space can be allocated accordingly
for the indirect area.  In the multi-processor case, this is achieved by
using a deterministic routing technique \cite{balancing} which first evenly
distributes messages across the network before completing the communication.
Messages are first sent to an arbitrary intermediary processor in a round-robin
fashion, then sent to their final destination by that intermediate processor.
Because messages are evenly distributed to an arbitrary intermediate processor,
this technique ensures balanced communication.

This technique is straightforward and works well to ensure balanced
communication, but in the context of PEMS incurs a large amount of overhead.
In order to be delivered, each message must be (in the worst case):
\begin{enumerate}
\item Sent over the network (by the sender)
\item Written to disk (by the intermediary)
\item Read from disk (by the intermediary)
\item Sent over the network (by the intermediary)
\item Written to disk (by the receiver, to the indirect area)
\item Read from disk (by the destination, from the indirect area)
\item Written to disk (by the destination, to its context)
\end{enumerate}

The multiple reads and writes of each message to disk, in particular, is a
significant amount of overhead due to the large cost of disk I/O.

\subsection{Allocation}
\label{improvements-alloc}

The simple memory allocation scheme used by PEMS1 has a serious limitation
for many programs: freeing memory is not possible.  Since only a pointer
to the end of all allocated memory is stored, there is no way to free a
particular chunk of allocated memory.

While some BSP-like algorithms allocate a large amount of memory initially
then use it throughout execution (such as the PSRS algorithm presented in
\S\ref{psrs-sec}), many have more dynamic memory allocation requirements.
PEMS1's basic allocator is not appropriate for algorithms that continuously
allocate and free chunks of memory, since memory comsumption will continue
to increase until available space is exhausted.

\subsection{Improvements}

Chapter~\ref{simulation} presents solutions to the shortcomings described in
this section, all of which are implemented in PEMS2.  These improvements depend
on two new fundamental aspects of the design: multi-core support and several
I/O drivers, presented in Chapter~\ref{multicore} and Chapter~\ref{io-drivers},
respectively.

\chapter{Overview of PEMS2}
\thispagestyle{empty}
\label{overview-ch}

\section{Software Design}

\begin{figure}[ht]
\begin{center}
	\resizebox{!}{0.6\textheight}{
	\includegraphics{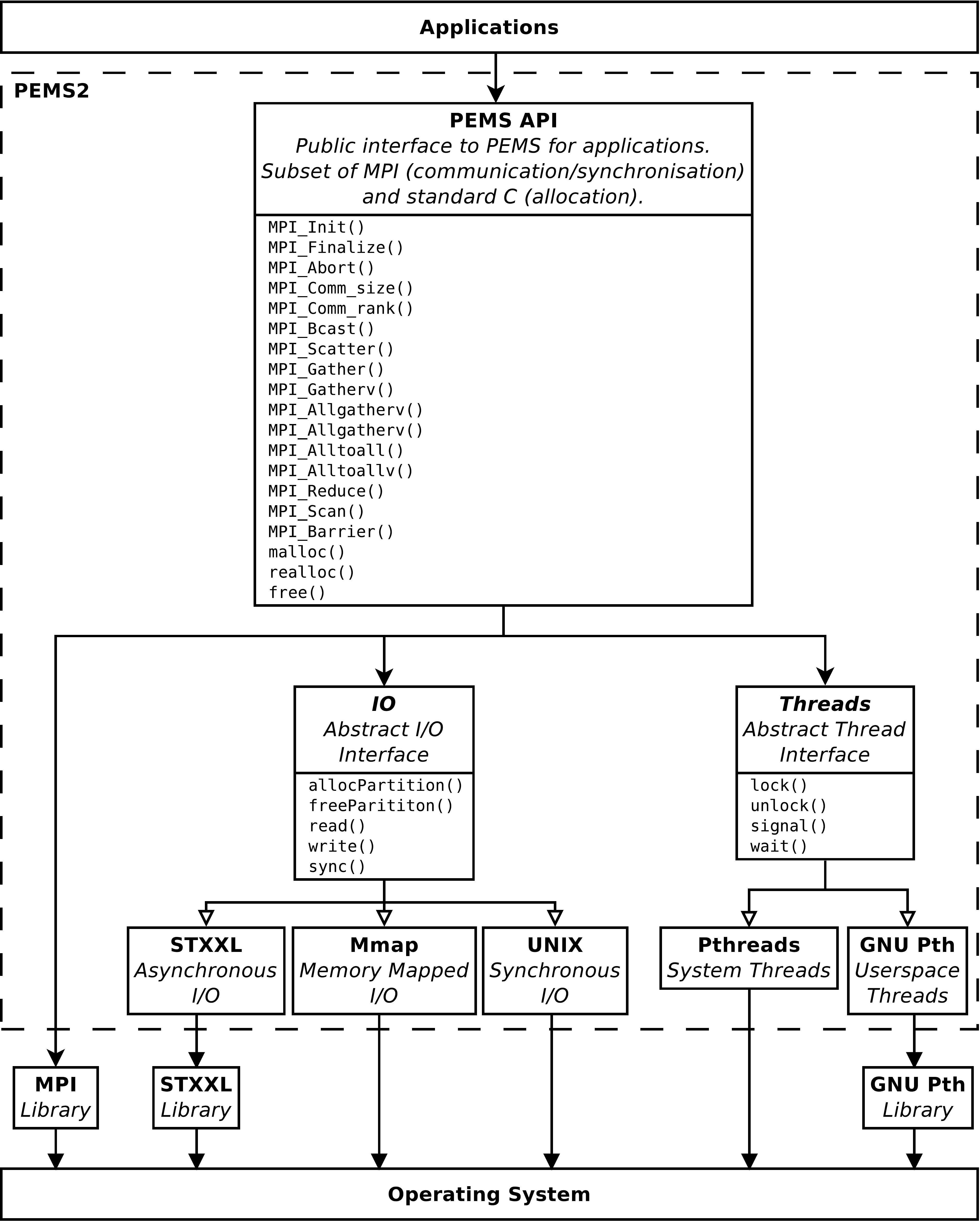}
	}
\caption{PEMS2 Design}
\label{pems2_design}
\end{center}
\end{figure}

Fig.~\ref{pems2_design} shows an overview of the PEMS2 design.

The most significant change from the more static architecture of PEMS1 is
the addition of abstract interfaces for I/O and threading.  All use of these
subsystems occurs through these relatively simple interfaces, which makes
the addition of new I/O and threading drivers to PEMS2 a straightforward
process with little impact on other components.

The original I/O (synchronous) and threading (user-space) implementations
from PEMS1 have been modified to fit within this framework.  Both remain
available for use as user options in PEMS2.

Two new I/O drivers have been implemented: Asynchronous I/O, which allows
PEMS to submit many I/O requests to the disk at once and resume computation
or communication while they are performed, is described in \S\ref{async}.
Memory-mapped I/O, which allows PEMS to only swap in the required portions
of a virtual processor context at each superstep, is described in \S\ref{mmap}.

A new threading driver based on POSIX threads has been added which
supports true concurrency, the implications of which are discussed in
Chapter~\ref{multicore}.

\clearpage
\section{Computational Model}

\begin{figure}[ht]
\begin{center}
	\resizebox{!}{0.3\textheight}{
	\includegraphics{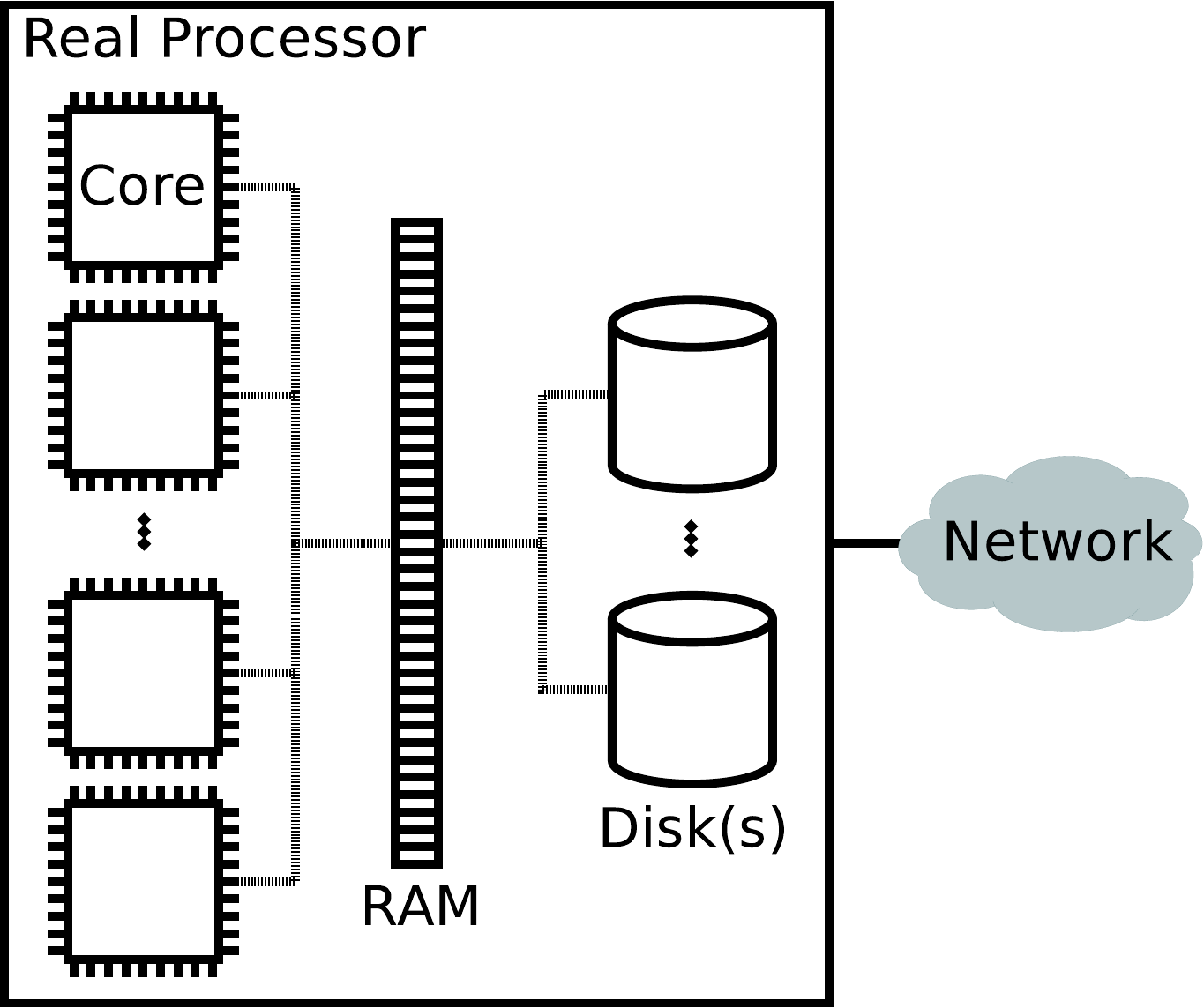}
	} \\
Example: $k=4$~Cores/Processor,
$D=2$~Disks
\end{center}
\caption{PEMS2 Computational Model}
\label{pems2_model}
\end{figure}

PEMS2 extends the EM-BSP \cite{dhthesis}\cite{bspem}\cite{emsimulation}
models shown in Fig.~\ref{pems1_model} with one or more ``cores'' per real
processor.  Each set of cores on a real processor access a single shared main
memory, and one or more disks.  The cluster of real processors is assumed
to be homogeneous, i.e.\ each real processor has $k$ cores and $D$ disks.
This extended model is shown in Fig.~\ref{pems2_model}.

Adding the ability for threads to execute concurrently is a relatively
straightforward modification to PEMS1 (replace the use of GNU Pth functions
with POSIX threads equivalents).  The difficulty in adding multi-core
support lies in the implications, e.g.\ more sophisticated synchronisation and
inter-thread communication methods must be used, and the relevant portions
of the system must be made thread-safe.  The details of how this has been
accomplished are discussed in Chapter~\ref{multicore}.

\chapter{Multi-Core Support}
\thispagestyle{empty}
\label{multicore}

PEMS1 supported only user-space threads via the GNU Pth library.  On single-core
machines this can be advantageous because user-space threads avoid the
overhead of context switching.  However, achieving true concurrency on a
multi-core machine requires the use of ``real'' system threads.

While it is possible to run several MPI processes concurrently on a single
multi-core machine, running a single process with a thread for each virtual
processor avoids the overhead of inter-process communication, synchronisation,
and context switching.  Using threads also allows for more effective parallel
disk I/O strategies since PEMS can control parallel access to disk(s) in
more flexible ways.

To achieve this, the threading system has been redesigned around a small
set of simple synchronisation primitives which are safe for both user-space
and kernel threads.  PEMS2 can use either user-space threads via GNU Pth,
or system threads via the POSIX Threads (``pthreads'') API.  In either
case there is a $1:1$ relationship between threads and virtual processors
regardless of the number of cores available.

The number of virtual processors that execute concurrently on a local real
processor is denoted $k$.  The user may choose any value for $k$ provided
$1 \le k \le \frac{v}{P}$.

\section{Memory Partitions}
\label{partitions}

Thread concurrency in PEMS2 is achieved by allocating $k$ separate memory
partitions (rather than the single partition used by PEMS1).  Thus,
$k$ separate threads may be swapped in at a given time and perform work
concurrently.  The user must ensure that $k\mu$ real memory is available
for these partitions.

A simple static mapping is used to assign threads to memory partitions:
thread $t$ uses partition $t\mod{k}$.  A dynamic mapping would be beneficial
in many respects, but this would have the effect of changing the address
of a given piece of virtual processor memory, thus invalidating pointers.
For example, if a virtual processor is swapped in at memory address 10,
a pointer to the first allocated piece of memory would have the value 10.
If that virtual processor was subsequently swapped in at memory address 20,
the pointer {\em should} have value 20, but still has value 10, thus memory
is corrupted.  Because of this, a dynamic mapping of contexts on disk to
memory partitions is not feasible within PEMS.

\section{Controlling Concurrency}

With the addition of system thread support, several virtual processors may
execute in parallel on a single real processor, taking advantage of multiple
cores.  This raises an issue when $\frac{v}{P} > k$ (which is generally the
case for any reasonable configuration): threads can run concurrently, thus
more than $k$ threads may attempt to run simultaneously.  However, only $k$
memory partitions are available.  PEMS itself can not explicitly schedule
threads $k$ at a time since the operating system scheduler is used.  Instead,
an exclusive lock (mutex) is associated with each of the $k$ partitions
in main memory.  A thread must obtain a lock on its memory partition before
executing any part of the simulated virtual processor's algorithm.  Therefore,
the number of virtual processors which can run concurrently on a single real
processor is at most $k$.

\section{Thread Synchronisation}
\label{synchronisation}

Superstep synchronisation, a simple barrier, is sufficient for collective
communication methods in which all processors participate as equals.  However,
there are many methods which have more complex synchronisation requirements.
A simple example of such a method is the broadcast, or {\tt Bcast}.  In a {\tt
Bcast}, a single virtual processor called the ``root'' sends a message to
every other virtual processor (see \S\ref{bcast-sec}).  Thus, other virtual
processors have to wait specifically for the root to perform some action.
While full superstep barriers could be used for this purpose, synchronisation
methods specifically designed for such cases can achieve better performance.
With a full barrier, each virtual processor waits for every other virtual
processor.  However, in a ``rooted'' case such as {\tt Bcast}, any virtual
processor that reaches the barrier after the root need not wait at all.
Since I/O is triggered by virtual processor execution (e.g.\ I/O will occur
when a virtual processor calls {\tt MPI\_Bcast}), this can be a significant
performance factor -- the sooner a thread passes a barrier, the sooner it
can submit further I/O requests, resulting in higher throughput and more
communication/computation overlap.

The collective communication algorithms presented here require three
styles of synchronisation (in addition to superstep barriers):

\begin{enumerate}
\item {\em Initial Synchronisation}: Wait for the first thread
\item {\em Rooted Synchronisation}: Wait for a specific ``root'' thread
\item {\em Final Synchronisation}: Wait for all other threads
\end{enumerate}

These operations are implemented to work with any number of threads running
at a time, swapping virtual processors in or out as required.

Since each thread holds its memory partition lock while executing, and other
inactive threads require the same partition, simply using a primitive signal
(e.g.\ that provided by pthreads) would result in a deadlock and/or missed
signals.  This is because primitive signals are not persistent, i.e.\ only
those threads waiting on a signal at the moment it fires are notified.
In PEMS2, a primitive signal with an associated counter and flag make up a
composite synchronisation structure.  This allows for synchronisation both
between threads which are currently swapped in (via the primitive signal)
and threads which are not (via the counter or flag).

The primitive signal is only used to synchronise the $k$ currently swapped in
threads, eliminating the possibility of deadlock.  The counter keeps track
of how many threads have reached the synchronisation barrier, and the flag
is used to signal an arbitrary condition (e.g.\ ``the root has finished'').

This composite signal structure is simply referred to as a ``signal''; which
is the main threading abstraction used to implement our synchronisation
primitives.

All functions described in this section are called while the thread holds
the lock on its memory partition.  Because swapping is generally the most
expensive operation performed by PEMS during a simulation, the goal of these
primitives is to swap only when necessary.  Run times stated for these methods
only consider time spent performing I/O, since no significant computation
takes place.

\clearpage
\subsection{Rooted Synchronisation}

Alg.~\ref{waitForRoot} {\sc EM-Wait-For-Root} waits for the root thread to
signal.  This is generally necessary for any rooted collective communication
method (e.g.\ {\tt Bcast}, {\tt Gather}).  Swapping to disk occurs only when
a thread is blocking the memory partition required by the root.  The return
value indicates whether the partition has been swapped out, which allows the
caller to only swap in/out again if necessary.  Only the non-root threads
call this function; the root thread must perform whatever work is required,
then signal (using Alg.~\ref{signalThreads}, {\sc EM-Signal-Threads}) to
unblock the other threads.

\begin{algorithm}[h]
	\KwData{$s$ (signal), $t$ (this thread ID), $r$ (root thread ID) $\vert$ $t \ne r$} 
	\KwResult{True iff thread was swapped out}
	\BlankLine
	result $\longleftarrow$ false\;
	s.lock()\;
	\If(\tcp*[h]{If the root has not already signalled}){s.flag = false}{
		$p_t \longleftarrow t$ mod $k$ \tcp*[h]{Current thread's partition}\;
		$p_r \longleftarrow r$ mod $k$ \tcp*[h]{Thread $r$'s partition}\;
		\If(\tcp*[h]{If $t$ and $r$ share a partition}){$p_t = p_r$}{
			\tcc{Yield to root}\;
			result $\longleftarrow$ true\;
			Swap out\;
			Unlock partition\;
		}
		s.wait() \tcp*[h]{Wait for root to signal}\;
		\If(\tcp*[h]{If $t$ and $r$ share a partition}){$p_t = p_r$}{
			\tcc{Yielded above, so re-lock partition}\;
			s.unlock() \tcp*[h]{Release signal lock to prevent deadlock}\;
			Lock partition\;
			s.lock()\;
		}
	}
	s.count $\longleftarrow$ s.count + 1\;
	\If(\tcp*[h]{If all non-root threads are finished waiting}){s.count = $\frac{v}{P}$}{
		\tcc{Reset signal}\;
		s.count $\longleftarrow 0$\;
		s.flag $\longleftarrow$ false\;
	}
	s.unlock()\;
	\Return result\;
	\caption{\sc EM-Wait-For-Root}
	\label{waitForRoot}
\end{algorithm}

\begin{lemma}
\label{wait-root-io}
Alg.~\ref{waitForRoot} takes $S\frac{v\mu}{PkB}$ time in the worst case.
\end{lemma}
\begin{proof}
The only possible I/O occurs at line 8, which is only executed by virtual
processors which share a memory partition with the root processor.  There are
$k$ partitions per real processor, shared by $\frac{v}{P}$ virtual processors,
thus $\frac{v}{Pk}$ virtual processors may perform I/O.  If a virtual
processor performs I/O, it swaps out once at line 8, resulting in $\mu$ I/O per
virtual processor that shares a partition with the root.  Since all virtual
processors that perform I/O share a memory partition, only one may be swapped
in at a given time, therefore no disk parallelism occurs in the worst case
when striping is not in use.
\end{proof}

Note that Lemma~\ref{wait-root-io} does not take disk striping into
consideration, i.e.\ it is assumed that each virtual processor is mapped to
a single disk.  If PEMS is being used on a configuration where all data
is striped across all disks, then all I/O is inherently fully parallel,
and therefore Alg.~\ref{waitForRoot} would take $S\frac{v\mu}{PkBD}$ time.

\subsection{Initial Synchronisation}

Implementations of several collective communication functions require an
arbitrary single thread to do some work (e.g.\ perform MPI communication) before
any other threads continue.  Alg.~\ref{firstThread} ({\sc EM-First-Thread}),
when called by all threads, will return true immediately if the caller is the
first thread, or otherwise block until the first thread has signalled and
return false.  Note that when true is returned the signal is still locked;
this allows the first thread to perform the necessary work while other
threads wait.  The first thread must signal (using Alg.~\ref{signalThreads}
with false as the ``lock'' parameter) when it has completed the work in
order to wake any waiting threads.

\begin{algorithm}[h]
	\KwData{$t$ (this thread ID), $s$ (signal)} 
	\KwResult{True iff caller is the first thread}
	\BlankLine
	s.lock()\;
	\If(\tcp*[h]{If this is the first thread}){s.count $= 0$}{
		\tcc{Keep lock and return true}
		s.flag $\longleftarrow$ false\;
		\Return true\;
	}
	s.count $\longleftarrow$ (s.count $+ 1$)$\mod \frac{v}{P}$\;
	\If(\tcp*[h]{If first thread has not finished}){s.flag = false}{
		s.wait()\;
	}
	\If(\tcp*[h]{If this is the last thread}){s.count $= 0$}{
		\tcc{Reset signal}
		s.flag $\longleftarrow$ false\;
	}
	s.unlock()\;
	\Return false\;
	\caption{\sc EM-First-Thread}
	\label{firstThread}
\end{algorithm}

\begin{lemma}
\label{first-io}
Alg.~\ref{firstThread} performs no I/O.
\end{lemma}

\subsection{Final Synchronisation}

Collective communication calls which collect data at a single root processor
(e.g.\ Gather) must wait for other threads to finish their work before
the results can be gathered and delivered to their final destination.
Alg.~\ref{allThreadsFinished} ({\sc EM-All-Threads-Finished}) along with
Alg.~\ref{waitThreads} ({\sc EM-Wait-Threads}) provides the required mechanism.
If true is returned, all threads have reached the call and the work may
be safely performed.  Whether or not a swap has occurred is passed as an
input/output parameter (e.g.\ a pointer) to allow cascading several calls
without performing unnecessary swaps: if true is passed for this parameter,
no swap will be performed.  Otherwise, if a swap is performed, the
parameter will be set to true to notify the caller.  

Like Alg.~\ref{firstThread} ({\sc EM-First-Thread}), if false is returned
the lock is not released.  When this happens the caller must call
Alg.~\ref{waitThreads} ({\sc EM-Wait-Threads}) which will block until all
threads have completed.

\begin{algorithm}[h]
	\KwData{$t$ (this thread ID), $s$ (signal), $w$ (whether swap has occurred)} 
	\KwResult{True iff the caller is last}
	\BlankLine
	result $\longleftarrow$ true\;
	last $\longleftarrow$ false\;
	s.lock()\;

	\If(\tcp*[h]{If this is the last thread}){s.count $= \frac{v}{P} - 1$}{
		\tcc{Signal others, reset signal and return true}
		s.count $\longleftarrow 0$\;
		s.broadcast()\;
		s.unlock()\;
		\Return true\;
	}\Else(\tcp*[h]{This is not the last thread}){
		s.count $\longleftarrow$ (s.count $+ 1$)$\mod \frac{v}{P}$\;
		\If(\tcp*[h]{If the last thread has not already finished}){$s.flag = true$}{
			\If(\tcp*[h]{If this thread hasn't already swapped out}){$w = false$}{
				\tcc{Swap out and notify caller}
				Swap out\;
				w $\longleftarrow$ true\;
			}
			\tcc{Wait for last thread to finish}
			Unlock partition\;
			s.wait()\;
			Lock partition\;
		}
		\If(\tcp*[h]{Last thread has finished}){s.flag = true}{
			s.unlock()\;
		}\Else(\tcp*[h]{This thread is blocking the last thread}){
			\tcc{Keep lock and return false}
			\Return false\;
		}
	}
	\Return result\;
	\caption{\sc EM-All-Threads-Finished}
	\label{allThreadsFinished}
\end{algorithm}

\begin{algorithm}[h]
	\KwData{$s$ (signal), $w$ (whether swap has occurred)} 
	\BlankLine
	\If(\tcp*[h]{If this thread hasn't been swapped out yet}){$w$ = false}{
		Swap out\;
		$w \longleftarrow$ true\;
	}
	
	\BlankLine
	\tcc{Yield partition and wait for signal}
	Unlock partition\;
	s.wait()\;
	Lock partition\;

	\BlankLine
	\tcc{Reset signal}
	s.flag = false\;
	s.count = 0\;
	s.unlock()\;
	\caption{\sc EM-Wait-Threads}
	\label{waitThreads}
\end{algorithm}

\begin{lemma}
\label{wait-threads-io}
Alg.~\ref{waitThreads} performs at most $v\mu$ I/O.
\end{lemma}
\begin{proof}
The only I/O performed is a swap out of size $\mu$, which is called
$v$ times in the worst case (once by each virtual processor).
\end{proof}

\subsection{Signalling}

Both Initial and Rooted synchronisation require a thread to signal the others
once some work has been performed.  Alg.~\ref{signalThreads} ({\sc
EM-Signal-Threads}) is used for this purpose in both cases.  Since these
cases have different locking semantics, whether the signal lock should be
taken is passed as a parameter (specifically: false must be passed in the
Initial case, and true in the Rooted case).

\begin{algorithm}[h]
	\KwData{$t$ (this thread ID), $s$ (signal), $l$ (whether to lock)} 
	\BlankLine
	\If{$l$ = true}{
		s.lock()\;
	}
	s.count $\longleftarrow$ (s.count $+ 1$)$\mod \frac{v}{P}$\;
	s.flag $\longleftarrow$ true \tcp*[h]{Set flag for threads yet to run}\;
	s.broadcast() \tcp*[h]{Signal the $k-1$ other currently running threads}\;
	s.unlock()\;
	\caption{\sc EM-Signal-Threads}
	\label{signalThreads}
\end{algorithm}

\chapter{New I/O Drivers}
\thispagestyle{empty}
\label{io-drivers}

\section{Asynchronous I/O}
\label{async}

\subsection{Background}

The UNIX system I/O used by PEMS1 is synchronous, i.e.\ a call to {\tt read}
or {\tt write} blocks until the I/O operation has finished.  In some cases
this is necessary because execution can not continue until I/O is finished,
typically because the buffers used are required for the next operation.
In other cases, however, there is useful work that can be safely performed
in parallel with the I/O operation.  In these cases, asynchronous I/O is
advantageous.  Asynchronous I/O allows an I/O request to be submitted with a
non-blocking call, and provides a separate mechanism to wait for completion.
This allows I/O to proceed in parallel with computation, improving overall
performance.

An additional benefit of asynchronous I/O is the ability to send many
I/O requests to the operating system (OS) at once.  With synchronous I/O,
this is not possible because all I/O requests block.  Asynchronous I/O,
however, allows submitting many requests in rapid succession, keeping the
OS and disk busy with I/O requests.  This is beneficial because the OS
attempts to schedule disk I/O optimally when several requests are pending.
Several algorithms exist for this purpose which yield better performance than
a trivial {\em First Come First Served} (FCFS) algorithm \cite{disksched}.
All modern commonly used operating systems include at least one disk
scheduling algorithm; Linux in particular provides several which may be
selected at runtime for a specific disk volume (see \S\ref{future}).

\subsection{Design}

PEMS2 uses the STXXL \cite{stxxl} file layer for asynchronous I/O.  This is
the lowest level abstraction in STXXL, essentially a portability layer for
asynchronous I/O with a more elegant interface than the operating system's API.
The scheduling and caching mechanisms in other layers of STXXL are not used
(see \S\ref{stxxl-sec} for additional discussion of STXXL).

The non-trivial modifications to PEMS required for asynchronous I/O are
concerned with waiting for the necessary I/O requests to finish.  All I/O in
PEMS is performed by some virtual processor, and written to / read from the
context of another virtual processor.  It is important that threads only wait
when necessary to avoid blocking other threads which could otherwise proceed.
Generally, the thread that initiated the I/O request is the only thread that
should wait.  Accordingly, PEMS2 has $k$ independent I/O request queues per
real processor, one for each local virtual processor that is swapped in.
Each virtual processor can make multiple I/O requests (e.g.\ during message
delivery) and explicitly wait for all, or some, of its own requests to finish
if necessary.  Otherwise, all requests are waited on at the next superstep
barrier before the virtual processor is swapped out.

\section{Memory Mapped I/O}
\label{mmap}

\subsection{Background}

The I/O approaches previously discussed (both synchronous and asynchronous)
have a major disadvantage for certain algorithms: at each virtual superstep,
the entire context of every virtual processor is swapped regardless of how
much data the algorithm actually uses.  In cases where the algorithm only
accesses a small portion of the data (e.g.\ sampling) this can result in a
very large amount of unnecessary I/O.  This can cause the I/O complexity
of the simulation to be far from optimal, particularly for algorithms
with many supersteps each of which do not access the majority of memory.
This problem can not be solved with explicit I/O (i.e.\ read/write calls)
because PEMS has no way of knowing which areas of memory are actually used
by the simulated algorithm.

Special API calls could be added to PEMS to address this problem, but
this conflicts with the goal of simulating generic BSP-like algorithms,
and would not be compatible with MPI.  Fortunately, there is a mechanism
available in all modern operating systems which can solve this problem:
memory mapped I/O.  Memory mapping is a facility which allows a file (or
other addressable resource) to be mapped onto a range of virtual memory and
used normally like any other region of memory, without actually reading the
entire file into physical memory.  Pages are swapped to/from disk by the OS
as necessary without any effort on behalf of the programmer.

The critical property of memory mapped I/O is that this page swapping is
performed by the OS kernel which, unlike ``userland'' code such as PEMS,
does know which areas of memory are accessed.  This allows PEMS to avoid
unnecessary swapping, since the kernel will only swap in/out those regions
of memory which are actually used by the simulated algorithm.  This implies
the cache behavior of the algorithm may also affect I/O performance --
algorithms with favourable memory access patterns will make use of the
kernel-managed cache more effectively, and achieve better performance
with memory mapped I/O.

Experiments in \S\ref{cgmlib} confirm experimentally that memory mapping
avoids a significant amount of I/O in some cases.

\subsection{Design}

When used with memory mapping, PEMS2 simply maps the entire used portion
of disk into memory.  Rather than allocate in-memory partitions and swap
in/out from/to disk, the simulated algorithm works directly with a range
of this mapped memory.  All other aspects of the simulation remain the
same, in particular, only $k$ virtual processors execute at a given time.
If suitable parameters are chosen such that $k\mu$ fits within physical memory
(as it must with explicit I/O), this ensures that the amount of virtual memory
used at any given time fits within physical memory, so thrashing is avoided.

Because memory-mapped disk regions are used in the same way as any other
region of memory, message delivery in PEMS2 with memory-mapped I/O is simply
a direct virtual memory copy (e.g.\ using {\tt memcpy}).  In degenerate cases
where the problem size is smaller than the available physical memory, this
effectively makes PEMS an in-memory multi-core MPI system.  This allows PEMS
to scale gracefully over a wide range of problem sizes from very small,
to the majority of physical memory, to much larger than physical memory.

\chapter{Simulation Enhancements}
\thispagestyle{empty}
\label{simulation}

\section{Swapping}

A straightforward implementation of many communication algorithms could perform
many complete swaps in a virtual superstep (i.e.\ a superstep in the simulated
algorithm), since virtual supersteps may be composed of several internal
supersteps (i.e.\ a superstep performed by PEMS).  A careful implementation,
however, can ensure that each virtual processor is completely swapped out and
completely swapped in only once per virtual superstep.  Thus, for explicit
I/O, $L \ge S\frac{2v\mu}{B}$\footnote{Note that $L$ in this thesis differs
from previous work on PEMS, see Appendix~\ref{parameters}}.

With the use of memory mapped I/O, supersteps cause no explicit I/O at all.
In this case the analysis of a simulated algorithm must take into consideration
any swapping I/O it would cause by accessing its own memory mapped partition.
Because of this generic bounds for a PEMS simulation using memory mapped
I/O can not be given, the analysis is specific to a particular algorithm.

\section{Message Delivery}
\label{delivery}

This section introduces a new communication strategy for PEMS which addresses
the limitations discussed in \S\ref{potential}.  For illustrative purposes,
the basic concept is first presented in the form of a simplified algorithm,
Alg.~\ref{alltoall-buf-seq}, which does not consider details such as block
alignment.

Message delivery in PEMS2 is discussed using {\tt Alltoallv} as an example.
Rather than write/read messages to a separate area on disk as in PEMS1,
all virtual processors record in a table where in their contexts they expect
to receive incoming messages.  Then, they deliver directly to other virtual
processors' contexts on disk.  Thus the additional communication area on disk
(and with it a significant amount of I/O and disk seeking) is eliminated.

\begin{algorithm}[h]
\KwData{$\mathcal{S}$ : Array of pointers to $v$ messages to send}
\KwData{$\mathcal{R}$ : Array of pointers to $v$ messages to receive}
\BlankLine
Let $\mathcal{T}$ be a shared $v$ x $v$ table of incoming message offsets
\BlankLine
\tcc{Store message offsets}
	Store incoming message offsets from $\mathcal{R}$ in $\mathcal{T}$\;
	Swap out\;
\BlankLine\tcc{Finished Internal Superstep 1}
\tcc{Begin Internal Superstep 2}\BlankLine
\tcc{Deliver messages}
	Swap in\;
	\ForEach{message $m_{\rho{\rightarrow}i}$ in $\mathcal{S}$}{
		Write $m_{\rho \rightarrow i}$ to $\mathcal{T}_{\rho \rightarrow i}$\;
	}
	Swap out\;
\BlankLine\tcc{Finished Virtual Superstep}\BlankLine
\caption{\sc Simple-Direct-Alltoallv}
\label{alltoall-buf-seq}
\end{algorithm}

This strategy avoids out-of-place message delivery, but is not an ideal
solution for two reasons: I/O operations are not necessarily block aligned, and
messages are written to disk and read again in cases where this can be avoided.

One possible approach to eliminate alignment issues is to simply use
buffered I/O.  However, the caching and copying inherent to buffered I/O is not
suitable for a system like PEMS which consumes as much main memory as possible.
PEMS1 resolved this by organizing all message data in an appropriate way in a
separate area on disk.  PEMS2 instead directly delivers the largest aligned
portion of a message possible, and keeps a cache of remaining blocks which
require ``cleaning up''.  The key observation is that for a given message, a
maximum of $2$ blocks may not be properly aligned (namely the first and last
block of the message\footnote{Note that while messages may be distributed
in any way across the context, an {\em individual} message is a contiguous
range}).  Since each virtual processor receives $v$ messages, a given virtual
processor must receive at most $2v$ unaligned blocks -- dramatically
less than the total message volume for a typical coarse-grained algorithm.

The overhead inherent in buffered I/O is due to the fact that a portion of
a block can not physically be written to disk.  For example, if the first
half of a block must be written, the contents of the second half must be in
memory so the complete block can be assembled and written to disk.  Therefore,
some sort of cache is required to avoid corrupting blocks when a partial block
is written.  We will need to emulate this behavior to achieve our goal, but
are able to do so more efficiently than the generic cache mechanism in the
kernel since we know precisely what the kernel must guess\footnote{``There
are only two hard problems in Computer Science: cache invalidation and naming
things.'' -- Phil Karlton}.

The general solution to this problem is trivial: simply read in the desired
block from the destination context, modify it, and write it back out again.
Unfortunately this is not sufficient for our purposes since two (or more,
in cases with very small messages) messages can overlap a single block,
which raises synchronisation issues when $k > 1$.  The overhead associated
with a read/write cycle is also undesirable.  Instead, we will cache the
blocks containing unaligned messages ends (``boundary blocks'') in memory
throughout the course of the Alltoallv call.  As virtual processors deliver
the bulk of their messages directly to their destinations, they update this
cache with the remaining fragments of the delivered messages.  Since this is
done when the relevant contexts are already swapped in, the read/write cycle
is avoided.  Finally, when the bulk of all messages have been delivered each
processor flushes the necessary boundary blocks from the cache in memory to
its context on disk and the algorithm completes.

\begin{figure}[ht]
\begin{center}
	\resizebox{0.5\textwidth}{!}{
	\includegraphics{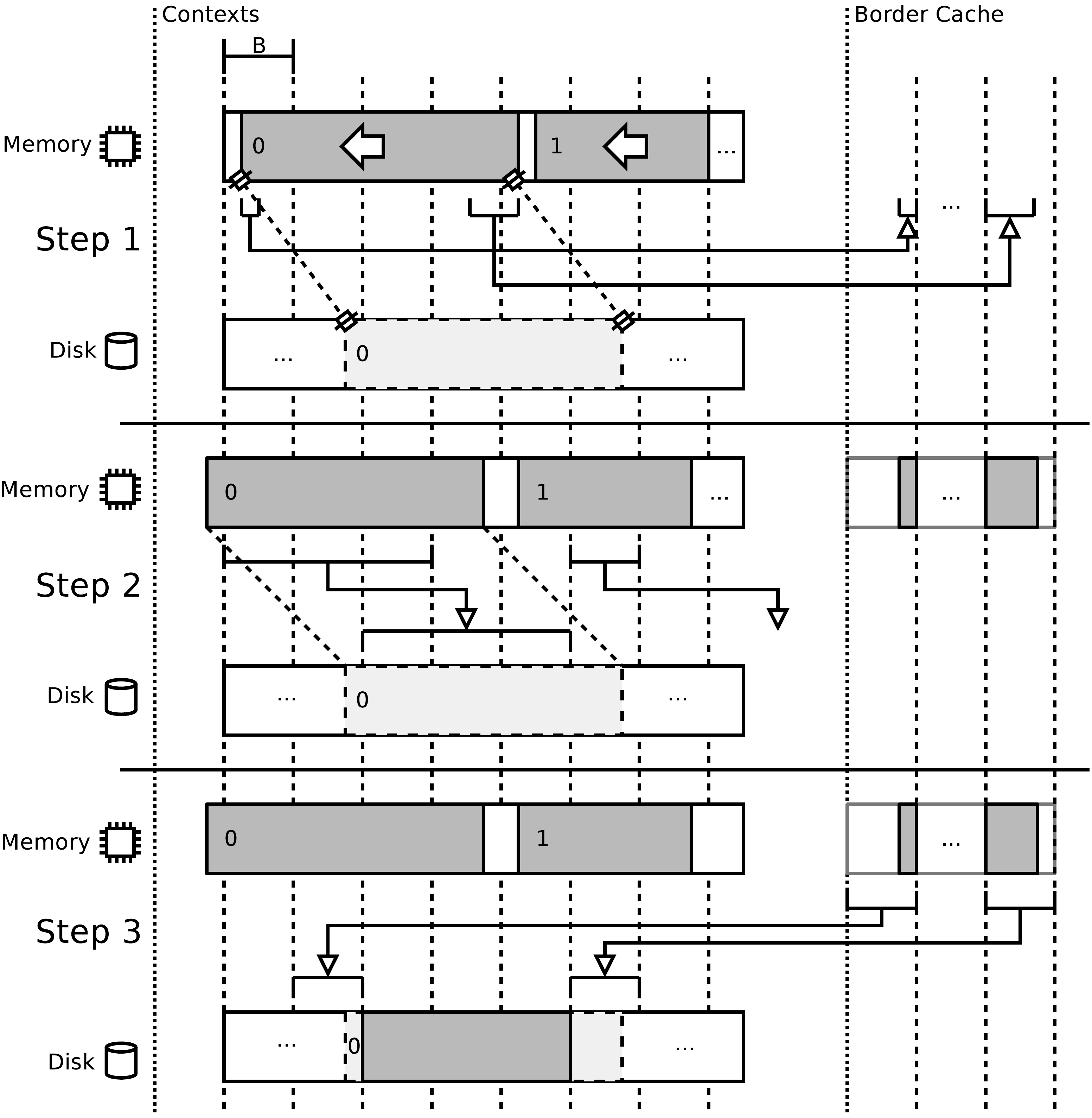}
	}
\caption{Direct Message Delivery}
\label{message-direct}
\end{center}
\end{figure}

Another issue arises with the direct delivery of messages in the absence
of buffered I/O: while the source and destination of each message contain
aligned regions of equal size, these regions may not have equivalent
alignment (i.e.\ their start offsets are not equivalent mod $B$).
Fig.~\ref{message-direct} illustrates such a case (the top and bottom
regions represent the sender's and receiver's contexts, respectively).
The largest aligned region within the message source does not correspond
directly to the largest block-aligned region within the message destination
because their alignment differs.  This is a problem because non-buffered I/O
requires that {\em all} offsets be block aligned, both in memory and on disk.
To resolve this we take advantage of the fact that the source context is
both in memory and on disk at write time, so we can destroy the context in
memory and avoid swapping out to prevent corruption.  We shift each message
in memory leftward so the regions in the sender and receiver align properly
(Step 1 in Fig.~\ref{message-direct}).  Thus aligned, these regions can
be delivered directly, and the remainder of the message is handled by the
boundary block cache in memory.

\section{Disk Space Reduction}
\label{s-disk-space}

In addition to reducing the volume of I/O performed, the elimination of the
indirect area significantly reduces the amount of disk space required to run
a given simulation, particularly with large numbers of virtual processors.
This allows a given system configuration to handle larger problem sizes.

In PEMS1, each real processor required $\frac{v\mu}{P}$ disk space for its
local virtual processor contexts, and $v\mu$ disk space for the indirect
area.  Note that the size of the indirect area increases with $v$ rather
than $\frac{v}{P}$.  This has the effect of increasing disk space when real
processors are added even if $\frac{v}{P}$ remains constant, which can be a
significant scalability problem.  The ideal strategy for scaling up a PEMS
simulation is to determine the parameters that fully utilize the resources
available to a single machine, then be able to easily add real processors
as necessary to reach the desired problem size.  An area on disk that scales
with $v$ rather than $\frac{v}{P}$ conflicts with this concept.  In practice
this makes scaling more tedious than necessary, since predicting the amount
of required disk space is more difficult.  Additionally, as experiments
in \S\ref{psrs-results} show, this large region of disk space can incur a
serious performance penalty when $\mu$ is large, due to both disk seek time
and file system overhead.

During the course of a simulation the disk is continually reading and writing
for swapping and message delivery.  As a result the disk head must constantly
seek across the entire region of disk space, including, in PEMS1, the huge
indirect area.  This occasionally had the counter-intuitive effect of making
the simulation {\em slower} when more RAM was added and $\mu$ correspondingly
increased, because the disk seek time dwarfed the time spent actually swapping
a given context.

To solve this problem, the improved simulation algorithms introduced in this
thesis eliminate the indirect disk area entirely, so the amount of disk space
required per real processor is precisely $\frac{v\mu}{P}$.  As a result,
real processors can be added to increase problem size without increasing
the disk space requirement for each real processor.  In practice, this makes
tuning a PEMS simulation much more manageable.

Fig.~\ref{disk-space} illustrates the difference in disk space consumption
between the two strategies.  Even for a modest $\frac{v}{P}$ the disk space
requirements for PEMS1 rapidly increase; in this case with 16 processors the
disk space required of a {\em single} real processor exceeds the {\em total}
problem size.  PEMS2, in contrast, only uses disk for virtual processor
contexts, so the amount of disk space required precisely matches the problem
size regardless of how many real processors are added.

\begin{figure}[ht]

\begin{center}
\begin{tabular}[]{ccccccc}
$p$ & $v$ & Required & PEMS1/Proc & PEMS1 & PEMS2/Proc & PEMS2 \\ \hline
1 & 8 & 16 GiB & 32 GiB & 32 GiB & 16 GiB & 16 GiB \\
2 & 16 & 32 GiB & 48 GiB & 96 GiB & 16 GiB & 32 GiB \\
4 & 32 & 64 GiB & 80 GiB & 320 GiB & 16 GiB & 64 GiB \\
8 & 64 & 128 GiB & 144 GiB & 1152 GiB & 16 GiB & 128 GiB \\
16 & 128 & 256 GiB & 272 GiB & 4352 GiB & 16 GiB & 256 GiB \\
\end{tabular}
\\ \bigskip
($\frac{v}{P} = 8$, $\mu = 2$ GiB)
\end{center}

\label{disk-space}
\caption{Disk Space Requirements}
\end{figure}

\section{Communication Buffer Size}
\label{comm-balance}

Due to the removal of the indirect message area, PEMS2 does not require
an upper bound on communication volume in order to allocate disk space, but
communication volume must still be bounded in many cases to avoid exceeding
the available communication buffer.  Note that one bound on communication
volume is inherent: each virtual processor can send at most $\mu$ data in
total, since each virtual processor has $\mu$ memory and messages must reside
in that memory before being sent.

Each virtual processor sends at most $v$ messages in a communication superstep.
The user may configure how many of these messages are sent at once using
the parameter $\alpha$.  By choosing an appropriate value for $\alpha$,
the user may ensure there is always sufficient buffer space to handle
communication.  This strategy removes the need for indirect routing as in
PEMS1 (see \S\ref{pems1-balancing}).  Performance is therefore improved since
each message is sent over the network precisely once, to its destination
real processor.

The amount of I/O performed by communication methods depends on the
size of messages.  To represent this in analytical results, the variable
$\omega$ is used to represent an arbitrary bound on virtual message size.
Specific values may be substituted for $\omega$ to find the run time for
a particular call, or a particular computational model.  For example, if a
{\tt Bcast} is performed where the message is simply a single 32-bit integer,
$\omega = 4$ bytes; for a CGM algorithm, $\omega = \Theta(\frac{N}{v})$; etc.

\section{Scheduling and Disk Parallelism}
\label{sched-par}

If each virtual processor is mapped to a single disk (i.e.\ striping or similar
techniques are not in use), the runtime of a collective communication method
depends on the order of execution of virtual processors.  This is because
virtual processors execute in synchronised rounds $k$ at a time, where each
round includes a single virtual processor mapped to each memory partition
($0{\ldots}k-1$).  However, this does not automatically imply that each round
contains a virtual processor mapped to each disk.  In the worst case,
only a single disk may be used despite several disks being available.
Fig.~\ref{part-disk-mapping} shows such a case: if the virtual processors
shown in bold (0, 4, and 8) are executed in a round, only disk 0 is used
for that round and thus disk parallelism is not exploited.

\begin{figure}[ht]
\begin{center}
\begin{tabular}[]{c|c|c}
Processor ($\rho$) & Memory Partition ($\rho \mod k$) & Disk ($\rho \mod D$) \\ \hline
\textbf 0 & \textbf 0 & \textbf 0 \\
1         & 1         & 1 \\
2         & 2         & 0 \\
3         & 0         & 1 \\
\textbf 4 & \textbf 1 & \textbf 0 \\
5         & 2         & 1 \\
6         & 0         & 0 \\
7         & 1         & 1 \\
\textbf 8 & \textbf 2 & \textbf 0 \\
\end{tabular}
\end{center}
\caption{Memory Partition and Disk Mapping ($k=3$, $D=2$)}
\label{part-disk-mapping}
\end{figure}

This problem must be addressed in order to precisely analyse the communication
functions in PEMS2 and applications built with them.  Restrictions
on $k$ and $D$ could solve the problem, but this approach is not realistic
since $k$ and $D$ reflect physical system characteristics.  Defining the
scheduler's behaviour such that these situations are avoided is more flexible,
and feasible to implement in practice.  Conveniently, a trivial scheduling
algorithm results in the desired behaviour: if virtual processors are
executed in ID order, then message delivery is distributed across all disks.
For example, with $k=3$ as in Fig.~\ref{part-disk-mapping}, processors 0, 1,
2 would execute in the first round, 3, 4, 5 in the next round, etc.  If $k \ge
D$, then clearly each round uses $D$ disks in parallel (since an increasing
sequence of $k$ integers mod $D$ contains all integers in $0 \dots D-1$ if
$k \ge D$).  If $k < D$, then this is not the case, and virtual processor
contexts should be distributed across disks to exploit disk parallelism.

When each virtual processor context is distributed across disks, {\em all}
disk I/O of sufficient size is fully parallel, so the scheduler behaviour
need not be defined and no restriction is required of $k$ and $D$.  In this
case, a separate restriction is necessary: individual reads and writes must
be large enough that they will be performed across all $D$ disks.  With a
straightforward round-robin block distribution strategy as used by striped RAID
systems and the STXXL block layer, this requires $\omega \ge BD$ for fully
parallel message delivery, and $\mu \ge BD$ for fully parallel swapping.
For any reasonable configuration, $\mu \gg BD$.  $\omega$ may be $< BD$,
but if this is the case messages are so small that full disk parallelism
is impossible (since disks can not perform transfers smaller than $B$),
so we will simply assume $\omega \ge BD$ to simplify analysis.

Def.~\ref{full-swap-par} summarises these conditions.

\begin{definition}[Fully Parallel Swapping]
If each virtual processor context resides on a single disk, $k \ge D$,
and virtual processors are scheduled in increasing order by ID, then PEMS2
performs all swapping I/O across all $D$ disks in parallel.

If each virtual processor context is distributed evenly across all disks
in a blockwise fashion, and $\mu \ge BD$, then PEMS2 performs all swapping
I/O across all $D$ disks in parallel.
\label{full-swap-par}
\end{definition}

Unfortunately, this behaviour conflicts with the potential swapping
optimisation described in \S\ref{improvements-swap} where $k$ swaps can be
avoided at each virtual superstep barrier.  Accordingly, the run times given
in this thesis do not include that optimisation.

\section{Allocation}
\label{allocation-sec}

When PEMS is initialised it first allocates all memory required by virtual
processors.  It then intercepts allocation requests from the simulated algorithm
and satisfies them by allocating the requested memory from this pool.

To fully support dynamic memory allocation and deallocation, PEMS2 uses a more
sophisticated allocation scheme than PEMS1 (see \S\ref{improvements-alloc}).
All virtual processor memory is still contained within a single region of
size $\mu$.  Unlike PEMS1, however, PEMS2 stores the offset and size of
each allocation.  This enables freeing of allocated memory, which can then
be reused by future allocations.

The allocation records are stored using a simple balanced binary search tree
in memory.  Since the number of allocations is relatively small and the
overhead of this data structure is not significant compared to disk I/O, a
more sophisticated structure would not likely show any significant improvement.

The allocation algorithm is simple: search from the lowest address until a
large enough free chunk is found, then split the start of this chunk into
a newly allocated area of appropriate size.

\begin{figure}[h]
\begin{center}
	\resizebox{0.4\textwidth}{!}{
	\includegraphics{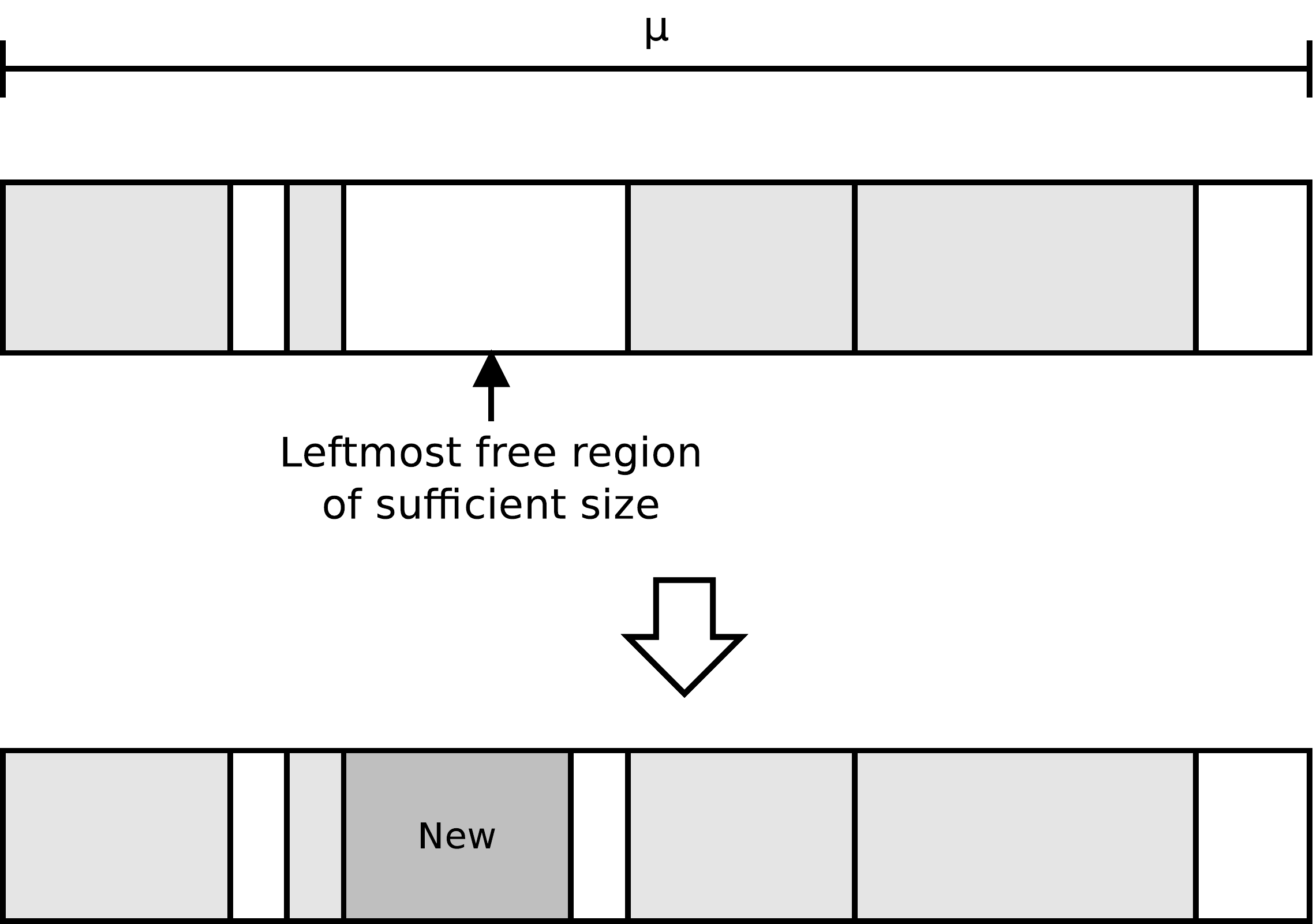}
	}
\end{center}
\caption{Memory Allocation in PEMS2}
\label{pems2-allocation-fig}
\end{figure}

Deallocation is also straightforward: remove the allocated chunk, and merge
with any adjacent free chunks.  If there are no adjacent free chunks, simply
record the area as deallocated.

\begin{figure}[h]
\begin{center}
	\resizebox{0.4\textwidth}{!}{
	\includegraphics{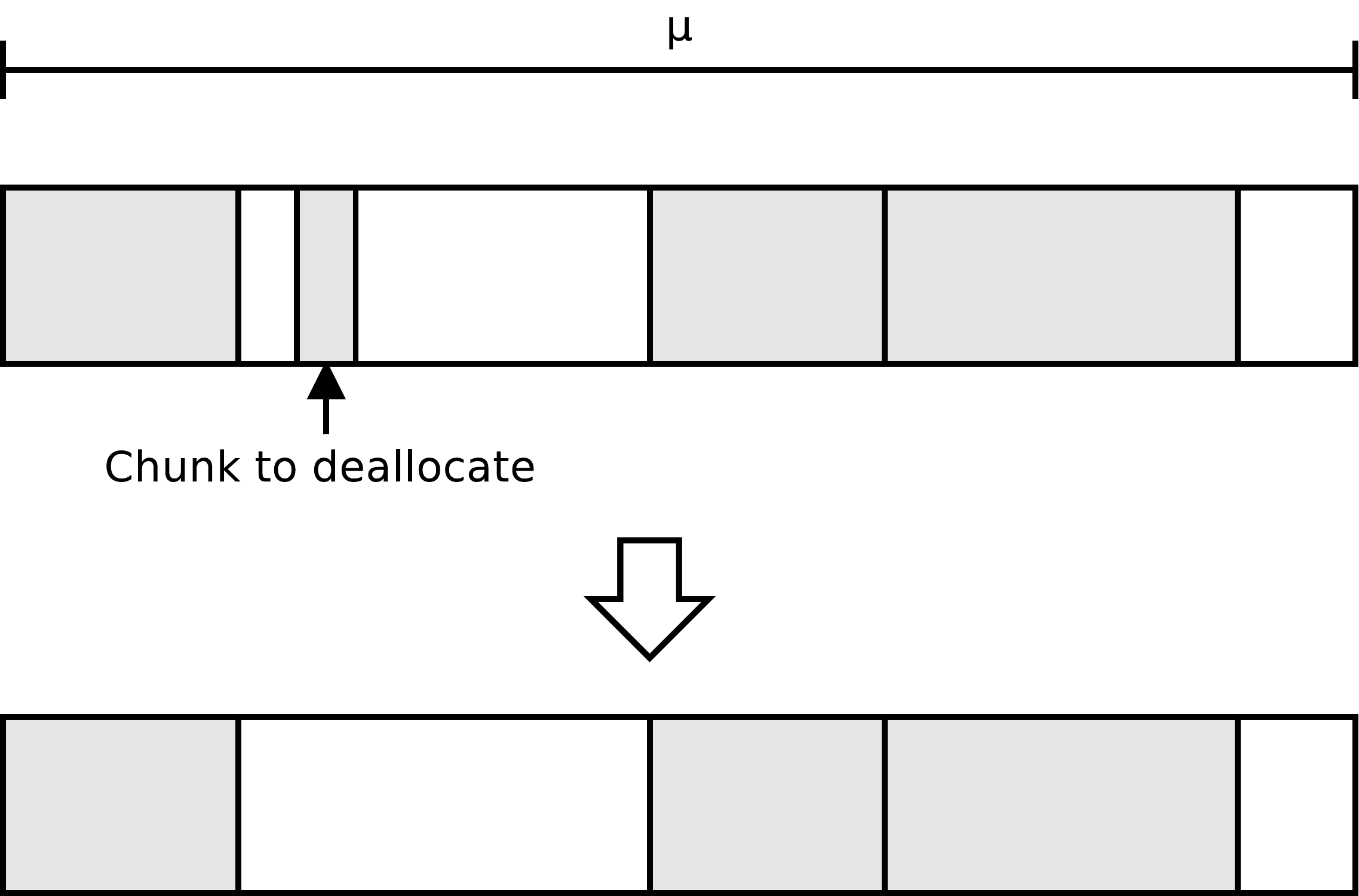}
	}
\end{center}
\caption{Memory Deallocation in PEMS2}
\label{pems2-deallocation-fig}
\end{figure}

More sophisticated strategies are of course possible; efficient allocation
with minimal fragmentation is a much researched problem.  In the context
of PEMS, however, the most important benefit of an allocator over the basic
design of PEMS1 is the ability to re-use deallocated memory, and avoid I/O
for currently unallocated memory regions.  The relatively simple allocator
presented here, though not optimal with respect to fragmentation, does
provide these two advantages.

The swapping related I/O function in PEMS2 have been modified to only swap
currently allocated regions of memory, rather than swap the entire partition
in a single read/write operation as in PEMS1.  As a result, programs which
free memory as soon as possible see improved performance due to less I/O.
For programs with very dynamic allocation behaviour, this can amount to a
significant reduction in I/O and total run time compared to the PEMS1 strategy.

\chapter{New and Improved Communication Algorithms}
\thispagestyle{empty}
\label{comm-algs}

\section{Alltoallv}

In an {\sc Alltoallv}, every virtual processor sends a message of arbitrary
size to every other virtual processor, thus $v^2$ messages are exchanged
in total.  {\sc Alltoallv} is the most powerful collective communication
operation implemented in PEMS that performs only communication (i.e.\ new
values are not computed as part of the operation).

Due to the complexity and size of the {\sc EM-Alltoallv} algorithm, the
single-processor and multi-processor versions are presented here separately.
These are referred to as {\sc EM-Alltoallv-Seq} and {\sc EM-Alltoallv-Par},
respectively.  The algorithm in general (i.e.\ for both single processor
and multi-processor cases) is referred to as {\sc EM-Alltoallv}.  Note the
implementation makes no such distinction and simply provides an implementation
of the {\tt MPI\_Alltoallv} function that works in both cases.

\begin{figure}[ht]
\begin{center}
	\resizebox{0.6\textwidth}{!}{
	\includegraphics{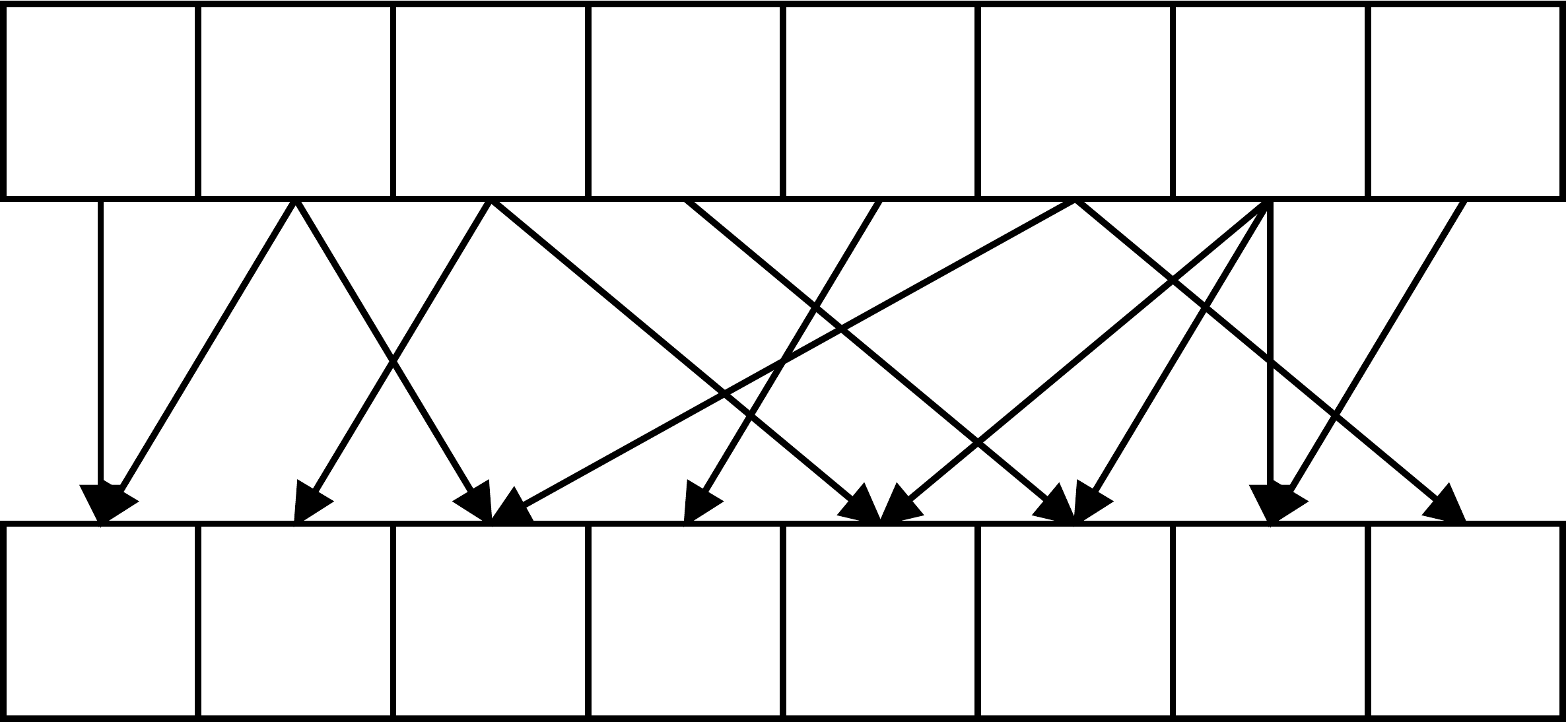}
	}
\caption{Alltoallv Operation}
\end{center}
\end{figure}

\clearpage
\subsection{Single Processor}

\subsubsection{Algorithm}

Alg.~\ref{alltoall-direct-seq} describes the single processor implementation of
{\tt Alltoallv} in PEMS2.  Note that all I/O (including swapping) is explicitly
performed, superstep barriers do not imply swapping.  This algorithm and the
others in this section perform fine-grained swapping, e.g.\ ``Swap message
in'' means the message (which resides in the virtual processor's context)
should be swapped in from disk to its usual location in the memory partition,
just as if the entire partition was swapped in.

The reader is encouraged to review the simpler implementations of {\tt
Alltoallv} (Alg.~\ref{simple-alltoall-seq} and Alg.~\ref{alltoall-buf-seq}),
since the algorithm given here solves the same problem in a similar but more
intricate way.

\begin{algorithm}[ht]
\KwData{$\mathcal{S}$ : Array of pointers to $v$ messages to send}
\KwData{$\mathcal{R}$ : Array of pointers to $v$ messages to receive}
\BlankLine
Let $\mathcal{T}$ be a shared $v$ x $v$ table of incoming message offsets\\
Let $\mathcal{E}$ be a shared array of $v$ execution states, all initially false\\
Let $\mathcal{M}$ be a cache of at most $2v^2$ border blocks ($2v$ per virtual processor)\\
\BlankLine
\tcc{Store message offsets and synchronise}
	Swap out everything except regions in $\mathcal{R}$\;
	Store incoming message offsets from $\mathcal{R}$ in $\mathcal{T}$
		\tcp*[h]{$\mathcal{T}_{* \rightarrow \rho}$ is valid}\;
	Set $\mathcal{E}_\rho$ to true\tcp*[h]{This thread has reached this point}\;
	Synchronise with the $k - 1$ other currently running threads\;
\BlankLine\tcc{Deliver messages if possible}
	\ForEach{message $m_{\rho{\rightarrow}i}$ in $\mathcal{S}$}{
		Update $\mathcal{M}$ with the start and end of this message\;
		\If(\tcp*[h]{Thread $i$ has recorded its offsets in $\mathcal{T}$}){$\mathcal{E}_i$ is true}{
			Align and deliver directly to $\mathcal{T}_{\rho{\rightarrow}i}$ on disk\;
		}
	}
\BlankLine\tcc{Finished Internal Superstep 1}
\tcc{Begin Internal Superstep 2}\BlankLine
\tcc{Deliver remaining messages}
	\ForEach{message $m_{\rho{\rightarrow}i}$ in $\mathcal{S}$ not delivered in superstep 1}{
		Swap message in\;
		Align and deliver directly to $\mathcal{T}_{\rho{\rightarrow}i}$ on disk\;
	}
\BlankLine\tcc{Finished Internal Superstep 2}
\tcc{Begin Internal Superstep 3}\BlankLine
\tcc{(Blocked I/O only) Flush border block cache}
	Flush border blocks in $\mathcal{M}$ to our context\;
\BlankLine\tcc{Finished Virtual Superstep}\BlankLine
\caption{{\sc EM-Alltoallv-Seq}}
\label{alltoall-direct-seq}
\end{algorithm}

\clearpage
\subsubsection{Analysis}

Similar disk parallelism issues arise in the analysis of {\sc EM-Alltoallv}
as those described in \S\ref{sched-par}, but with respect to message delivery
rather than swapping.  Since message delivery, like swapping, happens in
the same order as virtual processor execution, the same arguments used for
swapping (Def.~\ref{full-swap-par}) apply to message delivery as well.
Def.~\ref{full-delivery-par} summarises the necessary conditions for
communication methods that, like {\sc EM-Alltoallv}, perform message I/O
to/from all virtual processors.

\begin{definition}[Fully Parallel Message Delivery]
If each virtual processor context resides on a single disk, $k \ge D$,
and virtual processors are scheduled in increasing order by ID, then a
communication function which performs message I/O to/from all virtual
processors does so across all $D$ disks in parallel.

If each virtual processor context is distributed evenly across all disks in
a blockwise fashion, and $\omega \ge BD$, then a communication function which
performs message I/O to-from all virtual processors does so across all $D$
disks in parallel.
\label{full-delivery-par}
\end{definition}

\begin{definition}[Fully Parallel I/O]
A communication function has ``fully parallel I/O'' if it has both
fully parallel swapping (Def.~\ref{full-swap-par}) and fully parallel
message delivery (Def.~\ref{full-delivery-par})
\label{full-disk-par}
\end{definition}

\begin{lemma}
\label{alltoall-direct-seq-io}
When used with explicit I/O, {\sc EM-Alltoallv-Seq} performs
$v\mu + \frac{v^2 - vk}{2}\omega + 2v^2B$ I/O.
\end{lemma}
\begin{proof}
The fundamental difference between Alg.~\ref{alltoall-direct-seq} and
Alg.~\ref{simple-alltoall-seq} is that the amount of I/O performed by a
given virtual processor depends on how many virtual processors have finished
executing previously.\\
Let $\delta$ be the number of messages delivered directly on line 11.\\
Let $\iota$ be the number of messages delivered indirectly on line 14.\\
In lines $8{\ldots}10$, threads deliver directly to all threads that have
completed Internal Superstep 1.  Since threads execute in synchronised
rounds $k$ at a time, the first round of $k$ threads {\em each} deliver $k$
messages directly, the next round $2k$ (since $2k$ threads have now run),
the next round $3k$, etc.  Hence:
\begin{align*}
\delta	&= \displaystyle\sum_{i=1}^{\frac{v}{k}} ik^2 \\
		&= k^2 \left[\frac{v}{k}\left(\frac{\frac{v}{k}+1}{2}\right)\right] \\
		&= vk\left(\frac{\frac{v}{k}+1}{2}\right) \\
		&= \frac{v^2 + vk}{2} \\
\\
\iota	&= v^2 - \delta \\
\intertext{The remaining analysis is straightforward:}\\
I_{\text{seq}} &= I_{4} + I_{11} + I_{13..14} + I_{15} \\
&=	  \left( v\mu - v^2\omega \right)
	+ \left( \delta \omega \right)
	+ \left( 2\iota\omega \right)
	+ \left( 2v^2B \right) \\
&=	v\mu - v^2\omega + \delta\omega + 2v^2\omega - 2\delta\omega + 2v^2B \\
&=	v\mu + v^2\omega - \left(\frac{v^2 + vk}{2}\right)\omega + 2v^2B \\
&=	v\mu + \frac{v^2 - vk}{2}\omega + 2v^2B
\end{align*}
\end{proof}

\begin{corollary}[Improvement]
\label{alltoall-direct-seq-imp}
When used with explicit I/O, {\sc EM-Alltoallv-Seq} performs $2v\mu +
\frac{3v^2 + vk}{2}\omega - 2v^2B$ less message delivery I/O per virtual
superstep than Alg.~\ref{simple-alltoall-seq} ({\sc PEMS1-Alltoallv-Seq}).
\end{corollary}
\begin{proof}
\[
\begin{aligned}
\Delta{I}
	&= I_{\text{orig-seq}} - I_{\text{seq}} \\
	&= \left(3v\mu + 2v^2\omega\right)
	   - \left(v\mu + \frac{v^2 - vk}{2}\omega + 2v^2B\right) \\
	&= 2v\mu + \frac{3v^2 + vk}{2}\omega - 2v^2B \\
\end{aligned}
\]
\end{proof}

\begin{lemma}
\label{alltoall-direct-seq-mem}
{\sc EM-Alltoallv-Seq} uses at most $\frac{2v^2B}{P}$ shared buffer space.
\end{lemma}
\begin{proof}
The only buffer space used is for the block cache, when direct I/O is in use.
Each of the $\frac{v}{P}$ local virtual processors has 2 blocks in the cache for each of
its $v$ received messages.
\end{proof}

\begin{thm}
\label{alltoall-direct-seq-time}
Given fully parallel I/O (Def.~\ref{full-disk-par}), {\sc EM-Alltoallv-Seq}
takes $S\frac{v\mu}{BD} + G\frac{v^2 - vk}{2BD}\omega + G\frac{2v^2}{D} +
L$ time.
\end{thm}
\begin{proof}
Follows directly from Lem.~\ref{alltoall-direct-seq-io}, since {\sc
EM-Alltoallv-Seq} performs no network communication and no significant
computation.
\end{proof}

\subsubsection{Benchmarks}

\begin{figure}[ht]
\begin{center}
	\resizebox{0.9\textwidth}{!}{
	\input{alltoall-plot-tex}
	}
\caption{Single Processor {\sc EM-Alltoallv} Performance}
\label{em-alltoall-plot}
\end{center}
\end{figure}
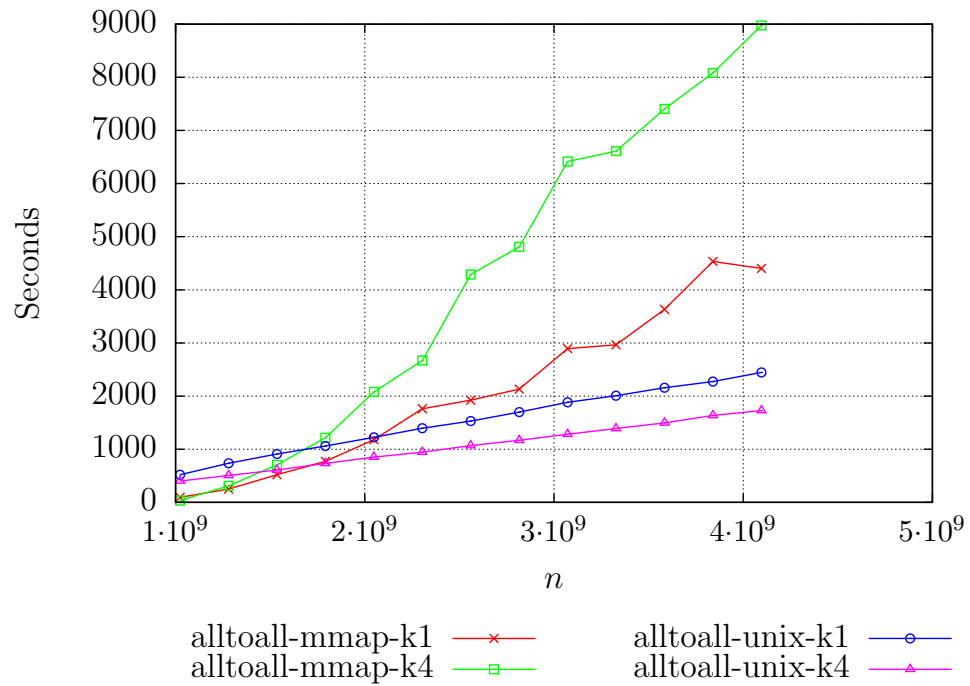

Fig.~\ref{em-alltoall-plot} shows the run time of a single call to {\sc
EM-Alltoall-Seq} for various numbers of 32-bit integers.  The x-axis represents
total problem size as a number of 32-bit integers, and the y-axis represents
total run time.  Times are shown for both memory-mapped (``mmap'') and explicit
(``unix'') I/O, for $k=1$ and $k=4$ cores (e.g.\ alltoall-mmap-k1 represents
memory-mapped I/O with 1 core).

No action is performed by the program other than a single {\sc Alltoallv}
on the complete data set.  Note in particular the performance improvement
seen with UNIX I/O when using 4 cores compared to using a single core.
Since the test program performs no significant computation, this shows that
the run time of {\sc EM-Alltoall-Seq} itself improves when $k$ increases,
as Thm.~\ref{alltoall-direct-seq-time} predicts (since the $vk$ term in the
message delivery time is subtracted).

The situation is reversed with memory mapped I/O, due to the overhead
of the operating system's cache mechanism.  This is not surprising: since
communication primitives in PEMS2 are carefully tuned to minimise I/O, explicit
I/O will always result in better performance for a trivial program that simply
calls a collective communication function once.  The potential benefit of
memory-mapped I/O is that the operating system's cache mechanism can avoid
a large amount of I/O for certain programs, but that is not the case here.

Note that the experiment shown in Fig.~\ref{em-alltoall-plot} is a trivial
program that does not represent a case where multi-core or memory-mapping are
expected to show much benefit.  Fig.~\ref{em-alltoall-plot} is not intended
to illustrate the improved performance of PEMS2, only that an improvement
is seen when using multiple cores in spite of the fact that no computation
is performed.  Experiments to illustrate the improvements in PEMS2 for
realistic use cases are shown in Chapter~\ref{experiments}.

\clearpage
\subsection{Multiple Processor}

\subsubsection{Algorithm}

Alg.~\ref{alltoall-direct-seq} describes the multiple processor implementation of
{\tt Alltoallv} in PEMS2.  Local message delivery occurs in an identical manner as
in the single processor case.  Remote messages are handled in the second internal
superstep, when all message destinations are known (since they have been recorded
in the shared table in the previous internal superstep).  Using this information,
virtual processors receive on behalf of their local peers and deliver directly
to their contexts on disk.

\begin{algorithm}[ht]
\KwData{$\mathcal{S}$ : Array of pointers to $v$ messages to send}
\KwData{$\mathcal{R}$ : Array of pointers to $v$ messages to receive}
\BlankLine
Let $\mathcal{T}$ be a shared $v$ x $\frac{v}{P}$ table of incoming message offsets\\
Let $\mathcal{E}$ be a shared array of $\frac{v}{P}$ execution states, all initially false\\
Let $\mathcal{M}$ be a cache of at most $\frac{2v^2}{P}$ border blocks ($2v$ per local thread)\\
\BlankLine
\tcc{Store message offsets and synchronise}
	Swap out everything except regions in $\mathcal{R}$\;
	Store incoming message offsets from $\mathcal{R}$ in $\mathcal{T}$
		\tcp*[h]{$\mathcal{T}_{* \rightarrow \rho}$ is now valid}\;
	Set $\mathcal{E}_\rho$ to true\tcp*[h]{This thread has reached this point}\;
	Synchronise with the $k - 1$ other currently executing local threads\;
\tcc{Deliver messages if possible}
	\ForEach{local message $m_{\rho{\rightarrow}i}$ in $\mathcal{S}$}{
		Update $\mathcal{M}$ with the start and end of this message\;
		\If(\tcp*[h]{Thread $i$ has recorded its offsets in $\mathcal{T}$}){$\mathcal{E}_i$ is true}{
			Align and deliver directly to $\mathcal{T}_{\rho{\rightarrow}i}$ on disk\;
		}
	}
\BlankLine\tcc{Finished Internal Superstep 1}
\tcc{Begin Internal Superstep 2}\BlankLine
\tcc{Deliver remaining messages}
	\ForEach{local message $m_{\rho{\rightarrow}i}$ in $\mathcal{S}$ not delivered in superstep 1}{
		Swap message in\;
		Align and deliver directly to $\mathcal{T}_{\rho{\rightarrow}i}$ on disk\;
	}
	Communicate using Alg.~\ref{alltoall-direct-par-comm}\;
	\ForEach{remote message $m_{i{\rightarrow}j}$ received}{
		Update $\mathcal{M}$ with the start and end of this message\;
		Align and deliver directly to $\mathcal{T}_{i{\rightarrow}j}$ on disk\;
	}
\BlankLine\tcc{Finished Internal Superstep 2}
\tcc{Begin Internal Superstep 3}\BlankLine
\tcc{(Blocked I/O only) Flush border blocks}
	Flush border blocks in $\mathcal{M}$ to our context\;
\BlankLine\tcc{Finished Virtual Superstep}\BlankLine
\caption{{\sc EM-Alltoallv-Par}}
\label{alltoall-direct-par}
\end{algorithm}

\begin{algorithm}[ht]
\KwData{$\mathcal{S}$ : Array of pointers to $v$ messages to send}
\KwData{$\mathcal{R}$ : Array of pointers to $v$ messages to receive}
\KwData{$\mathcal{T}$ : Shared $v$ x $\frac{v}{P}$ table of message offsets}
\BlankLine
Let $\mathcal{\alpha}$ be the network ``chunk size'' parameter\\
Let $\mathcal{B}$ be the shared communication buffer
\BlankLine
\ForEach{$i$ in 0, $\alpha, 2\alpha, 3\alpha, \dots, \frac{v}{P}-1$}{
	Assemble messages to threads $i \ldots i+\alpha-1$ on each real
	      processor contiguously in $\mathcal{B}$\;
	\If{this is the last of $k$ threads to reach this point}{
		Send/Receive assembled messages with {\tt MPI\_Alltoallv}\;
	}
	Deliver received messages using $\mathcal{T}$\;
}
\caption{{\sc EM-Alltoallv-Par-Comm}}
\label{alltoall-direct-par-comm}
\end{algorithm}

\subsubsection{Analysis}

\begin{lemma}
\label{em-alltoall-comm}
{\sc EM-Alltoallv-Par} takes $g\frac{\alpha{k}\omega}{b} + l\frac{v^2}{Pk\alpha}$
communication time, assuming $\alpha k\omega \geq b$, with $g$, $b$, and $l$ as in
the BSP model (see Appendix~\ref{conventions}).
\end{lemma}
\begin{proof}
Each virtual processor must send one message to each other virtual processor,
and all messages are sent directly to the real processor that hosts the
destination virtual processor.

All communication performed by {\sc EM-Alltoallv-Par} is performed by {\sc
EM-Alltoallv-Par-Comm}.  This algorithm sends messages from the ``round''
of $Pk$ currently executing virtual processors in ``chunks'' of size $\alpha$
(where $\alpha$ is a user-defined parameter indicating the number of messages
to send at once, $1 \leq \alpha < v$).

In each round, $k$ virtual processors are active on each real processor,
and each of these virtual processors sends $v$ messages over the
network\footnote{For simplicity of analysis, messages to local virtual
processors are included in this figure though in reality this is optimised
away and each virtual processor sends $v - \frac{v}{P}$ messages over
the network}.  Thus, each round consists of $\frac{v}{P\alpha}$ separate
$\alpha{k}\omega$-relations.

$\frac{v}{Pk}$ such rounds occur, therefore there are $\frac{v^2}{P^2k\alpha}$
such relations in total.
\end{proof}

\begin{lemma}
\label{alltoall-direct-par-io}
When used with explicit I/O, {\sc EM-Alltoallv-Par} performs
$
\frac{v\mu}{P}
	+ \left( \frac{v^2}{P} - v^2
	        + \frac{3v^2}{2P^2}
	        - \frac{kv}{2P}
	  \right) \omega
	+ 2v^2B
$ I/O.
\end{lemma}
\begin{proof}
The local message delivery of the parallel version of {\sc EM-Alltoallv-Seq}
is identical to that of {\sc EM-Alltoallv-Seq}, therefore this portion of
the analysis ($I_{4}$,  $I_{11}$, $I_{13..14}$, and $I_{15}$) is identical
except with $\frac{v}{P}$ local virtual processors rather than $v$.  The remaining
I/O is $I_{17..18} = \frac{v^2}{P}\omega$ for the received network messages.
\\
\[
\begin{aligned}
\delta	&= \displaystyle\sum_{i=1}^{\frac{v}{Pk}} ik^2 \\
		&= k^2 \left[\frac{v}{Pk}\left(\frac{\frac{v}{Pk}+1}{2}\right)\right] \\
		&= \frac{vk}{P}\left(\frac{v}{2Pk} + \frac{1}{2}\right) \\
		&= \frac{v}{P}\left(\frac{v}{2P} + \frac{k}{2}\right) \\
		&= \frac{1}{2}\left(\frac{v}{P}\right)^2 + \frac{k}{2}\left(\frac{v}{P}\right) \\
\\
\iota	&= \left(\frac{v}{P}\right)^2 - \delta \\
\\
I_{\text{seq}} &= I_{4} + I_{11} + I_{13..14} + I_{17..18} + I_{15} \\
&=	  \left( \frac{v\mu}{P} - v^2\omega \right)
	+ \left( \delta \omega \right)
	+ \left( 2\iota\omega \right)
	+ \left( \frac{v^2}{P}\omega \right)
	+ \left( 2v^2B \right) \\
&=	\frac{v\mu}{P} - v^2\omega - \delta\omega + 2\left(\frac{v}{P}\right)^2\omega
	+ \frac{v^2}{P}\omega + 2v^2B \\
&=	\frac{v\mu}{P} - v^2\omega
	- \frac{1}{2}\left(\frac{v}{P}\right)^2\omega
    - \frac{k}{2}\left(\frac{v}{P}\right)\omega
	+ 2\left(\frac{v}{P}\right)^2\omega
	+ \frac{v^2}{P}\omega + 2v^2B \\
&=	\frac{v\mu}{P}
	+ \left( \frac{v^2}{P}
	        + \frac{3v^2}{2P^2}
	        - \frac{kv}{2P}
			- v^2
	  \right) \omega
	+ 2v^2B \\
\end{aligned}
\]
\end{proof}

\begin{lemma}
\label{alltoall-direct-par-mem}
{\sc EM-Alltoallv-Par} uses at most $\frac{2v^2B}{P} + \alpha{k}\omega$ shared buffer space.
\end{lemma}
\begin{proof}
The size of the block cache is equivalent to the sequential case
(Lem.~\ref{alltoall-direct-seq-mem}).  Additional space is used by {\sc
EM-Alltoallv-Par-Comm} to assemble messages contiguously for communication.
The ``chunk size'', $\alpha$, is a user parameter which controls the amount
of buffer space used for this purpose: at most $\alpha{k}\omega$ buffer
space is used.
\end{proof}

\begin{thm}
\label{alltoall-direct-par-time}
Given fully parallel I/O (Def.~\ref{full-disk-par}), {\sc EM-Alltoallv-Par}
takes $
S\frac{v\mu}{PDB}
	+ G\left( \frac{v^2}{P}
	        + \frac{3v^2}{2P^2}
	        - \frac{kv}{2P}
			- v^2
	  \right) \frac{\omega}{PDB}
	+ G2v^2B
	+ g\frac{\alpha{k}\omega}{b} + l\frac{v^2}{Pk\alpha} + L$ time.
\end{thm}
\begin{proof}
Follows directly from Lem.~\ref{alltoall-direct-par-io} and
Lem.~\ref{alltoall-direct-par-comm}.
\end{proof}

\clearpage
\section{Bcast}
\label{bcast-sec}

In a {\sc Bcast} (broadcast), a single root virtual processor sends a single
message to every other virtual processor, i.e.\ every virtual processor
receives the same individual message.

\begin{figure}[ht]
\begin{center}
	\resizebox{0.6\textwidth}{!}{
	\includegraphics{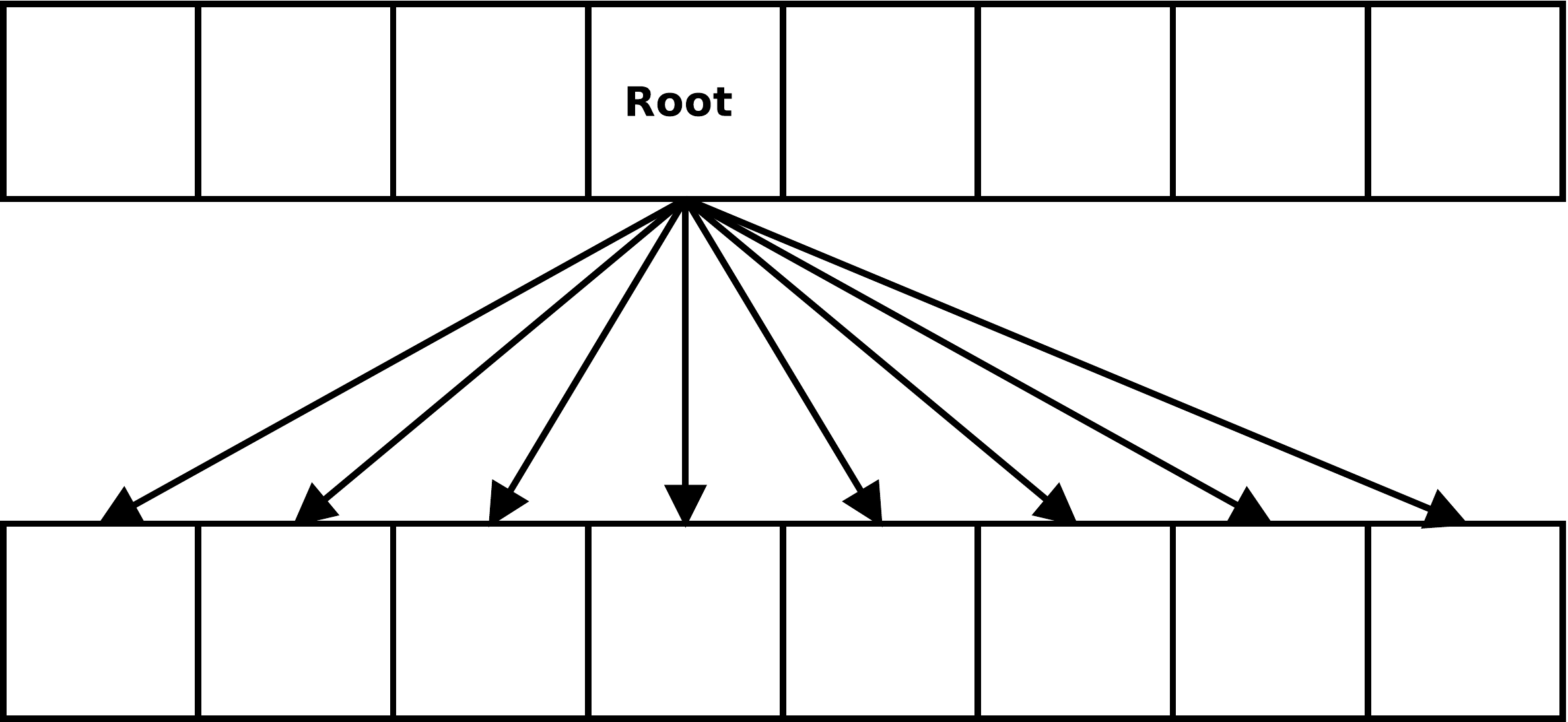}
	}
\caption{Bcast Operation}
\end{center}
\end{figure}

The PEMS2 implementation of Bcast uses rooted synchronisation on the real
processor that contains the root, and initial synchronisation on all other
real processors (see \S\ref{synchronisation}).

All threads on the same real processor as the root wait for the root to
write the message to the shared buffer.  They then copy the message from the shared
buffer to their receive buffer.

The message is delivered remotely using a single {\tt MPI\_Bcast}: the root
sends, and the first virtual processor to run on other real processors
receives into the shared buffer.  This receiving virtual processor then
signals, and other virtual processors copy from the shared buffer to their
receive buffers.

\clearpage
\subsection{Algorithm}

\begin{algorithm}
\KwData{$\mathcal{S}$ : Send buffer of size $\omega$ (valid only at root)}
\KwData{$\mathcal{R}$ : Receive buffer of size $\omega$}
\BlankLine
Let $\mathcal{B}$ be the initial portion of shared buffer of size $\omega$
\BlankLine
\tcc{Broadcast}
	\If{this is the root}{
		Copy $\mathcal{S}$ to $\mathcal{B}$\;
		{\sc EM-Signal-Threads}()\tcp*[h]{Signal that data is ready}\;
		\If{$P > 1$}{
			MPI\_Bcast from $\mathcal{S}$\tcp*[h]{Send to other real processors}\;
		}
	}\Else(\tcp*[h]{This is not the root}){
		\If{the root is on this real processor}{
			{\sc EM-Wait-For-Root}()\;
			Copy from shared buffer to $\mathcal{R}$\;
		}\Else(root is on another real processor){
			\If{$P > 1$ and {\sc EM-First-Thread()}}{
				MPI\_Bcast to $\mathcal{B}$ (receive from root)\;
				{\sc EM-Signal-Threads()}\;
			}
			{\sc EM-Wait-Threads()}\;
			Copy from $\mathcal{B}$ to $\mathcal{R}$
		}
	}
\BlankLine\tcc{Finished Virtual Superstep}\BlankLine
\caption{{\sc EM-Bcast}}
\label{bcast-alg}
\end{algorithm}

\clearpage
\subsection{Analysis}

\begin{lemma}
\label{em-bcast-io}
{\sc EM-Bcast} takes $S\frac{2v\mu}{PkB} + G\frac{v\omega}{PDB}$ time to
perform I/O (not including virtual superstep overhead).
\end{lemma}
\begin{proof}
{\sc EM-Wait-For-Root} takes $S\frac{2v\mu}{PkB}$ time
(Lem.~\ref{wait-root-io}), and each virtual processor delivers the buffer (of
size $\omega$) to its context.
\end{proof}

$ $ \\

\begin{lemma}
\label{em-bcast-comm}
{\sc EM-Bcast} performs a single network $\omega$-relation, where $\omega$
is the size of the buffer to broadcast.
\end{lemma}
\begin{proof}
{\tt MPI\_Bcast} is called exactly once with a buffer of size $\omega$.
\end{proof}

$ $ \\

\begin{thm}
\label{em-bcast-time}
{\sc EM-Bcast} takes $S\frac{2v\mu}{PkB} + G\frac{v\omega}{PDB} + g\frac{\omega}{b} + l + L$
time where $\omega$ is the size of the buffer to broadcast, assuming $v\omega > B$
and $\omega > b$.
\end{thm}
\begin{proof}
Follows directly from Lemma~\ref{em-bcast-comm} and Lemma~\ref{em-bcast-io}, since
extra swapping only occurs for a single virtual processor and message delivery
occurs for all virtual processors in parallel.
\end{proof}

\clearpage
\section{Gather}

In a {\sc Gather}, each virtual processor sends a message to the root virtual
processor.

\begin{figure}[ht]
\begin{center}
	\resizebox{0.6\textwidth}{!}{
	\includegraphics{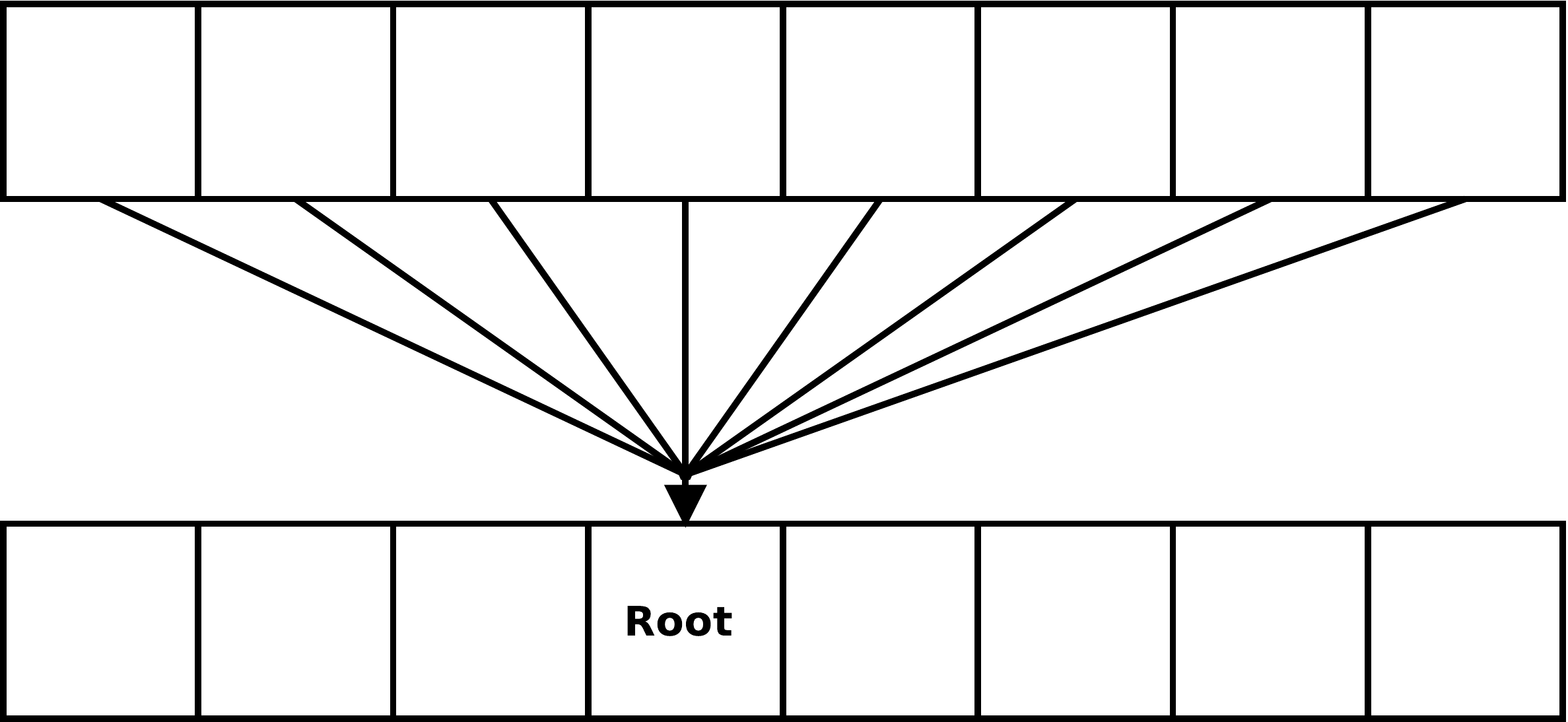}
	}
\caption{Gather Operation}
\end{center}
\end{figure}

The PEMS2 implementation of Gather uses final synchronisation (see
\S\ref{synchronisation}).  In both the single and multiple processor cases,
the gathered messages are assembled in the shared buffer before finally
being collected by the root virtual processor.

In the single processor case, the virtual processors simply copy to the
appropriate location in the shared buffer, then signal.  When all threads
have signalled, the root copies the result from the shared buffer to its
receive buffer and the operation is complete.

In the multiple processor case, each virtual processor participates in an
{\sc MPI\_Gather} to send data to the real processor which hosts the root.
When these communication rounds are completed, all gathered data resides
in the shared buffer at the real processor which hosts the root.  The root
virtual processor then copies this data to its receive buffer and the
operation is complete.

\subsection{Algorithm}

\begin{algorithm}
\KwData{$\mathcal{S}$ : Send buffer of size $\omega$}
\KwData{$\mathcal{R}$ : Receive buffer of size $v\omega$ (valid only at root)}
\BlankLine
Let $\mathcal{B}$ be the initial portion of shared buffer of size $v\omega$\;
Let $y$ = false (yielded, swapped out)\;
\If{$P > 1$}{
	\If{the root is on this real processor}{
		{\sc MPI\_Gather} $P$ ranks of current senders (receive)\;
		{\sc MPI\_Gather}($\mathcal{S}$, $\mathcal{B}$) (receive)\;
		\If{this is the root}{
			\If{not {\sc EM-All-Threads-Finished}($y$)}{
				{\sc EM-Wait-Threads}($y$)\;
			}
			Copy data from $\mathcal{B}$ to $\mathcal{R}$\;
		}\Else{
			{\sc EM-Thread-Finished}()\;
		}
	}\Else(\tcp*[h]{Root is not on this real processor}){
		{\sc MPI\_Gather} $P$ ranks of current senders (send)\;
		{\sc MPI\_Gather}($\mathcal{S}$) (send)\;
	}
}\Else(\tcp*[h]{Single processor}){
	\If{this is the root}{
		\If{not {\sc EM-All-Threads-Finished}($y$)}{
			{\sc EM-Wait-Threads}($y$)\;
		}
		\If{$y$}{Swap in $\mathcal{R}$}
		Copy $\mathcal{S}$ to $\mathcal{B}$\;
		Copy $\mathcal{B}$ to $\mathcal{R}$\;
	}\Else(\tcp*[h]{This is not the root}){
		Copy $\mathcal{S}$ to $\mathcal{B}$\;
		{\sc EM-Thread-Finished()}\;
	}
}
\BlankLine\tcc{Finished Virtual Superstep}\BlankLine
\caption{{\sc EM-Gather}}
\label{gather-alg}
\end{algorithm}

\clearpage
\subsection{Analysis}

\begin{lemma}
\label{em-gather-io}
{\sc EM-Gather} takes at most $S\frac{\mu + \omega}{BD}$ time to perform I/O
(not including virtual superstep overhead), assuming $\omega > B$.
\end{lemma}
\begin{proof}
In the multi-processor case, the root may swap its context out via {\sc
EM-Wait-Threads} ($\mu$ I/O since only the root calls this function).
In this case, $\mathcal{R}$ is not swapped in at line 10, so this copy is
actually a disk write ($\omega$ I/O), for a total of $\mu + \omega$ I/O
in the worst case.  Other virtual processors perform no additional I/O.
The single processor case clearly performs less I/O in the worst case.
\end{proof}

$ $ \\

\begin{lemma}
\label{em-gather-comm}
{\sc EM-Gather} takes $g\frac{v\omega}{Pb} + l\frac{v}{P}$ communication
time, assuming $\omega > b$.
\end{lemma}
\begin{proof}
{\sc EM-Gather} performs an {\sc MPI\_Gather} for $P$ threads at once,
where each real processor sends one message of size $\omega$ to the root.
Thus, {\sc EM-Gather} performs $\frac{v}{P}$ network $\omega$-relations.
The lemma follows directly from the definitions of $l$, $g$, and $b$ in the
BSP model.
\end{proof}

$ $ \\

\begin{thm}
\label{em-gather-time}
{\sc EM-Gather} takes $S\frac{\mu + \omega}{BD} + g\frac{v\omega}{Pb} +
l\frac{v}{P} + L$ time, assuming $\omega > b$ and $\omega > B$.
\end{thm}
\begin{proof}
Follows directly from Lem.~\ref{em-gather-io} and Lem.~\ref{em-gather-comm}.
\end{proof}

\clearpage
\section{Reduce}

A Reduce operation applies an associative and commutative\footnote{MPI
allows the user to define non-commutative operators, but PEMS currently
requires operators to be commutative} operator to $v$ values (one from
each virtual processor), placing the single result in a buffer on some
root virtual processor.  This operation is vectorized across $n$ values,
i.e.\ a single Reduce performs $n$ reductions of size $v$ resulting in $n$
values at the root.  (This definition corresponds to {\tt MPI\_Reduce},
which is not implemented by PEMS1).

A Reduce can be performed with less I/O and communication than an Alltoall,
since several values may be reduced to a single value before delivery.
On each real processor, for each of the $n$ reductions, {\sc EM-Reduce}
performs $k$ operations at a time in parallel into the shared buffer.  $\sigma$
must therefore be large enough to hold $kn$ values.  When all local threads
have finished, the final thread reduces these $k$ values to a single value,
and communicates that value to the root.  This requires 2 internal supersteps,
but only a single swap and a single network communication (if necessary).

\begin{figure}[h]
\begin{center}
	\resizebox{0.6\textwidth}{!}{
	\includegraphics{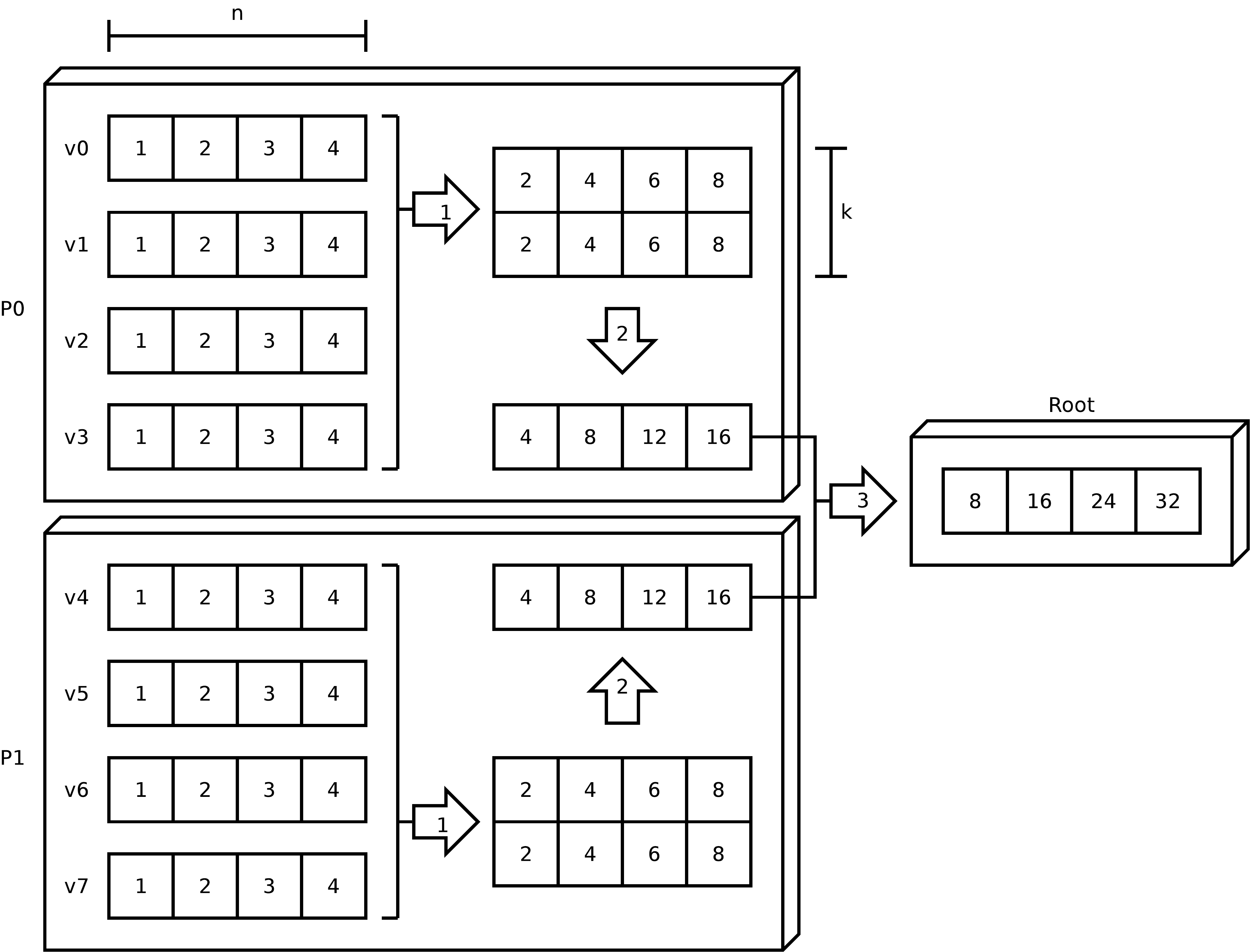}
	}
\caption{{\sc EM-Reduce} ($P=2$, $v=8$, $k=2$, and $n=4$)}
\label{em-reduce-diag}
\end{center}
\end{figure}

\clearpage
\subsection{Algorithm}

\begin{algorithm}[h]
\KwData{$\mathcal{S}$ : Array of $n$ values to send}
\KwData{$\mathcal{R}$ (root only) : Array of $n$ values for result}
\KwData{$r$ (root) : ID of root virtual processor}
\BlankLine
Let $\mathcal{B}$ be one of $k$ shared buffer portions of size $n$\;
\BlankLine
\tcc{Partially reduce local data}
	Swap out\;
	Reduce $S$ into $\mathcal{B}$\;
\BlankLine\tcc{Finished Internal Superstep 1}
\tcc{Begin Internal Superstep 2}\BlankLine
\tcc{Merge partial reductions}
	\If{$\rho = r$ or $\rho$ is thread 0 on a different real processor than $r$}{
		Reduce $kn$ values in shared buffer to $n$ local results\;
		\If{$P>1$}{
			{\tt MPI\_Reduce} $n$ local results to $r$'s real processor\;
		}
	}\If{$\rho = r$}{
		\If{$P>1$}{
			{\tt MPI\_Reduce} $Pn$ values from the network into $\mathcal{R}$\;
		}
		Swap $R$ out to partition on disk\;
	}
\BlankLine\tcc{Finished Virtual Superstep}\BlankLine
\caption{{\sc EM-Reduce}}
\label{em-reduce}
\end{algorithm}

\subsection{Analysis}

\begin{lemma}
\label{em-reduce-cpu}
{\sc EM-Reduce} takes $\frac{nv}{Pk} + nk$ computation time to
reduce all real processor's local values to a local result.
\end{lemma}
\begin{proof}
All operations are performed on vectors of size $n$.  For each of these
$n$ elements: Each real processor first reduces $\frac{v}{P}$ values on
$k$ cores in parallel, resulting in $k$ intermediate values (Step 1 in
Fig.~\ref{em-reduce-diag}), which takes $\frac{v}{Pk}$ time.  Then, these
$k$ intermediate local results are combined by application of the reduction
operator (Step 2 in Fig.~\ref{em-reduce-diag}), which takes $k$ time.
Thus, it takes $\frac{v}{Pk} + k$ time to reduce a vector of size $1$,
or $\frac{nv}{Pk} + nk$ time to reduce a vector of size $n$.
\end{proof}

\begin{lemma}
\label{em-reduce-io}
{\sc EM-Reduce} takes $G\frac{n\omega}{B}$ time to perform I/O
(not including virtual superstep overhead).
\end{lemma}
\begin{proof}
The root processor delivers the final result of size $n\omega$ to its context
on disk.  Precisely one swap occurs per virtual processor, which is accounted
for by by $L$ if necessary.
\end{proof}

When $P>1$, {\sc EM-Reduce} performs a single network {\tt MPI\_Reduce}
operation.  The precise communication and computation time may vary
between MPI implementations, but we can find reasonable bounds by assuming
the MPI implementation is at least as good as the ``obvious'' algorithm, as
described in Lem.~\ref{mpi-reduce-time}.

\begin{figure}[ht]
\begin{center}
	\resizebox{0.6\textwidth}{!}{
	\includegraphics{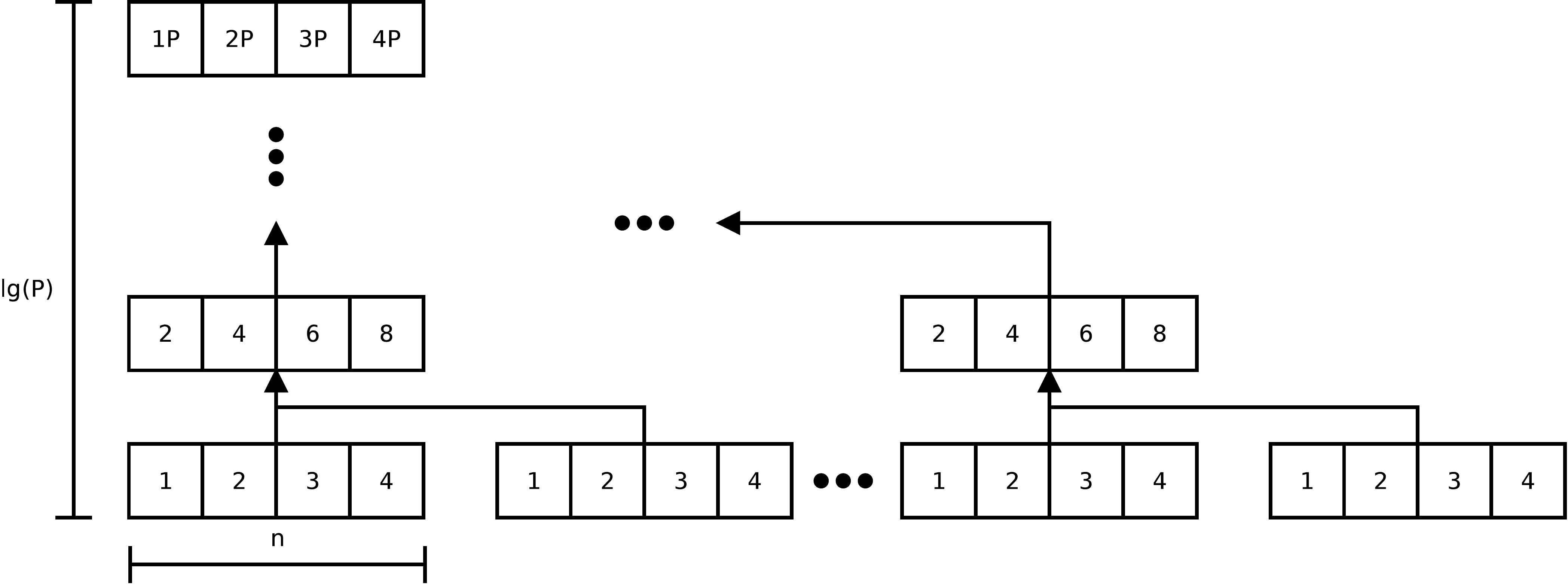}
	}
\caption{Logarithmic {\sc MPI\_Reduce}}
\label{mpi-reduce}
\end{center}
\end{figure}

\begin{lemma}
\label{mpi-reduce-time}
A reasonable {\tt MPI\_Reduce} implementation on a switched network reduces
$nP$ values across $P$ processors (Step 3 in Fig.~\ref{em-reduce-diag}) in
$n\lg(P) + g\frac{n\omega\lg(P)}{b} + l\lg(P)$ time, assuming $n\omega\lg(P)
\ge b$.
\end{lemma}
\begin{proof}
{\tt MPI\_Reduce} can be implemented as a parallel tree reduction to achieve
logarithmic time, as shown in Fig.~\ref{mpi-reduce}.  The result is computed
as $\lg(P)$ parallel partial reductions.  Each partial reduction combines
two vectors of length $n$ ($n\lg(P)$ total time), and is the result of
sending a single vector of $n$ values each of size $\omega$ over the network
($g\lg(P){\frac{n\omega}{b}}$ total time).
\end{proof}

\begin{thm}
\label{em-reduce-time}
{\sc EM-Reduce} takes $G\frac{n\omega}{B} + g\frac{n\omega\lg(P)}{b} +
l\lg(P) + n\lg(P) + \frac{nv}{Pk} + nk + L$ time.
\end{thm}
\begin{proof}
Follows directly from Lem.~\ref{em-reduce-cpu}, Lem.~\ref{em-reduce-io},
and Lem.~\ref{mpi-reduce-time}.
\end{proof}

\clearpage
\section{Summary}

\begin{figure}[h]
\begin{center}
\begin{tabular}[ht]{cccc}
Operation & Buffer Space \\ \hline
Bcast
	& $\omega$ \\
Gather
	& $v\omega$ \\
Reduce
	& $kn$ \\
Alltoallv-Seq
	& $\frac{2v^2B}{P}$ \\
Alltoallv-Par
	& $\frac{2v^2B}{P} + \alpha{k}\omega$ \\
\end{tabular}
\caption{Communication Algorithm Buffer Space}
\end{center}
\end{figure}

\begin{figure}[h]
\begin{center}
\begin{tabular}[ht]{cccc}
Operation & Time \\ \hline
Bcast
	& $S\frac{2v\mu}{PkB} + G\frac{v\omega}{PDB} + g\frac{\omega}{b} + l + L$ \\
Gather
	& $S\frac{\mu + \omega}{BD} + g\frac{v\omega}{Pb} + l\frac{v}{P} + L$ \\
Reduce
	& $G\frac{n\omega}{B} + g\frac{n\omega\lg(P)}{b} + l\lg(P) + n\lg(P) + \frac{nv}{Pk} + nk + L$ \\
Alltoallv-Seq
	& $S\frac{v\mu}{BD} + G\frac{v^2 - vk}{2BD}\omega + G\frac{2v^2}{D}$ + L \\
Alltoallv-Par
	& $S\frac{v\mu}{PDB}
	+ G\left( \frac{v^2}{P}
	        + \frac{3v^2}{2P^2}
	        - \frac{kv}{2P}
			- v^2
	  \right) \frac{\omega}{PDB}
	+ G2v^2B
	+ g\frac{\alpha{k}\omega}{b} + l\frac{v^2}{Pk\alpha} + L$ \\
\end{tabular}
\caption{Communication Algorithm Run Time}
\end{center}
\end{figure}

\clearpage

\chapter{Experiments}
\thispagestyle{empty}
\label{experiments}

\section{Experimental Setup}

All experiments in this chapter were performed on the HPCVL cluster described
in detail in Appendix~\ref{setup}.

\section{Plot Style}

Labels for plot lines show the program name, PEMS version, I/O style,
and number of processors.  For example, ``PSRS PEMS2 (mmap) P=2'' refers
to the PSRS algorithm running on PEMS2 with mmap I/O on 2 real processors.
The three types of I/O style referred to are shown in Fig.~\ref{io-styles}.
\begin{figure}[ht]
\begin{center}
\begin{tabular}{l|l}
Label & I/O Style \\ \hline
unix & Synchronous UNIX I/O \\
stxxl-file & Asynchronous STXXL File I/O (\S\ref{async}) \\
mmap & Memory Mapped I/O (\S\ref{mmap}) \\
\end{tabular}
\label{io-styles}
\caption{PEMS2 I/O Styles}
\end{center}
\end{figure}

The label ``stxxl'' refers to STXXL's included sorting algorithm, run on
a single processor (the program does not support distributed processors).
This data is included on all plots to provide a consistent baseline for
comparison of other results.

Variables shown below the plot (e.g.\ $\mu$ or $\frac{n}{v}$) apply to all
runs shown in that plot, with the exception of the ``stxxl'' data.

\clearpage
\section{Sorting}
\label{psrs-sec}

The well-known {\em Parallel Sorting by Regular Sampling} \cite{psrs}
(PSRS) algorithm is a good candidate for use with PEMS and explicit I/O
due to its coarse granularity and small, constant number of supersteps.
Alg.~\ref{psrs-alg} shows a simple high-level description of this algorithm,
with calls to collective communication functions (i.e.\ calls that result in
I/O via PEMS) shown in bold.

\subsection{Algorithm}

\begin{algorithm}[ht]
\KwData{$\mathcal{D}$ (data) : Array of size $\frac{n}{v}$}
\BlankLine
Sort $\mathcal{D}$\;
Choose $v$ equally spaced splitters in $\mathcal{D}$\;
{\bf Gather} all $v^2$ splitters at root\;
Sort all $v^2$ splitters at the root\;
{\bf Bcast} splitters evenly to all processors (each receives $v$ splitters)\;
Locate splitters in (sorted) $\mathcal{D}$\;
Compute the number of elements in $\mathcal{D}$ in each bucket\;
{\bf Alltoall} bucket sizes (each sends/receives $v$ sizes)\;
{\bf Alltoallv} buckets to final destination processor\;
Merge received buckets\;
\caption{\sc PSRS}
\label{psrs-alg}
\end{algorithm}

\subsection{Analysis}

Let $\pi$ be the size of an integer used for counts.\\
Let $\epsilon$ be the size of an individual data element.

Alg.~\ref{psrs-alg} consists of four supersteps, the first three of which
communicate only counts of a fixed size.

The remaining call, Alltoallv, does all the work of distributing data among
processors.  The message sizes in this step therefore depends on how balanced
the global partitioning is.  The partitioning scheme used in PSRS guarantees
the balancing is within a factor of 2 of an ideal partitioning \cite{psrs}
We can assume, then, that the worst case virtual message size for this final
step is $\omega \le \frac{2n\epsilon}{v^2}$.

\begin{center}
\begin{tabular}[ht]{ccc}
Operation & Size & Time \\ \hline
Gather    & $v\pi$ &
	$S\frac{\mu + 2v\pi}{BD} + g\frac{kv\pi}{b} + l\frac{v}{Pk} + L$ \\
Bcast     & $v\pi$ &
	$S\frac{2v\mu}{PkB} + G\frac{v^2\pi}{PDB} + g\frac{v\pi}{b} + l + L$ \\
Alltoall & $v\pi$ &
	$ O\left(S\frac{v\mu}{PDB}
	+ G\frac{v^3\pi}{PDB}
	+ g\frac{\alpha{k}v\pi}{b} + l\frac{v^2}{Pk\alpha} + L\right)$ \\
Alltoallv  & $\frac{2n\epsilon}{v^2}$ &
	$O\left(S\frac{v\mu}{PDB}
	+ G\frac{2n\epsilon\omega}{PDB}
	+ g\frac{\alpha{k}2n\epsilon}{v^2b} + l\frac{v^2}{Pk\alpha} + L\right)$ \\
\end{tabular}
\end{center}

\subsection{Performance}
\label{psrs-results}

\subsubsection{PEMS1 vs. PEMS2 Time (Scaling $v$)}
\label{vs_small}

To allow direct comparison with previous experimental results for
PEMS1 \cite{mnthesis}, these experiments use a small virtual processor context
size, $\mu = 64$ MiB, with an additional $64$ MiB for the shared buffer.
Because direct I/O is used the operating system does not use extra RAM
for caching, i.e.\ performance is as if the system had only this amount of
RAM and is unaffected by additional memory which goes unused\footnote{This
has been verified by monitoring resource consumption and running identical
experiments on machines with reduced RAM}.

The context size, $\mu$, remains constant for all runs, while the problem
size is increased via $v$.  This is the ideal way to scale PEMS: choose
the memory parameters suitably for the available hardware, then scale $v$
up to reach the desired problem size.

The PEMS1 and PEMS2 programs are identical, and experiments were run on the
same machines with identical configuration and PEMS parameters.

As the figures in this section show, PEMS2 is significantly faster than
PEMS1, particularly with several real processors.  When run with 8 processors,
PEMS1 is still not competitive with STXXL, taking over twice as long.  PEMS2,
however, is faster than STXXL with 8 processors, and very close in speed
with 4.  PEMS2 also scales better than PEMS1, with a slope nearly identical
to that of STXXL.  In contrast, the performance gap between PEMS1 and STXXL
gets larger as the problem size increases.

\input{improvement-tex}

\clearpage
\subsubsection{PEMS1 vs. PEMS2 Speedup}
\label{vs_speedup}

Fig.~\ref{pems1-pems2-speedup-plot} shows the relative speedup of the
experiments in the previous section where $n = 4$ billion.  The speedup
shown is relative to the sequential execution of the same system, i.e.\ PEMS1
speedup is relative to PEMS1 with $P=1$ and PEMS2 speedup is relative to
PEMS2 with $P=1$.

This figure illustrates that PEMS2 performance improves as real processors
are added significantly more than that of PEMS1.

\input{pems1-pems2-speedup-tex}

\subsubsection{PEMS1 vs. PEMS2 Time (Scaling $\mu$)}
\label{vs_mu}

The results in \S\ref{vs_small} confirm the hypothesis that always delivering
directly to virtual processor contexts on disk results in improved performance
compared to delivering indirectly via a separate area on disk.  However,
the parameters used here are not realistic -- modern machines have {\em far}
more RAM than 128 MiB.  While these smaller runs can be extrapolated to
estimate the performance of runs using more RAM (as mentioned in previous
work \cite{mnthesis}\cite{experimentswith}), when comparing PEMS1 and PEMS2,
the context size has an impact on performance due to the differing disk
layouts and delivery strategies.

As described in \S\ref{pems1_delivery}, a significant motivation factor
for the new direct delivery strategy was the reduction of disk seeking.
Because virtual processor contexts and the message area reside on separate
areas of disk in PEMS1, during a simulation the disk must seek back and forth
between these (possibly very distant) areas in order to perform swapping and
delivery.  This effect should increase as $\mu$ increases, since increasing
$\mu$ increases the distance from a context to the indirect area, as well
as the distance between each region of the indirect area itself.

Fig.~\ref{mu-plot} shows the results of an experiment with the PSRS algorithm
that confirms this observation.  In this experiment, the context size ($\mu$)
increases, while $v$ remains constant.  Thus there is an equal number of
virtual processors for every run, but each virtual processor handles a larger
number of elements.  The results clearly show that PEMS2 scales significantly
better than PEMS1 with respect to increasing $\mu$.  This is an important
observation, since experiments using small contexts do not illustrate this
aspect of performance.  Because modern machines have several GiB of memory,
the experiments in the previous section (and similar experiments in the PEMS1
literature) do not realistically reflect the performance of PEMS when used
for practical purposes, where it is desirable to use as much RAM as possible
to maximise performance.

\input{improvement-mu-tex}

\clearpage
\subsubsection{PEMS2 Large Runs}

The dramatically different slopes in Fig.~\ref{mu-plot} suggest that PEMS2
should be more suited to exploiting the full capabilities of modern machines.
In order to investigate this, the experiments in this section use a more
realistic context size, $\mu = 1$ GiB.  Each machine has 4 cores ($k =
4$), thus $4$ GiB RAM per machine is used for virtual processor
contexts.  An additional GiB is used for the shared buffer, for a total of
$5$ GiB\footnote{Of course, a small amount of additional memory is used for
internal control structures, thread stacks, the operating system, etc.}

These runs illustrate performance with much larger problem sizes: up to
roughly 32 billion ($32\cdot10^9$) 32-bit integers; or about 119 GiB of data.
Because the PSRS algorithm requires twice the space in order to sort, as well
as additional space for counts, this amounts to well over 200 GiB of data;
significantly larger than the total amount of physical memory available on
the machines used.

The improved performance of PEMS2 with larger context sizes can be seen
by comparing performance to STXXL with the small context results in
\S\ref{vs_small} (the STXXL data is identical).  In those experiments
with small contexts, PEMS2 only surpasses STXXL performance at $P=8$
(Fig.~\ref{v-p8}).  In the experiments here with larger contexts, PEMS2
surpasses STXXL performance at $P=4$ (Fig.~\ref{psrs-p4}), and is very close
when $P=2$ (Fig.~\ref{psrs-p2}), a significant improvement.

Direct comparison of individual runs with similar problem sizes illustrates
the improvement clearly.  For example, with $P=8$, PEMS2 with small
contexts takes 1441 seconds to sort 4 billion elements (Fig.~\ref{v-p8}).
PEMS2 with large contexts takes only 704 seconds to sort 4 billion elements
(Fig.~\ref{psrs-p8}), more than twice as fast as the comparable run with
small contexts.  PEMS1 with small contexts takes 3925 seconds to sort 4
billion elements (Fig.~\ref{v-p8}), making PEMS2 with large contexts more
than 5 times as fast.  Considering the fact that PEMS2 scales better than
PEMS1 with respect to context size (Fig.~\ref{mu-plot}), it is clear that
PEMS2 is a significant improvement over PEMS1 for practical applications
which utilize the resources available on modern hardware.

Performance is best, and most predictable, when using UNIX I/O.  Memory-mapped
I/O performs significantly worse, though this is not surprising since PSRS
is not a good choice of algorithm for memory mapping: most memory is used in
all steps, so caching provides little benefit but a large amount of overhead.
More surprisingly, asynchronous STXXL I/O does not outperform the synchronous
UNIX I/O with the exception of a few runs.  If the $n=32$ billion data point
is ignored, the asynchronous performance for $P=8$ looks promising; it may
be the case that further optimisation of the implementation to increase I/O,
computation, and communication overlap will allow asynchronous I/O to show
a consistent improvement over synchronous I/O.

\input{psrs-plot-p1-tex}

\input{psrs-plot-p2-tex}

\input{psrs-plot-p4-tex}

\input{psrs-plot-p8-tex}

\clearpage
\subsubsection{Internal Benchmarks}

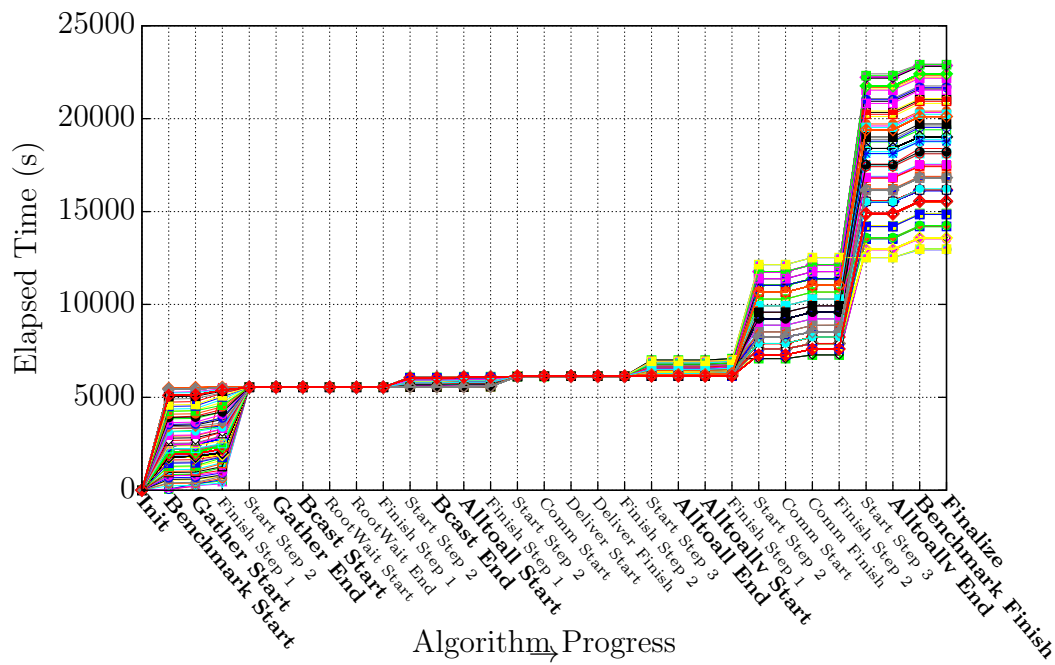
\begin{figure}[h]
\begin{center}
	\input{final_psrs-stxxl-file-internal-v-tex}
\\ \footnotesize (Each line represents elapsed time for a single virtual processor)
\end{center}
\caption{PSRS PEMS2 (stxxl-file) Elapsed Time Per Thread}
\label{psrs-internal-stxxl-file}
\end{figure}

PEMS includes a more fine-grained benchmarking system which plots the time
of a program's execution at each superstep barrier.  This can be used to
precisely investigate the run time of a program and see which sections of code
spend the most time.  This data is shown in Fig.~\ref{psrs-internal-unix},
Fig.~\ref{psrs-internal-mmap}, and Fig.~\ref{psrs-internal-stxxl-file} for
a PSRS run using unix, mmap, and stxxl-file I/O, respectively.  These plots
clearly illustrate where time is consumed among the 4 communication calls
in the PSRS algorithm.

Each plot shows a single PSRS execution on a single real processor (runs are
using 2 real processors and the same parameters as in the previous section,
but each plot shows only a single real processor).  Each line represents the
elapsed time of a single thread.  Because $\frac{v}{P}$ is relatively high
for these runs, there are many lines on these plots and distinguishing each
individually is difficult.  However, it is the overall trend and distribution
that is interesting in these plots, not the time taken by any particular
thread.

These plots illustrate the fundamental performance difference between explicit
and memory mapped IO.  The UNIX and STXXL plots have similar structures -
time increases in jumps at each superstep, roughly corresponding to the amount
of I/O performed in that superstep.  MMap, however, differs significantly:
elapsed time is nearly flat until the final Alltoallv.  This illustrates the
effect of caching and benefits of memory mapping - the first 3 steps deal
with only splitter data, thus a small amount of data is accessed each step.
This allows the cache to work effectively, keeping the splitter data in
memory and avoiding I/O.  The advantage of memory-mapped I/O in PEMS is
clearly visible: these steps take almost no time at all, because no swapping
I/O is performed.  The final {\tt Alltoallv}, however, moves all the data to
its final location, accessing the majority of the virtual processor's memory.
This data is large and not cached, so a large amount of I/O is performed.
This last step which accesses and moves the majority of memory on every
virtual processor causes PSRS to not see much overall performance advantage
with memory-mapped I/O.  Other algorithms which communicate in smaller chunks
would see a significant improvement in overall runtime, as the first 3 steps
in Fig.~\ref{psrs-internal-mmap} illustrate.  Thus, memory mapping expands
the scope of PEMS: with explicit I/O as in PEMS1, only algorithms like PSRS
with a small number of very coarse supersteps are appropriate.  PEMS2 with
memory mapped I/O, however, can see good performance with algorithms that
use a large number of supersteps and finer grained communication, since a
superstep barrier no longer forces a complete swap.

\begin{figure}[h]
\begin{center}
	\input{final_psrs-unix-internal-v-tex}
\\ \footnotesize (Each line represents elapsed time for a single virtual processor)
\end{center}
\caption{PSRS PEMS2 (unix) Elapsed Time Per Thread}
\label{psrs-internal-unix}
\end{figure}

\begin{figure}[h]
\begin{center}
	\input{final_psrs-mmap-internal-v-tex}
\\ \footnotesize (Each line represents elapsed time for a single virtual processor)
\end{center}
\caption{PSRS PEMS2 (mmap) Elapsed Time Per Thread}
\label{psrs-internal-mmap}
\end{figure}

\clearpage
\section{CGMLib}
\label{cgmlib}

CGMLib/CGMGraph \cite{cgmlib} (here collectively referred to as ``CGMLib'') is
an implementation of several CGM algorithms and associated utilities.  CGMLib is a
high-level object based C++ library implemented on top of MPI, which implements several
communication methods:
\begin{description}
\item[\tt oneToAllBCast(int source, CommObjectList \&data)]
	Broadcast the list {\tt data} from processor number {\tt source}
	to all processors.
\item[\tt allToOneGather(int target, CommObjectList \&data)]
	Gather the lists {\tt data} from all processors to processor number
	{\tt target}.
\item[\tt hRelation(CommObjectList \&data, int *ns)]
	Perform an h-Relation on the lists {\tt data} using the integer
	array {\tt ns} to indicate for each processor which list objects
	are to be sent to which processor.
\item[\tt allToAllBCast(CommObjectList \&data)]
	Every processor broadcasts its list {\tt data} to all other processors.
\item[\tt arrayBalancing(CommObjectList \&data, int expectedN=-1)]
	Shift the list elements between the lists {\tt data} such that every
	processor contains the same number of elements.
\item[\tt partitionCGM(int groupId)]
	Partition the CGM into groups indicated by {\tt groupId}. All
	subsequent communication operations, such as the ones listed above,
	operate within the respective processor's group only.
\item[\tt unPartitionCGM()] Undo the previous partition operation.
\end{description}

These communication methods are implemented using MPI collective communication
methods.  All methods are supported by PEMS excluding {\tt partitionCGM}
and {\tt unPartitionCGM}, which depend on the {\tt MPI\_Comm\_split} call
which is not currently implemented by PEMS.

CGMLib also provides additional utilities, such as communication and computation
benchmarking, a system for routing data requests between processors, and commonly
useful algorithms such as sorting, prefix sum, and list ranking.

\subsection{Sort}

The sorting algorithm included in CGMLib is a simple deterministic
parallel sample sort, based on PSRS \cite{psrs} and techniques described
in \cite{cgmlibsort}.  The figures in this section show the performance of
this sort under PEMS.

Though CGMLib Sort is similar to PSRS, performance under PEMS differs
dramatically from the straightforward PSRS MPI implementation presented
in \S\ref{psrs-sec}.  Unfortunately, characteristics of CGMLib and PEMS
interact in ways that limit the problem size achievable for a given system.
In particular, the CGMLib sort allocates much more memory.  In the context
for which CGMLib was originally designed (direct execution on a cluster using
MPI) this does not significantly impact performance.  However, when explicit
I/O is used in PEMS, the amount of memory allocated has a very large impact
on performance since this dictates the amount of swapping I/O performed.
The problem is amplified by the fact that the CGMLib communication primitives
typically use several MPI communication functions, which results in a larger
number of swaps each superstep.

Though $n$ is much smaller than the PSRS results from \S\ref{psrs-sec},
because of a larger constant factor of memory consumption, the runs in this
section represent a large problem in terms of the amount of data handled
by PEMS.  The largest runs reach the limit of available disk space on our
test configuration.  Importantly, though $n$ itself is not very large from
an EM perspective, the actual amount of allocated memory used is well in
excess of the total amount of available system RAM.

\input{cgm-sort-plot-p1-tex}

\input{cgm-sort-plot-p2-tex}

\input{cgm-sort-plot-p4-tex}

\subsection{Prefix Sum}

The CGMLib Prefix Sum application finds the inclusive prefix sum of an array
distributed across all processors.

The inclusive prefix sum of an array $[a_0, a_1, \ldots, a_{n-1}]$ is the
array $[a_0, a_0 + a_1, \ldots, a_0 + a_1 + \ldots + a_{n-1}]$, i.e.\ each
element in the result is the sum of that element and all previous elements.
For example, the prefix sum of $[1, 2, 3, 4]$ is $[1, 3, 6, 10]$.

This application shows similar performance to CGMLib Sort.  This is expected,
since both algorithm perform a small constant amount of communication.

\input{cgm-psum-plot-p1-tex}

\input{cgm-psum-plot-p2-tex}

\input{cgm-psum-plot-p4-tex}

\clearpage
\subsection{Euler Tour}

The CGMLib Euler Tour application\cite{bsp-etour} finds the Euler Tour of a forest.

\begin{figure}[h]
\begin{center}
	\resizebox{0.3\textwidth}{!}{
	\includegraphics{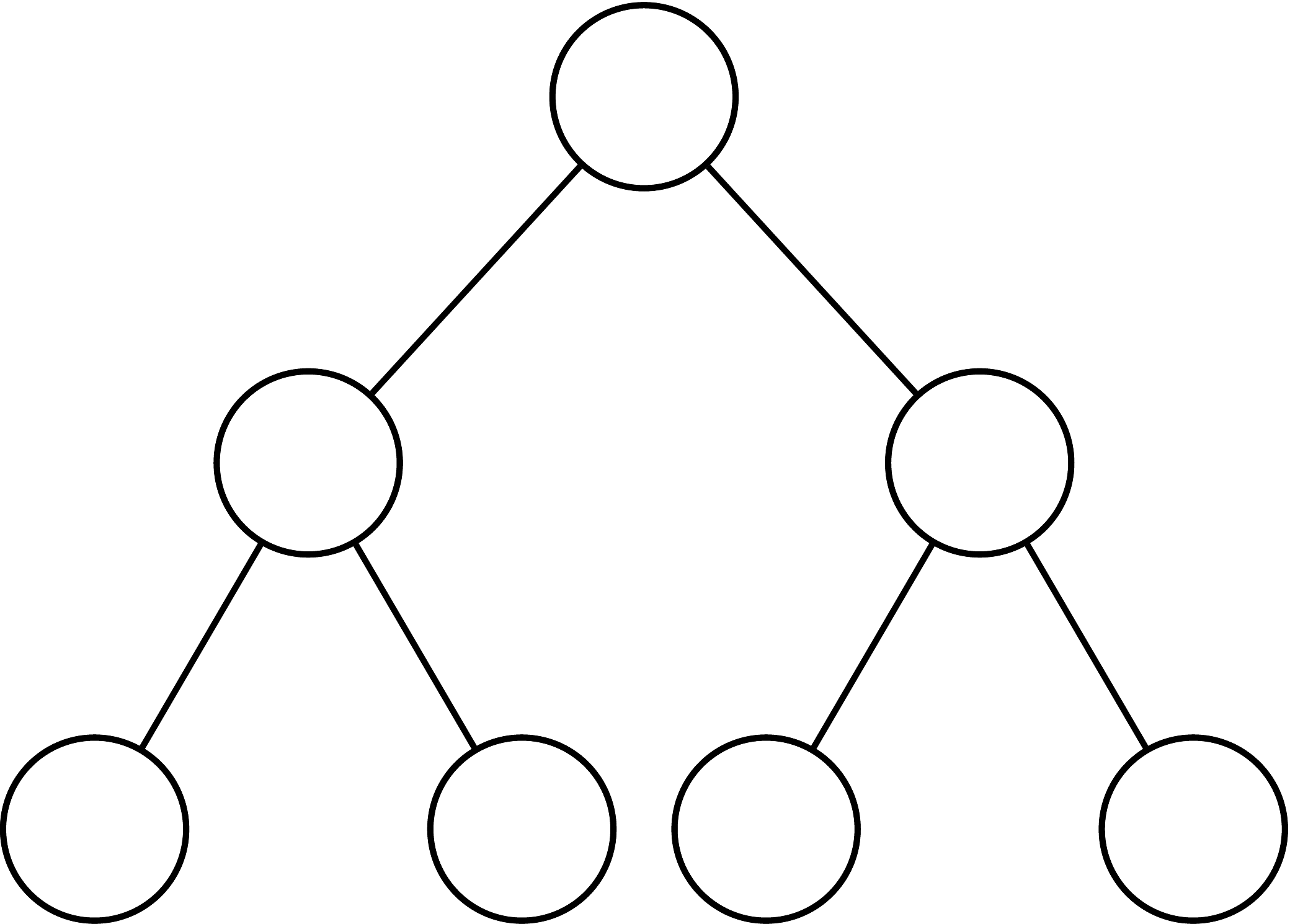}
	}
\end{center}
\caption{Euler Tour Input}
\label{euler-tree-start-fig}
\end{figure}

\begin{figure}[h]
\begin{center}
	\resizebox{0.3\textwidth}{!}{
	\includegraphics{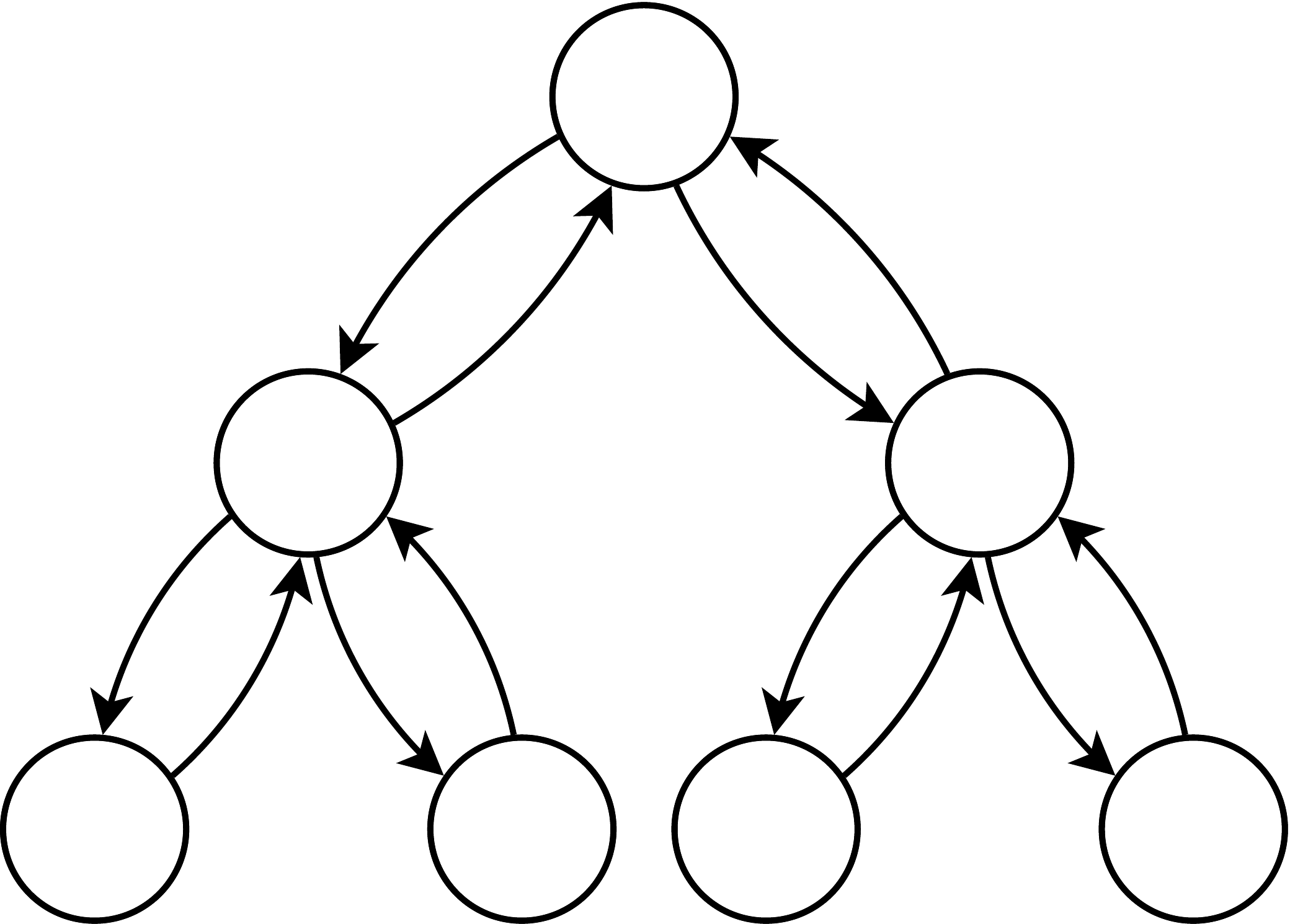}
	}
\end{center}
\caption{Euler Tour Input (Doubled Edges)}
\label{euler-tree-doubled-fig}
\end{figure}

\begin{figure}[h]
\begin{center}
	\resizebox{0.3\textwidth}{!}{
	\includegraphics{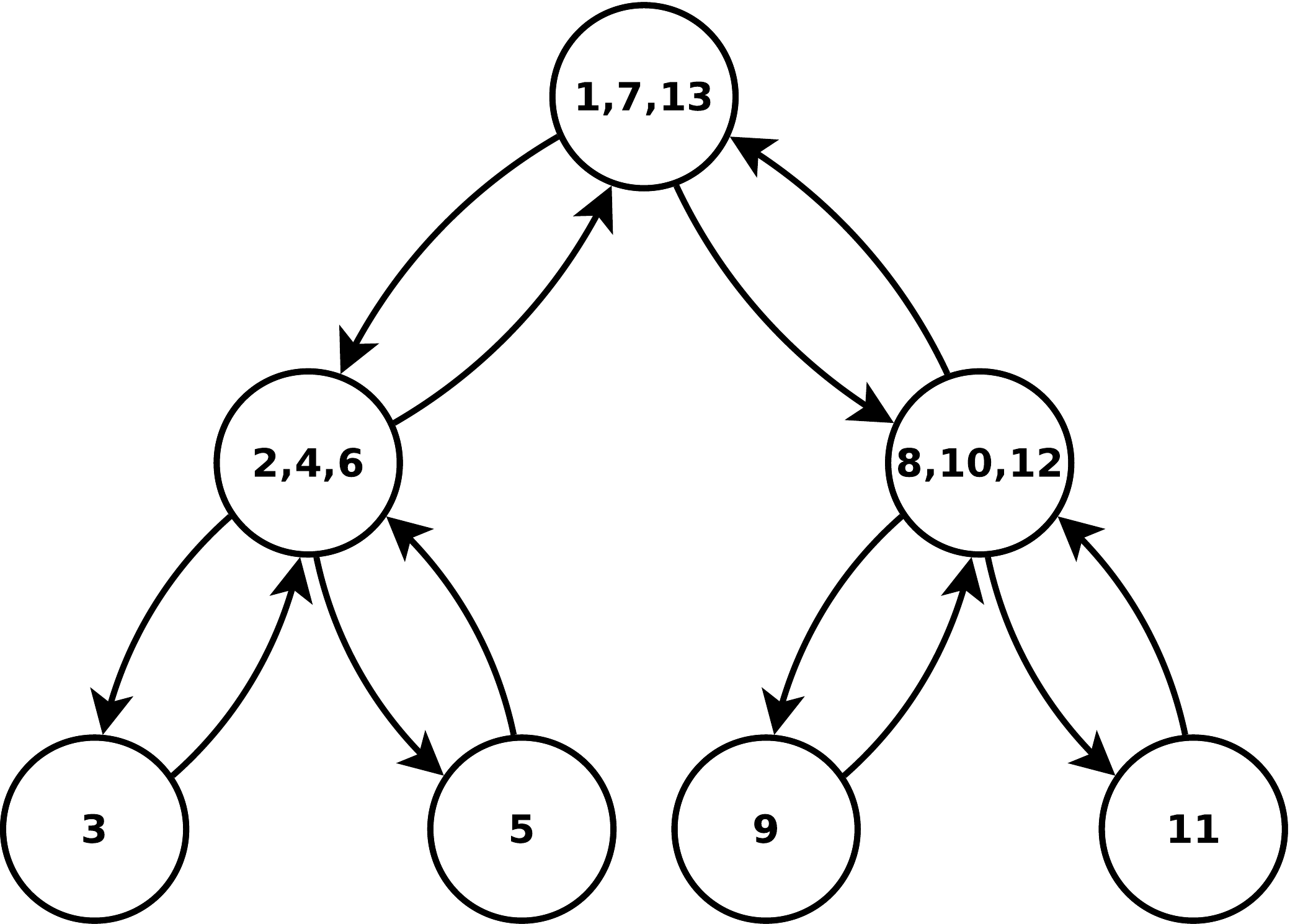}
	}
\end{center}
\caption{Euler Tour Solution}
\label{euler-tree-solved-fig}
\end{figure}

The Euler Tour of a graph is a path which traverses every edge of the
graph exactly once and returns to the starting point.  In order to apply
this problem to a tree, or a forest (a collection of trees), each edge is
doubled.  Figures \ref{euler-tree-start-fig}, \ref{euler-tree-doubled-fig},
and \ref{euler-tree-solved-fig} show example input, transormed input, and
output for this problem, respectively.  The labels on nodes in
\ref{euler-tree-solved-fig} represent the order each node is visited in the
Euler Tour.

\clearpage

This application is significantly more complex than CGMLib Sort and CGMLib
Prefix Sum, and uses several other facilities of CGMLib (including sorting
and list ranking).  Fig.~\ref{cgm-etour} shows the performance of the CGMLib
Euler Tour application with memory mapped I/O.  Here, $n$ refers to the
total number of trees in the forest, each of which contains $n^2$ nodes.

\input{cgm-etour-plot-tex}

\clearpage
\subsection{CGMLib + PEMS Conclusions}

Though the high constant factor of memory consumption prevents the CGMLib sort
from being competitive with the simpler PSRS implementation, the results do
show favourable scalability characteristics.  It is likely that improvements
to the PEMS memory allocator to reduce fragmentation, reductions in CGMLib's
memory usage, and other improvements would result in a significant reduction
in constant overhead and make PEMS+CGMLib a competitive solution.

A positive outcome of these experiments can be seen in the results where
{\tt mmap} I/O is used.  The problems described above are directly related
to the use of explicit I/O -- more allocated memory translates to more I/O.
Memory mapped I/O, however, avoids this problem (as described in detail
in \S\ref{mmap}).  CGMLib, then, provides an ideal example of where the
new memory mapped I/O capability of PEMS can be advantageous.  As can be
seen in the results, the CGMLib applications perform dramatically better
with memory mapped I/O compared to Unix and STXXL I/O.  This is because
the large amount of allocated memory is not entirely used in each superstep,
allowing the Operating System's cache to do a good job of keeping required
data in memory across supersteps, avoiding disk I/O.  This confirms that
memory mapping is an effective strategy in PEMS for improving the performance
of algorithms with certain characteristics.

\chapter{Conclusions}
\thispagestyle{empty}
\label{conclusions-ch}

This thesis presents PEMS2, an improved version of PEMS (Parallel External
Memory System).  PEMS can be used to execute BSP-like algorithms implemented as
MPI programs with massive data sets larger than main memory by utilising disk.

PEMS2 incorporates most of the future work mentioned in the literature
associated with PEMS1 \cite{mnthesis}.  In particular, PEMS2 adds multi-core
support, asynchronous I/O, and reduces disk requirements.  Beyond this, a new
message delivery strategy has been introduced, and the implementation heavily
reworked to a more flexible and easy to use form, with MPI compatibility
and a run-time configuration system which allows simple experimentation with
any algorithm.

Experiments show that PEMS2 performs significantly better than PEMS1,
particularly when using the full resources of modern hardware.

\section{Future Work}
\label{future}

Since PEMS2 is simple to use with existing MPI code, much of the interesting
work to be done based on this thesis is experimentation with various algorithms
and configurations.  It is hoped that PEMS2 will prove useful in practice
to other researchers and practitioners interested in very large problems.

However, there are several potential avenues of investigation related to
PEMS itself:
\begin{itemize}
\item Further MPI Compatibility: Unfortunately, many existing MPI programs are
not actually BSP or BSP-like algorithms.  It may be possible to implement
non-collective communication functions (e.g.\ {\tt MPI\_Send} and {\tt
MPI\_Recv}) in PEMS.  However, because these functions do not adhere to the
superstep model (i.e.\ they are not {\em collective} communication methods),
implementation in PEMS may be difficult.  If possible, though, this would
increase the number of readily available PEMS compatible programs dramatically.
While there may be negative performance implications with algorithms that
use these functions, the fact that a great deal of MPI programs use them
is undeniable.  Such algorithms, if well designed, could be practical EM
solutions if PEMS had efficient support for non-collective communication.
\item Asynchronous and Memory-Mapped I/O: Further investigation is required to
evaluate the effectiveness of these modifications.  The experiments in this
thesis do not fully investigate the full potential of these new styles of I/O.
While experiments with CGMLib have shown memory mapping to be a useful strategy,
results for asynchronous I/O have (perhaps surprisingly) not shown much advantage.
\item Dynamic $\alpha$: The communication ``chunk size'' parameter used by
Alltoallv, $\alpha$, is currently specified as a user parameter.  This could be
made dynamic, so $\alpha$ is as large as possible for each communication (e.g.\
an Alltoallv with very small messages would only perform a single network communication).
\item Fully Asynchronous Design: Lack of significant performance increases seen
when using asynchronous STXXL I/O suggest that synchronisation in PEMS limits
performance.  A fully asynchronous design where both network communication
and disk delivery are handled by a separate ``controller'' thread could
alleviate this problem.  In such a design, virtual processors would first
record their message destinations much like the current Alltoallv design
in PEMS2 (see \S\ref{delivery}).  Rather than delivering in synchronised
rounds, however, messages would be sent over the network immediately.
The controller thread on the receiving real processor would receive the
message and immediately write it to the correct location on disk.  This way,
delivery of messages does not require synchronisation between the sending
and receiving virtual processors, thus communication, computation, and I/O
overlap would be significantly increased.  Such a design would also be more
appropriate for implementing non-collective communication methods like {\tt
MPI\_Send} and {\tt MPI\_Recv}.
\item Multi-Core and Multi-Disk: Due to hardware limitations, the multiple-disk
capabilities of PEMS2 have not been tested.  It is likely that the use of
several cores and several disk will show significant performance improvements
since several virtual processors running at once can make full use of up to
$k$ disks (or more, depending on the type of I/O used).
\item Disk Scheduling: Recent versions of Linux include several disk
scheduling algorithms.  It would be interesting to investigate the impact
these have on the performance of PEMS.
\item CGMLib: The scalability shown by CGMLib+PEMS2 experiments in
\S\ref{cgmlib} is promising, but absolute performance is hindered by excessive
copying and memory consumption.  Improving these characteristics of CGMLib
would make the combination of CGMLib and PEMS2 a more practical solution,
and due to the impressive breadth of functionality available in CGMLib,
simplify the development of many advanced EM algorithms with PEMS2.
\item New Architectures: There are interesting similarities between disk-based
models such as those used in this thesis and increasingly popular special
purpose multi-core (e.g.\ the Cell BE) and General Purpose Graphics Processing
Unit (GPGPU) architectures.  Both have a fast local memory store, and a slower
external memory store.  Transferring data between these two stores is a key
factor in performance.  PEMS, with some modifications, may allow suitable
BSP algorithms to run on these new architectures with good performance.
The implementation currently contains a ``mem'' I/O driver which simply uses
allocated memory and does no I/O at all.  This driver shows good performance
and multi-core speedup (but of course can not scale beyond the limits of RAM).
This illustrates that PEMS2 is not inherently tied to disk I/O, and
adapting PEMS2's strategy to novel architectures is an interesting avenue for
future research.
\end{itemize}

\bibliographystyle{plain}
\bibliography{bibliography}

\appendix

\chapter{Availability of PEMS2}
\thispagestyle{empty}
\label{availability}

PEMS is freely available online at {\tt http://pems.sourceforge.net/}.
PEMS is Free / Open Source software licensed under the GNU General Public
License (GPL).  Essentially this means PEMS is free to use, modify, and
distribute, but all derivative works must also be released with source code
under the GPL.  The PEMS2 implementation is designed to be simple to use
and extend (e.g.\ use with applications is trivial due to MPI compatibility,
adding new I/O drivers is straightforward).  Use, experimentation, and
modification is encouraged; if licensing is an issue please contact the
authors.  Contact information is available at the PEMS website.

PEMS can be compiled and installed using the typical process for UNIX software:
{\tt ./configure; make; sudo make install}.  Run {\tt ./configure --help}
for a summary of the available compile-time options which can be passed to
{\tt ./configure}.

Using PEMS with MPI programs is as simple as using any other MPI
implementation.  No source code modifications are necessary.  There are two
ways of doing so:
\begin{enumerate}
\item PEMS uses the {\tt pkg-config} system for compiler and linker flags.
The command {\tt pkg-config --cflags pems2} will return the compiler flags
required for building against PEMS2, and {\tt pkg-config --libs pems2}
will return the linker flags required for linking against PEMS2.
\item Like many MPI implementation, PEMS ships with compiler wrapper scripts to
automatically add the necessary compiler and linker flags.  Simply replacing
uses of {\tt mpicc} and {\tt mpic++} with {\tt pemscc} and {\tt pemsc++},
respectively, will build an MPI program against pems.  Most MPI programs ship
with a Makefile where this modification can be easily made.
\end{enumerate}

\chapter{Conventions}
\thispagestyle{empty}
\label{conventions}

\section{Terminology}

\begin{description}
\item[real processor] A single physical computer which runs a single process
(of possibly many threads) and may have several \textbf{cores} which share
main memory.

\item[virtual processor] A processor in the simulated bulk-synchronous
algorithm.

\item[thread] The implementation of a virtual processor.  While there is
a $1:1$ correspondence between threads and virtual processors, the two are
not identical -- a thread performs work internal to the PEMS implementation
in addition to the work of the simulated virtual processor.

\item[context] The memory of a virtual processor, which may exist on
disk or in main memory depending on whether or not the virtual processor
is currently swapped in.

\item[memory partition] A context-sized block of real main memory into
which a context is swapped.  Unlike contexts, all memory partitions fit into
main memory at once.

\item[internal superstep] A superstep performed internally by PEMS
(e.g.\ as part of a multi-superstep communication method).

\item[virtual superstep] A superstep performed by the simulated algorithm (the
simulation of a virtual superstep may require several internal supersteps).

\item[swap] The process of writing/reading an entire context to/from disk.
The more specific terms \textbf{swap in} and \textbf{swap out} are used
where necessary.
\end{description}

\section{Notation}
\label{notation}

\begin{center}
\begin{tabular}[ht]{ll}
$n$ & = Size of the problem to be solved \\

$t$ & = a thread ID which is a {\em local} identifier for a thread
($0 \le t < \frac{v}{P}$). \\

$\rho$ & = the current virtual processor's {\em global} ID ($0 \le \rho < v$). \\

$m_{i \rightarrow j}$ & = a message sent from virtual processor $i$ to
virtual processor $j$. \\

$\llfloor{x}\rrfloor$ & = $x$ rounded down to the nearest disk block boundary. \\

$\llceil{x}\rrceil$ & = $x$ rounded up to the nearest disk block boundary. \\

$\llfloor{r}\rrceil$ & = the smallest block aligned region containing range $r$ \\

$\llceil{r}\rrfloor$ & = the largest aligned region within range $r$ \\

\end{tabular}
\end{center}

\section{Simulation Parameters}

A PEMS simulation has the following run-time parameters:

\begin{center}
\begin{tabular}[ht]{ll}
$P$      & = Number of real processors \\
$\mu$    & = Memory size of a single virtual processor \\
$D$      & = Number of disks per real processor ($D \geq 1$) \\
$v$      & = Total number of virtual processors (v $\geq P$) \\
$k$      & = Number of concurrent threads per real processor ($k \leq \frac{v}{P}$) \\
$\sigma$ & = Size of the ``shared buffer'' in main memory
\end{tabular}
\end{center}

\section{System Parameters}
\label{parameters}

For performance analysis we make use of the following variables,  which are
essentially those of the BSP* model (in lowercase) with analogous variables
(in uppercase) to represent EM performance:

\begin{center}
\begin{tabular}[ht]{ll}
$b$ & = Minimum size of a network message to achieve rated throughput (BSP*)\\
$g$ & = Time to deliver a network packet of size $b$, or 0 if $P=1$ (BSP*) \\
$l$ & = Overhead of a single network superstep (BSP*) \\
$B$ & = Size of a single disk block (EM) \\
$G$ & = Time to write/read a single block of size $B$ to/from disk (EM) \\
$L$ & = Overhead of a single virtual superstep (EM) \\
$S$ & = Time to {\em swap} a single block of size $B$ to/from disk (EM)
\end{tabular}
\end{center}

$L$ represents the constant overhead of a virtual superstep, including
any synchronisation time and swapping I/O.  $L$ may vary depending
on the I/O system in use.  Note that, due to the introduction of
memory-mapped I/O, this is in contrast to previous work related to PEMS
\cite{dhthesis}\cite{bspem}\cite{emsimulation}\cite{mnthesis}\cite{experimentswith}
where $L$ does not include any I/O.

$S$ is used to separate terms representing swap I/O from terms representing
message delivery I/O.  $S$ is identical to $G$ when using explicit I/O (i.e.\
UNIX or STXXL), and $0$ by definition when using memory mapped I/O.

$g$ represents parallel network performance given a fully connected
network, i.e.\ each processor can send messages directly to each other
processor.  When two processors communicate, network bandwidth between other
processors is not affected.  This is true for switched ethernet networks,
but not true for ethernet hubs.  Hubs are not suitable for high performance
computing/networking, but this is not a concern with modern hardware since
switches have reduced in price so much that even low-end consumer hardware
is typically switched.

\chapter{Methodology}
\thispagestyle{empty}
\label{methodology}

\section{Hardware/Software Configuration}
\label{setup}

Experiments were run on the HPCVL Beowulf cluster at Carleton University.
Each node has 2 dual-core\footnote{i.e.\ 4 cores total} AMD Opteron 2214
processors at 2.2 GHz with 1 MiB cache per core, 8 GiB RAM, and a single 200
GiB disk.  Nodes are interconnected with a high-end gigabit ethernet switch.

Linux 2.6.28 was used, with the ext4 filesystem and standard I/O scheduling.
All code was compiled with GCC 4.1.1.

\section{File Systems}
\label{filesystems}

EM algorithms, including PEMS, attempt to optimise disk access for
performance reasons.  Locality of reference and favourable access patterns
(e.g.\ linear sweeps) provide the best performance, since disk seeking
is extremely expensive.  However, in practice most applications running
on a modern operating system are not actually accessing disk directly --
disk access is provided via a file system.  This has the implication that
a linear sweep in code may not actually translate to a linear sweep on disk
due to file system fragmentation (files are generally not guaranteed to be
contiguous ranges of blocks).  This can result in unpredictable performance.

Thankfully, some file systems take this into consideration and provide
facilities for allocating large areas of disk.  The new default file system
for Linux, ext4, includes this ability (support for ``extents'').  PEMS2 makes
use of this functionality on systems modern enough to support this feature.
All experiments in this thesis explicitly allocate disk via this mechanism.

Fig.~\ref{ext3vs4} shows the potential impact a fragmented filesystem can have.
In this experiment, all parameters remain constant {\em including $n$} and
only $\mu$ is increased.  That is, more disk space is used, but the actual
problem size remains constant.  Notice ext4 (with extents) has consistent
performance regardless of the space used, but ext3 (without extents) degrades
in performance severely as more space is used.  Further experiments have shown
that, without extents, disk performance can be very unpredictable.

Note that Fig.~\ref{ext3vs4} illustrates a particular pathological case.
In particular, it is possible (with luck) to achieve nearly\footnote{Using
extents actually reduces filesystem overhead in general, since blocks are
much larger and less tree traversal is required.  Thus, using extents can
yield a performance improvement even on a completely unfragmented filesystem}
equivalent performance without explicitly allocating disk.  However, unless
the file system is entirely empty, this is extremely unlikely for large files.

Those working with external memory algorithms should be careful to choose
an appropriate filesystem, and make use of explicit disk space allocation
routines ({\tt fallocate} or {\tt posix\_fallocate} in Linux) to ensure good,
predictable performance.

\input{ext-tex}

\chapter{MPI Compatibility}
\thispagestyle{empty}
\label{MPI-funcs}

Fig.~\ref{mpi_functions} shows the subset of MPI implemented by PEMS2.
Additionally, {\tt malloc}, {\tt realloc}, and {\tt free} are wrapped by PEMS
to allocate memory in the virtual processor context rather than system RAM.

\begin{figure}[ht]
\begin{center}
{\tt
\begin{itemize}
	\item MPI\_Allgather
	\item MPI\_Allgatherv
	\item MPI\_Allreduce
	\item MPI\_Alltoall
	\item MPI\_Alltoallv
	\item MPI\_Bcast
	\item MPI\_Gather
	\item MPI\_Gatherv
	\item MPI\_Reduce
	\item MPI\_Scatter
	\item MPI\_Barrier
	\item MPI\_Wtime
	\item MPI\_Init
	\item MPI\_Finalize
	\item MPI\_Abort
	\item MPI\_Comm\_rank
	\item MPI\_Comm\_size
\end{itemize}
}
\end{center}
\caption{Supported MPI Functions}
\label{mpi_functions}
\end{figure}

\end{document}

%% file: alltoall-plot-tex.tex
\begingroup
  \makeatletter
  \providecommand\color[2][]{%
    \GenericError{(gnuplot) \space\space\space\@spaces}{%
      Package color not loaded in conjunction with
      terminal option `colourtext'%
    }{See the gnuplot documentation for explanation.%
    }{Either use 'blacktext' in gnuplot or load the package
      color.sty in LaTeX.}%
    \renewcommand\color[2][]{}%
  }%
  \providecommand\includegraphics[2][]{%
    \GenericError{(gnuplot) \space\space\space\@spaces}{%
      Package graphicx or graphics not loaded%
    }{See the gnuplot documentation for explanation.%
    }{The gnuplot epslatex terminal needs graphicx.sty or graphics.sty.}%
    \renewcommand\includegraphics[2][]{}%
  }%
  \providecommand\rotatebox[2]{#2}%
  \@ifundefined{ifGPcolor}{%
    \newif\ifGPcolor
    \GPcolortrue
  }{}%
  \@ifundefined{ifGPblacktext}{%
    \newif\ifGPblacktext
    \GPblacktexttrue
  }{}%
  \let\gplgaddtomacro\g@addto@macro
  \gdef\gplbacktext{}%
  \gdef\gplfronttext{}%
  \makeatother
  \ifGPblacktext
    \def\colorrgb#1{}%
    \def\colorgray#1{}%
  \else
    \ifGPcolor
      \def\colorrgb#1{\color[rgb]{#1}}%
      \def\colorgray#1{\color[gray]{#1}}%
      \expandafter\def\csname LTw\endcsname{\color{white}}%
      \expandafter\def\csname LTb\endcsname{\color{black}}%
      \expandafter\def\csname LTa\endcsname{\color{black}}%
      \expandafter\def\csname LT0\endcsname{\color[rgb]{1,0,0}}%
      \expandafter\def\csname LT1\endcsname{\color[rgb]{0,1,0}}%
      \expandafter\def\csname LT2\endcsname{\color[rgb]{0,0,1}}%
      \expandafter\def\csname LT3\endcsname{\color[rgb]{1,0,1}}%
      \expandafter\def\csname LT4\endcsname{\color[rgb]{0,1,1}}%
      \expandafter\def\csname LT5\endcsname{\color[rgb]{1,1,0}}%
      \expandafter\def\csname LT6\endcsname{\color[rgb]{0,0,0}}%
      \expandafter\def\csname LT7\endcsname{\color[rgb]{1,0.3,0}}%
      \expandafter\def\csname LT8\endcsname{\color[rgb]{0.5,0.5,0.5}}%
    \else
      \def\colorrgb#1{\color{black}}%
      \def\colorgray#1{\color[gray]{#1}}%
      \expandafter\def\csname LTw\endcsname{\color{white}}%
      \expandafter\def\csname LTb\endcsname{\color{black}}%
      \expandafter\def\csname LTa\endcsname{\color{black}}%
      \expandafter\def\csname LT0\endcsname{\color{black}}%
      \expandafter\def\csname LT1\endcsname{\color{black}}%
      \expandafter\def\csname LT2\endcsname{\color{black}}%
      \expandafter\def\csname LT3\endcsname{\color{black}}%
      \expandafter\def\csname LT4\endcsname{\color{black}}%
      \expandafter\def\csname LT5\endcsname{\color{black}}%
      \expandafter\def\csname LT6\endcsname{\color{black}}%
      \expandafter\def\csname LT7\endcsname{\color{black}}%
      \expandafter\def\csname LT8\endcsname{\color{black}}%
    \fi
  \fi
  \setlength{\unitlength}{0.0500bp}%
  \begin{picture}(7200.00,5040.00)%
    \gplgaddtomacro\gplbacktext{%
      \csname LTb\endcsname%
      \put(1342,1364){\makebox(0,0)[r]{\strut{} 0}}%
      \csname LTb\endcsname%
      \put(1342,1743){\makebox(0,0)[r]{\strut{} 1000}}%
      \csname LTb\endcsname%
      \put(1342,2122){\makebox(0,0)[r]{\strut{} 2000}}%
      \csname LTb\endcsname%
      \put(1342,2501){\makebox(0,0)[r]{\strut{} 3000}}%
      \csname LTb\endcsname%
      \put(1342,2880){\makebox(0,0)[r]{\strut{} 4000}}%
      \csname LTb\endcsname%
      \put(1342,3260){\makebox(0,0)[r]{\strut{} 5000}}%
      \csname LTb\endcsname%
      \put(1342,3639){\makebox(0,0)[r]{\strut{} 6000}}%
      \csname LTb\endcsname%
      \put(1342,4018){\makebox(0,0)[r]{\strut{} 7000}}%
      \csname LTb\endcsname%
      \put(1342,4397){\makebox(0,0)[r]{\strut{} 8000}}%
      \csname LTb\endcsname%
      \put(1342,4776){\makebox(0,0)[r]{\strut{} 9000}}%
      \csname LTb\endcsname%
      \put(1474,1144){\makebox(0,0){\strut{}\small$1{\cdot}10^{9}$}}%
      \csname LTb\endcsname%
      \put(2823,1144){\makebox(0,0){\strut{}\small$2{\cdot}10^{9}$}}%
      \csname LTb\endcsname%
      \put(4172,1144){\makebox(0,0){\strut{}\small$3{\cdot}10^{9}$}}%
      \csname LTb\endcsname%
      \put(5521,1144){\makebox(0,0){\strut{}\small$4{\cdot}10^{9}$}}%
      \csname LTb\endcsname%
      \put(6870,1144){\makebox(0,0){\strut{}\small$5{\cdot}10^{9}$}}%
      \put(440,3070){\rotatebox{90}{\makebox(0,0){\strut{}Seconds}}}%
      \put(4172,814){\makebox(0,0){\strut{}$n$}}%
    }%
    \gplgaddtomacro\gplfronttext{%
      \csname LTb\endcsname%
      \put(3317,393){\makebox(0,0)[r]{\strut{}alltoall-mmap-k1}}%
      \csname LTb\endcsname%
      \put(3317,173){\makebox(0,0)[r]{\strut{}alltoall-mmap-k4}}%
      \csname LTb\endcsname%
      \put(6284,393){\makebox(0,0)[r]{\strut{}alltoall-unix-k1}}%
      \csname LTb\endcsname%
      \put(6284,173){\makebox(0,0)[r]{\strut{}alltoall-unix-k4}}%
    }%
    \gplbacktext
    \put(0,0){\includegraphics{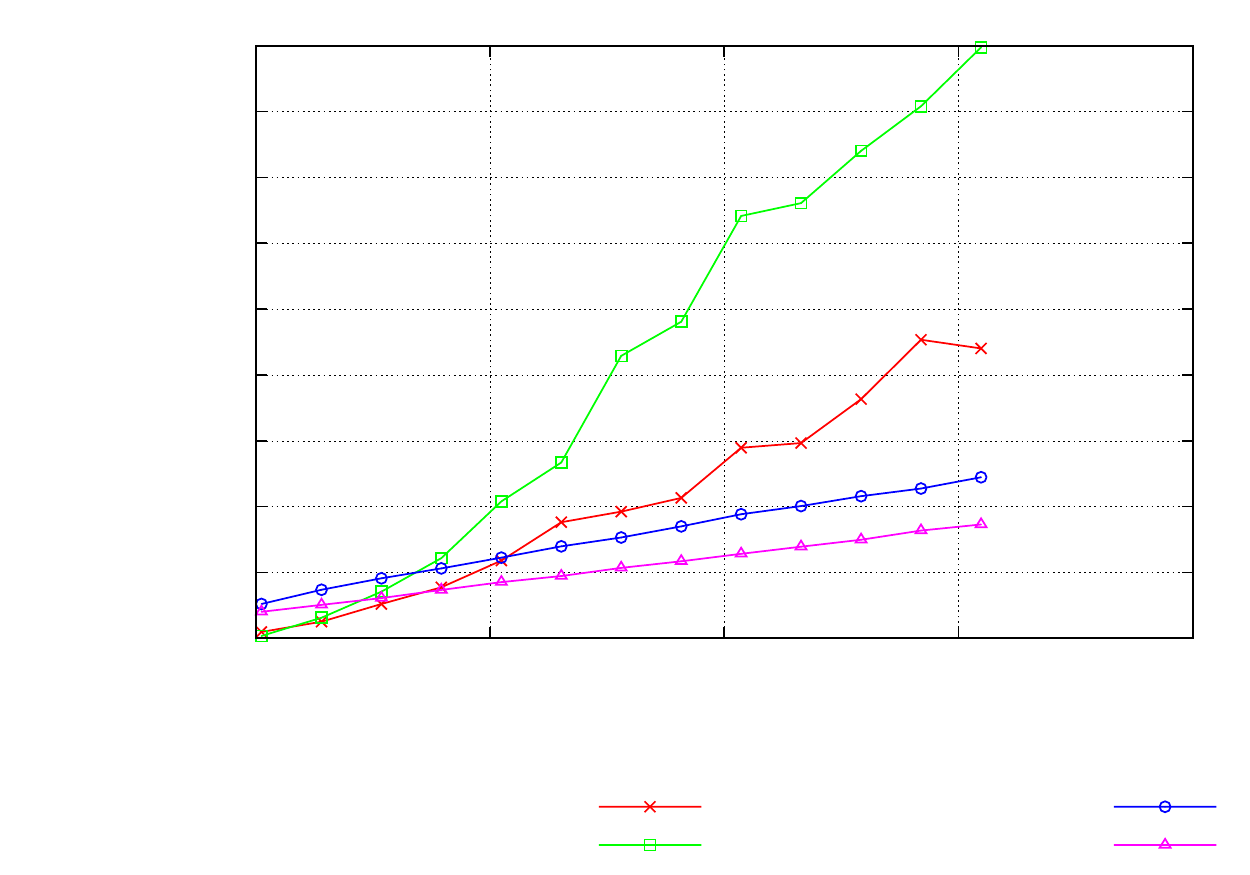}}%
    \gplfronttext
  \end{picture}%
\endgroup

%% file: improvement-tex.tex
\begin{figure}[ht]
	\begin{center}
	\resizebox{!}{0.4\textheight}{
		\input{v-p1-tex}
	}
	\\
$\frac{n}{v} = 8000000$
,  $P = 1$
,  $\mu = 64$ {\small MiB}

	\caption{PEMS1 vs. PEMS2 (PSRS, P=1)}
	\label{v-p1}
	\end{center}
\end{figure}

\begin{figure}[ht]
	\begin{center}
	\resizebox{!}{0.4\textheight}{
		\input{v-p2-tex}
	}
	\\
$\frac{n}{v} = 8000000$
,  $P = 2$
,  $\mu = 64$ {\small MiB}

	\caption{PEMS1 vs. PEMS2 (PSRS, P=2)}
	\label{v-p2}
	\end{center}
\end{figure}

\begin{figure}[ht]
	\begin{center}
	\resizebox{!}{0.4\textheight}{
		\input{v-p4-tex}
	}
	\\
$\frac{n}{v} = 8000000$
,  $P = 4$
,  $\mu = 64$ {\small MiB}

	\caption{PEMS1 vs. PEMS2 (PSRS, P=4)}
	\label{v-p4}
	\end{center}
\end{figure}

\begin{figure}[ht]
	\begin{center}
	\resizebox{!}{0.4\textheight}{
		\input{v-p8-tex}
	}
	\\
$\frac{n}{v} = 8000000$
,  $P = 8$
,  $\mu = 64$ {\small MiB}

	\caption{PEMS1 vs. PEMS2 (PSRS, P=8)}
	\label{v-p8}
	\end{center}
\end{figure}

%% file: v-p1-tex.tex
\begingroup
  \makeatletter
  \providecommand\color[2][]{%
    \GenericError{(gnuplot) \space\space\space\@spaces}{%
      Package color not loaded in conjunction with
      terminal option `colourtext'%
    }{See the gnuplot documentation for explanation.%
    }{Either use 'blacktext' in gnuplot or load the package
      color.sty in LaTeX.}%
    \renewcommand\color[2][]{}%
  }%
  \providecommand\includegraphics[2][]{%
    \GenericError{(gnuplot) \space\space\space\@spaces}{%
      Package graphicx or graphics not loaded%
    }{See the gnuplot documentation for explanation.%
    }{The gnuplot epslatex terminal needs graphicx.sty or graphics.sty.}%
    \renewcommand\includegraphics[2][]{}%
  }%
  \providecommand\rotatebox[2]{#2}%
  \@ifundefined{ifGPcolor}{%
    \newif\ifGPcolor
    \GPcolortrue
  }{}%
  \@ifundefined{ifGPblacktext}{%
    \newif\ifGPblacktext
    \GPblacktexttrue
  }{}%
  \let\gplgaddtomacro\g@addto@macro
  \gdef\gplbacktext{}%
  \gdef\gplfronttext{}%
  \makeatother
  \ifGPblacktext
    \def\colorrgb#1{}%
    \def\colorgray#1{}%
  \else
    \ifGPcolor
      \def\colorrgb#1{\color[rgb]{#1}}%
      \def\colorgray#1{\color[gray]{#1}}%
      \expandafter\def\csname LTw\endcsname{\color{white}}%
      \expandafter\def\csname LTb\endcsname{\color{black}}%
      \expandafter\def\csname LTa\endcsname{\color{black}}%
      \expandafter\def\csname LT0\endcsname{\color[rgb]{1,0,0}}%
      \expandafter\def\csname LT1\endcsname{\color[rgb]{0,1,0}}%
      \expandafter\def\csname LT2\endcsname{\color[rgb]{0,0,1}}%
      \expandafter\def\csname LT3\endcsname{\color[rgb]{1,0,1}}%
      \expandafter\def\csname LT4\endcsname{\color[rgb]{0,1,1}}%
      \expandafter\def\csname LT5\endcsname{\color[rgb]{1,1,0}}%
      \expandafter\def\csname LT6\endcsname{\color[rgb]{0,0,0}}%
      \expandafter\def\csname LT7\endcsname{\color[rgb]{1,0.3,0}}%
      \expandafter\def\csname LT8\endcsname{\color[rgb]{0.5,0.5,0.5}}%
    \else
      \def\colorrgb#1{\color{black}}%
      \def\colorgray#1{\color[gray]{#1}}%
      \expandafter\def\csname LTw\endcsname{\color{white}}%
      \expandafter\def\csname LTb\endcsname{\color{black}}%
      \expandafter\def\csname LTa\endcsname{\color{black}}%
      \expandafter\def\csname LT0\endcsname{\color{black}}%
      \expandafter\def\csname LT1\endcsname{\color{black}}%
      \expandafter\def\csname LT2\endcsname{\color{black}}%
      \expandafter\def\csname LT3\endcsname{\color{black}}%
      \expandafter\def\csname LT4\endcsname{\color{black}}%
      \expandafter\def\csname LT5\endcsname{\color{black}}%
      \expandafter\def\csname LT6\endcsname{\color{black}}%
      \expandafter\def\csname LT7\endcsname{\color{black}}%
      \expandafter\def\csname LT8\endcsname{\color{black}}%
    \fi
  \fi
  \setlength{\unitlength}{0.0500bp}%
  \begin{picture}(7200.00,5040.00)%
    \gplgaddtomacro\gplbacktext{%
      \csname LTb\endcsname%
      \put(1474,1584){\makebox(0,0)[r]{\strut{} 0}}%
      \csname LTb\endcsname%
      \put(1474,1903){\makebox(0,0)[r]{\strut{} 1000}}%
      \csname LTb\endcsname%
      \put(1474,2222){\makebox(0,0)[r]{\strut{} 2000}}%
      \csname LTb\endcsname%
      \put(1474,2542){\makebox(0,0)[r]{\strut{} 3000}}%
      \csname LTb\endcsname%
      \put(1474,2861){\makebox(0,0)[r]{\strut{} 4000}}%
      \csname LTb\endcsname%
      \put(1474,3180){\makebox(0,0)[r]{\strut{} 5000}}%
      \csname LTb\endcsname%
      \put(1474,3499){\makebox(0,0)[r]{\strut{} 6000}}%
      \csname LTb\endcsname%
      \put(1474,3818){\makebox(0,0)[r]{\strut{} 7000}}%
      \csname LTb\endcsname%
      \put(1474,4138){\makebox(0,0)[r]{\strut{} 8000}}%
      \csname LTb\endcsname%
      \put(1474,4457){\makebox(0,0)[r]{\strut{} 9000}}%
      \csname LTb\endcsname%
      \put(1474,4776){\makebox(0,0)[r]{\strut{} 10000}}%
      \csname LTb\endcsname%
      \put(1606,1364){\makebox(0,0){\strut{}\small$0{\cdot}10^{0}$}}%
      \csname LTb\endcsname%
      \put(2659,1364){\makebox(0,0){\strut{}\small$1{\cdot}10^{9}$}}%
      \csname LTb\endcsname%
      \put(3712,1364){\makebox(0,0){\strut{}\small$2{\cdot}10^{9}$}}%
      \csname LTb\endcsname%
      \put(4764,1364){\makebox(0,0){\strut{}\small$3{\cdot}10^{9}$}}%
      \csname LTb\endcsname%
      \put(5817,1364){\makebox(0,0){\strut{}\small$4{\cdot}10^{9}$}}%
      \csname LTb\endcsname%
      \put(6870,1364){\makebox(0,0){\strut{}\small$5{\cdot}10^{9}$}}%
      \put(440,3180){\rotatebox{90}{\makebox(0,0){\strut{}Seconds}}}%
      \put(4238,1034){\makebox(0,0){\strut{}$n$}}%
    }%
    \gplgaddtomacro\gplfronttext{%
      \csname LTb\endcsname%
      \put(5197,613){\makebox(0,0)[r]{\strut{}PSRS PEMS1 (unix) P=1}}%
      \csname LTb\endcsname%
      \put(5197,393){\makebox(0,0)[r]{\strut{}PSRS PEMS2 (unix) P=1}}%
      \csname LTb\endcsname%
      \put(5197,173){\makebox(0,0)[r]{\strut{}stxxl}}%
    }%
    \gplbacktext
    \put(0,0){\includegraphics{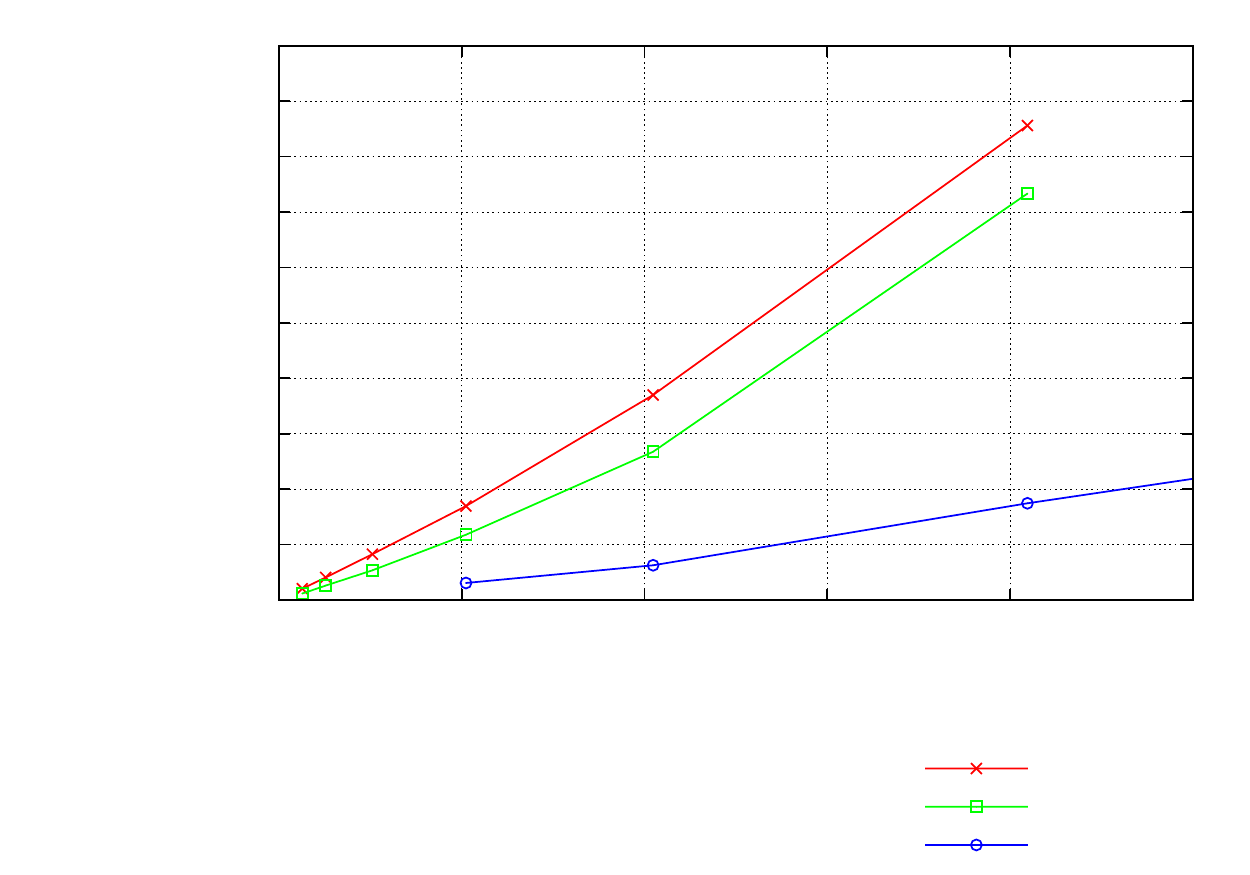}}%
    \gplfronttext
  \end{picture}%
\endgroup

%% file: v-p2-tex.tex
\begingroup
  \makeatletter
  \providecommand\color[2][]{%
    \GenericError{(gnuplot) \space\space\space\@spaces}{%
      Package color not loaded in conjunction with
      terminal option `colourtext'%
    }{See the gnuplot documentation for explanation.%
    }{Either use 'blacktext' in gnuplot or load the package
      color.sty in LaTeX.}%
    \renewcommand\color[2][]{}%
  }%
  \providecommand\includegraphics[2][]{%
    \GenericError{(gnuplot) \space\space\space\@spaces}{%
      Package graphicx or graphics not loaded%
    }{See the gnuplot documentation for explanation.%
    }{The gnuplot epslatex terminal needs graphicx.sty or graphics.sty.}%
    \renewcommand\includegraphics[2][]{}%
  }%
  \providecommand\rotatebox[2]{#2}%
  \@ifundefined{ifGPcolor}{%
    \newif\ifGPcolor
    \GPcolortrue
  }{}%
  \@ifundefined{ifGPblacktext}{%
    \newif\ifGPblacktext
    \GPblacktexttrue
  }{}%
  \let\gplgaddtomacro\g@addto@macro
  \gdef\gplbacktext{}%
  \gdef\gplfronttext{}%
  \makeatother
  \ifGPblacktext
    \def\colorrgb#1{}%
    \def\colorgray#1{}%
  \else
    \ifGPcolor
      \def\colorrgb#1{\color[rgb]{#1}}%
      \def\colorgray#1{\color[gray]{#1}}%
      \expandafter\def\csname LTw\endcsname{\color{white}}%
      \expandafter\def\csname LTb\endcsname{\color{black}}%
      \expandafter\def\csname LTa\endcsname{\color{black}}%
      \expandafter\def\csname LT0\endcsname{\color[rgb]{1,0,0}}%
      \expandafter\def\csname LT1\endcsname{\color[rgb]{0,1,0}}%
      \expandafter\def\csname LT2\endcsname{\color[rgb]{0,0,1}}%
      \expandafter\def\csname LT3\endcsname{\color[rgb]{1,0,1}}%
      \expandafter\def\csname LT4\endcsname{\color[rgb]{0,1,1}}%
      \expandafter\def\csname LT5\endcsname{\color[rgb]{1,1,0}}%
      \expandafter\def\csname LT6\endcsname{\color[rgb]{0,0,0}}%
      \expandafter\def\csname LT7\endcsname{\color[rgb]{1,0.3,0}}%
      \expandafter\def\csname LT8\endcsname{\color[rgb]{0.5,0.5,0.5}}%
    \else
      \def\colorrgb#1{\color{black}}%
      \def\colorgray#1{\color[gray]{#1}}%
      \expandafter\def\csname LTw\endcsname{\color{white}}%
      \expandafter\def\csname LTb\endcsname{\color{black}}%
      \expandafter\def\csname LTa\endcsname{\color{black}}%
      \expandafter\def\csname LT0\endcsname{\color{black}}%
      \expandafter\def\csname LT1\endcsname{\color{black}}%
      \expandafter\def\csname LT2\endcsname{\color{black}}%
      \expandafter\def\csname LT3\endcsname{\color{black}}%
      \expandafter\def\csname LT4\endcsname{\color{black}}%
      \expandafter\def\csname LT5\endcsname{\color{black}}%
      \expandafter\def\csname LT6\endcsname{\color{black}}%
      \expandafter\def\csname LT7\endcsname{\color{black}}%
      \expandafter\def\csname LT8\endcsname{\color{black}}%
    \fi
  \fi
  \setlength{\unitlength}{0.0500bp}%
  \begin{picture}(7200.00,5040.00)%
    \gplgaddtomacro\gplbacktext{%
      \csname LTb\endcsname%
      \put(1474,1584){\makebox(0,0)[r]{\strut{} 0}}%
      \csname LTb\endcsname%
      \put(1474,1903){\makebox(0,0)[r]{\strut{} 1000}}%
      \csname LTb\endcsname%
      \put(1474,2222){\makebox(0,0)[r]{\strut{} 2000}}%
      \csname LTb\endcsname%
      \put(1474,2542){\makebox(0,0)[r]{\strut{} 3000}}%
      \csname LTb\endcsname%
      \put(1474,2861){\makebox(0,0)[r]{\strut{} 4000}}%
      \csname LTb\endcsname%
      \put(1474,3180){\makebox(0,0)[r]{\strut{} 5000}}%
      \csname LTb\endcsname%
      \put(1474,3499){\makebox(0,0)[r]{\strut{} 6000}}%
      \csname LTb\endcsname%
      \put(1474,3818){\makebox(0,0)[r]{\strut{} 7000}}%
      \csname LTb\endcsname%
      \put(1474,4138){\makebox(0,0)[r]{\strut{} 8000}}%
      \csname LTb\endcsname%
      \put(1474,4457){\makebox(0,0)[r]{\strut{} 9000}}%
      \csname LTb\endcsname%
      \put(1474,4776){\makebox(0,0)[r]{\strut{} 10000}}%
      \csname LTb\endcsname%
      \put(1606,1364){\makebox(0,0){\strut{}\small$0{\cdot}10^{0}$}}%
      \csname LTb\endcsname%
      \put(2659,1364){\makebox(0,0){\strut{}\small$1{\cdot}10^{9}$}}%
      \csname LTb\endcsname%
      \put(3712,1364){\makebox(0,0){\strut{}\small$2{\cdot}10^{9}$}}%
      \csname LTb\endcsname%
      \put(4764,1364){\makebox(0,0){\strut{}\small$3{\cdot}10^{9}$}}%
      \csname LTb\endcsname%
      \put(5817,1364){\makebox(0,0){\strut{}\small$4{\cdot}10^{9}$}}%
      \csname LTb\endcsname%
      \put(6870,1364){\makebox(0,0){\strut{}\small$5{\cdot}10^{9}$}}%
      \put(440,3180){\rotatebox{90}{\makebox(0,0){\strut{}Seconds}}}%
      \put(4238,1034){\makebox(0,0){\strut{}$n$}}%
    }%
    \gplgaddtomacro\gplfronttext{%
      \csname LTb\endcsname%
      \put(5197,613){\makebox(0,0)[r]{\strut{}PSRS PEMS1 (unix) P=2}}%
      \csname LTb\endcsname%
      \put(5197,393){\makebox(0,0)[r]{\strut{}PSRS PEMS2 (unix) P=2}}%
      \csname LTb\endcsname%
      \put(5197,173){\makebox(0,0)[r]{\strut{}stxxl}}%
    }%
    \gplbacktext
    \put(0,0){\includegraphics{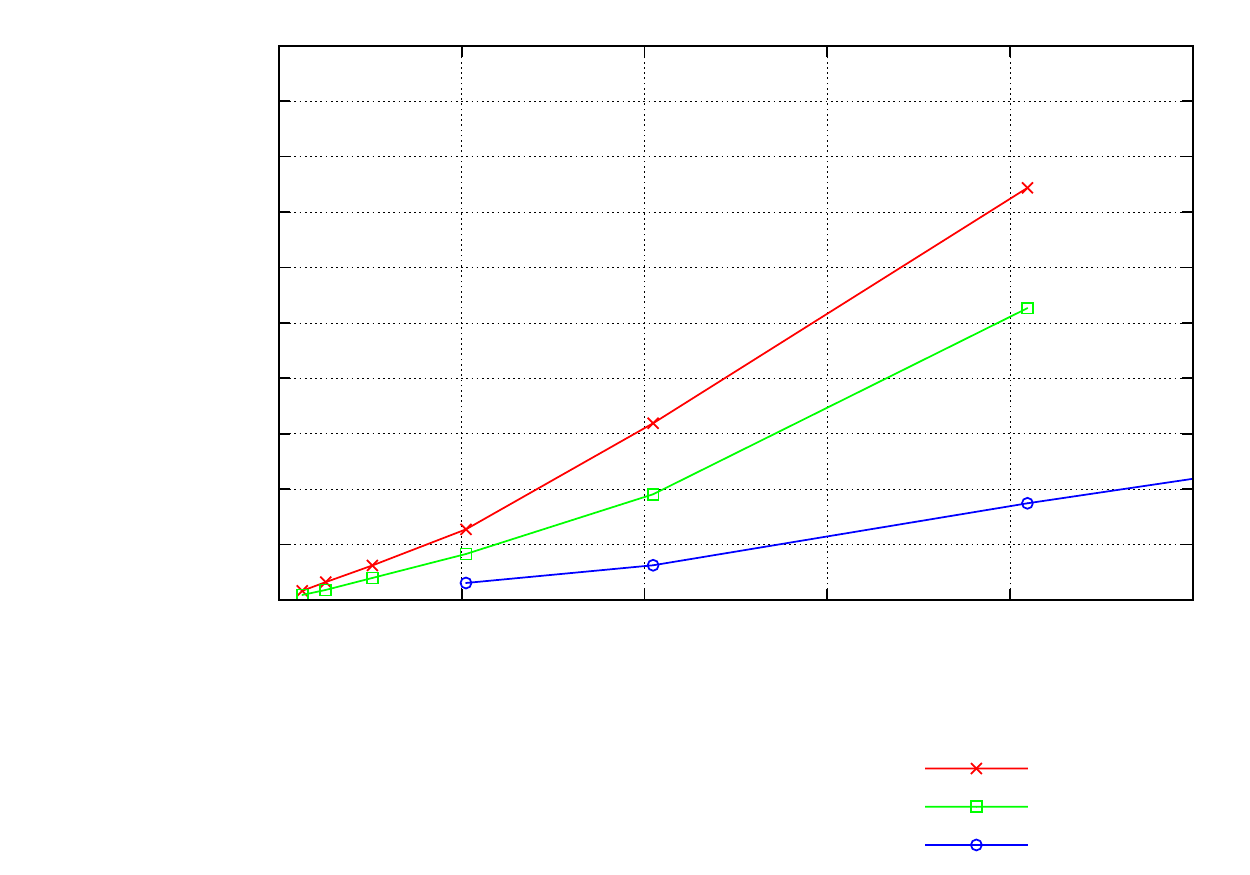}}%
    \gplfronttext
  \end{picture}%
\endgroup

%% file: v-p4-tex.tex
\begingroup
  \makeatletter
  \providecommand\color[2][]{%
    \GenericError{(gnuplot) \space\space\space\@spaces}{%
      Package color not loaded in conjunction with
      terminal option `colourtext'%
    }{See the gnuplot documentation for explanation.%
    }{Either use 'blacktext' in gnuplot or load the package
      color.sty in LaTeX.}%
    \renewcommand\color[2][]{}%
  }%
  \providecommand\includegraphics[2][]{%
    \GenericError{(gnuplot) \space\space\space\@spaces}{%
      Package graphicx or graphics not loaded%
    }{See the gnuplot documentation for explanation.%
    }{The gnuplot epslatex terminal needs graphicx.sty or graphics.sty.}%
    \renewcommand\includegraphics[2][]{}%
  }%
  \providecommand\rotatebox[2]{#2}%
  \@ifundefined{ifGPcolor}{%
    \newif\ifGPcolor
    \GPcolortrue
  }{}%
  \@ifundefined{ifGPblacktext}{%
    \newif\ifGPblacktext
    \GPblacktexttrue
  }{}%
  \let\gplgaddtomacro\g@addto@macro
  \gdef\gplbacktext{}%
  \gdef\gplfronttext{}%
  \makeatother
  \ifGPblacktext
    \def\colorrgb#1{}%
    \def\colorgray#1{}%
  \else
    \ifGPcolor
      \def\colorrgb#1{\color[rgb]{#1}}%
      \def\colorgray#1{\color[gray]{#1}}%
      \expandafter\def\csname LTw\endcsname{\color{white}}%
      \expandafter\def\csname LTb\endcsname{\color{black}}%
      \expandafter\def\csname LTa\endcsname{\color{black}}%
      \expandafter\def\csname LT0\endcsname{\color[rgb]{1,0,0}}%
      \expandafter\def\csname LT1\endcsname{\color[rgb]{0,1,0}}%
      \expandafter\def\csname LT2\endcsname{\color[rgb]{0,0,1}}%
      \expandafter\def\csname LT3\endcsname{\color[rgb]{1,0,1}}%
      \expandafter\def\csname LT4\endcsname{\color[rgb]{0,1,1}}%
      \expandafter\def\csname LT5\endcsname{\color[rgb]{1,1,0}}%
      \expandafter\def\csname LT6\endcsname{\color[rgb]{0,0,0}}%
      \expandafter\def\csname LT7\endcsname{\color[rgb]{1,0.3,0}}%
      \expandafter\def\csname LT8\endcsname{\color[rgb]{0.5,0.5,0.5}}%
    \else
      \def\colorrgb#1{\color{black}}%
      \def\colorgray#1{\color[gray]{#1}}%
      \expandafter\def\csname LTw\endcsname{\color{white}}%
      \expandafter\def\csname LTb\endcsname{\color{black}}%
      \expandafter\def\csname LTa\endcsname{\color{black}}%
      \expandafter\def\csname LT0\endcsname{\color{black}}%
      \expandafter\def\csname LT1\endcsname{\color{black}}%
      \expandafter\def\csname LT2\endcsname{\color{black}}%
      \expandafter\def\csname LT3\endcsname{\color{black}}%
      \expandafter\def\csname LT4\endcsname{\color{black}}%
      \expandafter\def\csname LT5\endcsname{\color{black}}%
      \expandafter\def\csname LT6\endcsname{\color{black}}%
      \expandafter\def\csname LT7\endcsname{\color{black}}%
      \expandafter\def\csname LT8\endcsname{\color{black}}%
    \fi
  \fi
  \setlength{\unitlength}{0.0500bp}%
  \begin{picture}(7200.00,5040.00)%
    \gplgaddtomacro\gplbacktext{%
      \csname LTb\endcsname%
      \put(1474,1584){\makebox(0,0)[r]{\strut{} 0}}%
      \csname LTb\endcsname%
      \put(1474,1903){\makebox(0,0)[r]{\strut{} 1000}}%
      \csname LTb\endcsname%
      \put(1474,2222){\makebox(0,0)[r]{\strut{} 2000}}%
      \csname LTb\endcsname%
      \put(1474,2542){\makebox(0,0)[r]{\strut{} 3000}}%
      \csname LTb\endcsname%
      \put(1474,2861){\makebox(0,0)[r]{\strut{} 4000}}%
      \csname LTb\endcsname%
      \put(1474,3180){\makebox(0,0)[r]{\strut{} 5000}}%
      \csname LTb\endcsname%
      \put(1474,3499){\makebox(0,0)[r]{\strut{} 6000}}%
      \csname LTb\endcsname%
      \put(1474,3818){\makebox(0,0)[r]{\strut{} 7000}}%
      \csname LTb\endcsname%
      \put(1474,4138){\makebox(0,0)[r]{\strut{} 8000}}%
      \csname LTb\endcsname%
      \put(1474,4457){\makebox(0,0)[r]{\strut{} 9000}}%
      \csname LTb\endcsname%
      \put(1474,4776){\makebox(0,0)[r]{\strut{} 10000}}%
      \csname LTb\endcsname%
      \put(1606,1364){\makebox(0,0){\strut{}\small$0{\cdot}10^{0}$}}%
      \csname LTb\endcsname%
      \put(2659,1364){\makebox(0,0){\strut{}\small$1{\cdot}10^{9}$}}%
      \csname LTb\endcsname%
      \put(3712,1364){\makebox(0,0){\strut{}\small$2{\cdot}10^{9}$}}%
      \csname LTb\endcsname%
      \put(4764,1364){\makebox(0,0){\strut{}\small$3{\cdot}10^{9}$}}%
      \csname LTb\endcsname%
      \put(5817,1364){\makebox(0,0){\strut{}\small$4{\cdot}10^{9}$}}%
      \csname LTb\endcsname%
      \put(6870,1364){\makebox(0,0){\strut{}\small$5{\cdot}10^{9}$}}%
      \put(440,3180){\rotatebox{90}{\makebox(0,0){\strut{}Seconds}}}%
      \put(4238,1034){\makebox(0,0){\strut{}$n$}}%
    }%
    \gplgaddtomacro\gplfronttext{%
      \csname LTb\endcsname%
      \put(5197,613){\makebox(0,0)[r]{\strut{}PSRS PEMS1 (unix) P=4}}%
      \csname LTb\endcsname%
      \put(5197,393){\makebox(0,0)[r]{\strut{}PSRS PEMS2 (unix) P=4}}%
      \csname LTb\endcsname%
      \put(5197,173){\makebox(0,0)[r]{\strut{}stxxl}}%
    }%
    \gplbacktext
    \put(0,0){\includegraphics{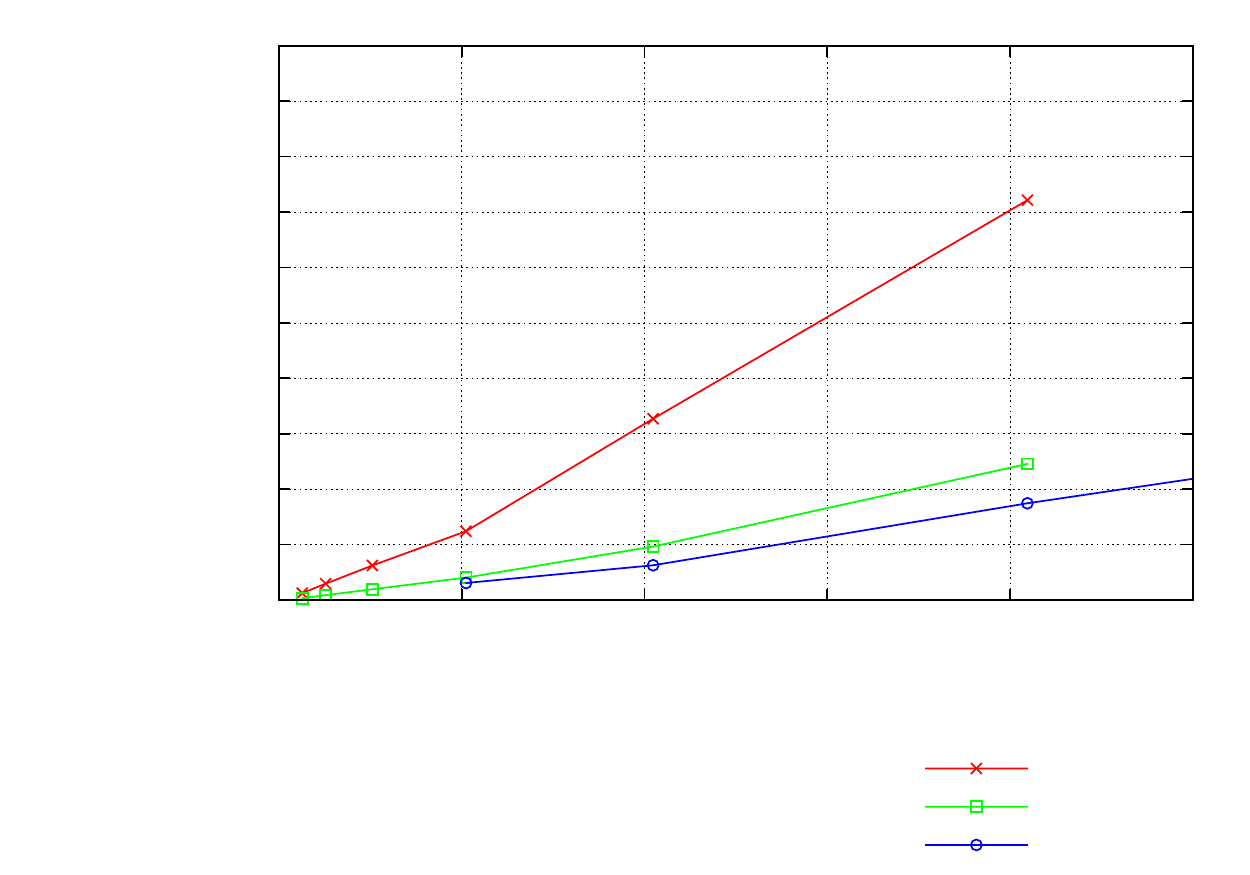}}%
    \gplfronttext
  \end{picture}%
\endgroup

%% file: v-p8-tex.tex
\begingroup
  \makeatletter
  \providecommand\color[2][]{%
    \GenericError{(gnuplot) \space\space\space\@spaces}{%
      Package color not loaded in conjunction with
      terminal option `colourtext'%
    }{See the gnuplot documentation for explanation.%
    }{Either use 'blacktext' in gnuplot or load the package
      color.sty in LaTeX.}%
    \renewcommand\color[2][]{}%
  }%
  \providecommand\includegraphics[2][]{%
    \GenericError{(gnuplot) \space\space\space\@spaces}{%
      Package graphicx or graphics not loaded%
    }{See the gnuplot documentation for explanation.%
    }{The gnuplot epslatex terminal needs graphicx.sty or graphics.sty.}%
    \renewcommand\includegraphics[2][]{}%
  }%
  \providecommand\rotatebox[2]{#2}%
  \@ifundefined{ifGPcolor}{%
    \newif\ifGPcolor
    \GPcolortrue
  }{}%
  \@ifundefined{ifGPblacktext}{%
    \newif\ifGPblacktext
    \GPblacktexttrue
  }{}%
  \let\gplgaddtomacro\g@addto@macro
  \gdef\gplbacktext{}%
  \gdef\gplfronttext{}%
  \makeatother
  \ifGPblacktext
    \def\colorrgb#1{}%
    \def\colorgray#1{}%
  \else
    \ifGPcolor
      \def\colorrgb#1{\color[rgb]{#1}}%
      \def\colorgray#1{\color[gray]{#1}}%
      \expandafter\def\csname LTw\endcsname{\color{white}}%
      \expandafter\def\csname LTb\endcsname{\color{black}}%
      \expandafter\def\csname LTa\endcsname{\color{black}}%
      \expandafter\def\csname LT0\endcsname{\color[rgb]{1,0,0}}%
      \expandafter\def\csname LT1\endcsname{\color[rgb]{0,1,0}}%
      \expandafter\def\csname LT2\endcsname{\color[rgb]{0,0,1}}%
      \expandafter\def\csname LT3\endcsname{\color[rgb]{1,0,1}}%
      \expandafter\def\csname LT4\endcsname{\color[rgb]{0,1,1}}%
      \expandafter\def\csname LT5\endcsname{\color[rgb]{1,1,0}}%
      \expandafter\def\csname LT6\endcsname{\color[rgb]{0,0,0}}%
      \expandafter\def\csname LT7\endcsname{\color[rgb]{1,0.3,0}}%
      \expandafter\def\csname LT8\endcsname{\color[rgb]{0.5,0.5,0.5}}%
    \else
      \def\colorrgb#1{\color{black}}%
      \def\colorgray#1{\color[gray]{#1}}%
      \expandafter\def\csname LTw\endcsname{\color{white}}%
      \expandafter\def\csname LTb\endcsname{\color{black}}%
      \expandafter\def\csname LTa\endcsname{\color{black}}%
      \expandafter\def\csname LT0\endcsname{\color{black}}%
      \expandafter\def\csname LT1\endcsname{\color{black}}%
      \expandafter\def\csname LT2\endcsname{\color{black}}%
      \expandafter\def\csname LT3\endcsname{\color{black}}%
      \expandafter\def\csname LT4\endcsname{\color{black}}%
      \expandafter\def\csname LT5\endcsname{\color{black}}%
      \expandafter\def\csname LT6\endcsname{\color{black}}%
      \expandafter\def\csname LT7\endcsname{\color{black}}%
      \expandafter\def\csname LT8\endcsname{\color{black}}%
    \fi
  \fi
  \setlength{\unitlength}{0.0500bp}%
  \begin{picture}(7200.00,5040.00)%
    \gplgaddtomacro\gplbacktext{%
      \csname LTb\endcsname%
      \put(1474,1584){\makebox(0,0)[r]{\strut{} 0}}%
      \csname LTb\endcsname%
      \put(1474,1903){\makebox(0,0)[r]{\strut{} 1000}}%
      \csname LTb\endcsname%
      \put(1474,2222){\makebox(0,0)[r]{\strut{} 2000}}%
      \csname LTb\endcsname%
      \put(1474,2542){\makebox(0,0)[r]{\strut{} 3000}}%
      \csname LTb\endcsname%
      \put(1474,2861){\makebox(0,0)[r]{\strut{} 4000}}%
      \csname LTb\endcsname%
      \put(1474,3180){\makebox(0,0)[r]{\strut{} 5000}}%
      \csname LTb\endcsname%
      \put(1474,3499){\makebox(0,0)[r]{\strut{} 6000}}%
      \csname LTb\endcsname%
      \put(1474,3818){\makebox(0,0)[r]{\strut{} 7000}}%
      \csname LTb\endcsname%
      \put(1474,4138){\makebox(0,0)[r]{\strut{} 8000}}%
      \csname LTb\endcsname%
      \put(1474,4457){\makebox(0,0)[r]{\strut{} 9000}}%
      \csname LTb\endcsname%
      \put(1474,4776){\makebox(0,0)[r]{\strut{} 10000}}%
      \csname LTb\endcsname%
      \put(1606,1364){\makebox(0,0){\strut{}\small$0{\cdot}10^{0}$}}%
      \csname LTb\endcsname%
      \put(2659,1364){\makebox(0,0){\strut{}\small$1{\cdot}10^{9}$}}%
      \csname LTb\endcsname%
      \put(3712,1364){\makebox(0,0){\strut{}\small$2{\cdot}10^{9}$}}%
      \csname LTb\endcsname%
      \put(4764,1364){\makebox(0,0){\strut{}\small$3{\cdot}10^{9}$}}%
      \csname LTb\endcsname%
      \put(5817,1364){\makebox(0,0){\strut{}\small$4{\cdot}10^{9}$}}%
      \csname LTb\endcsname%
      \put(6870,1364){\makebox(0,0){\strut{}\small$5{\cdot}10^{9}$}}%
      \put(440,3180){\rotatebox{90}{\makebox(0,0){\strut{}Seconds}}}%
      \put(4238,1034){\makebox(0,0){\strut{}$n$}}%
    }%
    \gplgaddtomacro\gplfronttext{%
      \csname LTb\endcsname%
      \put(5197,613){\makebox(0,0)[r]{\strut{}PSRS PEMS1 (unix) P=8}}%
      \csname LTb\endcsname%
      \put(5197,393){\makebox(0,0)[r]{\strut{}PSRS PEMS2 (unix) P=8}}%
      \csname LTb\endcsname%
      \put(5197,173){\makebox(0,0)[r]{\strut{}stxxl}}%
    }%
    \gplbacktext
    \put(0,0){\includegraphics{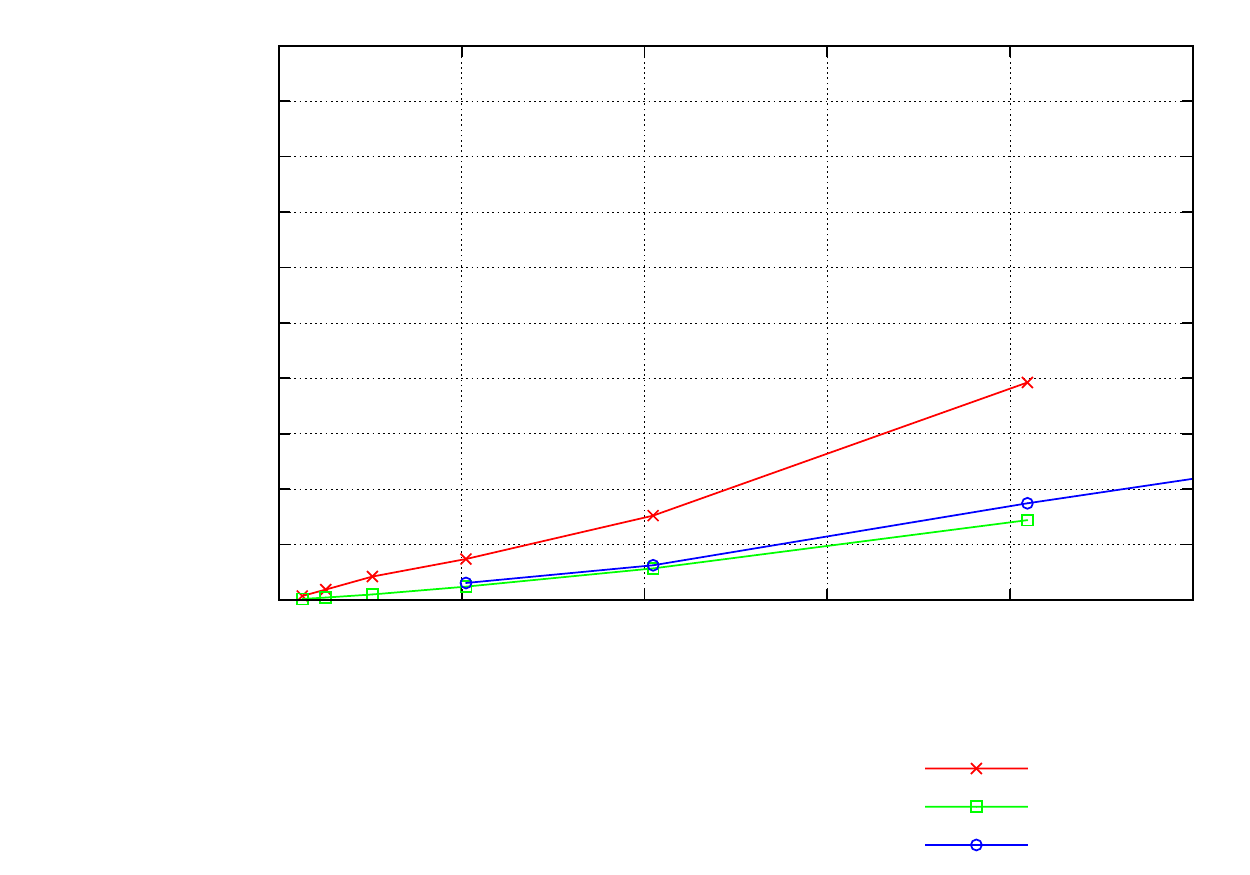}}%
    \gplfronttext
  \end{picture}%
\endgroup

%% file: pems1-pems2-speedup-tex.tex
\begin{figure}[ht]
	\begin{center}
	\resizebox{!}{0.4\textheight}{
		\input{pems1-pems2-speedup-plot-tex}
	}
	\\
n = 4096000000
,  v = 512
,  $\frac{n}{v} = 8000000$
,  $\mu = 64$ {\small MiB}

	\caption{PEMS1 vs. PEMS2 PSRS Relative Speedup}
	\label{pems1-pems2-speedup-plot}
	\end{center}
\end{figure}
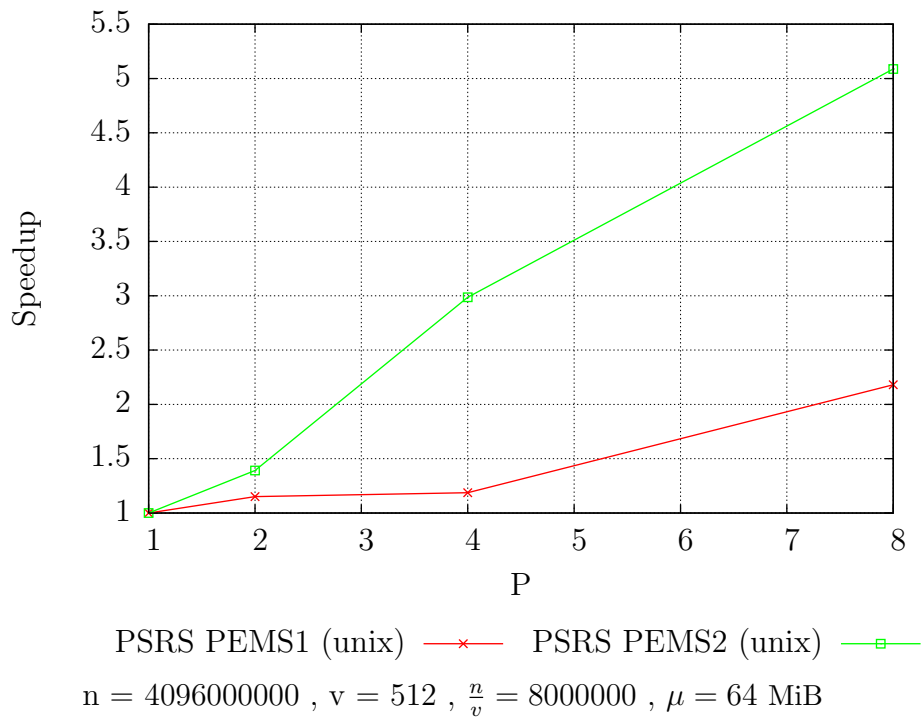

%% file: pems1-pems2-speedup-plot-tex.tex
\begingroup
  \makeatletter
  \providecommand\color[2][]{%
    \GenericError{(gnuplot) \space\space\space\@spaces}{%
      Package color not loaded in conjunction with
      terminal option `colourtext'%
    }{See the gnuplot documentation for explanation.%
    }{Either use 'blacktext' in gnuplot or load the package
      color.sty in LaTeX.}%
    \renewcommand\color[2][]{}%
  }%
  \providecommand\includegraphics[2][]{%
    \GenericError{(gnuplot) \space\space\space\@spaces}{%
      Package graphicx or graphics not loaded%
    }{See the gnuplot documentation for explanation.%
    }{The gnuplot epslatex terminal needs graphicx.sty or graphics.sty.}%
    \renewcommand\includegraphics[2][]{}%
  }%
  \providecommand\rotatebox[2]{#2}%
  \@ifundefined{ifGPcolor}{%
    \newif\ifGPcolor
    \GPcolortrue
  }{}%
  \@ifundefined{ifGPblacktext}{%
    \newif\ifGPblacktext
    \GPblacktexttrue
  }{}%
  \let\gplgaddtomacro\g@addto@macro
  \gdef\gplbacktext{}%
  \gdef\gplfronttext{}%
  \makeatother
  \ifGPblacktext
    \def\colorrgb#1{}%
    \def\colorgray#1{}%
  \else
    \ifGPcolor
      \def\colorrgb#1{\color[rgb]{#1}}%
      \def\colorgray#1{\color[gray]{#1}}%
      \expandafter\def\csname LTw\endcsname{\color{white}}%
      \expandafter\def\csname LTb\endcsname{\color{black}}%
      \expandafter\def\csname LTa\endcsname{\color{black}}%
      \expandafter\def\csname LT0\endcsname{\color[rgb]{1,0,0}}%
      \expandafter\def\csname LT1\endcsname{\color[rgb]{0,1,0}}%
      \expandafter\def\csname LT2\endcsname{\color[rgb]{0,0,1}}%
      \expandafter\def\csname LT3\endcsname{\color[rgb]{1,0,1}}%
      \expandafter\def\csname LT4\endcsname{\color[rgb]{0,1,1}}%
      \expandafter\def\csname LT5\endcsname{\color[rgb]{1,1,0}}%
      \expandafter\def\csname LT6\endcsname{\color[rgb]{0,0,0}}%
      \expandafter\def\csname LT7\endcsname{\color[rgb]{1,0.3,0}}%
      \expandafter\def\csname LT8\endcsname{\color[rgb]{0.5,0.5,0.5}}%
    \else
      \def\colorrgb#1{\color{black}}%
      \def\colorgray#1{\color[gray]{#1}}%
      \expandafter\def\csname LTw\endcsname{\color{white}}%
      \expandafter\def\csname LTb\endcsname{\color{black}}%
      \expandafter\def\csname LTa\endcsname{\color{black}}%
      \expandafter\def\csname LT0\endcsname{\color{black}}%
      \expandafter\def\csname LT1\endcsname{\color{black}}%
      \expandafter\def\csname LT2\endcsname{\color{black}}%
      \expandafter\def\csname LT3\endcsname{\color{black}}%
      \expandafter\def\csname LT4\endcsname{\color{black}}%
      \expandafter\def\csname LT5\endcsname{\color{black}}%
      \expandafter\def\csname LT6\endcsname{\color{black}}%
      \expandafter\def\csname LT7\endcsname{\color{black}}%
      \expandafter\def\csname LT8\endcsname{\color{black}}%
    \fi
  \fi
  \setlength{\unitlength}{0.0500bp}%
  \begin{picture}(7200.00,5040.00)%
    \gplgaddtomacro\gplbacktext{%
      \csname LTb\endcsname%
      \put(1210,1144){\makebox(0,0)[r]{\strut{} 1}}%
      \csname LTb\endcsname%
      \put(1210,1548){\makebox(0,0)[r]{\strut{} 1.5}}%
      \csname LTb\endcsname%
      \put(1210,1951){\makebox(0,0)[r]{\strut{} 2}}%
      \csname LTb\endcsname%
      \put(1210,2355){\makebox(0,0)[r]{\strut{} 2.5}}%
      \csname LTb\endcsname%
      \put(1210,2758){\makebox(0,0)[r]{\strut{} 3}}%
      \csname LTb\endcsname%
      \put(1210,3162){\makebox(0,0)[r]{\strut{} 3.5}}%
      \csname LTb\endcsname%
      \put(1210,3565){\makebox(0,0)[r]{\strut{} 4}}%
      \csname LTb\endcsname%
      \put(1210,3969){\makebox(0,0)[r]{\strut{} 4.5}}%
      \csname LTb\endcsname%
      \put(1210,4372){\makebox(0,0)[r]{\strut{} 5}}%
      \csname LTb\endcsname%
      \put(1210,4776){\makebox(0,0)[r]{\strut{} 5.5}}%
      \csname LTb\endcsname%
      \put(1342,924){\makebox(0,0){\strut{} 1}}%
      \csname LTb\endcsname%
      \put(2132,924){\makebox(0,0){\strut{} 2}}%
      \csname LTb\endcsname%
      \put(2921,924){\makebox(0,0){\strut{} 3}}%
      \csname LTb\endcsname%
      \put(3711,924){\makebox(0,0){\strut{} 4}}%
      \csname LTb\endcsname%
      \put(4501,924){\makebox(0,0){\strut{} 5}}%
      \csname LTb\endcsname%
      \put(5291,924){\makebox(0,0){\strut{} 6}}%
      \csname LTb\endcsname%
      \put(6080,924){\makebox(0,0){\strut{} 7}}%
      \csname LTb\endcsname%
      \put(6870,924){\makebox(0,0){\strut{} 8}}%
      \put(440,2960){\rotatebox{90}{\makebox(0,0){\strut{}Speedup}}}%
      \put(4106,594){\makebox(0,0){\strut{}P}}%
    }%
    \gplgaddtomacro\gplfronttext{%
      \csname LTb\endcsname%
      \put(3251,173){\makebox(0,0)[r]{\strut{}PSRS PEMS1 (unix)}}%
      \csname LTb\endcsname%
      \put(6350,173){\makebox(0,0)[r]{\strut{}PSRS PEMS2 (unix)}}%
    }%
    \gplbacktext
    \put(0,0){\includegraphics{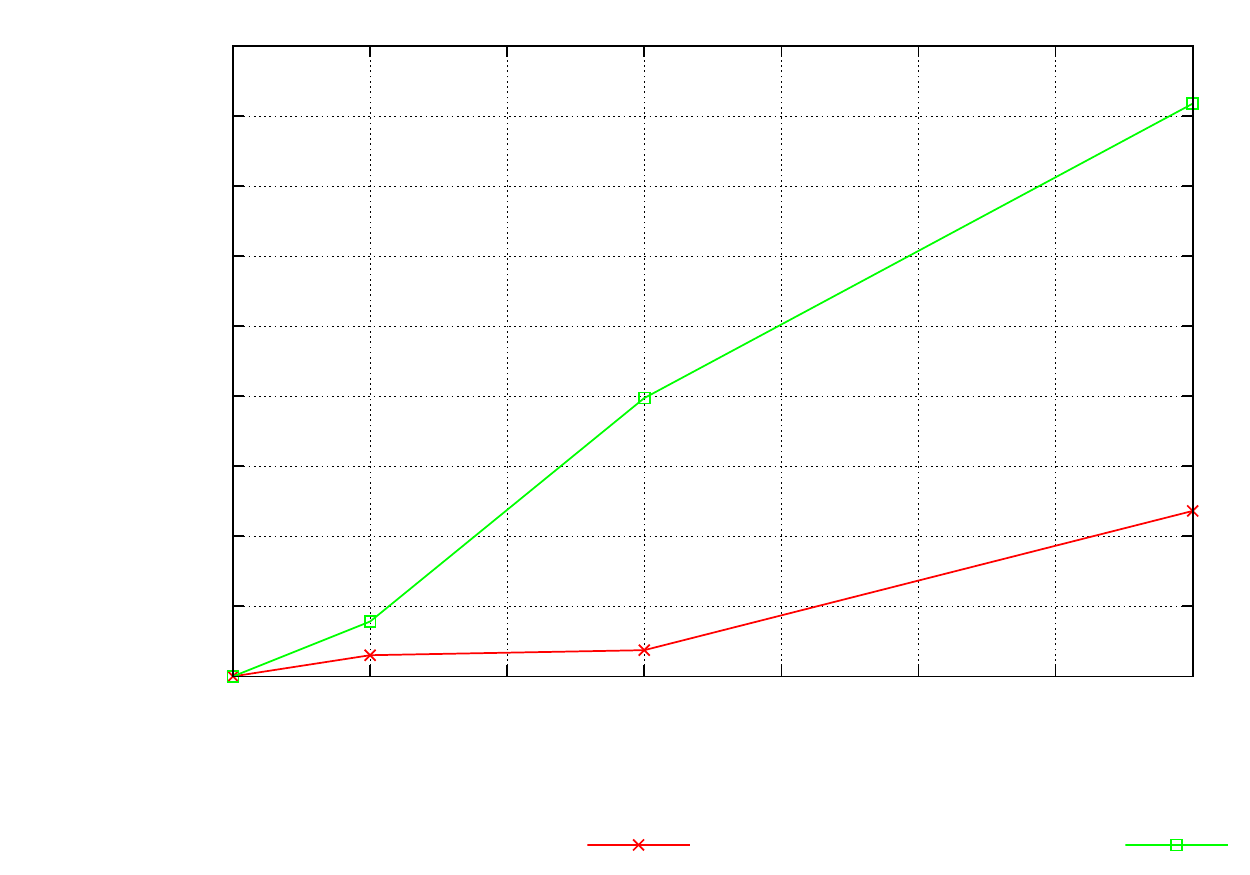}}%
    \gplfronttext
  \end{picture}%
\endgroup

%% file: improvement-mu-tex.tex
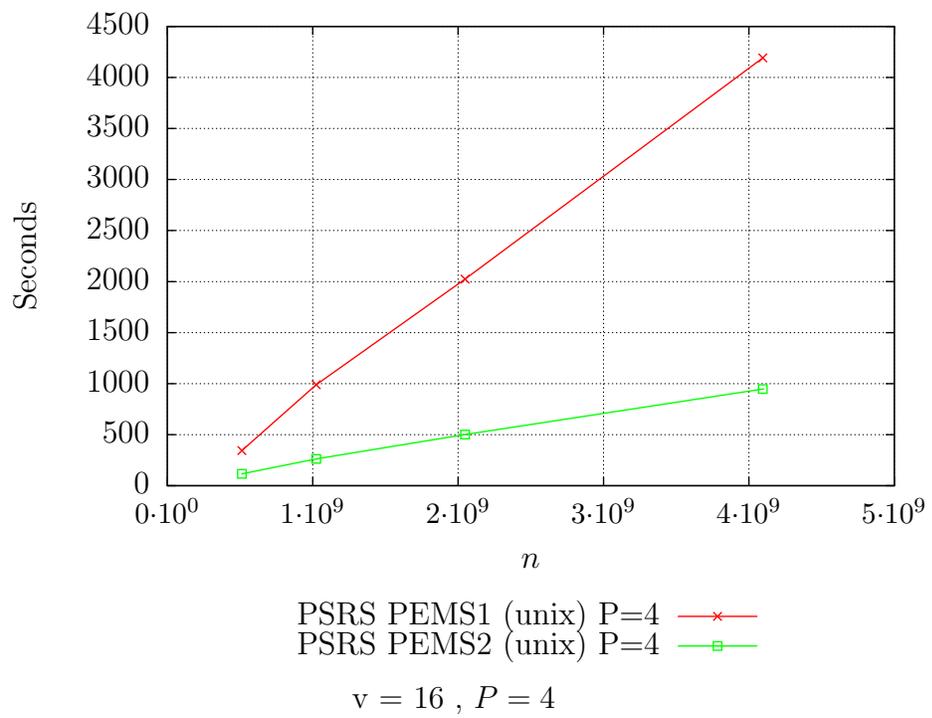
\begin{figure}[ht]
	\begin{center}
	\resizebox{!}{0.4\textheight}{
		\input{mu-plot-tex}
	}
	\\
v = 16
,  $P = 4$

	\caption{Increasing Context Size with Constant $v$}
	\label{mu-plot}
	\end{center}
\end{figure}

%% file: mu-plot-tex.tex
\begingroup
  \makeatletter
  \providecommand\color[2][]{%
    \GenericError{(gnuplot) \space\space\space\@spaces}{%
      Package color not loaded in conjunction with
      terminal option `colourtext'%
    }{See the gnuplot documentation for explanation.%
    }{Either use 'blacktext' in gnuplot or load the package
      color.sty in LaTeX.}%
    \renewcommand\color[2][]{}%
  }%
  \providecommand\includegraphics[2][]{%
    \GenericError{(gnuplot) \space\space\space\@spaces}{%
      Package graphicx or graphics not loaded%
    }{See the gnuplot documentation for explanation.%
    }{The gnuplot epslatex terminal needs graphicx.sty or graphics.sty.}%
    \renewcommand\includegraphics[2][]{}%
  }%
  \providecommand\rotatebox[2]{#2}%
  \@ifundefined{ifGPcolor}{%
    \newif\ifGPcolor
    \GPcolortrue
  }{}%
  \@ifundefined{ifGPblacktext}{%
    \newif\ifGPblacktext
    \GPblacktexttrue
  }{}%
  \let\gplgaddtomacro\g@addto@macro
  \gdef\gplbacktext{}%
  \gdef\gplfronttext{}%
  \makeatother
  \ifGPblacktext
    \def\colorrgb#1{}%
    \def\colorgray#1{}%
  \else
    \ifGPcolor
      \def\colorrgb#1{\color[rgb]{#1}}%
      \def\colorgray#1{\color[gray]{#1}}%
      \expandafter\def\csname LTw\endcsname{\color{white}}%
      \expandafter\def\csname LTb\endcsname{\color{black}}%
      \expandafter\def\csname LTa\endcsname{\color{black}}%
      \expandafter\def\csname LT0\endcsname{\color[rgb]{1,0,0}}%
      \expandafter\def\csname LT1\endcsname{\color[rgb]{0,1,0}}%
      \expandafter\def\csname LT2\endcsname{\color[rgb]{0,0,1}}%
      \expandafter\def\csname LT3\endcsname{\color[rgb]{1,0,1}}%
      \expandafter\def\csname LT4\endcsname{\color[rgb]{0,1,1}}%
      \expandafter\def\csname LT5\endcsname{\color[rgb]{1,1,0}}%
      \expandafter\def\csname LT6\endcsname{\color[rgb]{0,0,0}}%
      \expandafter\def\csname LT7\endcsname{\color[rgb]{1,0.3,0}}%
      \expandafter\def\csname LT8\endcsname{\color[rgb]{0.5,0.5,0.5}}%
    \else
      \def\colorrgb#1{\color{black}}%
      \def\colorgray#1{\color[gray]{#1}}%
      \expandafter\def\csname LTw\endcsname{\color{white}}%
      \expandafter\def\csname LTb\endcsname{\color{black}}%
      \expandafter\def\csname LTa\endcsname{\color{black}}%
      \expandafter\def\csname LT0\endcsname{\color{black}}%
      \expandafter\def\csname LT1\endcsname{\color{black}}%
      \expandafter\def\csname LT2\endcsname{\color{black}}%
      \expandafter\def\csname LT3\endcsname{\color{black}}%
      \expandafter\def\csname LT4\endcsname{\color{black}}%
      \expandafter\def\csname LT5\endcsname{\color{black}}%
      \expandafter\def\csname LT6\endcsname{\color{black}}%
      \expandafter\def\csname LT7\endcsname{\color{black}}%
      \expandafter\def\csname LT8\endcsname{\color{black}}%
    \fi
  \fi
  \setlength{\unitlength}{0.0500bp}%
  \begin{picture}(7200.00,5040.00)%
    \gplgaddtomacro\gplbacktext{%
      \csname LTb\endcsname%
      \put(1342,1364){\makebox(0,0)[r]{\strut{} 0}}%
      \csname LTb\endcsname%
      \put(1342,1743){\makebox(0,0)[r]{\strut{} 500}}%
      \csname LTb\endcsname%
      \put(1342,2122){\makebox(0,0)[r]{\strut{} 1000}}%
      \csname LTb\endcsname%
      \put(1342,2501){\makebox(0,0)[r]{\strut{} 1500}}%
      \csname LTb\endcsname%
      \put(1342,2880){\makebox(0,0)[r]{\strut{} 2000}}%
      \csname LTb\endcsname%
      \put(1342,3260){\makebox(0,0)[r]{\strut{} 2500}}%
      \csname LTb\endcsname%
      \put(1342,3639){\makebox(0,0)[r]{\strut{} 3000}}%
      \csname LTb\endcsname%
      \put(1342,4018){\makebox(0,0)[r]{\strut{} 3500}}%
      \csname LTb\endcsname%
      \put(1342,4397){\makebox(0,0)[r]{\strut{} 4000}}%
      \csname LTb\endcsname%
      \put(1342,4776){\makebox(0,0)[r]{\strut{} 4500}}%
      \csname LTb\endcsname%
      \put(1474,1144){\makebox(0,0){\strut{}\small$0{\cdot}10^{0}$}}%
      \csname LTb\endcsname%
      \put(2553,1144){\makebox(0,0){\strut{}\small$1{\cdot}10^{9}$}}%
      \csname LTb\endcsname%
      \put(3632,1144){\makebox(0,0){\strut{}\small$2{\cdot}10^{9}$}}%
      \csname LTb\endcsname%
      \put(4712,1144){\makebox(0,0){\strut{}\small$3{\cdot}10^{9}$}}%
      \csname LTb\endcsname%
      \put(5791,1144){\makebox(0,0){\strut{}\small$4{\cdot}10^{9}$}}%
      \csname LTb\endcsname%
      \put(6870,1144){\makebox(0,0){\strut{}\small$5{\cdot}10^{9}$}}%
      \put(440,3070){\rotatebox{90}{\makebox(0,0){\strut{}Seconds}}}%
      \put(4172,814){\makebox(0,0){\strut{}$n$}}%
    }%
    \gplgaddtomacro\gplfronttext{%
      \csname LTb\endcsname%
      \put(5131,393){\makebox(0,0)[r]{\strut{}PSRS PEMS1 (unix) P=4}}%
      \csname LTb\endcsname%
      \put(5131,173){\makebox(0,0)[r]{\strut{}PSRS PEMS2 (unix) P=4}}%
    }%
    \gplbacktext
    \put(0,0){\includegraphics{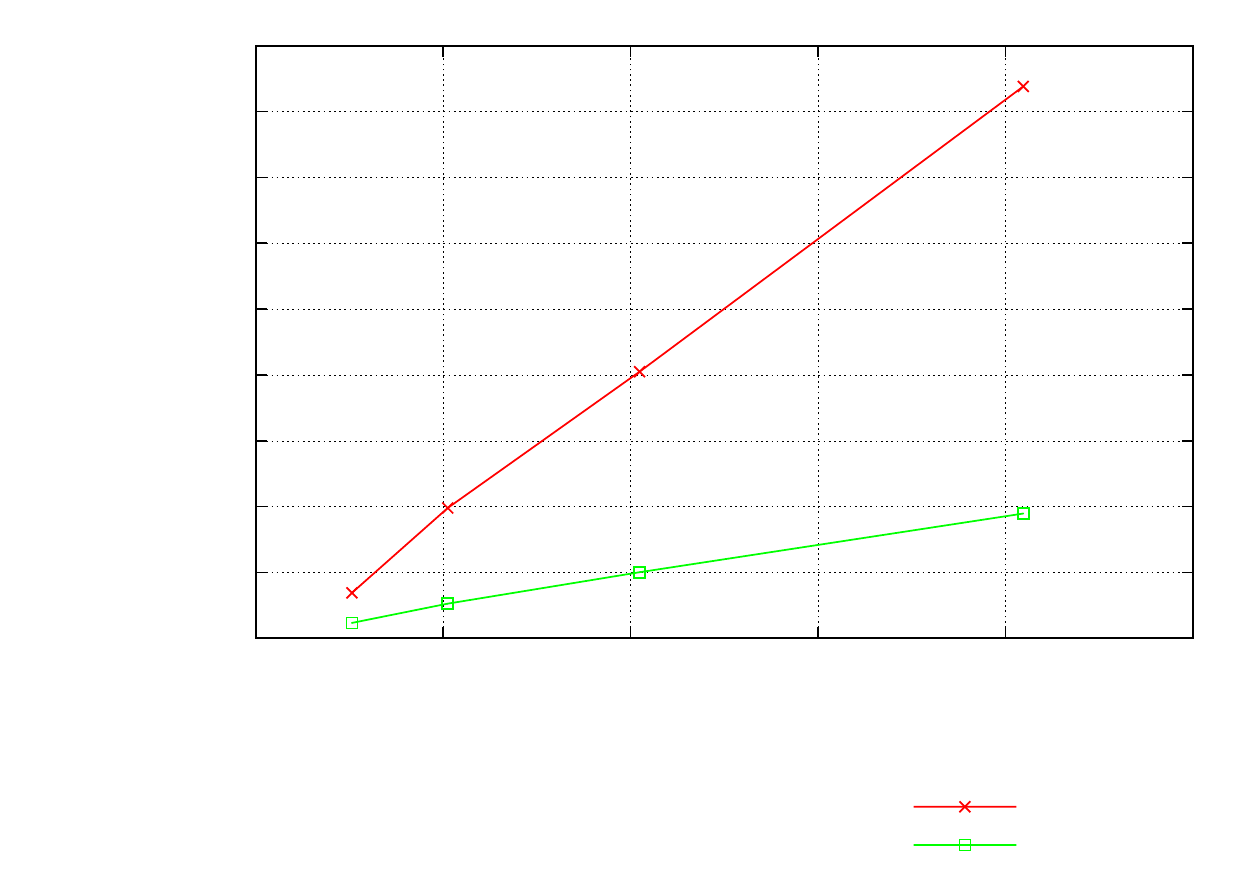}}%
    \gplfronttext
  \end{picture}%
\endgroup

%% file: psrs-plot-p1-tex.tex
\begin{figure}[ht]
	\begin{center}
	\resizebox{!}{0.4\textheight}{
		\input{psrs-p1-tex}
	}
	\\
$\frac{n}{v} = 128000000$
,  $P = 1$
,  k = 4
,  $\mu = 1024$ {\small MiB}

	\caption{PSRS PEMS2 P=1}
	\label{psrs-p1}
	\end{center}
\end{figure}

%% file: psrs-p1-tex.tex
\begingroup
  \makeatletter
  \providecommand\color[2][]{%
    \GenericError{(gnuplot) \space\space\space\@spaces}{%
      Package color not loaded in conjunction with
      terminal option `colourtext'%
    }{See the gnuplot documentation for explanation.%
    }{Either use 'blacktext' in gnuplot or load the package
      color.sty in LaTeX.}%
    \renewcommand\color[2][]{}%
  }%
  \providecommand\includegraphics[2][]{%
    \GenericError{(gnuplot) \space\space\space\@spaces}{%
      Package graphicx or graphics not loaded%
    }{See the gnuplot documentation for explanation.%
    }{The gnuplot epslatex terminal needs graphicx.sty or graphics.sty.}%
    \renewcommand\includegraphics[2][]{}%
  }%
  \providecommand\rotatebox[2]{#2}%
  \@ifundefined{ifGPcolor}{%
    \newif\ifGPcolor
    \GPcolortrue
  }{}%
  \@ifundefined{ifGPblacktext}{%
    \newif\ifGPblacktext
    \GPblacktexttrue
  }{}%
  \let\gplgaddtomacro\g@addto@macro
  \gdef\gplbacktext{}%
  \gdef\gplfronttext{}%
  \makeatother
  \ifGPblacktext
    \def\colorrgb#1{}%
    \def\colorgray#1{}%
  \else
    \ifGPcolor
      \def\colorrgb#1{\color[rgb]{#1}}%
      \def\colorgray#1{\color[gray]{#1}}%
      \expandafter\def\csname LTw\endcsname{\color{white}}%
      \expandafter\def\csname LTb\endcsname{\color{black}}%
      \expandafter\def\csname LTa\endcsname{\color{black}}%
      \expandafter\def\csname LT0\endcsname{\color[rgb]{1,0,0}}%
      \expandafter\def\csname LT1\endcsname{\color[rgb]{0,1,0}}%
      \expandafter\def\csname LT2\endcsname{\color[rgb]{0,0,1}}%
      \expandafter\def\csname LT3\endcsname{\color[rgb]{1,0,1}}%
      \expandafter\def\csname LT4\endcsname{\color[rgb]{0,1,1}}%
      \expandafter\def\csname LT5\endcsname{\color[rgb]{1,1,0}}%
      \expandafter\def\csname LT6\endcsname{\color[rgb]{0,0,0}}%
      \expandafter\def\csname LT7\endcsname{\color[rgb]{1,0.3,0}}%
      \expandafter\def\csname LT8\endcsname{\color[rgb]{0.5,0.5,0.5}}%
    \else
      \def\colorrgb#1{\color{black}}%
      \def\colorgray#1{\color[gray]{#1}}%
      \expandafter\def\csname LTw\endcsname{\color{white}}%
      \expandafter\def\csname LTb\endcsname{\color{black}}%
      \expandafter\def\csname LTa\endcsname{\color{black}}%
      \expandafter\def\csname LT0\endcsname{\color{black}}%
      \expandafter\def\csname LT1\endcsname{\color{black}}%
      \expandafter\def\csname LT2\endcsname{\color{black}}%
      \expandafter\def\csname LT3\endcsname{\color{black}}%
      \expandafter\def\csname LT4\endcsname{\color{black}}%
      \expandafter\def\csname LT5\endcsname{\color{black}}%
      \expandafter\def\csname LT6\endcsname{\color{black}}%
      \expandafter\def\csname LT7\endcsname{\color{black}}%
      \expandafter\def\csname LT8\endcsname{\color{black}}%
    \fi
  \fi
  \setlength{\unitlength}{0.0500bp}%
  \begin{picture}(7200.00,5040.00)%
    \gplgaddtomacro\gplbacktext{%
      \csname LTb\endcsname%
      \put(1474,1804){\makebox(0,0)[r]{\strut{} 0}}%
      \csname LTb\endcsname%
      \put(1474,2229){\makebox(0,0)[r]{\strut{} 10000}}%
      \csname LTb\endcsname%
      \put(1474,2653){\makebox(0,0)[r]{\strut{} 20000}}%
      \csname LTb\endcsname%
      \put(1474,3078){\makebox(0,0)[r]{\strut{} 30000}}%
      \csname LTb\endcsname%
      \put(1474,3502){\makebox(0,0)[r]{\strut{} 40000}}%
      \csname LTb\endcsname%
      \put(1474,3927){\makebox(0,0)[r]{\strut{} 50000}}%
      \csname LTb\endcsname%
      \put(1474,4351){\makebox(0,0)[r]{\strut{} 60000}}%
      \csname LTb\endcsname%
      \put(1474,4776){\makebox(0,0)[r]{\strut{} 70000}}%
      \csname LTb\endcsname%
      \put(1606,1584){\makebox(0,0){\strut{} 0}}%
      \csname LTb\endcsname%
      \put(2358,1584){\makebox(0,0){\strut{} 5}}%
      \csname LTb\endcsname%
      \put(3110,1584){\makebox(0,0){\strut{} 10}}%
      \csname LTb\endcsname%
      \put(3862,1584){\makebox(0,0){\strut{} 15}}%
      \csname LTb\endcsname%
      \put(4614,1584){\makebox(0,0){\strut{} 20}}%
      \csname LTb\endcsname%
      \put(5366,1584){\makebox(0,0){\strut{} 25}}%
      \csname LTb\endcsname%
      \put(6118,1584){\makebox(0,0){\strut{} 30}}%
      \csname LTb\endcsname%
      \put(6870,1584){\makebox(0,0){\strut{} 35}}%
      \put(440,3290){\rotatebox{90}{\makebox(0,0){\strut{}Seconds}}}%
      \put(4238,1254){\makebox(0,0){\strut{}$n$ (billions)}}%
    }%
    \gplgaddtomacro\gplfronttext{%
      \csname LTb\endcsname%
      \put(5593,833){\makebox(0,0)[r]{\strut{}PSRS PEMS2 (mmap) P=1}}%
      \csname LTb\endcsname%
      \put(5593,613){\makebox(0,0)[r]{\strut{}PSRS PEMS2 (stxxl-file) P=1}}%
      \csname LTb\endcsname%
      \put(5593,393){\makebox(0,0)[r]{\strut{}PSRS PEMS2 (unix) P=1}}%
      \csname LTb\endcsname%
      \put(5593,173){\makebox(0,0)[r]{\strut{}stxxl}}%
    }%
    \gplbacktext
    \put(0,0){\includegraphics{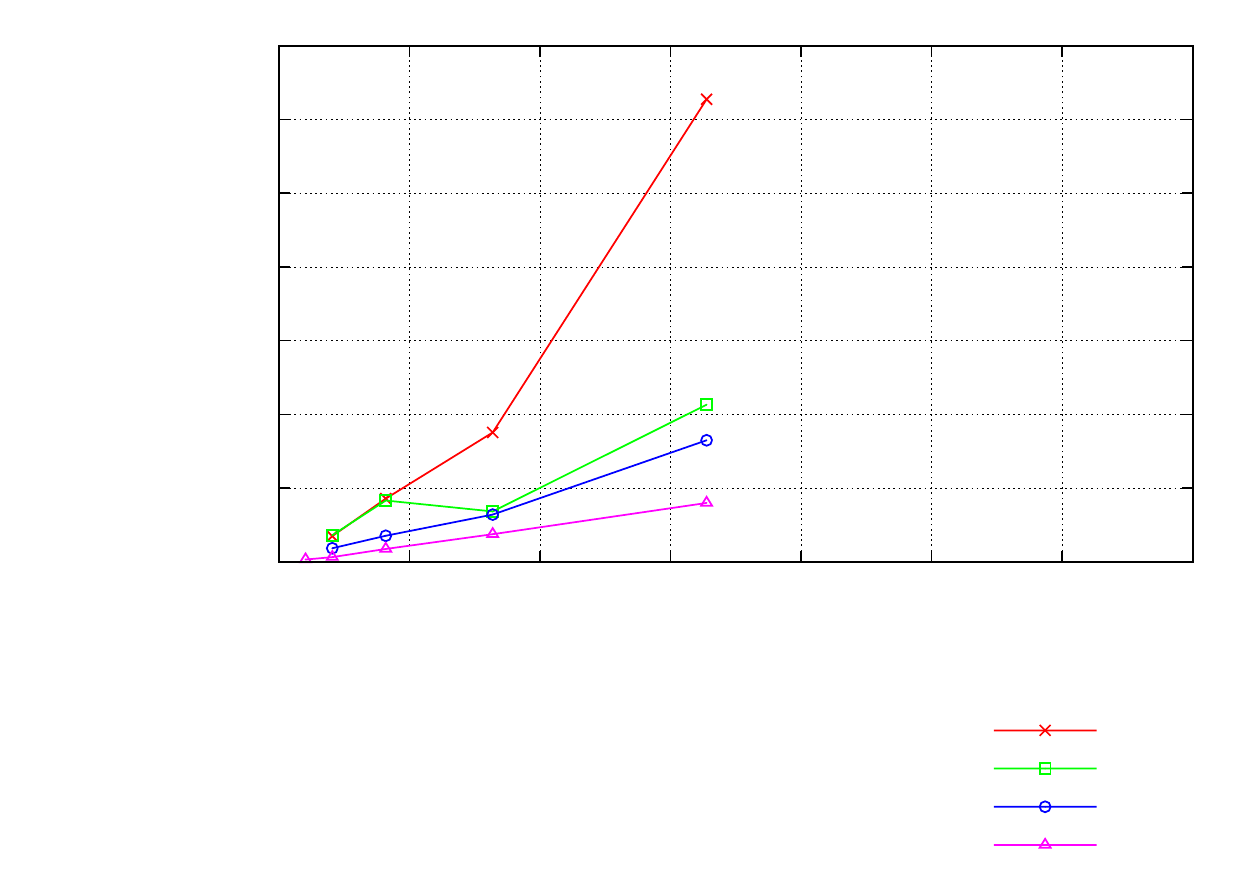}}%
    \gplfronttext
  \end{picture}%
\endgroup

%% file: psrs-plot-p2-tex.tex
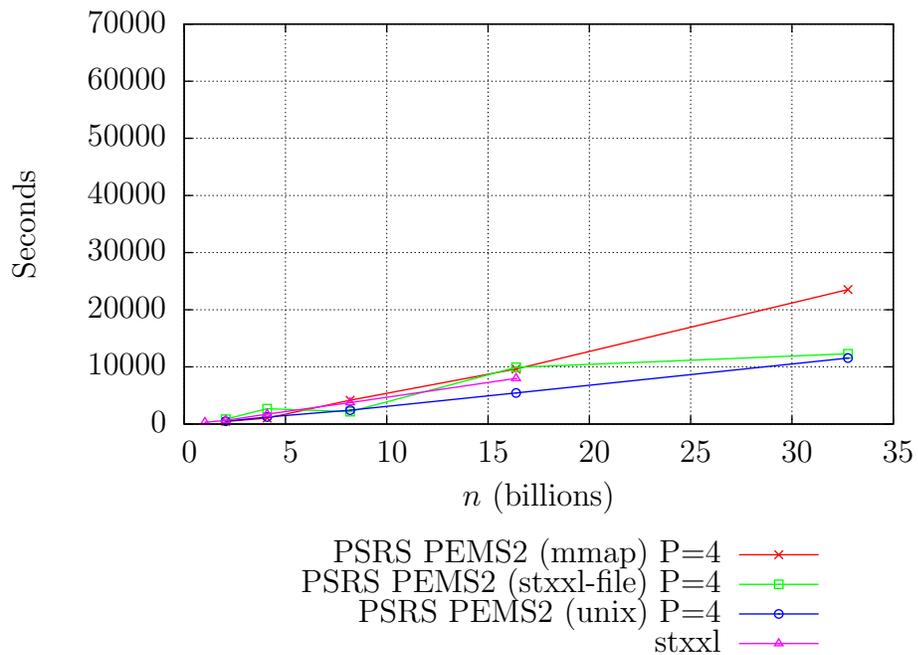
\begin{figure}[ht]
	\begin{center}
	\resizebox{!}{0.4\textheight}{
		\input{psrs-p2-tex}
	}
	\\
$\frac{n}{v} = 128000000$
,  $P = 2$
,  k = 4
,  $\mu = 1024$ {\small MiB}

	\caption{PSRS PEMS2 P=2}
	\label{psrs-p2}
	\end{center}
\end{figure}

%% file: psrs-p2-tex.tex
\begingroup
  \makeatletter
  \providecommand\color[2][]{%
    \GenericError{(gnuplot) \space\space\space\@spaces}{%
      Package color not loaded in conjunction with
      terminal option `colourtext'%
    }{See the gnuplot documentation for explanation.%
    }{Either use 'blacktext' in gnuplot or load the package
      color.sty in LaTeX.}%
    \renewcommand\color[2][]{}%
  }%
  \providecommand\includegraphics[2][]{%
    \GenericError{(gnuplot) \space\space\space\@spaces}{%
      Package graphicx or graphics not loaded%
    }{See the gnuplot documentation for explanation.%
    }{The gnuplot epslatex terminal needs graphicx.sty or graphics.sty.}%
    \renewcommand\includegraphics[2][]{}%
  }%
  \providecommand\rotatebox[2]{#2}%
  \@ifundefined{ifGPcolor}{%
    \newif\ifGPcolor
    \GPcolortrue
  }{}%
  \@ifundefined{ifGPblacktext}{%
    \newif\ifGPblacktext
    \GPblacktexttrue
  }{}%
  \let\gplgaddtomacro\g@addto@macro
  \gdef\gplbacktext{}%
  \gdef\gplfronttext{}%
  \makeatother
  \ifGPblacktext
    \def\colorrgb#1{}%
    \def\colorgray#1{}%
  \else
    \ifGPcolor
      \def\colorrgb#1{\color[rgb]{#1}}%
      \def\colorgray#1{\color[gray]{#1}}%
      \expandafter\def\csname LTw\endcsname{\color{white}}%
      \expandafter\def\csname LTb\endcsname{\color{black}}%
      \expandafter\def\csname LTa\endcsname{\color{black}}%
      \expandafter\def\csname LT0\endcsname{\color[rgb]{1,0,0}}%
      \expandafter\def\csname LT1\endcsname{\color[rgb]{0,1,0}}%
      \expandafter\def\csname LT2\endcsname{\color[rgb]{0,0,1}}%
      \expandafter\def\csname LT3\endcsname{\color[rgb]{1,0,1}}%
      \expandafter\def\csname LT4\endcsname{\color[rgb]{0,1,1}}%
      \expandafter\def\csname LT5\endcsname{\color[rgb]{1,1,0}}%
      \expandafter\def\csname LT6\endcsname{\color[rgb]{0,0,0}}%
      \expandafter\def\csname LT7\endcsname{\color[rgb]{1,0.3,0}}%
      \expandafter\def\csname LT8\endcsname{\color[rgb]{0.5,0.5,0.5}}%
    \else
      \def\colorrgb#1{\color{black}}%
      \def\colorgray#1{\color[gray]{#1}}%
      \expandafter\def\csname LTw\endcsname{\color{white}}%
      \expandafter\def\csname LTb\endcsname{\color{black}}%
      \expandafter\def\csname LTa\endcsname{\color{black}}%
      \expandafter\def\csname LT0\endcsname{\color{black}}%
      \expandafter\def\csname LT1\endcsname{\color{black}}%
      \expandafter\def\csname LT2\endcsname{\color{black}}%
      \expandafter\def\csname LT3\endcsname{\color{black}}%
      \expandafter\def\csname LT4\endcsname{\color{black}}%
      \expandafter\def\csname LT5\endcsname{\color{black}}%
      \expandafter\def\csname LT6\endcsname{\color{black}}%
      \expandafter\def\csname LT7\endcsname{\color{black}}%
      \expandafter\def\csname LT8\endcsname{\color{black}}%
    \fi
  \fi
  \setlength{\unitlength}{0.0500bp}%
  \begin{picture}(7200.00,5040.00)%
    \gplgaddtomacro\gplbacktext{%
      \csname LTb\endcsname%
      \put(1474,1804){\makebox(0,0)[r]{\strut{} 0}}%
      \csname LTb\endcsname%
      \put(1474,2229){\makebox(0,0)[r]{\strut{} 10000}}%
      \csname LTb\endcsname%
      \put(1474,2653){\makebox(0,0)[r]{\strut{} 20000}}%
      \csname LTb\endcsname%
      \put(1474,3078){\makebox(0,0)[r]{\strut{} 30000}}%
      \csname LTb\endcsname%
      \put(1474,3502){\makebox(0,0)[r]{\strut{} 40000}}%
      \csname LTb\endcsname%
      \put(1474,3927){\makebox(0,0)[r]{\strut{} 50000}}%
      \csname LTb\endcsname%
      \put(1474,4351){\makebox(0,0)[r]{\strut{} 60000}}%
      \csname LTb\endcsname%
      \put(1474,4776){\makebox(0,0)[r]{\strut{} 70000}}%
      \csname LTb\endcsname%
      \put(1606,1584){\makebox(0,0){\strut{} 0}}%
      \csname LTb\endcsname%
      \put(2358,1584){\makebox(0,0){\strut{} 5}}%
      \csname LTb\endcsname%
      \put(3110,1584){\makebox(0,0){\strut{} 10}}%
      \csname LTb\endcsname%
      \put(3862,1584){\makebox(0,0){\strut{} 15}}%
      \csname LTb\endcsname%
      \put(4614,1584){\makebox(0,0){\strut{} 20}}%
      \csname LTb\endcsname%
      \put(5366,1584){\makebox(0,0){\strut{} 25}}%
      \csname LTb\endcsname%
      \put(6118,1584){\makebox(0,0){\strut{} 30}}%
      \csname LTb\endcsname%
      \put(6870,1584){\makebox(0,0){\strut{} 35}}%
      \put(440,3290){\rotatebox{90}{\makebox(0,0){\strut{}Seconds}}}%
      \put(4238,1254){\makebox(0,0){\strut{}$n$ (billions)}}%
    }%
    \gplgaddtomacro\gplfronttext{%
      \csname LTb\endcsname%
      \put(5593,833){\makebox(0,0)[r]{\strut{}PSRS PEMS2 (mmap) P=2}}%
      \csname LTb\endcsname%
      \put(5593,613){\makebox(0,0)[r]{\strut{}PSRS PEMS2 (stxxl-file) P=2}}%
      \csname LTb\endcsname%
      \put(5593,393){\makebox(0,0)[r]{\strut{}PSRS PEMS2 (unix) P=2}}%
      \csname LTb\endcsname%
      \put(5593,173){\makebox(0,0)[r]{\strut{}stxxl}}%
    }%
    \gplbacktext
    \put(0,0){\includegraphics{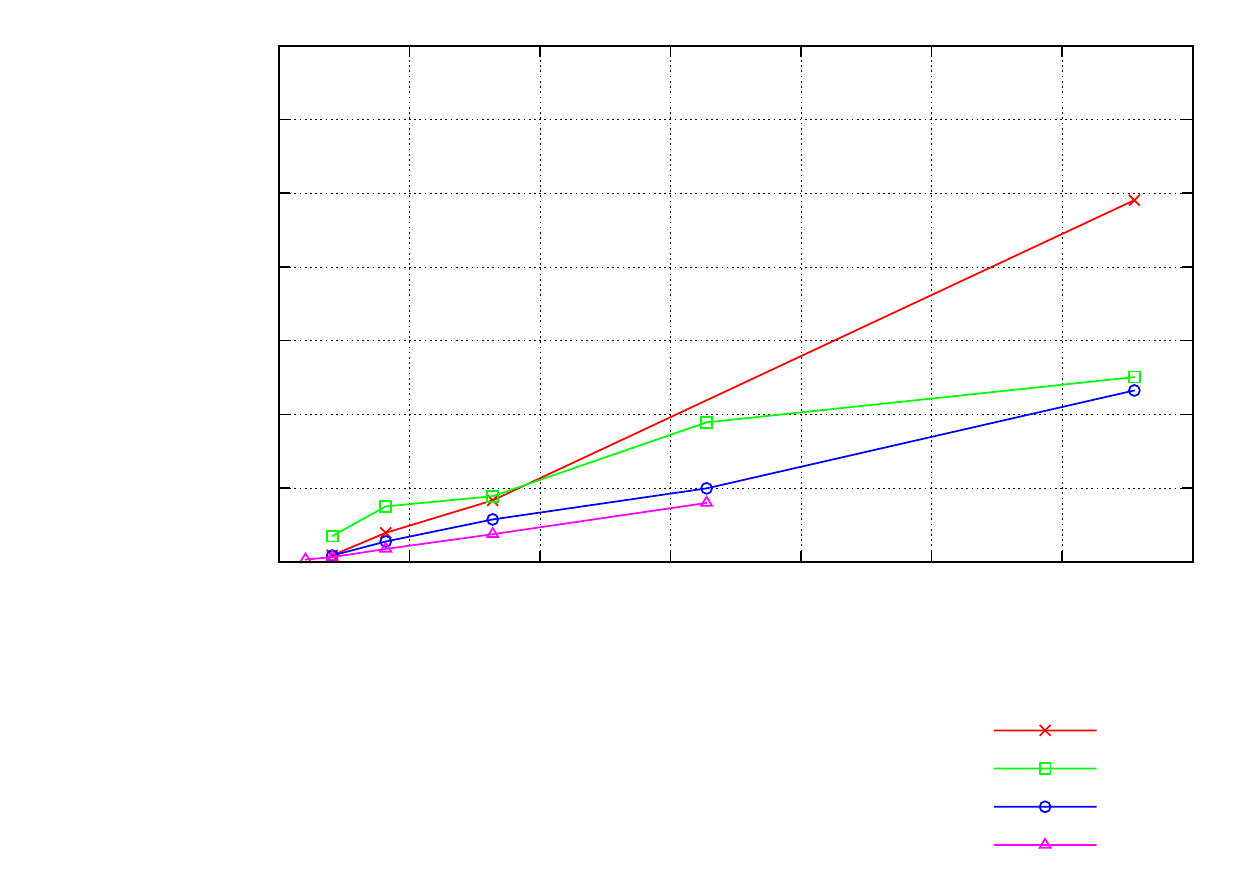}}%
    \gplfronttext
  \end{picture}%
\endgroup

%% file: psrs-plot-p4-tex.tex
\begin{figure}[ht]
	\begin{center}
	\resizebox{!}{0.4\textheight}{
		\input{psrs-p4-tex}
	}
	\\
$\frac{n}{v} = 128000000$
,  $P = 4$
,  k = 4
,  $\mu = 1024$ {\small MiB}

	\caption{PSRS PEMS2 P=4}
	\label{psrs-p4}
	\end{center}
\end{figure}

%% file: psrs-p4-tex.tex
\begingroup
  \makeatletter
  \providecommand\color[2][]{%
    \GenericError{(gnuplot) \space\space\space\@spaces}{%
      Package color not loaded in conjunction with
      terminal option `colourtext'%
    }{See the gnuplot documentation for explanation.%
    }{Either use 'blacktext' in gnuplot or load the package
      color.sty in LaTeX.}%
    \renewcommand\color[2][]{}%
  }%
  \providecommand\includegraphics[2][]{%
    \GenericError{(gnuplot) \space\space\space\@spaces}{%
      Package graphicx or graphics not loaded%
    }{See the gnuplot documentation for explanation.%
    }{The gnuplot epslatex terminal needs graphicx.sty or graphics.sty.}%
    \renewcommand\includegraphics[2][]{}%
  }%
  \providecommand\rotatebox[2]{#2}%
  \@ifundefined{ifGPcolor}{%
    \newif\ifGPcolor
    \GPcolortrue
  }{}%
  \@ifundefined{ifGPblacktext}{%
    \newif\ifGPblacktext
    \GPblacktexttrue
  }{}%
  \let\gplgaddtomacro\g@addto@macro
  \gdef\gplbacktext{}%
  \gdef\gplfronttext{}%
  \makeatother
  \ifGPblacktext
    \def\colorrgb#1{}%
    \def\colorgray#1{}%
  \else
    \ifGPcolor
      \def\colorrgb#1{\color[rgb]{#1}}%
      \def\colorgray#1{\color[gray]{#1}}%
      \expandafter\def\csname LTw\endcsname{\color{white}}%
      \expandafter\def\csname LTb\endcsname{\color{black}}%
      \expandafter\def\csname LTa\endcsname{\color{black}}%
      \expandafter\def\csname LT0\endcsname{\color[rgb]{1,0,0}}%
      \expandafter\def\csname LT1\endcsname{\color[rgb]{0,1,0}}%
      \expandafter\def\csname LT2\endcsname{\color[rgb]{0,0,1}}%
      \expandafter\def\csname LT3\endcsname{\color[rgb]{1,0,1}}%
      \expandafter\def\csname LT4\endcsname{\color[rgb]{0,1,1}}%
      \expandafter\def\csname LT5\endcsname{\color[rgb]{1,1,0}}%
      \expandafter\def\csname LT6\endcsname{\color[rgb]{0,0,0}}%
      \expandafter\def\csname LT7\endcsname{\color[rgb]{1,0.3,0}}%
      \expandafter\def\csname LT8\endcsname{\color[rgb]{0.5,0.5,0.5}}%
    \else
      \def\colorrgb#1{\color{black}}%
      \def\colorgray#1{\color[gray]{#1}}%
      \expandafter\def\csname LTw\endcsname{\color{white}}%
      \expandafter\def\csname LTb\endcsname{\color{black}}%
      \expandafter\def\csname LTa\endcsname{\color{black}}%
      \expandafter\def\csname LT0\endcsname{\color{black}}%
      \expandafter\def\csname LT1\endcsname{\color{black}}%
      \expandafter\def\csname LT2\endcsname{\color{black}}%
      \expandafter\def\csname LT3\endcsname{\color{black}}%
      \expandafter\def\csname LT4\endcsname{\color{black}}%
      \expandafter\def\csname LT5\endcsname{\color{black}}%
      \expandafter\def\csname LT6\endcsname{\color{black}}%
      \expandafter\def\csname LT7\endcsname{\color{black}}%
      \expandafter\def\csname LT8\endcsname{\color{black}}%
    \fi
  \fi
  \setlength{\unitlength}{0.0500bp}%
  \begin{picture}(7200.00,5040.00)%
    \gplgaddtomacro\gplbacktext{%
      \csname LTb\endcsname%
      \put(1474,1804){\makebox(0,0)[r]{\strut{} 0}}%
      \csname LTb\endcsname%
      \put(1474,2229){\makebox(0,0)[r]{\strut{} 10000}}%
      \csname LTb\endcsname%
      \put(1474,2653){\makebox(0,0)[r]{\strut{} 20000}}%
      \csname LTb\endcsname%
      \put(1474,3078){\makebox(0,0)[r]{\strut{} 30000}}%
      \csname LTb\endcsname%
      \put(1474,3502){\makebox(0,0)[r]{\strut{} 40000}}%
      \csname LTb\endcsname%
      \put(1474,3927){\makebox(0,0)[r]{\strut{} 50000}}%
      \csname LTb\endcsname%
      \put(1474,4351){\makebox(0,0)[r]{\strut{} 60000}}%
      \csname LTb\endcsname%
      \put(1474,4776){\makebox(0,0)[r]{\strut{} 70000}}%
      \csname LTb\endcsname%
      \put(1606,1584){\makebox(0,0){\strut{} 0}}%
      \csname LTb\endcsname%
      \put(2358,1584){\makebox(0,0){\strut{} 5}}%
      \csname LTb\endcsname%
      \put(3110,1584){\makebox(0,0){\strut{} 10}}%
      \csname LTb\endcsname%
      \put(3862,1584){\makebox(0,0){\strut{} 15}}%
      \csname LTb\endcsname%
      \put(4614,1584){\makebox(0,0){\strut{} 20}}%
      \csname LTb\endcsname%
      \put(5366,1584){\makebox(0,0){\strut{} 25}}%
      \csname LTb\endcsname%
      \put(6118,1584){\makebox(0,0){\strut{} 30}}%
      \csname LTb\endcsname%
      \put(6870,1584){\makebox(0,0){\strut{} 35}}%
      \put(440,3290){\rotatebox{90}{\makebox(0,0){\strut{}Seconds}}}%
      \put(4238,1254){\makebox(0,0){\strut{}$n$ (billions)}}%
    }%
    \gplgaddtomacro\gplfronttext{%
      \csname LTb\endcsname%
      \put(5593,833){\makebox(0,0)[r]{\strut{}PSRS PEMS2 (mmap) P=4}}%
      \csname LTb\endcsname%
      \put(5593,613){\makebox(0,0)[r]{\strut{}PSRS PEMS2 (stxxl-file) P=4}}%
      \csname LTb\endcsname%
      \put(5593,393){\makebox(0,0)[r]{\strut{}PSRS PEMS2 (unix) P=4}}%
      \csname LTb\endcsname%
      \put(5593,173){\makebox(0,0)[r]{\strut{}stxxl}}%
    }%
    \gplbacktext
    \put(0,0){\includegraphics{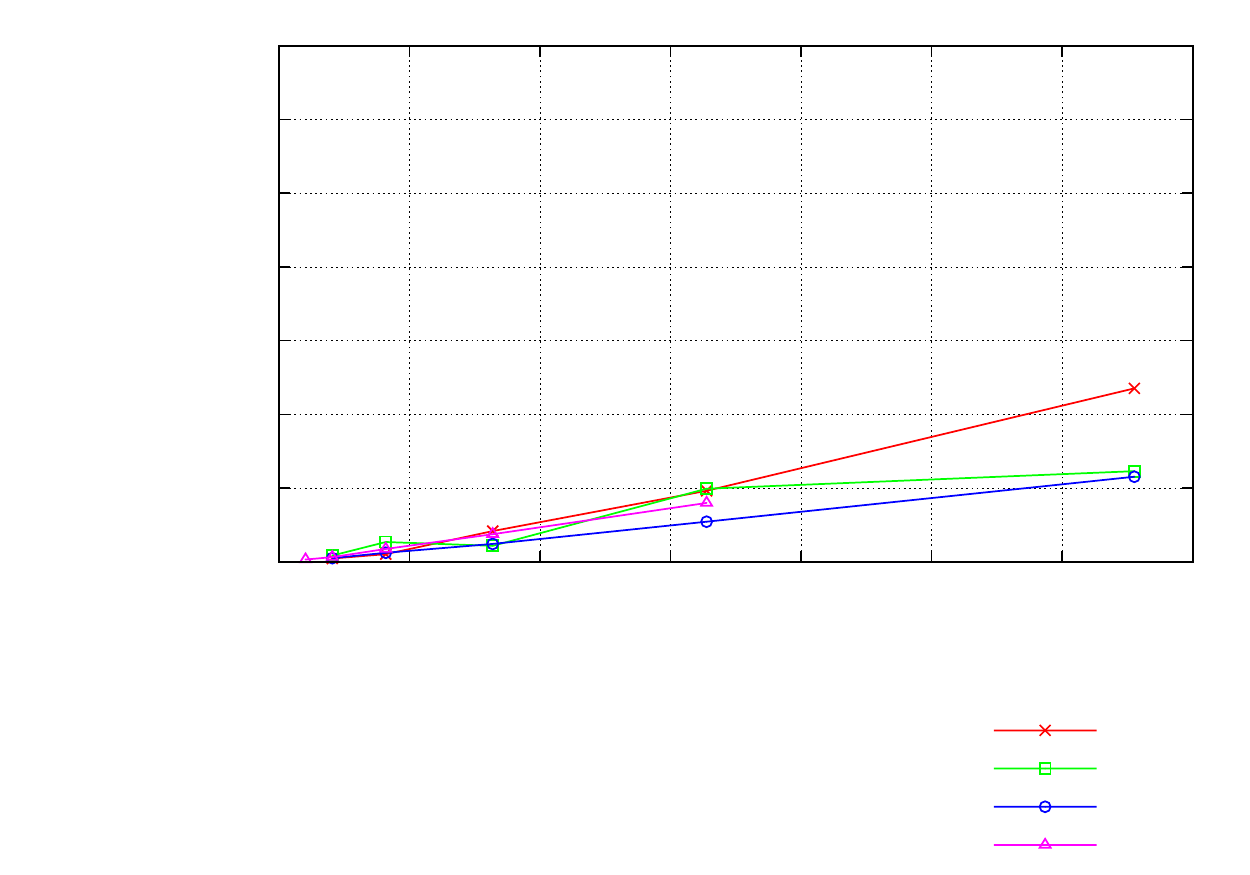}}%
    \gplfronttext
  \end{picture}%
\endgroup

%% file: psrs-plot-p8-tex.tex
\begin{figure}[ht]
	\begin{center}
	\resizebox{!}{0.4\textheight}{
		\input{psrs-p8-tex}
	}
	\\
$\frac{n}{v} = 128000000$
,  $P = 8$
,  $\mu = 1024$ {\small MiB}

	\caption{PSRS PEMS2 P=8}
	\label{psrs-p8}
	\end{center}
\end{figure}

%% file: psrs-p8-tex.tex
\begingroup
  \makeatletter
  \providecommand\color[2][]{%
    \GenericError{(gnuplot) \space\space\space\@spaces}{%
      Package color not loaded in conjunction with
      terminal option `colourtext'%
    }{See the gnuplot documentation for explanation.%
    }{Either use 'blacktext' in gnuplot or load the package
      color.sty in LaTeX.}%
    \renewcommand\color[2][]{}%
  }%
  \providecommand\includegraphics[2][]{%
    \GenericError{(gnuplot) \space\space\space\@spaces}{%
      Package graphicx or graphics not loaded%
    }{See the gnuplot documentation for explanation.%
    }{The gnuplot epslatex terminal needs graphicx.sty or graphics.sty.}%
    \renewcommand\includegraphics[2][]{}%
  }%
  \providecommand\rotatebox[2]{#2}%
  \@ifundefined{ifGPcolor}{%
    \newif\ifGPcolor
    \GPcolortrue
  }{}%
  \@ifundefined{ifGPblacktext}{%
    \newif\ifGPblacktext
    \GPblacktexttrue
  }{}%
  \let\gplgaddtomacro\g@addto@macro
  \gdef\gplbacktext{}%
  \gdef\gplfronttext{}%
  \makeatother
  \ifGPblacktext
    \def\colorrgb#1{}%
    \def\colorgray#1{}%
  \else
    \ifGPcolor
      \def\colorrgb#1{\color[rgb]{#1}}%
      \def\colorgray#1{\color[gray]{#1}}%
      \expandafter\def\csname LTw\endcsname{\color{white}}%
      \expandafter\def\csname LTb\endcsname{\color{black}}%
      \expandafter\def\csname LTa\endcsname{\color{black}}%
      \expandafter\def\csname LT0\endcsname{\color[rgb]{1,0,0}}%
      \expandafter\def\csname LT1\endcsname{\color[rgb]{0,1,0}}%
      \expandafter\def\csname LT2\endcsname{\color[rgb]{0,0,1}}%
      \expandafter\def\csname LT3\endcsname{\color[rgb]{1,0,1}}%
      \expandafter\def\csname LT4\endcsname{\color[rgb]{0,1,1}}%
      \expandafter\def\csname LT5\endcsname{\color[rgb]{1,1,0}}%
      \expandafter\def\csname LT6\endcsname{\color[rgb]{0,0,0}}%
      \expandafter\def\csname LT7\endcsname{\color[rgb]{1,0.3,0}}%
      \expandafter\def\csname LT8\endcsname{\color[rgb]{0.5,0.5,0.5}}%
    \else
      \def\colorrgb#1{\color{black}}%
      \def\colorgray#1{\color[gray]{#1}}%
      \expandafter\def\csname LTw\endcsname{\color{white}}%
      \expandafter\def\csname LTb\endcsname{\color{black}}%
      \expandafter\def\csname LTa\endcsname{\color{black}}%
      \expandafter\def\csname LT0\endcsname{\color{black}}%
      \expandafter\def\csname LT1\endcsname{\color{black}}%
      \expandafter\def\csname LT2\endcsname{\color{black}}%
      \expandafter\def\csname LT3\endcsname{\color{black}}%
      \expandafter\def\csname LT4\endcsname{\color{black}}%
      \expandafter\def\csname LT5\endcsname{\color{black}}%
      \expandafter\def\csname LT6\endcsname{\color{black}}%
      \expandafter\def\csname LT7\endcsname{\color{black}}%
      \expandafter\def\csname LT8\endcsname{\color{black}}%
    \fi
  \fi
  \setlength{\unitlength}{0.0500bp}%
  \begin{picture}(7200.00,5040.00)%
    \gplgaddtomacro\gplbacktext{%
      \csname LTb\endcsname%
      \put(1474,1804){\makebox(0,0)[r]{\strut{} 0}}%
      \csname LTb\endcsname%
      \put(1474,2229){\makebox(0,0)[r]{\strut{} 10000}}%
      \csname LTb\endcsname%
      \put(1474,2653){\makebox(0,0)[r]{\strut{} 20000}}%
      \csname LTb\endcsname%
      \put(1474,3078){\makebox(0,0)[r]{\strut{} 30000}}%
      \csname LTb\endcsname%
      \put(1474,3502){\makebox(0,0)[r]{\strut{} 40000}}%
      \csname LTb\endcsname%
      \put(1474,3927){\makebox(0,0)[r]{\strut{} 50000}}%
      \csname LTb\endcsname%
      \put(1474,4351){\makebox(0,0)[r]{\strut{} 60000}}%
      \csname LTb\endcsname%
      \put(1474,4776){\makebox(0,0)[r]{\strut{} 70000}}%
      \csname LTb\endcsname%
      \put(1606,1584){\makebox(0,0){\strut{} 0}}%
      \csname LTb\endcsname%
      \put(2358,1584){\makebox(0,0){\strut{} 5}}%
      \csname LTb\endcsname%
      \put(3110,1584){\makebox(0,0){\strut{} 10}}%
      \csname LTb\endcsname%
      \put(3862,1584){\makebox(0,0){\strut{} 15}}%
      \csname LTb\endcsname%
      \put(4614,1584){\makebox(0,0){\strut{} 20}}%
      \csname LTb\endcsname%
      \put(5366,1584){\makebox(0,0){\strut{} 25}}%
      \csname LTb\endcsname%
      \put(6118,1584){\makebox(0,0){\strut{} 30}}%
      \csname LTb\endcsname%
      \put(6870,1584){\makebox(0,0){\strut{} 35}}%
      \put(440,3290){\rotatebox{90}{\makebox(0,0){\strut{}Seconds}}}%
      \put(4238,1254){\makebox(0,0){\strut{}$n$ (billions)}}%
    }%
    \gplgaddtomacro\gplfronttext{%
      \csname LTb\endcsname%
      \put(5593,833){\makebox(0,0)[r]{\strut{}PSRS PEMS2 (mmap) P=8}}%
      \csname LTb\endcsname%
      \put(5593,613){\makebox(0,0)[r]{\strut{}PSRS PEMS2 (stxxl-file) P=8}}%
      \csname LTb\endcsname%
      \put(5593,393){\makebox(0,0)[r]{\strut{}PSRS PEMS2 (unix) P=8}}%
      \csname LTb\endcsname%
      \put(5593,173){\makebox(0,0)[r]{\strut{}stxxl}}%
    }%
    \gplbacktext
    \put(0,0){\includegraphics{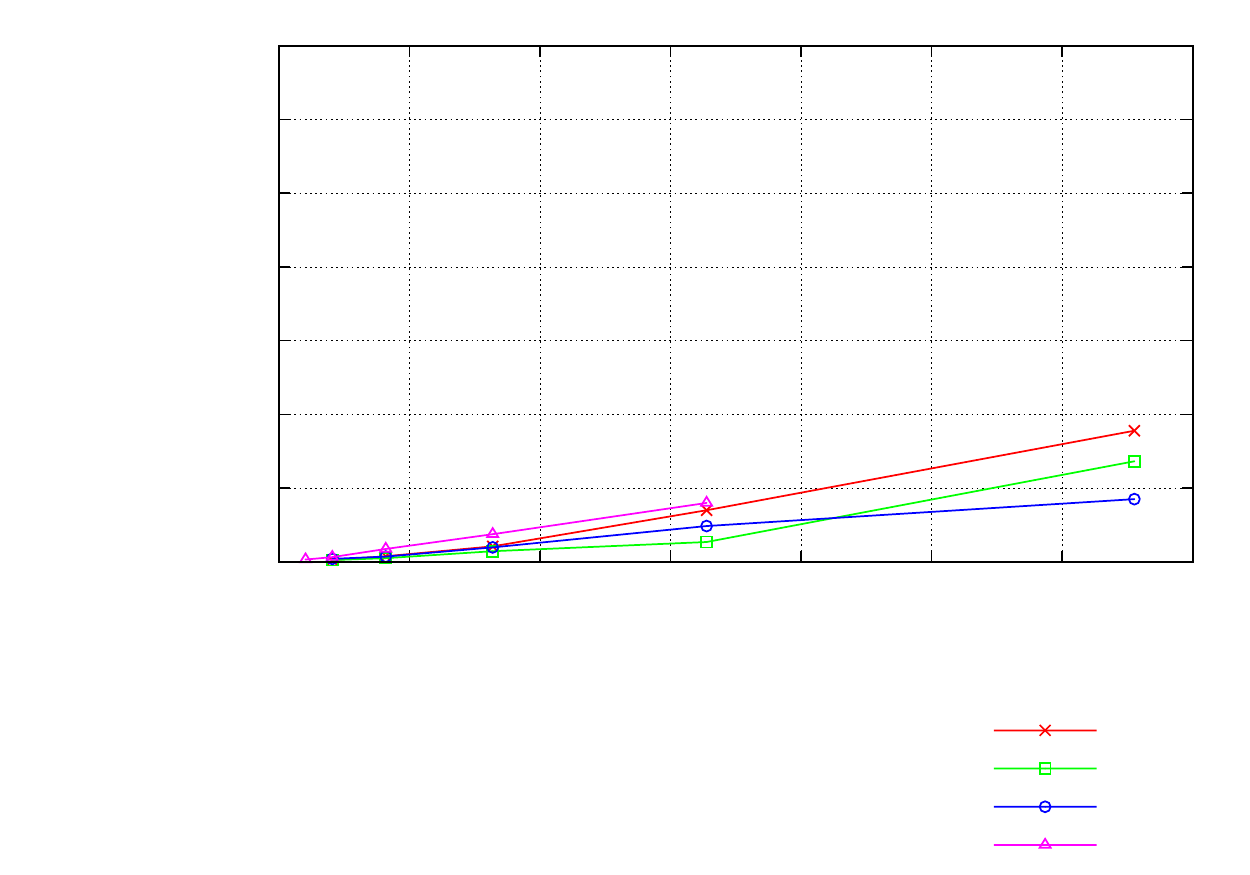}}%
    \gplfronttext
  \end{picture}%
\endgroup

%% file: final_psrs-unix-internal-v-tex.tex
\begingroup
  \makeatletter
  \providecommand\color[2][]{%
    \GenericError{(gnuplot) \space\space\space\@spaces}{%
      Package color not loaded in conjunction with
      terminal option `colourtext'%
    }{See the gnuplot documentation for explanation.%
    }{Either use 'blacktext' in gnuplot or load the package
      color.sty in LaTeX.}%
    \renewcommand\color[2][]{}%
  }%
  \providecommand\includegraphics[2][]{%
    \GenericError{(gnuplot) \space\space\space\@spaces}{%
      Package graphicx or graphics not loaded%
    }{See the gnuplot documentation for explanation.%
    }{The gnuplot epslatex terminal needs graphicx.sty or graphics.sty.}%
    \renewcommand\includegraphics[2][]{}%
  }%
  \providecommand\rotatebox[2]{#2}%
  \@ifundefined{ifGPcolor}{%
    \newif\ifGPcolor
    \GPcolortrue
  }{}%
  \@ifundefined{ifGPblacktext}{%
    \newif\ifGPblacktext
    \GPblacktexttrue
  }{}%
  \let\gplgaddtomacro\g@addto@macro
  \gdef\gplbacktext{}%
  \gdef\gplfronttext{}%
  \makeatother
  \ifGPblacktext
    \def\colorrgb#1{}%
    \def\colorgray#1{}%
  \else
    \ifGPcolor
      \def\colorrgb#1{\color[rgb]{#1}}%
      \def\colorgray#1{\color[gray]{#1}}%
      \expandafter\def\csname LTw\endcsname{\color{white}}%
      \expandafter\def\csname LTb\endcsname{\color{black}}%
      \expandafter\def\csname LTa\endcsname{\color{black}}%
      \expandafter\def\csname LT0\endcsname{\color[rgb]{1,0,0}}%
      \expandafter\def\csname LT1\endcsname{\color[rgb]{0,1,0}}%
      \expandafter\def\csname LT2\endcsname{\color[rgb]{0,0,1}}%
      \expandafter\def\csname LT3\endcsname{\color[rgb]{1,0,1}}%
      \expandafter\def\csname LT4\endcsname{\color[rgb]{0,1,1}}%
      \expandafter\def\csname LT5\endcsname{\color[rgb]{1,1,0}}%
      \expandafter\def\csname LT6\endcsname{\color[rgb]{0,0,0}}%
      \expandafter\def\csname LT7\endcsname{\color[rgb]{1,0.3,0}}%
      \expandafter\def\csname LT8\endcsname{\color[rgb]{0.5,0.5,0.5}}%
    \else
      \def\colorrgb#1{\color{black}}%
      \def\colorgray#1{\color[gray]{#1}}%
      \expandafter\def\csname LTw\endcsname{\color{white}}%
      \expandafter\def\csname LTb\endcsname{\color{black}}%
      \expandafter\def\csname LTa\endcsname{\color{black}}%
      \expandafter\def\csname LT0\endcsname{\color{black}}%
      \expandafter\def\csname LT1\endcsname{\color{black}}%
      \expandafter\def\csname LT2\endcsname{\color{black}}%
      \expandafter\def\csname LT3\endcsname{\color{black}}%
      \expandafter\def\csname LT4\endcsname{\color{black}}%
      \expandafter\def\csname LT5\endcsname{\color{black}}%
      \expandafter\def\csname LT6\endcsname{\color{black}}%
      \expandafter\def\csname LT7\endcsname{\color{black}}%
      \expandafter\def\csname LT8\endcsname{\color{black}}%
    \fi
  \fi
  \setlength{\unitlength}{0.0500bp}%
  \begin{picture}(7200.00,5040.00)%
    \gplgaddtomacro\gplbacktext{%
      \csname LTb\endcsname%
      \put(968,1440){\makebox(0,0)[r]{\strut{} 0}}%
      \csname LTb\endcsname%
      \put(968,2141){\makebox(0,0)[r]{\strut{} 5000}}%
      \csname LTb\endcsname%
      \put(968,2842){\makebox(0,0)[r]{\strut{} 10000}}%
      \csname LTb\endcsname%
      \put(968,3542){\makebox(0,0)[r]{\strut{} 15000}}%
      \csname LTb\endcsname%
      \put(968,4243){\makebox(0,0)[r]{\strut{} 20000}}%
      \csname LTb\endcsname%
      \put(968,4944){\makebox(0,0)[r]{\strut{} 25000}}%
      \csname LTb\endcsname%
      \put(1016,1392){\rotatebox{-50}{\makebox(0,0)[l]{\strut{}\scriptsize \textbf{Init}}}}%
      \csname LTb\endcsname%
      \put(1218,1392){\rotatebox{-50}{\makebox(0,0)[l]{\strut{}\scriptsize \textbf{Benchmark Start}}}}%
      \csname LTb\endcsname%
      \put(1420,1392){\rotatebox{-50}{\makebox(0,0)[l]{\strut{}\scriptsize \textbf{Gather~Start}}}}%
      \csname LTb\endcsname%
      \put(1622,1392){\rotatebox{-50}{\makebox(0,0)[l]{\strut{}\tiny Finish Step~1}}}%
      \csname LTb\endcsname%
      \put(1825,1392){\rotatebox{-50}{\makebox(0,0)[l]{\strut{}\tiny Start Step~2}}}%
      \csname LTb\endcsname%
      \put(2027,1392){\rotatebox{-50}{\makebox(0,0)[l]{\strut{}\scriptsize \textbf{Gather~End}}}}%
      \csname LTb\endcsname%
      \put(2229,1392){\rotatebox{-50}{\makebox(0,0)[l]{\strut{}\scriptsize \textbf{Bcast~Start}}}}%
      \csname LTb\endcsname%
      \put(2431,1392){\rotatebox{-50}{\makebox(0,0)[l]{\strut{}\tiny RootWait Start}}}%
      \csname LTb\endcsname%
      \put(2633,1392){\rotatebox{-50}{\makebox(0,0)[l]{\strut{}\tiny RootWait End}}}%
      \csname LTb\endcsname%
      \put(2835,1392){\rotatebox{-50}{\makebox(0,0)[l]{\strut{}\tiny Finish Step~1}}}%
      \csname LTb\endcsname%
      \put(3037,1392){\rotatebox{-50}{\makebox(0,0)[l]{\strut{}\tiny Start Step~2}}}%
      \csname LTb\endcsname%
      \put(3239,1392){\rotatebox{-50}{\makebox(0,0)[l]{\strut{}\scriptsize \textbf{Bcast~End}}}}%
      \csname LTb\endcsname%
      \put(3442,1392){\rotatebox{-50}{\makebox(0,0)[l]{\strut{}\scriptsize \textbf{Alltoall~Start}}}}%
      \csname LTb\endcsname%
      \put(3644,1392){\rotatebox{-50}{\makebox(0,0)[l]{\strut{}\tiny Finish Step~1}}}%
      \csname LTb\endcsname%
      \put(3846,1392){\rotatebox{-50}{\makebox(0,0)[l]{\strut{}\tiny Start Step~2}}}%
      \csname LTb\endcsname%
      \put(4048,1392){\rotatebox{-50}{\makebox(0,0)[l]{\strut{}\tiny Comm Start}}}%
      \csname LTb\endcsname%
      \put(4250,1392){\rotatebox{-50}{\makebox(0,0)[l]{\strut{}\tiny Deliver Start}}}%
      \csname LTb\endcsname%
      \put(4452,1392){\rotatebox{-50}{\makebox(0,0)[l]{\strut{}\tiny Deliver Finish}}}%
      \csname LTb\endcsname%
      \put(4654,1392){\rotatebox{-50}{\makebox(0,0)[l]{\strut{}\tiny Finish Step~2}}}%
      \csname LTb\endcsname%
      \put(4857,1392){\rotatebox{-50}{\makebox(0,0)[l]{\strut{}\tiny Start Step~3}}}%
      \csname LTb\endcsname%
      \put(5059,1392){\rotatebox{-50}{\makebox(0,0)[l]{\strut{}\scriptsize \textbf{Alltoall~End}}}}%
      \csname LTb\endcsname%
      \put(5261,1392){\rotatebox{-50}{\makebox(0,0)[l]{\strut{}\scriptsize \textbf{Alltoallv~Start}}}}%
      \csname LTb\endcsname%
      \put(5463,1392){\rotatebox{-50}{\makebox(0,0)[l]{\strut{}\tiny Finish Step~1}}}%
      \csname LTb\endcsname%
      \put(5665,1392){\rotatebox{-50}{\makebox(0,0)[l]{\strut{}\tiny Start Step~2}}}%
      \csname LTb\endcsname%
      \put(5867,1392){\rotatebox{-50}{\makebox(0,0)[l]{\strut{}\tiny Comm Start}}}%
      \csname LTb\endcsname%
      \put(6069,1392){\rotatebox{-50}{\makebox(0,0)[l]{\strut{}\tiny Comm Finish}}}%
      \csname LTb\endcsname%
      \put(6271,1392){\rotatebox{-50}{\makebox(0,0)[l]{\strut{}\tiny Finish Step~2}}}%
      \csname LTb\endcsname%
      \put(6474,1392){\rotatebox{-50}{\makebox(0,0)[l]{\strut{}\tiny Start Step~3}}}%
      \csname LTb\endcsname%
      \put(6676,1392){\rotatebox{-50}{\makebox(0,0)[l]{\strut{}\scriptsize \textbf{Alltoallv~End}}}}%
      \csname LTb\endcsname%
      \put(6878,1392){\rotatebox{-50}{\makebox(0,0)[l]{\strut{}\scriptsize \textbf{Benchmark Finish}}}}%
      \csname LTb\endcsname%
      \put(7080,1392){\rotatebox{-50}{\makebox(0,0)[l]{\strut{}\scriptsize \textbf{Finalize}}}}%
      \put(160,3192){\rotatebox{90}{\makebox(0,0){\strut{}Elapsed Time (s)}}}%
      \put(4048,280){\makebox(0,0){\strut{}Algorithm Progress}}%
      \put(4048,200){\makebox(0,0){\strut{}$\rightarrow$}}%
    }%
    \gplgaddtomacro\gplfronttext{%
    }%
    \gplbacktext
    \put(0,0){\includegraphics{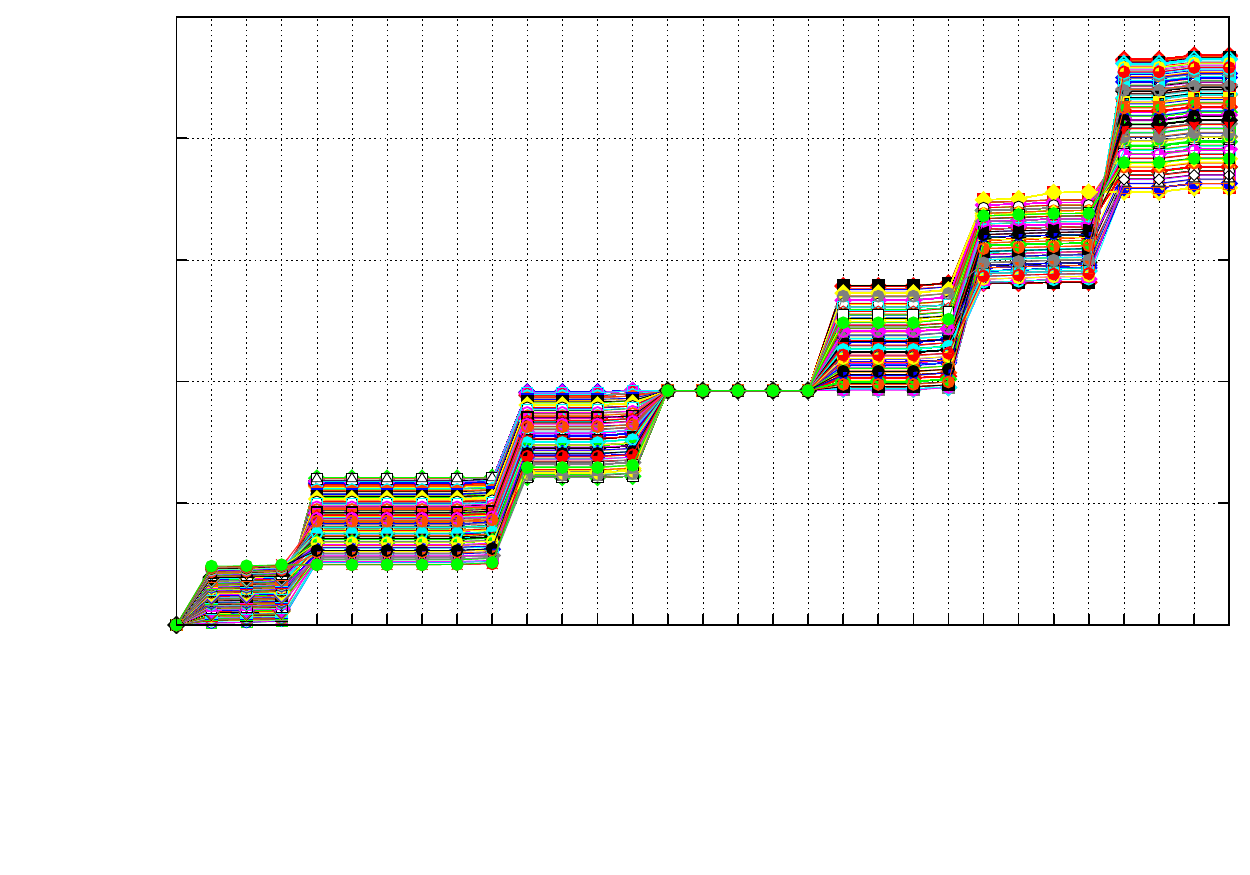}}%
    \gplfronttext
  \end{picture}%
\endgroup

%% file: final_psrs-stxxl-file-internal-v-tex.tex
\begingroup
  \makeatletter
  \providecommand\color[2][]{%
    \GenericError{(gnuplot) \space\space\space\@spaces}{%
      Package color not loaded in conjunction with
      terminal option `colourtext'%
    }{See the gnuplot documentation for explanation.%
    }{Either use 'blacktext' in gnuplot or load the package
      color.sty in LaTeX.}%
    \renewcommand\color[2][]{}%
  }%
  \providecommand\includegraphics[2][]{%
    \GenericError{(gnuplot) \space\space\space\@spaces}{%
      Package graphicx or graphics not loaded%
    }{See the gnuplot documentation for explanation.%
    }{The gnuplot epslatex terminal needs graphicx.sty or graphics.sty.}%
    \renewcommand\includegraphics[2][]{}%
  }%
  \providecommand\rotatebox[2]{#2}%
  \@ifundefined{ifGPcolor}{%
    \newif\ifGPcolor
    \GPcolortrue
  }{}%
  \@ifundefined{ifGPblacktext}{%
    \newif\ifGPblacktext
    \GPblacktexttrue
  }{}%
  \let\gplgaddtomacro\g@addto@macro
  \gdef\gplbacktext{}%
  \gdef\gplfronttext{}%
  \makeatother
  \ifGPblacktext
    \def\colorrgb#1{}%
    \def\colorgray#1{}%
  \else
    \ifGPcolor
      \def\colorrgb#1{\color[rgb]{#1}}%
      \def\colorgray#1{\color[gray]{#1}}%
      \expandafter\def\csname LTw\endcsname{\color{white}}%
      \expandafter\def\csname LTb\endcsname{\color{black}}%
      \expandafter\def\csname LTa\endcsname{\color{black}}%
      \expandafter\def\csname LT0\endcsname{\color[rgb]{1,0,0}}%
      \expandafter\def\csname LT1\endcsname{\color[rgb]{0,1,0}}%
      \expandafter\def\csname LT2\endcsname{\color[rgb]{0,0,1}}%
      \expandafter\def\csname LT3\endcsname{\color[rgb]{1,0,1}}%
      \expandafter\def\csname LT4\endcsname{\color[rgb]{0,1,1}}%
      \expandafter\def\csname LT5\endcsname{\color[rgb]{1,1,0}}%
      \expandafter\def\csname LT6\endcsname{\color[rgb]{0,0,0}}%
      \expandafter\def\csname LT7\endcsname{\color[rgb]{1,0.3,0}}%
      \expandafter\def\csname LT8\endcsname{\color[rgb]{0.5,0.5,0.5}}%
    \else
      \def\colorrgb#1{\color{black}}%
      \def\colorgray#1{\color[gray]{#1}}%
      \expandafter\def\csname LTw\endcsname{\color{white}}%
      \expandafter\def\csname LTb\endcsname{\color{black}}%
      \expandafter\def\csname LTa\endcsname{\color{black}}%
      \expandafter\def\csname LT0\endcsname{\color{black}}%
      \expandafter\def\csname LT1\endcsname{\color{black}}%
      \expandafter\def\csname LT2\endcsname{\color{black}}%
      \expandafter\def\csname LT3\endcsname{\color{black}}%
      \expandafter\def\csname LT4\endcsname{\color{black}}%
      \expandafter\def\csname LT5\endcsname{\color{black}}%
      \expandafter\def\csname LT6\endcsname{\color{black}}%
      \expandafter\def\csname LT7\endcsname{\color{black}}%
      \expandafter\def\csname LT8\endcsname{\color{black}}%
    \fi
  \fi
  \setlength{\unitlength}{0.0500bp}%
  \begin{picture}(7200.00,5040.00)%
    \gplgaddtomacro\gplbacktext{%
      \csname LTb\endcsname%
      \put(968,1440){\makebox(0,0)[r]{\strut{} 0}}%
      \csname LTb\endcsname%
      \put(968,1790){\makebox(0,0)[r]{\strut{} 2000}}%
      \csname LTb\endcsname%
      \put(968,2141){\makebox(0,0)[r]{\strut{} 4000}}%
      \csname LTb\endcsname%
      \put(968,2491){\makebox(0,0)[r]{\strut{} 6000}}%
      \csname LTb\endcsname%
      \put(968,2842){\makebox(0,0)[r]{\strut{} 8000}}%
      \csname LTb\endcsname%
      \put(968,3192){\makebox(0,0)[r]{\strut{} 10000}}%
      \csname LTb\endcsname%
      \put(968,3542){\makebox(0,0)[r]{\strut{} 12000}}%
      \csname LTb\endcsname%
      \put(968,3893){\makebox(0,0)[r]{\strut{} 14000}}%
      \csname LTb\endcsname%
      \put(968,4243){\makebox(0,0)[r]{\strut{} 16000}}%
      \csname LTb\endcsname%
      \put(968,4594){\makebox(0,0)[r]{\strut{} 18000}}%
      \csname LTb\endcsname%
      \put(968,4944){\makebox(0,0)[r]{\strut{} 20000}}%
      \csname LTb\endcsname%
      \put(1016,1392){\rotatebox{-50}{\makebox(0,0)[l]{\strut{}\scriptsize \textbf{Init}}}}%
      \csname LTb\endcsname%
      \put(1218,1392){\rotatebox{-50}{\makebox(0,0)[l]{\strut{}\scriptsize \textbf{Benchmark Start}}}}%
      \csname LTb\endcsname%
      \put(1420,1392){\rotatebox{-50}{\makebox(0,0)[l]{\strut{}\scriptsize \textbf{Gather~Start}}}}%
      \csname LTb\endcsname%
      \put(1622,1392){\rotatebox{-50}{\makebox(0,0)[l]{\strut{}\tiny Finish Step~1}}}%
      \csname LTb\endcsname%
      \put(1825,1392){\rotatebox{-50}{\makebox(0,0)[l]{\strut{}\tiny Start Step~2}}}%
      \csname LTb\endcsname%
      \put(2027,1392){\rotatebox{-50}{\makebox(0,0)[l]{\strut{}\scriptsize \textbf{Gather~End}}}}%
      \csname LTb\endcsname%
      \put(2229,1392){\rotatebox{-50}{\makebox(0,0)[l]{\strut{}\scriptsize \textbf{Bcast~Start}}}}%
      \csname LTb\endcsname%
      \put(2431,1392){\rotatebox{-50}{\makebox(0,0)[l]{\strut{}\tiny RootWait Start}}}%
      \csname LTb\endcsname%
      \put(2633,1392){\rotatebox{-50}{\makebox(0,0)[l]{\strut{}\tiny RootWait End}}}%
      \csname LTb\endcsname%
      \put(2835,1392){\rotatebox{-50}{\makebox(0,0)[l]{\strut{}\tiny Finish Step~1}}}%
      \csname LTb\endcsname%
      \put(3037,1392){\rotatebox{-50}{\makebox(0,0)[l]{\strut{}\tiny Start Step~2}}}%
      \csname LTb\endcsname%
      \put(3239,1392){\rotatebox{-50}{\makebox(0,0)[l]{\strut{}\scriptsize \textbf{Bcast~End}}}}%
      \csname LTb\endcsname%
      \put(3442,1392){\rotatebox{-50}{\makebox(0,0)[l]{\strut{}\scriptsize \textbf{Alltoall~Start}}}}%
      \csname LTb\endcsname%
      \put(3644,1392){\rotatebox{-50}{\makebox(0,0)[l]{\strut{}\tiny Finish Step~1}}}%
      \csname LTb\endcsname%
      \put(3846,1392){\rotatebox{-50}{\makebox(0,0)[l]{\strut{}\tiny Start Step~2}}}%
      \csname LTb\endcsname%
      \put(4048,1392){\rotatebox{-50}{\makebox(0,0)[l]{\strut{}\tiny Comm Start}}}%
      \csname LTb\endcsname%
      \put(4250,1392){\rotatebox{-50}{\makebox(0,0)[l]{\strut{}\tiny Deliver Start}}}%
      \csname LTb\endcsname%
      \put(4452,1392){\rotatebox{-50}{\makebox(0,0)[l]{\strut{}\tiny Deliver Finish}}}%
      \csname LTb\endcsname%
      \put(4654,1392){\rotatebox{-50}{\makebox(0,0)[l]{\strut{}\tiny Finish Step~2}}}%
      \csname LTb\endcsname%
      \put(4857,1392){\rotatebox{-50}{\makebox(0,0)[l]{\strut{}\tiny Start Step~3}}}%
      \csname LTb\endcsname%
      \put(5059,1392){\rotatebox{-50}{\makebox(0,0)[l]{\strut{}\scriptsize \textbf{Alltoall~End}}}}%
      \csname LTb\endcsname%
      \put(5261,1392){\rotatebox{-50}{\makebox(0,0)[l]{\strut{}\scriptsize \textbf{Alltoallv~Start}}}}%
      \csname LTb\endcsname%
      \put(5463,1392){\rotatebox{-50}{\makebox(0,0)[l]{\strut{}\tiny Finish Step~1}}}%
      \csname LTb\endcsname%
      \put(5665,1392){\rotatebox{-50}{\makebox(0,0)[l]{\strut{}\tiny Start Step~2}}}%
      \csname LTb\endcsname%
      \put(5867,1392){\rotatebox{-50}{\makebox(0,0)[l]{\strut{}\tiny Comm Start}}}%
      \csname LTb\endcsname%
      \put(6069,1392){\rotatebox{-50}{\makebox(0,0)[l]{\strut{}\tiny Comm Finish}}}%
      \csname LTb\endcsname%
      \put(6271,1392){\rotatebox{-50}{\makebox(0,0)[l]{\strut{}\tiny Finish Step~2}}}%
      \csname LTb\endcsname%
      \put(6474,1392){\rotatebox{-50}{\makebox(0,0)[l]{\strut{}\tiny Start Step~3}}}%
      \csname LTb\endcsname%
      \put(6676,1392){\rotatebox{-50}{\makebox(0,0)[l]{\strut{}\scriptsize \textbf{Alltoallv~End}}}}%
      \csname LTb\endcsname%
      \put(6878,1392){\rotatebox{-50}{\makebox(0,0)[l]{\strut{}\scriptsize \textbf{Benchmark Finish}}}}%
      \csname LTb\endcsname%
      \put(7080,1392){\rotatebox{-50}{\makebox(0,0)[l]{\strut{}\scriptsize \textbf{Finalize}}}}%
      \put(160,3192){\rotatebox{90}{\makebox(0,0){\strut{}Elapsed Time (s)}}}%
      \put(4048,280){\makebox(0,0){\strut{}Algorithm Progress}}%
      \put(4048,200){\makebox(0,0){\strut{}$\rightarrow$}}%
    }%
    \gplgaddtomacro\gplfronttext{%
    }%
    \gplbacktext
    \put(0,0){\includegraphics{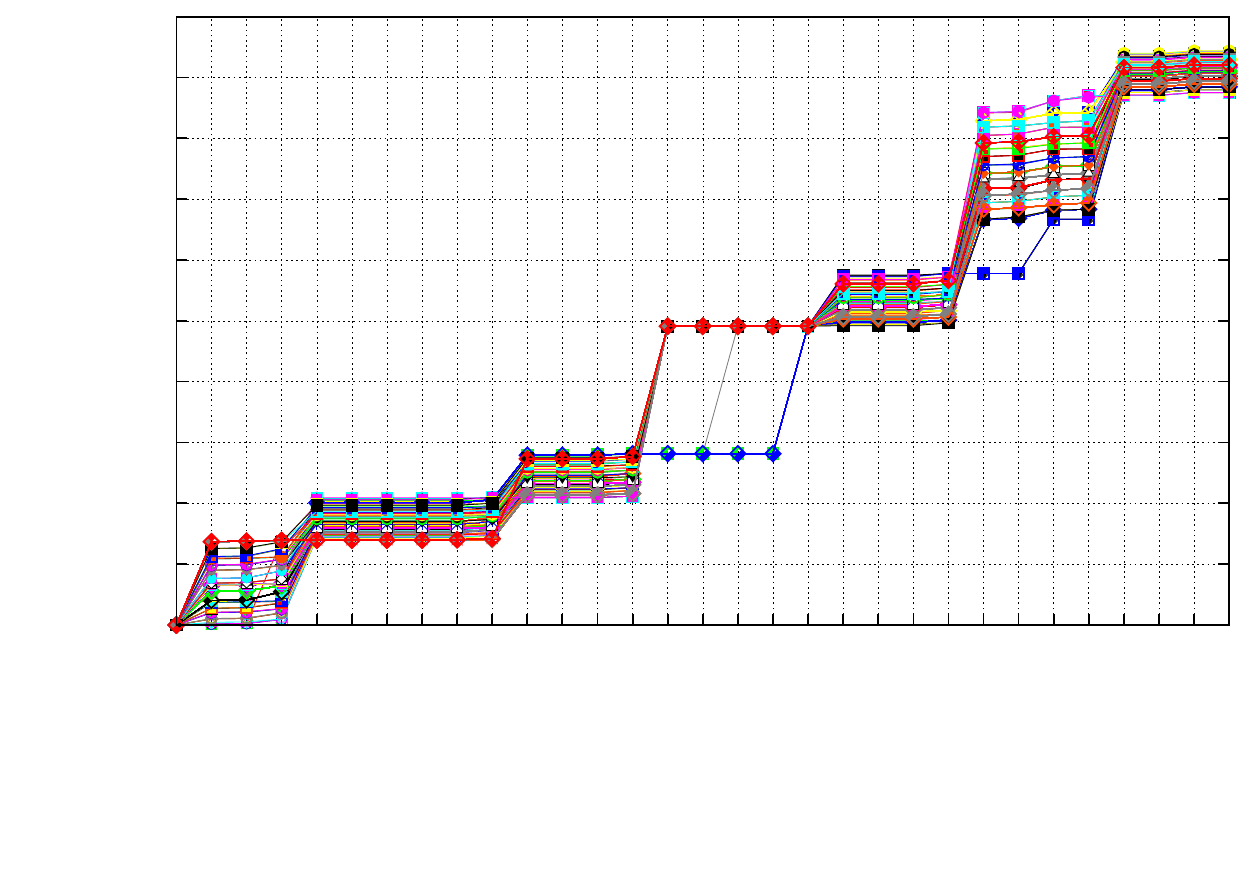}}%
    \gplfronttext
  \end{picture}%
\endgroup

%% file: final_psrs-mmap-internal-v-tex.tex
\begingroup
  \makeatletter
  \providecommand\color[2][]{%
    \GenericError{(gnuplot) \space\space\space\@spaces}{%
      Package color not loaded in conjunction with
      terminal option `colourtext'%
    }{See the gnuplot documentation for explanation.%
    }{Either use 'blacktext' in gnuplot or load the package
      color.sty in LaTeX.}%
    \renewcommand\color[2][]{}%
  }%
  \providecommand\includegraphics[2][]{%
    \GenericError{(gnuplot) \space\space\space\@spaces}{%
      Package graphicx or graphics not loaded%
    }{See the gnuplot documentation for explanation.%
    }{The gnuplot epslatex terminal needs graphicx.sty or graphics.sty.}%
    \renewcommand\includegraphics[2][]{}%
  }%
  \providecommand\rotatebox[2]{#2}%
  \@ifundefined{ifGPcolor}{%
    \newif\ifGPcolor
    \GPcolortrue
  }{}%
  \@ifundefined{ifGPblacktext}{%
    \newif\ifGPblacktext
    \GPblacktexttrue
  }{}%
  \let\gplgaddtomacro\g@addto@macro
  \gdef\gplbacktext{}%
  \gdef\gplfronttext{}%
  \makeatother
  \ifGPblacktext
    \def\colorrgb#1{}%
    \def\colorgray#1{}%
  \else
    \ifGPcolor
      \def\colorrgb#1{\color[rgb]{#1}}%
      \def\colorgray#1{\color[gray]{#1}}%
      \expandafter\def\csname LTw\endcsname{\color{white}}%
      \expandafter\def\csname LTb\endcsname{\color{black}}%
      \expandafter\def\csname LTa\endcsname{\color{black}}%
      \expandafter\def\csname LT0\endcsname{\color[rgb]{1,0,0}}%
      \expandafter\def\csname LT1\endcsname{\color[rgb]{0,1,0}}%
      \expandafter\def\csname LT2\endcsname{\color[rgb]{0,0,1}}%
      \expandafter\def\csname LT3\endcsname{\color[rgb]{1,0,1}}%
      \expandafter\def\csname LT4\endcsname{\color[rgb]{0,1,1}}%
      \expandafter\def\csname LT5\endcsname{\color[rgb]{1,1,0}}%
      \expandafter\def\csname LT6\endcsname{\color[rgb]{0,0,0}}%
      \expandafter\def\csname LT7\endcsname{\color[rgb]{1,0.3,0}}%
      \expandafter\def\csname LT8\endcsname{\color[rgb]{0.5,0.5,0.5}}%
    \else
      \def\colorrgb#1{\color{black}}%
      \def\colorgray#1{\color[gray]{#1}}%
      \expandafter\def\csname LTw\endcsname{\color{white}}%
      \expandafter\def\csname LTb\endcsname{\color{black}}%
      \expandafter\def\csname LTa\endcsname{\color{black}}%
      \expandafter\def\csname LT0\endcsname{\color{black}}%
      \expandafter\def\csname LT1\endcsname{\color{black}}%
      \expandafter\def\csname LT2\endcsname{\color{black}}%
      \expandafter\def\csname LT3\endcsname{\color{black}}%
      \expandafter\def\csname LT4\endcsname{\color{black}}%
      \expandafter\def\csname LT5\endcsname{\color{black}}%
      \expandafter\def\csname LT6\endcsname{\color{black}}%
      \expandafter\def\csname LT7\endcsname{\color{black}}%
      \expandafter\def\csname LT8\endcsname{\color{black}}%
    \fi
  \fi
  \setlength{\unitlength}{0.0500bp}%
  \begin{picture}(7200.00,5040.00)%
    \gplgaddtomacro\gplbacktext{%
      \csname LTb\endcsname%
      \put(968,1440){\makebox(0,0)[r]{\strut{} 0}}%
      \csname LTb\endcsname%
      \put(968,2141){\makebox(0,0)[r]{\strut{} 5000}}%
      \csname LTb\endcsname%
      \put(968,2842){\makebox(0,0)[r]{\strut{} 10000}}%
      \csname LTb\endcsname%
      \put(968,3542){\makebox(0,0)[r]{\strut{} 15000}}%
      \csname LTb\endcsname%
      \put(968,4243){\makebox(0,0)[r]{\strut{} 20000}}%
      \csname LTb\endcsname%
      \put(968,4944){\makebox(0,0)[r]{\strut{} 25000}}%
      \csname LTb\endcsname%
      \put(1016,1392){\rotatebox{-50}{\makebox(0,0)[l]{\strut{}\scriptsize \textbf{Init}}}}%
      \csname LTb\endcsname%
      \put(1218,1392){\rotatebox{-50}{\makebox(0,0)[l]{\strut{}\scriptsize \textbf{Benchmark Start}}}}%
      \csname LTb\endcsname%
      \put(1420,1392){\rotatebox{-50}{\makebox(0,0)[l]{\strut{}\scriptsize \textbf{Gather~Start}}}}%
      \csname LTb\endcsname%
      \put(1622,1392){\rotatebox{-50}{\makebox(0,0)[l]{\strut{}\tiny Finish Step~1}}}%
      \csname LTb\endcsname%
      \put(1825,1392){\rotatebox{-50}{\makebox(0,0)[l]{\strut{}\tiny Start Step~2}}}%
      \csname LTb\endcsname%
      \put(2027,1392){\rotatebox{-50}{\makebox(0,0)[l]{\strut{}\scriptsize \textbf{Gather~End}}}}%
      \csname LTb\endcsname%
      \put(2229,1392){\rotatebox{-50}{\makebox(0,0)[l]{\strut{}\scriptsize \textbf{Bcast~Start}}}}%
      \csname LTb\endcsname%
      \put(2431,1392){\rotatebox{-50}{\makebox(0,0)[l]{\strut{}\tiny RootWait Start}}}%
      \csname LTb\endcsname%
      \put(2633,1392){\rotatebox{-50}{\makebox(0,0)[l]{\strut{}\tiny RootWait End}}}%
      \csname LTb\endcsname%
      \put(2835,1392){\rotatebox{-50}{\makebox(0,0)[l]{\strut{}\tiny Finish Step~1}}}%
      \csname LTb\endcsname%
      \put(3037,1392){\rotatebox{-50}{\makebox(0,0)[l]{\strut{}\tiny Start Step~2}}}%
      \csname LTb\endcsname%
      \put(3239,1392){\rotatebox{-50}{\makebox(0,0)[l]{\strut{}\scriptsize \textbf{Bcast~End}}}}%
      \csname LTb\endcsname%
      \put(3442,1392){\rotatebox{-50}{\makebox(0,0)[l]{\strut{}\scriptsize \textbf{Alltoall~Start}}}}%
      \csname LTb\endcsname%
      \put(3644,1392){\rotatebox{-50}{\makebox(0,0)[l]{\strut{}\tiny Finish Step~1}}}%
      \csname LTb\endcsname%
      \put(3846,1392){\rotatebox{-50}{\makebox(0,0)[l]{\strut{}\tiny Start Step~2}}}%
      \csname LTb\endcsname%
      \put(4048,1392){\rotatebox{-50}{\makebox(0,0)[l]{\strut{}\tiny Comm Start}}}%
      \csname LTb\endcsname%
      \put(4250,1392){\rotatebox{-50}{\makebox(0,0)[l]{\strut{}\tiny Deliver Start}}}%
      \csname LTb\endcsname%
      \put(4452,1392){\rotatebox{-50}{\makebox(0,0)[l]{\strut{}\tiny Deliver Finish}}}%
      \csname LTb\endcsname%
      \put(4654,1392){\rotatebox{-50}{\makebox(0,0)[l]{\strut{}\tiny Finish Step~2}}}%
      \csname LTb\endcsname%
      \put(4857,1392){\rotatebox{-50}{\makebox(0,0)[l]{\strut{}\tiny Start Step~3}}}%
      \csname LTb\endcsname%
      \put(5059,1392){\rotatebox{-50}{\makebox(0,0)[l]{\strut{}\scriptsize \textbf{Alltoall~End}}}}%
      \csname LTb\endcsname%
      \put(5261,1392){\rotatebox{-50}{\makebox(0,0)[l]{\strut{}\scriptsize \textbf{Alltoallv~Start}}}}%
      \csname LTb\endcsname%
      \put(5463,1392){\rotatebox{-50}{\makebox(0,0)[l]{\strut{}\tiny Finish Step~1}}}%
      \csname LTb\endcsname%
      \put(5665,1392){\rotatebox{-50}{\makebox(0,0)[l]{\strut{}\tiny Start Step~2}}}%
      \csname LTb\endcsname%
      \put(5867,1392){\rotatebox{-50}{\makebox(0,0)[l]{\strut{}\tiny Comm Start}}}%
      \csname LTb\endcsname%
      \put(6069,1392){\rotatebox{-50}{\makebox(0,0)[l]{\strut{}\tiny Comm Finish}}}%
      \csname LTb\endcsname%
      \put(6271,1392){\rotatebox{-50}{\makebox(0,0)[l]{\strut{}\tiny Finish Step~2}}}%
      \csname LTb\endcsname%
      \put(6474,1392){\rotatebox{-50}{\makebox(0,0)[l]{\strut{}\tiny Start Step~3}}}%
      \csname LTb\endcsname%
      \put(6676,1392){\rotatebox{-50}{\makebox(0,0)[l]{\strut{}\scriptsize \textbf{Alltoallv~End}}}}%
      \csname LTb\endcsname%
      \put(6878,1392){\rotatebox{-50}{\makebox(0,0)[l]{\strut{}\scriptsize \textbf{Benchmark Finish}}}}%
      \csname LTb\endcsname%
      \put(7080,1392){\rotatebox{-50}{\makebox(0,0)[l]{\strut{}\scriptsize \textbf{Finalize}}}}%
      \put(160,3192){\rotatebox{90}{\makebox(0,0){\strut{}Elapsed Time (s)}}}%
      \put(4048,280){\makebox(0,0){\strut{}Algorithm Progress}}%
      \put(4048,200){\makebox(0,0){\strut{}$\rightarrow$}}%
    }%
    \gplgaddtomacro\gplfronttext{%
    }%
    \gplbacktext
    \put(0,0){\includegraphics{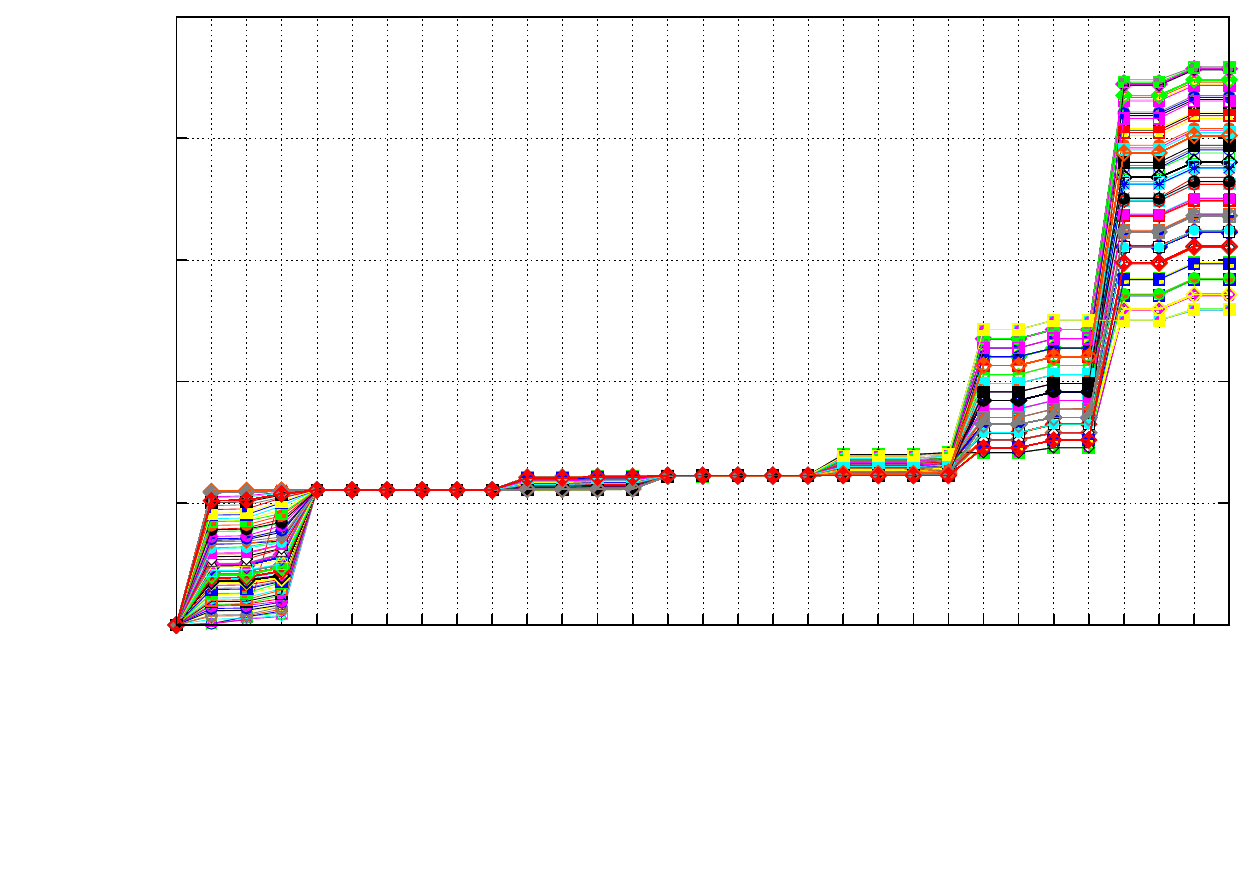}}%
    \gplfronttext
  \end{picture}%
\endgroup

%% file: cgm-sort-plot-p1-tex.tex
\begin{figure}[ht]
	\begin{center}
	\resizebox{!}{0.4\textheight}{
		\input{cgm-sort-p1-tex}
	}
	\\
$\frac{n}{v} = 5120000$
,  $P = 1$
,  k = 4
,  $\mu = 1400$ {\small MiB}

	\caption{CGMLib Sort PEMS2 P=1}
	\label{cgm-sort-p1}
	\end{center}
\end{figure}

%% file: cgm-sort-p1-tex.tex
\begingroup
  \makeatletter
  \providecommand\color[2][]{%
    \GenericError{(gnuplot) \space\space\space\@spaces}{%
      Package color not loaded in conjunction with
      terminal option `colourtext'%
    }{See the gnuplot documentation for explanation.%
    }{Either use 'blacktext' in gnuplot or load the package
      color.sty in LaTeX.}%
    \renewcommand\color[2][]{}%
  }%
  \providecommand\includegraphics[2][]{%
    \GenericError{(gnuplot) \space\space\space\@spaces}{%
      Package graphicx or graphics not loaded%
    }{See the gnuplot documentation for explanation.%
    }{The gnuplot epslatex terminal needs graphicx.sty or graphics.sty.}%
    \renewcommand\includegraphics[2][]{}%
  }%
  \providecommand\rotatebox[2]{#2}%
  \@ifundefined{ifGPcolor}{%
    \newif\ifGPcolor
    \GPcolortrue
  }{}%
  \@ifundefined{ifGPblacktext}{%
    \newif\ifGPblacktext
    \GPblacktexttrue
  }{}%
  \let\gplgaddtomacro\g@addto@macro
  \gdef\gplbacktext{}%
  \gdef\gplfronttext{}%
  \makeatother
  \ifGPblacktext
    \def\colorrgb#1{}%
    \def\colorgray#1{}%
  \else
    \ifGPcolor
      \def\colorrgb#1{\color[rgb]{#1}}%
      \def\colorgray#1{\color[gray]{#1}}%
      \expandafter\def\csname LTw\endcsname{\color{white}}%
      \expandafter\def\csname LTb\endcsname{\color{black}}%
      \expandafter\def\csname LTa\endcsname{\color{black}}%
      \expandafter\def\csname LT0\endcsname{\color[rgb]{1,0,0}}%
      \expandafter\def\csname LT1\endcsname{\color[rgb]{0,1,0}}%
      \expandafter\def\csname LT2\endcsname{\color[rgb]{0,0,1}}%
      \expandafter\def\csname LT3\endcsname{\color[rgb]{1,0,1}}%
      \expandafter\def\csname LT4\endcsname{\color[rgb]{0,1,1}}%
      \expandafter\def\csname LT5\endcsname{\color[rgb]{1,1,0}}%
      \expandafter\def\csname LT6\endcsname{\color[rgb]{0,0,0}}%
      \expandafter\def\csname LT7\endcsname{\color[rgb]{1,0.3,0}}%
      \expandafter\def\csname LT8\endcsname{\color[rgb]{0.5,0.5,0.5}}%
    \else
      \def\colorrgb#1{\color{black}}%
      \def\colorgray#1{\color[gray]{#1}}%
      \expandafter\def\csname LTw\endcsname{\color{white}}%
      \expandafter\def\csname LTb\endcsname{\color{black}}%
      \expandafter\def\csname LTa\endcsname{\color{black}}%
      \expandafter\def\csname LT0\endcsname{\color{black}}%
      \expandafter\def\csname LT1\endcsname{\color{black}}%
      \expandafter\def\csname LT2\endcsname{\color{black}}%
      \expandafter\def\csname LT3\endcsname{\color{black}}%
      \expandafter\def\csname LT4\endcsname{\color{black}}%
      \expandafter\def\csname LT5\endcsname{\color{black}}%
      \expandafter\def\csname LT6\endcsname{\color{black}}%
      \expandafter\def\csname LT7\endcsname{\color{black}}%
      \expandafter\def\csname LT8\endcsname{\color{black}}%
    \fi
  \fi
  \setlength{\unitlength}{0.0500bp}%
  \begin{picture}(7200.00,5040.00)%
    \gplgaddtomacro\gplbacktext{%
      \csname LTb\endcsname%
      \put(1474,1584){\makebox(0,0)[r]{\strut{} 0}}%
      \csname LTb\endcsname%
      \put(1474,2116){\makebox(0,0)[r]{\strut{} 5000}}%
      \csname LTb\endcsname%
      \put(1474,2648){\makebox(0,0)[r]{\strut{} 10000}}%
      \csname LTb\endcsname%
      \put(1474,3180){\makebox(0,0)[r]{\strut{} 15000}}%
      \csname LTb\endcsname%
      \put(1474,3712){\makebox(0,0)[r]{\strut{} 20000}}%
      \csname LTb\endcsname%
      \put(1474,4244){\makebox(0,0)[r]{\strut{} 25000}}%
      \csname LTb\endcsname%
      \put(1474,4776){\makebox(0,0)[r]{\strut{} 30000}}%
      \csname LTb\endcsname%
      \put(1606,1364){\makebox(0,0){\strut{} 80}}%
      \csname LTb\endcsname%
      \put(2191,1364){\makebox(0,0){\strut{} 90}}%
      \csname LTb\endcsname%
      \put(2776,1364){\makebox(0,0){\strut{} 100}}%
      \csname LTb\endcsname%
      \put(3361,1364){\makebox(0,0){\strut{} 110}}%
      \csname LTb\endcsname%
      \put(3946,1364){\makebox(0,0){\strut{} 120}}%
      \csname LTb\endcsname%
      \put(4530,1364){\makebox(0,0){\strut{} 130}}%
      \csname LTb\endcsname%
      \put(5115,1364){\makebox(0,0){\strut{} 140}}%
      \csname LTb\endcsname%
      \put(5700,1364){\makebox(0,0){\strut{} 150}}%
      \csname LTb\endcsname%
      \put(6285,1364){\makebox(0,0){\strut{} 160}}%
      \csname LTb\endcsname%
      \put(6870,1364){\makebox(0,0){\strut{} 170}}%
      \put(440,3180){\rotatebox{90}{\makebox(0,0){\strut{}Seconds}}}%
      \put(4238,1034){\makebox(0,0){\strut{}$n$ (millions)}}%
    }%
    \gplgaddtomacro\gplfronttext{%
      \csname LTb\endcsname%
      \put(6055,613){\makebox(0,0)[r]{\strut{}CGMLib Sort PEMS2 (mmap) P=1}}%
      \csname LTb\endcsname%
      \put(6055,393){\makebox(0,0)[r]{\strut{}CGMLib Sort PEMS2 (stxxl-file) P=1}}%
      \csname LTb\endcsname%
      \put(6055,173){\makebox(0,0)[r]{\strut{}CGMLib Sort PEMS2 (unix) P=1}}%
    }%
    \gplbacktext
    \put(0,0){\includegraphics{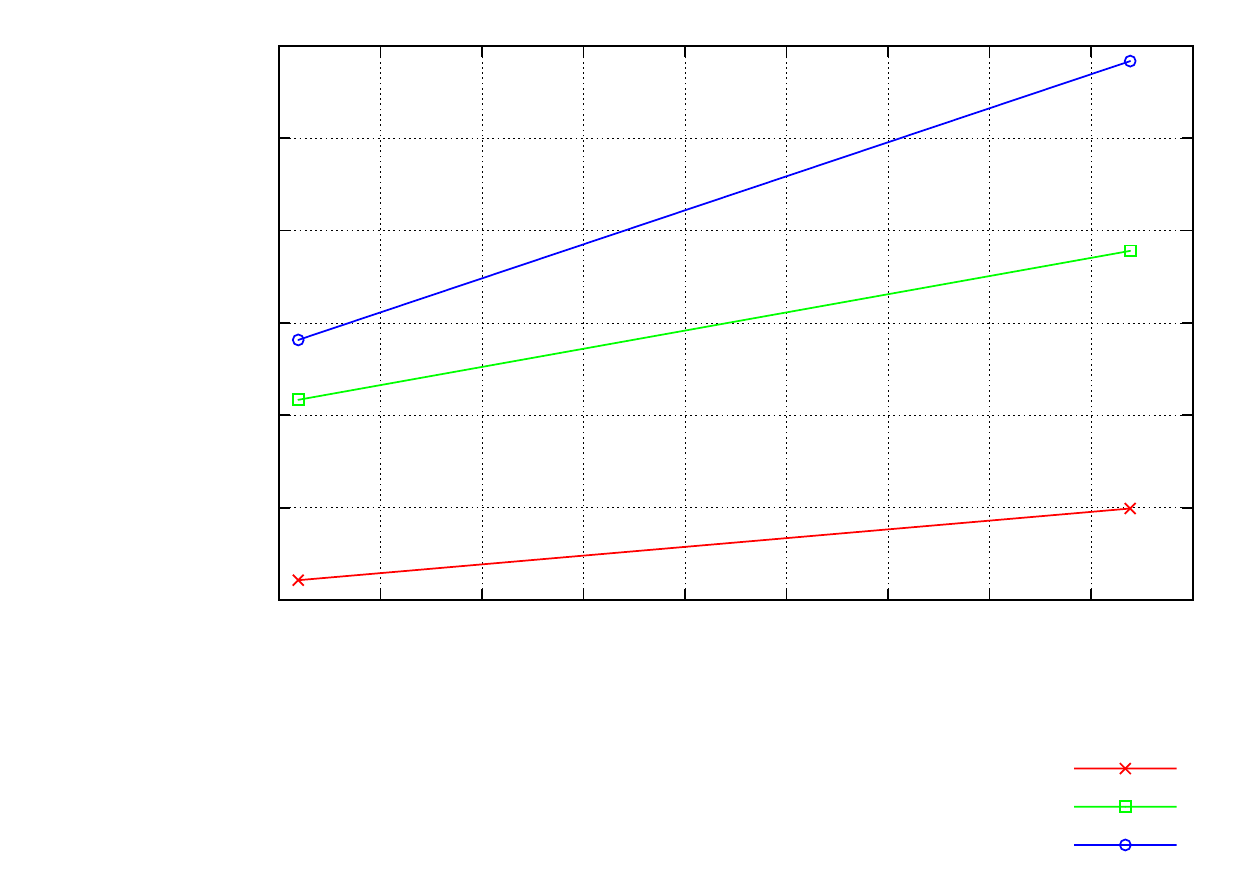}}%
    \gplfronttext
  \end{picture}%
\endgroup

%% file: cgm-sort-plot-p2-tex.tex
\begin{figure}[ht]
	\begin{center}
	\resizebox{!}{0.4\textheight}{
		\input{cgm-sort-p2-tex}
	}
	\\
$\frac{n}{v} = 5120000$
,  $P = 2$
,  k = 4
,  $\mu = 1400$ {\small MiB}

	\caption{CGMLib Sort PEMS2 P=2}
	\label{cgm-sort-p2}
	\end{center}
\end{figure}

%% file: cgm-sort-p2-tex.tex
\begingroup
  \makeatletter
  \providecommand\color[2][]{%
    \GenericError{(gnuplot) \space\space\space\@spaces}{%
      Package color not loaded in conjunction with
      terminal option `colourtext'%
    }{See the gnuplot documentation for explanation.%
    }{Either use 'blacktext' in gnuplot or load the package
      color.sty in LaTeX.}%
    \renewcommand\color[2][]{}%
  }%
  \providecommand\includegraphics[2][]{%
    \GenericError{(gnuplot) \space\space\space\@spaces}{%
      Package graphicx or graphics not loaded%
    }{See the gnuplot documentation for explanation.%
    }{The gnuplot epslatex terminal needs graphicx.sty or graphics.sty.}%
    \renewcommand\includegraphics[2][]{}%
  }%
  \providecommand\rotatebox[2]{#2}%
  \@ifundefined{ifGPcolor}{%
    \newif\ifGPcolor
    \GPcolortrue
  }{}%
  \@ifundefined{ifGPblacktext}{%
    \newif\ifGPblacktext
    \GPblacktexttrue
  }{}%
  \let\gplgaddtomacro\g@addto@macro
  \gdef\gplbacktext{}%
  \gdef\gplfronttext{}%
  \makeatother
  \ifGPblacktext
    \def\colorrgb#1{}%
    \def\colorgray#1{}%
  \else
    \ifGPcolor
      \def\colorrgb#1{\color[rgb]{#1}}%
      \def\colorgray#1{\color[gray]{#1}}%
      \expandafter\def\csname LTw\endcsname{\color{white}}%
      \expandafter\def\csname LTb\endcsname{\color{black}}%
      \expandafter\def\csname LTa\endcsname{\color{black}}%
      \expandafter\def\csname LT0\endcsname{\color[rgb]{1,0,0}}%
      \expandafter\def\csname LT1\endcsname{\color[rgb]{0,1,0}}%
      \expandafter\def\csname LT2\endcsname{\color[rgb]{0,0,1}}%
      \expandafter\def\csname LT3\endcsname{\color[rgb]{1,0,1}}%
      \expandafter\def\csname LT4\endcsname{\color[rgb]{0,1,1}}%
      \expandafter\def\csname LT5\endcsname{\color[rgb]{1,1,0}}%
      \expandafter\def\csname LT6\endcsname{\color[rgb]{0,0,0}}%
      \expandafter\def\csname LT7\endcsname{\color[rgb]{1,0.3,0}}%
      \expandafter\def\csname LT8\endcsname{\color[rgb]{0.5,0.5,0.5}}%
    \else
      \def\colorrgb#1{\color{black}}%
      \def\colorgray#1{\color[gray]{#1}}%
      \expandafter\def\csname LTw\endcsname{\color{white}}%
      \expandafter\def\csname LTb\endcsname{\color{black}}%
      \expandafter\def\csname LTa\endcsname{\color{black}}%
      \expandafter\def\csname LT0\endcsname{\color{black}}%
      \expandafter\def\csname LT1\endcsname{\color{black}}%
      \expandafter\def\csname LT2\endcsname{\color{black}}%
      \expandafter\def\csname LT3\endcsname{\color{black}}%
      \expandafter\def\csname LT4\endcsname{\color{black}}%
      \expandafter\def\csname LT5\endcsname{\color{black}}%
      \expandafter\def\csname LT6\endcsname{\color{black}}%
      \expandafter\def\csname LT7\endcsname{\color{black}}%
      \expandafter\def\csname LT8\endcsname{\color{black}}%
    \fi
  \fi
  \setlength{\unitlength}{0.0500bp}%
  \begin{picture}(7200.00,5040.00)%
    \gplgaddtomacro\gplbacktext{%
      \csname LTb\endcsname%
      \put(1474,1584){\makebox(0,0)[r]{\strut{} 0}}%
      \csname LTb\endcsname%
      \put(1474,2222){\makebox(0,0)[r]{\strut{} 5000}}%
      \csname LTb\endcsname%
      \put(1474,2861){\makebox(0,0)[r]{\strut{} 10000}}%
      \csname LTb\endcsname%
      \put(1474,3499){\makebox(0,0)[r]{\strut{} 15000}}%
      \csname LTb\endcsname%
      \put(1474,4138){\makebox(0,0)[r]{\strut{} 20000}}%
      \csname LTb\endcsname%
      \put(1474,4776){\makebox(0,0)[r]{\strut{} 25000}}%
      \csname LTb\endcsname%
      \put(1606,1364){\makebox(0,0){\strut{} 50}}%
      \csname LTb\endcsname%
      \put(2483,1364){\makebox(0,0){\strut{} 100}}%
      \csname LTb\endcsname%
      \put(3361,1364){\makebox(0,0){\strut{} 150}}%
      \csname LTb\endcsname%
      \put(4238,1364){\makebox(0,0){\strut{} 200}}%
      \csname LTb\endcsname%
      \put(5115,1364){\makebox(0,0){\strut{} 250}}%
      \csname LTb\endcsname%
      \put(5993,1364){\makebox(0,0){\strut{} 300}}%
      \csname LTb\endcsname%
      \put(6870,1364){\makebox(0,0){\strut{} 350}}%
      \put(440,3180){\rotatebox{90}{\makebox(0,0){\strut{}Seconds}}}%
      \put(4238,1034){\makebox(0,0){\strut{}$n$ (millions)}}%
    }%
    \gplgaddtomacro\gplfronttext{%
      \csname LTb\endcsname%
      \put(6055,613){\makebox(0,0)[r]{\strut{}CGMLib Sort PEMS2 (mmap) P=2}}%
      \csname LTb\endcsname%
      \put(6055,393){\makebox(0,0)[r]{\strut{}CGMLib Sort PEMS2 (stxxl-file) P=2}}%
      \csname LTb\endcsname%
      \put(6055,173){\makebox(0,0)[r]{\strut{}CGMLib Sort PEMS2 (unix) P=2}}%
    }%
    \gplbacktext
    \put(0,0){\includegraphics{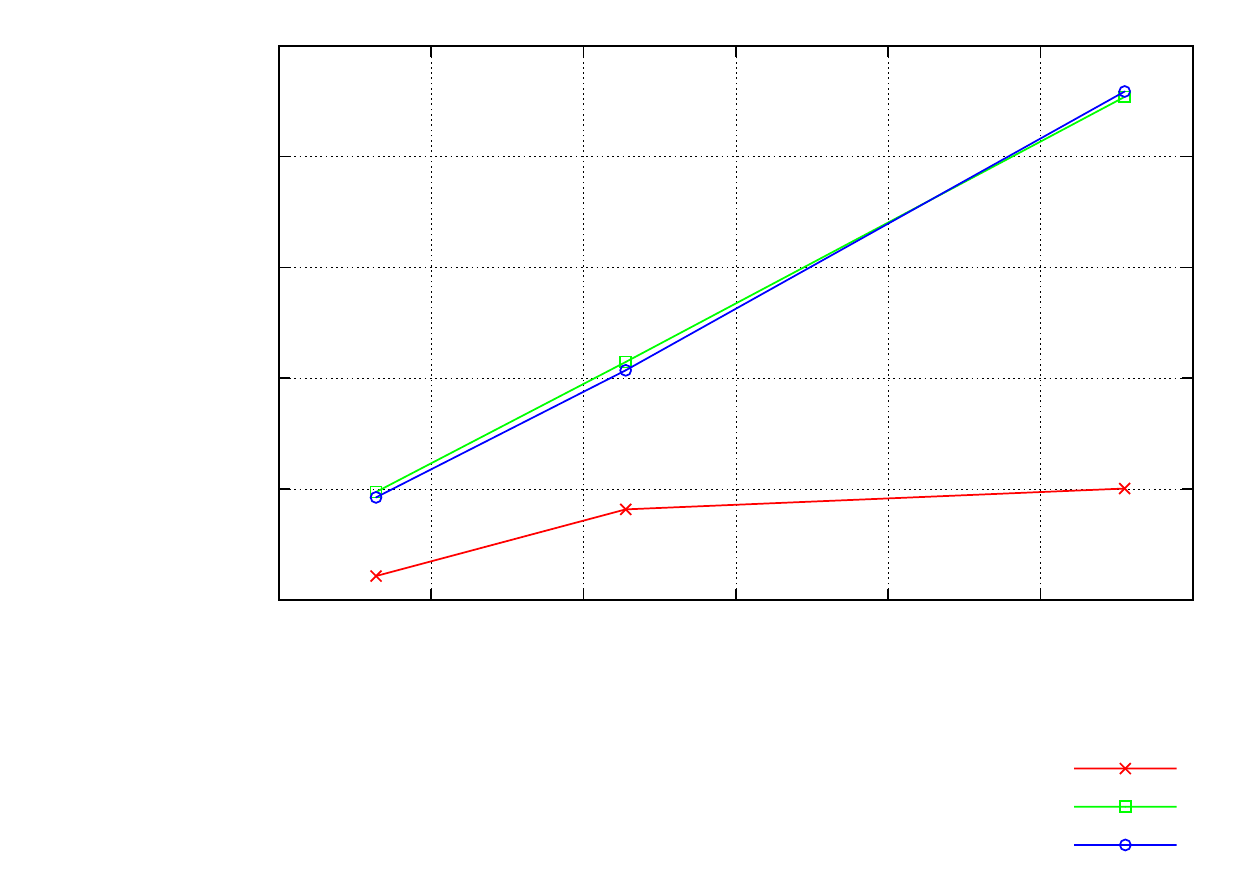}}%
    \gplfronttext
  \end{picture}%
\endgroup

%% file: cgm-sort-plot-p4-tex.tex
\begin{figure}[ht]
	\begin{center}
	\resizebox{!}{0.4\textheight}{
		\input{cgm-sort-p4-tex}
	}
	\\
$\frac{n}{v} = 5120000$
,  $P = 4$
,  k = 4

	\caption{CGMLib Sort PEMS2 P=4}
	\label{cgm-sort-p4}
	\end{center}
\end{figure}

%% file: cgm-sort-p4-tex.tex
\begingroup
  \makeatletter
  \providecommand\color[2][]{%
    \GenericError{(gnuplot) \space\space\space\@spaces}{%
      Package color not loaded in conjunction with
      terminal option `colourtext'%
    }{See the gnuplot documentation for explanation.%
    }{Either use 'blacktext' in gnuplot or load the package
      color.sty in LaTeX.}%
    \renewcommand\color[2][]{}%
  }%
  \providecommand\includegraphics[2][]{%
    \GenericError{(gnuplot) \space\space\space\@spaces}{%
      Package graphicx or graphics not loaded%
    }{See the gnuplot documentation for explanation.%
    }{The gnuplot epslatex terminal needs graphicx.sty or graphics.sty.}%
    \renewcommand\includegraphics[2][]{}%
  }%
  \providecommand\rotatebox[2]{#2}%
  \@ifundefined{ifGPcolor}{%
    \newif\ifGPcolor
    \GPcolortrue
  }{}%
  \@ifundefined{ifGPblacktext}{%
    \newif\ifGPblacktext
    \GPblacktexttrue
  }{}%
  \let\gplgaddtomacro\g@addto@macro
  \gdef\gplbacktext{}%
  \gdef\gplfronttext{}%
  \makeatother
  \ifGPblacktext
    \def\colorrgb#1{}%
    \def\colorgray#1{}%
  \else
    \ifGPcolor
      \def\colorrgb#1{\color[rgb]{#1}}%
      \def\colorgray#1{\color[gray]{#1}}%
      \expandafter\def\csname LTw\endcsname{\color{white}}%
      \expandafter\def\csname LTb\endcsname{\color{black}}%
      \expandafter\def\csname LTa\endcsname{\color{black}}%
      \expandafter\def\csname LT0\endcsname{\color[rgb]{1,0,0}}%
      \expandafter\def\csname LT1\endcsname{\color[rgb]{0,1,0}}%
      \expandafter\def\csname LT2\endcsname{\color[rgb]{0,0,1}}%
      \expandafter\def\csname LT3\endcsname{\color[rgb]{1,0,1}}%
      \expandafter\def\csname LT4\endcsname{\color[rgb]{0,1,1}}%
      \expandafter\def\csname LT5\endcsname{\color[rgb]{1,1,0}}%
      \expandafter\def\csname LT6\endcsname{\color[rgb]{0,0,0}}%
      \expandafter\def\csname LT7\endcsname{\color[rgb]{1,0.3,0}}%
      \expandafter\def\csname LT8\endcsname{\color[rgb]{0.5,0.5,0.5}}%
    \else
      \def\colorrgb#1{\color{black}}%
      \def\colorgray#1{\color[gray]{#1}}%
      \expandafter\def\csname LTw\endcsname{\color{white}}%
      \expandafter\def\csname LTb\endcsname{\color{black}}%
      \expandafter\def\csname LTa\endcsname{\color{black}}%
      \expandafter\def\csname LT0\endcsname{\color{black}}%
      \expandafter\def\csname LT1\endcsname{\color{black}}%
      \expandafter\def\csname LT2\endcsname{\color{black}}%
      \expandafter\def\csname LT3\endcsname{\color{black}}%
      \expandafter\def\csname LT4\endcsname{\color{black}}%
      \expandafter\def\csname LT5\endcsname{\color{black}}%
      \expandafter\def\csname LT6\endcsname{\color{black}}%
      \expandafter\def\csname LT7\endcsname{\color{black}}%
      \expandafter\def\csname LT8\endcsname{\color{black}}%
    \fi
  \fi
  \setlength{\unitlength}{0.0500bp}%
  \begin{picture}(7200.00,5040.00)%
    \gplgaddtomacro\gplbacktext{%
      \csname LTb\endcsname%
      \put(1474,1584){\makebox(0,0)[r]{\strut{} 0}}%
      \csname LTb\endcsname%
      \put(1474,2116){\makebox(0,0)[r]{\strut{} 5000}}%
      \csname LTb\endcsname%
      \put(1474,2648){\makebox(0,0)[r]{\strut{} 10000}}%
      \csname LTb\endcsname%
      \put(1474,3180){\makebox(0,0)[r]{\strut{} 15000}}%
      \csname LTb\endcsname%
      \put(1474,3712){\makebox(0,0)[r]{\strut{} 20000}}%
      \csname LTb\endcsname%
      \put(1474,4244){\makebox(0,0)[r]{\strut{} 25000}}%
      \csname LTb\endcsname%
      \put(1474,4776){\makebox(0,0)[r]{\strut{} 30000}}%
      \csname LTb\endcsname%
      \put(1606,1364){\makebox(0,0){\strut{} 0}}%
      \csname LTb\endcsname%
      \put(2358,1364){\makebox(0,0){\strut{} 100}}%
      \csname LTb\endcsname%
      \put(3110,1364){\makebox(0,0){\strut{} 200}}%
      \csname LTb\endcsname%
      \put(3862,1364){\makebox(0,0){\strut{} 300}}%
      \csname LTb\endcsname%
      \put(4614,1364){\makebox(0,0){\strut{} 400}}%
      \csname LTb\endcsname%
      \put(5366,1364){\makebox(0,0){\strut{} 500}}%
      \csname LTb\endcsname%
      \put(6118,1364){\makebox(0,0){\strut{} 600}}%
      \csname LTb\endcsname%
      \put(6870,1364){\makebox(0,0){\strut{} 700}}%
      \put(440,3180){\rotatebox{90}{\makebox(0,0){\strut{}Seconds}}}%
      \put(4238,1034){\makebox(0,0){\strut{}$n$ (millions)}}%
    }%
    \gplgaddtomacro\gplfronttext{%
      \csname LTb\endcsname%
      \put(6055,613){\makebox(0,0)[r]{\strut{}CGMLib Sort PEMS2 (mmap) P=4}}%
      \csname LTb\endcsname%
      \put(6055,393){\makebox(0,0)[r]{\strut{}CGMLib Sort PEMS2 (stxxl-file) P=4}}%
      \csname LTb\endcsname%
      \put(6055,173){\makebox(0,0)[r]{\strut{}CGMLib Sort PEMS2 (unix) P=4}}%
    }%
    \gplbacktext
    \put(0,0){\includegraphics{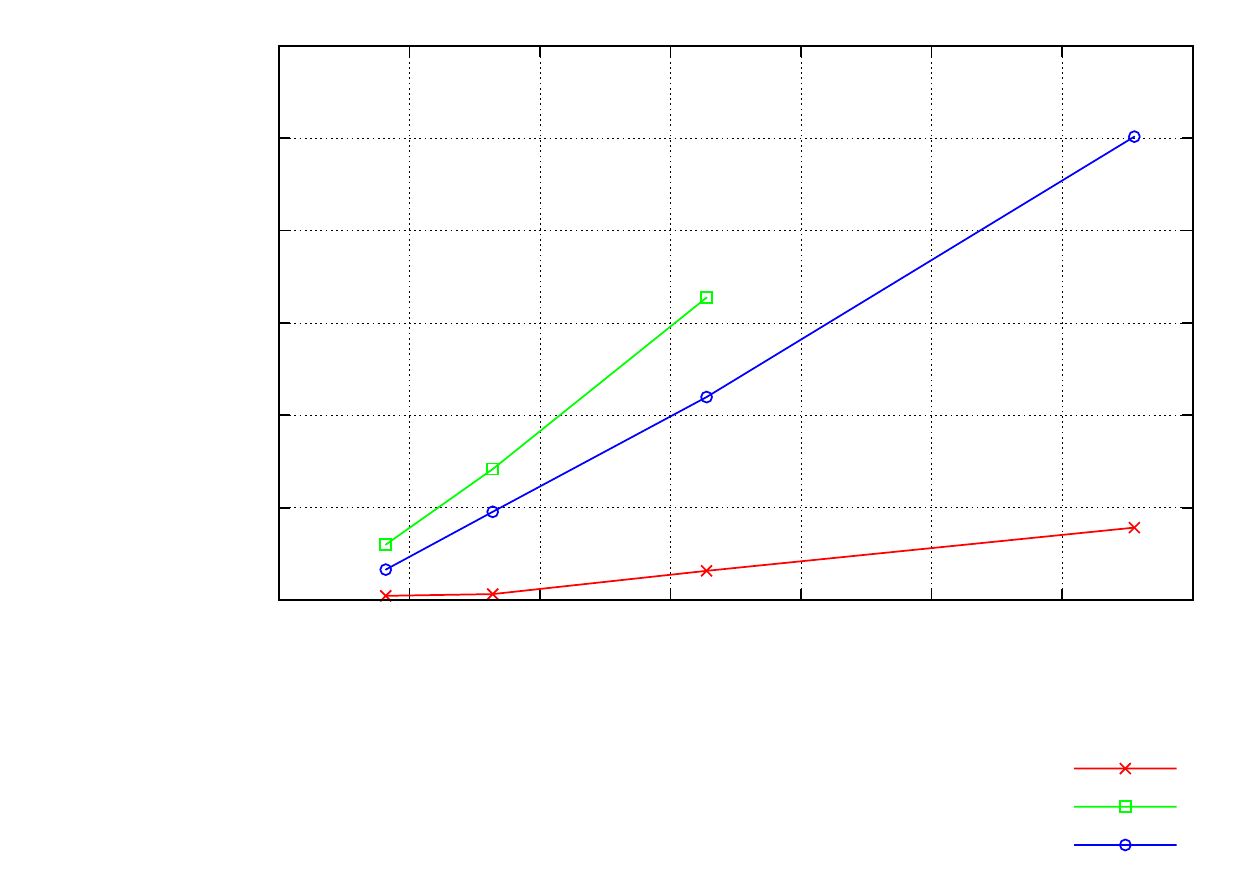}}%
    \gplfronttext
  \end{picture}%
\endgroup

%% file: cgm-psum-plot-p1-tex.tex
\begin{figure}[ht]
	\begin{center}
	\resizebox{!}{0.4\textheight}{
		\input{cgm-psum-p1-tex}
	}
	\\
$\frac{n}{v} = 5120000$
,  $P = 1$
,  k = 4
,  $\mu = 1400$ {\small MiB}

	\caption{CGMLib Prefix Sum PEMS2 P=1}
	\label{cgm-psum-p1}
	\end{center}
\end{figure}

%% file: cgm-psum-p1-tex.tex
\begingroup
  \makeatletter
  \providecommand\color[2][]{%
    \GenericError{(gnuplot) \space\space\space\@spaces}{%
      Package color not loaded in conjunction with
      terminal option `colourtext'%
    }{See the gnuplot documentation for explanation.%
    }{Either use 'blacktext' in gnuplot or load the package
      color.sty in LaTeX.}%
    \renewcommand\color[2][]{}%
  }%
  \providecommand\includegraphics[2][]{%
    \GenericError{(gnuplot) \space\space\space\@spaces}{%
      Package graphicx or graphics not loaded%
    }{See the gnuplot documentation for explanation.%
    }{The gnuplot epslatex terminal needs graphicx.sty or graphics.sty.}%
    \renewcommand\includegraphics[2][]{}%
  }%
  \providecommand\rotatebox[2]{#2}%
  \@ifundefined{ifGPcolor}{%
    \newif\ifGPcolor
    \GPcolortrue
  }{}%
  \@ifundefined{ifGPblacktext}{%
    \newif\ifGPblacktext
    \GPblacktexttrue
  }{}%
  \let\gplgaddtomacro\g@addto@macro
  \gdef\gplbacktext{}%
  \gdef\gplfronttext{}%
  \makeatother
  \ifGPblacktext
    \def\colorrgb#1{}%
    \def\colorgray#1{}%
  \else
    \ifGPcolor
      \def\colorrgb#1{\color[rgb]{#1}}%
      \def\colorgray#1{\color[gray]{#1}}%
      \expandafter\def\csname LTw\endcsname{\color{white}}%
      \expandafter\def\csname LTb\endcsname{\color{black}}%
      \expandafter\def\csname LTa\endcsname{\color{black}}%
      \expandafter\def\csname LT0\endcsname{\color[rgb]{1,0,0}}%
      \expandafter\def\csname LT1\endcsname{\color[rgb]{0,1,0}}%
      \expandafter\def\csname LT2\endcsname{\color[rgb]{0,0,1}}%
      \expandafter\def\csname LT3\endcsname{\color[rgb]{1,0,1}}%
      \expandafter\def\csname LT4\endcsname{\color[rgb]{0,1,1}}%
      \expandafter\def\csname LT5\endcsname{\color[rgb]{1,1,0}}%
      \expandafter\def\csname LT6\endcsname{\color[rgb]{0,0,0}}%
      \expandafter\def\csname LT7\endcsname{\color[rgb]{1,0.3,0}}%
      \expandafter\def\csname LT8\endcsname{\color[rgb]{0.5,0.5,0.5}}%
    \else
      \def\colorrgb#1{\color{black}}%
      \def\colorgray#1{\color[gray]{#1}}%
      \expandafter\def\csname LTw\endcsname{\color{white}}%
      \expandafter\def\csname LTb\endcsname{\color{black}}%
      \expandafter\def\csname LTa\endcsname{\color{black}}%
      \expandafter\def\csname LT0\endcsname{\color{black}}%
      \expandafter\def\csname LT1\endcsname{\color{black}}%
      \expandafter\def\csname LT2\endcsname{\color{black}}%
      \expandafter\def\csname LT3\endcsname{\color{black}}%
      \expandafter\def\csname LT4\endcsname{\color{black}}%
      \expandafter\def\csname LT5\endcsname{\color{black}}%
      \expandafter\def\csname LT6\endcsname{\color{black}}%
      \expandafter\def\csname LT7\endcsname{\color{black}}%
      \expandafter\def\csname LT8\endcsname{\color{black}}%
    \fi
  \fi
  \setlength{\unitlength}{0.0500bp}%
  \begin{picture}(7200.00,5040.00)%
    \gplgaddtomacro\gplbacktext{%
      \csname LTb\endcsname%
      \put(1342,1584){\makebox(0,0)[r]{\strut{} 0}}%
      \csname LTb\endcsname%
      \put(1342,2116){\makebox(0,0)[r]{\strut{} 1000}}%
      \csname LTb\endcsname%
      \put(1342,2648){\makebox(0,0)[r]{\strut{} 2000}}%
      \csname LTb\endcsname%
      \put(1342,3180){\makebox(0,0)[r]{\strut{} 3000}}%
      \csname LTb\endcsname%
      \put(1342,3712){\makebox(0,0)[r]{\strut{} 4000}}%
      \csname LTb\endcsname%
      \put(1342,4244){\makebox(0,0)[r]{\strut{} 5000}}%
      \csname LTb\endcsname%
      \put(1342,4776){\makebox(0,0)[r]{\strut{} 6000}}%
      \csname LTb\endcsname%
      \put(1474,1364){\makebox(0,0){\strut{} 80}}%
      \csname LTb\endcsname%
      \put(2074,1364){\makebox(0,0){\strut{} 90}}%
      \csname LTb\endcsname%
      \put(2673,1364){\makebox(0,0){\strut{} 100}}%
      \csname LTb\endcsname%
      \put(3273,1364){\makebox(0,0){\strut{} 110}}%
      \csname LTb\endcsname%
      \put(3872,1364){\makebox(0,0){\strut{} 120}}%
      \csname LTb\endcsname%
      \put(4472,1364){\makebox(0,0){\strut{} 130}}%
      \csname LTb\endcsname%
      \put(5071,1364){\makebox(0,0){\strut{} 140}}%
      \csname LTb\endcsname%
      \put(5671,1364){\makebox(0,0){\strut{} 150}}%
      \csname LTb\endcsname%
      \put(6270,1364){\makebox(0,0){\strut{} 160}}%
      \csname LTb\endcsname%
      \put(6870,1364){\makebox(0,0){\strut{} 170}}%
      \put(440,3180){\rotatebox{90}{\makebox(0,0){\strut{}Seconds}}}%
      \put(4172,1034){\makebox(0,0){\strut{}$n$ (millions)}}%
    }%
    \gplgaddtomacro\gplfronttext{%
      \csname LTb\endcsname%
      \put(6385,613){\makebox(0,0)[r]{\strut{}CGMLib Prefix Sum PEMS2 (mmap) P=1}}%
      \csname LTb\endcsname%
      \put(6385,393){\makebox(0,0)[r]{\strut{}CGMLib Prefix Sum PEMS2 (stxxl-file) P=1}}%
      \csname LTb\endcsname%
      \put(6385,173){\makebox(0,0)[r]{\strut{}CGMLib Prefix Sum PEMS2 (unix) P=1}}%
    }%
    \gplbacktext
    \put(0,0){\includegraphics{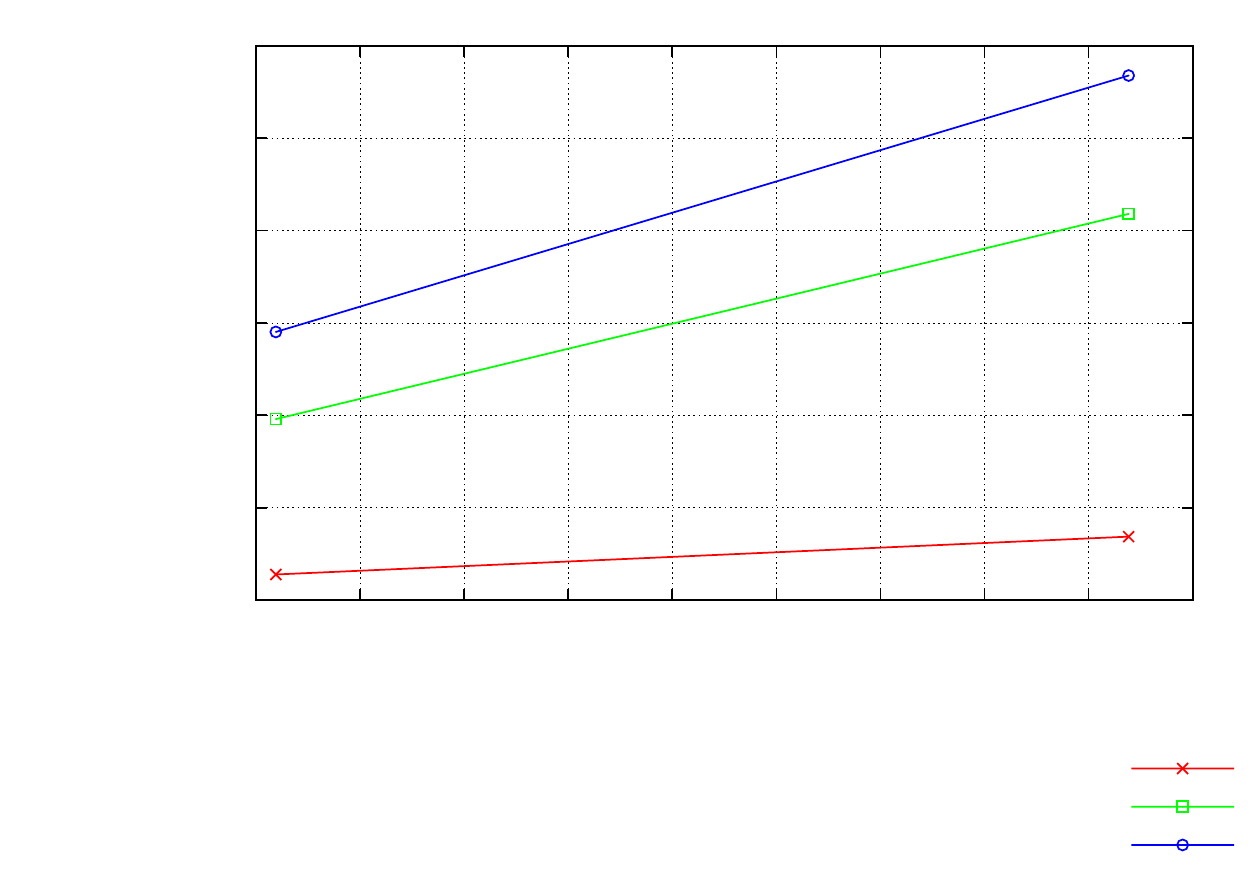}}%
    \gplfronttext
  \end{picture}%
\endgroup

%% file: cgm-psum-plot-p2-tex.tex
\begin{figure}[ht]
	\begin{center}
	\resizebox{!}{0.4\textheight}{
		\input{cgm-psum-p2-tex}
	}
	\\
$\frac{n}{v} = 5120000$
,  $P = 2$
,  k = 4
,  $\mu = 1400$ {\small MiB}

	\caption{CGMLib Prefix Sum PEMS2 P=2}
	\label{cgm-psum-p2}
	\end{center}
\end{figure}

%% file: cgm-psum-p2-tex.tex
\begingroup
  \makeatletter
  \providecommand\color[2][]{%
    \GenericError{(gnuplot) \space\space\space\@spaces}{%
      Package color not loaded in conjunction with
      terminal option `colourtext'%
    }{See the gnuplot documentation for explanation.%
    }{Either use 'blacktext' in gnuplot or load the package
      color.sty in LaTeX.}%
    \renewcommand\color[2][]{}%
  }%
  \providecommand\includegraphics[2][]{%
    \GenericError{(gnuplot) \space\space\space\@spaces}{%
      Package graphicx or graphics not loaded%
    }{See the gnuplot documentation for explanation.%
    }{The gnuplot epslatex terminal needs graphicx.sty or graphics.sty.}%
    \renewcommand\includegraphics[2][]{}%
  }%
  \providecommand\rotatebox[2]{#2}%
  \@ifundefined{ifGPcolor}{%
    \newif\ifGPcolor
    \GPcolortrue
  }{}%
  \@ifundefined{ifGPblacktext}{%
    \newif\ifGPblacktext
    \GPblacktexttrue
  }{}%
  \let\gplgaddtomacro\g@addto@macro
  \gdef\gplbacktext{}%
  \gdef\gplfronttext{}%
  \makeatother
  \ifGPblacktext
    \def\colorrgb#1{}%
    \def\colorgray#1{}%
  \else
    \ifGPcolor
      \def\colorrgb#1{\color[rgb]{#1}}%
      \def\colorgray#1{\color[gray]{#1}}%
      \expandafter\def\csname LTw\endcsname{\color{white}}%
      \expandafter\def\csname LTb\endcsname{\color{black}}%
      \expandafter\def\csname LTa\endcsname{\color{black}}%
      \expandafter\def\csname LT0\endcsname{\color[rgb]{1,0,0}}%
      \expandafter\def\csname LT1\endcsname{\color[rgb]{0,1,0}}%
      \expandafter\def\csname LT2\endcsname{\color[rgb]{0,0,1}}%
      \expandafter\def\csname LT3\endcsname{\color[rgb]{1,0,1}}%
      \expandafter\def\csname LT4\endcsname{\color[rgb]{0,1,1}}%
      \expandafter\def\csname LT5\endcsname{\color[rgb]{1,1,0}}%
      \expandafter\def\csname LT6\endcsname{\color[rgb]{0,0,0}}%
      \expandafter\def\csname LT7\endcsname{\color[rgb]{1,0.3,0}}%
      \expandafter\def\csname LT8\endcsname{\color[rgb]{0.5,0.5,0.5}}%
    \else
      \def\colorrgb#1{\color{black}}%
      \def\colorgray#1{\color[gray]{#1}}%
      \expandafter\def\csname LTw\endcsname{\color{white}}%
      \expandafter\def\csname LTb\endcsname{\color{black}}%
      \expandafter\def\csname LTa\endcsname{\color{black}}%
      \expandafter\def\csname LT0\endcsname{\color{black}}%
      \expandafter\def\csname LT1\endcsname{\color{black}}%
      \expandafter\def\csname LT2\endcsname{\color{black}}%
      \expandafter\def\csname LT3\endcsname{\color{black}}%
      \expandafter\def\csname LT4\endcsname{\color{black}}%
      \expandafter\def\csname LT5\endcsname{\color{black}}%
      \expandafter\def\csname LT6\endcsname{\color{black}}%
      \expandafter\def\csname LT7\endcsname{\color{black}}%
      \expandafter\def\csname LT8\endcsname{\color{black}}%
    \fi
  \fi
  \setlength{\unitlength}{0.0500bp}%
  \begin{picture}(7200.00,5040.00)%
    \gplgaddtomacro\gplbacktext{%
      \csname LTb\endcsname%
      \put(1342,1584){\makebox(0,0)[r]{\strut{} 500}}%
      \csname LTb\endcsname%
      \put(1342,2116){\makebox(0,0)[r]{\strut{} 1000}}%
      \csname LTb\endcsname%
      \put(1342,2648){\makebox(0,0)[r]{\strut{} 1500}}%
      \csname LTb\endcsname%
      \put(1342,3180){\makebox(0,0)[r]{\strut{} 2000}}%
      \csname LTb\endcsname%
      \put(1342,3712){\makebox(0,0)[r]{\strut{} 2500}}%
      \csname LTb\endcsname%
      \put(1342,4244){\makebox(0,0)[r]{\strut{} 3000}}%
      \csname LTb\endcsname%
      \put(1342,4776){\makebox(0,0)[r]{\strut{} 3500}}%
      \csname LTb\endcsname%
      \put(1474,1364){\makebox(0,0){\strut{} 50}}%
      \csname LTb\endcsname%
      \put(2373,1364){\makebox(0,0){\strut{} 100}}%
      \csname LTb\endcsname%
      \put(3273,1364){\makebox(0,0){\strut{} 150}}%
      \csname LTb\endcsname%
      \put(4172,1364){\makebox(0,0){\strut{} 200}}%
      \csname LTb\endcsname%
      \put(5071,1364){\makebox(0,0){\strut{} 250}}%
      \csname LTb\endcsname%
      \put(5971,1364){\makebox(0,0){\strut{} 300}}%
      \csname LTb\endcsname%
      \put(6870,1364){\makebox(0,0){\strut{} 350}}%
      \put(440,3180){\rotatebox{90}{\makebox(0,0){\strut{}Seconds}}}%
      \put(4172,1034){\makebox(0,0){\strut{}$n$ (millions)}}%
    }%
    \gplgaddtomacro\gplfronttext{%
      \csname LTb\endcsname%
      \put(6385,613){\makebox(0,0)[r]{\strut{}CGMLib Prefix Sum PEMS2 (mmap) P=2}}%
      \csname LTb\endcsname%
      \put(6385,393){\makebox(0,0)[r]{\strut{}CGMLib Prefix Sum PEMS2 (stxxl-file) P=2}}%
      \csname LTb\endcsname%
      \put(6385,173){\makebox(0,0)[r]{\strut{}CGMLib Prefix Sum PEMS2 (unix) P=2}}%
    }%
    \gplbacktext
    \put(0,0){\includegraphics{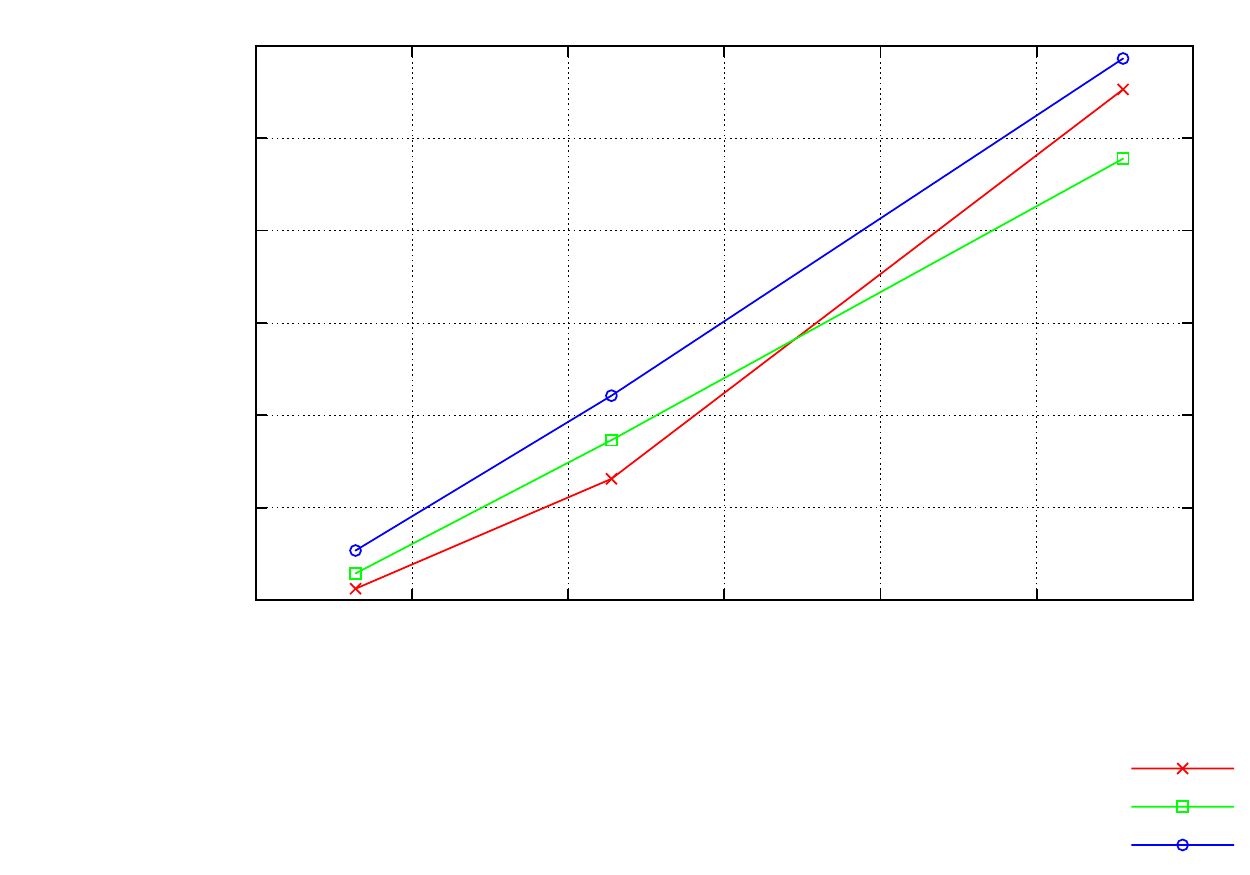}}%
    \gplfronttext
  \end{picture}%
\endgroup

%% file: cgm-psum-plot-p4-tex.tex
\begin{figure}[ht]
	\begin{center}
	\resizebox{!}{0.4\textheight}{
		\input{cgm-psum-p4-tex}
	}
	\\
$\frac{n}{v} = 5120000$
,  $P = 4$
,  k = 4

	\caption{CGMLib Prefix Sum PEMS2 P=4}
	\label{cgm-psum-p4}
	\end{center}
\end{figure}

%% file: cgm-psum-p4-tex.tex
\begingroup
  \makeatletter
  \providecommand\color[2][]{%
    \GenericError{(gnuplot) \space\space\space\@spaces}{%
      Package color not loaded in conjunction with
      terminal option `colourtext'%
    }{See the gnuplot documentation for explanation.%
    }{Either use 'blacktext' in gnuplot or load the package
      color.sty in LaTeX.}%
    \renewcommand\color[2][]{}%
  }%
  \providecommand\includegraphics[2][]{%
    \GenericError{(gnuplot) \space\space\space\@spaces}{%
      Package graphicx or graphics not loaded%
    }{See the gnuplot documentation for explanation.%
    }{The gnuplot epslatex terminal needs graphicx.sty or graphics.sty.}%
    \renewcommand\includegraphics[2][]{}%
  }%
  \providecommand\rotatebox[2]{#2}%
  \@ifundefined{ifGPcolor}{%
    \newif\ifGPcolor
    \GPcolortrue
  }{}%
  \@ifundefined{ifGPblacktext}{%
    \newif\ifGPblacktext
    \GPblacktexttrue
  }{}%
  \let\gplgaddtomacro\g@addto@macro
  \gdef\gplbacktext{}%
  \gdef\gplfronttext{}%
  \makeatother
  \ifGPblacktext
    \def\colorrgb#1{}%
    \def\colorgray#1{}%
  \else
    \ifGPcolor
      \def\colorrgb#1{\color[rgb]{#1}}%
      \def\colorgray#1{\color[gray]{#1}}%
      \expandafter\def\csname LTw\endcsname{\color{white}}%
      \expandafter\def\csname LTb\endcsname{\color{black}}%
      \expandafter\def\csname LTa\endcsname{\color{black}}%
      \expandafter\def\csname LT0\endcsname{\color[rgb]{1,0,0}}%
      \expandafter\def\csname LT1\endcsname{\color[rgb]{0,1,0}}%
      \expandafter\def\csname LT2\endcsname{\color[rgb]{0,0,1}}%
      \expandafter\def\csname LT3\endcsname{\color[rgb]{1,0,1}}%
      \expandafter\def\csname LT4\endcsname{\color[rgb]{0,1,1}}%
      \expandafter\def\csname LT5\endcsname{\color[rgb]{1,1,0}}%
      \expandafter\def\csname LT6\endcsname{\color[rgb]{0,0,0}}%
      \expandafter\def\csname LT7\endcsname{\color[rgb]{1,0.3,0}}%
      \expandafter\def\csname LT8\endcsname{\color[rgb]{0.5,0.5,0.5}}%
    \else
      \def\colorrgb#1{\color{black}}%
      \def\colorgray#1{\color[gray]{#1}}%
      \expandafter\def\csname LTw\endcsname{\color{white}}%
      \expandafter\def\csname LTb\endcsname{\color{black}}%
      \expandafter\def\csname LTa\endcsname{\color{black}}%
      \expandafter\def\csname LT0\endcsname{\color{black}}%
      \expandafter\def\csname LT1\endcsname{\color{black}}%
      \expandafter\def\csname LT2\endcsname{\color{black}}%
      \expandafter\def\csname LT3\endcsname{\color{black}}%
      \expandafter\def\csname LT4\endcsname{\color{black}}%
      \expandafter\def\csname LT5\endcsname{\color{black}}%
      \expandafter\def\csname LT6\endcsname{\color{black}}%
      \expandafter\def\csname LT7\endcsname{\color{black}}%
      \expandafter\def\csname LT8\endcsname{\color{black}}%
    \fi
  \fi
  \setlength{\unitlength}{0.0500bp}%
  \begin{picture}(7200.00,5040.00)%
    \gplgaddtomacro\gplbacktext{%
      \csname LTb\endcsname%
      \put(1342,1584){\makebox(0,0)[r]{\strut{} 0}}%
      \csname LTb\endcsname%
      \put(1342,1939){\makebox(0,0)[r]{\strut{} 1000}}%
      \csname LTb\endcsname%
      \put(1342,2293){\makebox(0,0)[r]{\strut{} 2000}}%
      \csname LTb\endcsname%
      \put(1342,2648){\makebox(0,0)[r]{\strut{} 3000}}%
      \csname LTb\endcsname%
      \put(1342,3003){\makebox(0,0)[r]{\strut{} 4000}}%
      \csname LTb\endcsname%
      \put(1342,3357){\makebox(0,0)[r]{\strut{} 5000}}%
      \csname LTb\endcsname%
      \put(1342,3712){\makebox(0,0)[r]{\strut{} 6000}}%
      \csname LTb\endcsname%
      \put(1342,4067){\makebox(0,0)[r]{\strut{} 7000}}%
      \csname LTb\endcsname%
      \put(1342,4421){\makebox(0,0)[r]{\strut{} 8000}}%
      \csname LTb\endcsname%
      \put(1342,4776){\makebox(0,0)[r]{\strut{} 9000}}%
      \csname LTb\endcsname%
      \put(1474,1364){\makebox(0,0){\strut{} 0}}%
      \csname LTb\endcsname%
      \put(2074,1364){\makebox(0,0){\strut{} 100}}%
      \csname LTb\endcsname%
      \put(2673,1364){\makebox(0,0){\strut{} 200}}%
      \csname LTb\endcsname%
      \put(3273,1364){\makebox(0,0){\strut{} 300}}%
      \csname LTb\endcsname%
      \put(3872,1364){\makebox(0,0){\strut{} 400}}%
      \csname LTb\endcsname%
      \put(4472,1364){\makebox(0,0){\strut{} 500}}%
      \csname LTb\endcsname%
      \put(5071,1364){\makebox(0,0){\strut{} 600}}%
      \csname LTb\endcsname%
      \put(5671,1364){\makebox(0,0){\strut{} 700}}%
      \csname LTb\endcsname%
      \put(6270,1364){\makebox(0,0){\strut{} 800}}%
      \csname LTb\endcsname%
      \put(6870,1364){\makebox(0,0){\strut{} 900}}%
      \put(440,3180){\rotatebox{90}{\makebox(0,0){\strut{}Seconds}}}%
      \put(4172,1034){\makebox(0,0){\strut{}$n$ (millions)}}%
    }%
    \gplgaddtomacro\gplfronttext{%
      \csname LTb\endcsname%
      \put(6385,613){\makebox(0,0)[r]{\strut{}CGMLib Prefix Sum PEMS2 (mmap) P=4}}%
      \csname LTb\endcsname%
      \put(6385,393){\makebox(0,0)[r]{\strut{}CGMLib Prefix Sum PEMS2 (stxxl-file) P=4}}%
      \csname LTb\endcsname%
      \put(6385,173){\makebox(0,0)[r]{\strut{}CGMLib Prefix Sum PEMS2 (unix) P=4}}%
    }%
    \gplbacktext
    \put(0,0){\includegraphics{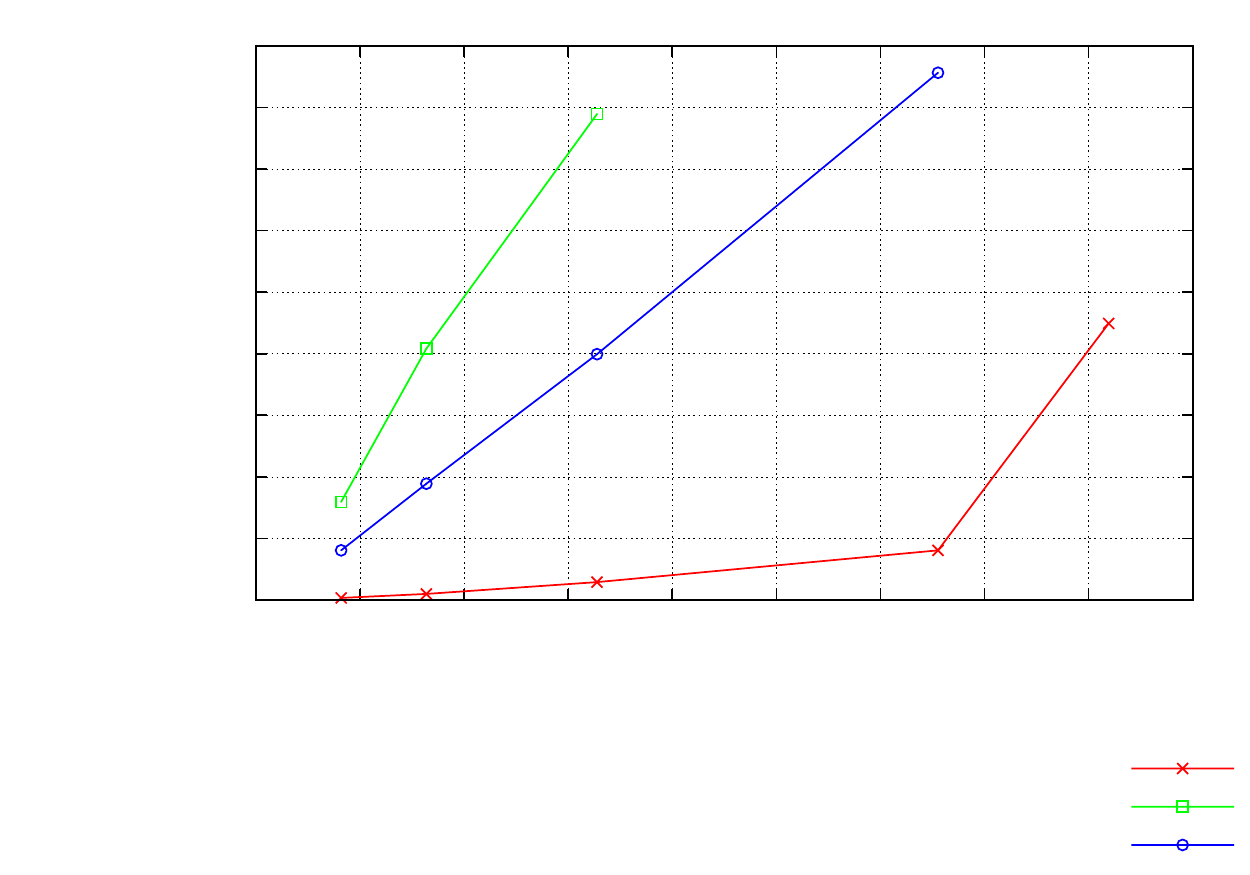}}%
    \gplfronttext
  \end{picture}%
\endgroup

%% file: cgm-etour-plot-tex.tex
\begin{figure}[ht]
	\begin{center}
	\resizebox{!}{0.4\textheight}{
		\input{cgm-etour-tex}
	}
	\\
v = 16
,  $P = 4$
,  $\frac{v}{P} = 4$
,  k = 4
,  io = mmap

	\caption{CGMLib Euler Tour}
	\label{cgm-etour}
	\end{center}
\end{figure}

%% file: cgm-etour-tex.tex
\begingroup
  \makeatletter
  \providecommand\color[2][]{%
    \GenericError{(gnuplot) \space\space\space\@spaces}{%
      Package color not loaded in conjunction with
      terminal option `colourtext'%
    }{See the gnuplot documentation for explanation.%
    }{Either use 'blacktext' in gnuplot or load the package
      color.sty in LaTeX.}%
    \renewcommand\color[2][]{}%
  }%
  \providecommand\includegraphics[2][]{%
    \GenericError{(gnuplot) \space\space\space\@spaces}{%
      Package graphicx or graphics not loaded%
    }{See the gnuplot documentation for explanation.%
    }{The gnuplot epslatex terminal needs graphicx.sty or graphics.sty.}%
    \renewcommand\includegraphics[2][]{}%
  }%
  \providecommand\rotatebox[2]{#2}%
  \@ifundefined{ifGPcolor}{%
    \newif\ifGPcolor
    \GPcolortrue
  }{}%
  \@ifundefined{ifGPblacktext}{%
    \newif\ifGPblacktext
    \GPblacktexttrue
  }{}%
  \let\gplgaddtomacro\g@addto@macro
  \gdef\gplbacktext{}%
  \gdef\gplfronttext{}%
  \makeatother
  \ifGPblacktext
    \def\colorrgb#1{}%
    \def\colorgray#1{}%
  \else
    \ifGPcolor
      \def\colorrgb#1{\color[rgb]{#1}}%
      \def\colorgray#1{\color[gray]{#1}}%
      \expandafter\def\csname LTw\endcsname{\color{white}}%
      \expandafter\def\csname LTb\endcsname{\color{black}}%
      \expandafter\def\csname LTa\endcsname{\color{black}}%
      \expandafter\def\csname LT0\endcsname{\color[rgb]{1,0,0}}%
      \expandafter\def\csname LT1\endcsname{\color[rgb]{0,1,0}}%
      \expandafter\def\csname LT2\endcsname{\color[rgb]{0,0,1}}%
      \expandafter\def\csname LT3\endcsname{\color[rgb]{1,0,1}}%
      \expandafter\def\csname LT4\endcsname{\color[rgb]{0,1,1}}%
      \expandafter\def\csname LT5\endcsname{\color[rgb]{1,1,0}}%
      \expandafter\def\csname LT6\endcsname{\color[rgb]{0,0,0}}%
      \expandafter\def\csname LT7\endcsname{\color[rgb]{1,0.3,0}}%
      \expandafter\def\csname LT8\endcsname{\color[rgb]{0.5,0.5,0.5}}%
    \else
      \def\colorrgb#1{\color{black}}%
      \def\colorgray#1{\color[gray]{#1}}%
      \expandafter\def\csname LTw\endcsname{\color{white}}%
      \expandafter\def\csname LTb\endcsname{\color{black}}%
      \expandafter\def\csname LTa\endcsname{\color{black}}%
      \expandafter\def\csname LT0\endcsname{\color{black}}%
      \expandafter\def\csname LT1\endcsname{\color{black}}%
      \expandafter\def\csname LT2\endcsname{\color{black}}%
      \expandafter\def\csname LT3\endcsname{\color{black}}%
      \expandafter\def\csname LT4\endcsname{\color{black}}%
      \expandafter\def\csname LT5\endcsname{\color{black}}%
      \expandafter\def\csname LT6\endcsname{\color{black}}%
      \expandafter\def\csname LT7\endcsname{\color{black}}%
      \expandafter\def\csname LT8\endcsname{\color{black}}%
    \fi
  \fi
  \setlength{\unitlength}{0.0500bp}%
  \begin{picture}(7200.00,5040.00)%
    \gplgaddtomacro\gplbacktext{%
      \csname LTb\endcsname%
      \put(1210,1144){\makebox(0,0)[r]{\strut{} 0}}%
      \csname LTb\endcsname%
      \put(1210,1548){\makebox(0,0)[r]{\strut{} 100}}%
      \csname LTb\endcsname%
      \put(1210,1951){\makebox(0,0)[r]{\strut{} 200}}%
      \csname LTb\endcsname%
      \put(1210,2355){\makebox(0,0)[r]{\strut{} 300}}%
      \csname LTb\endcsname%
      \put(1210,2758){\makebox(0,0)[r]{\strut{} 400}}%
      \csname LTb\endcsname%
      \put(1210,3162){\makebox(0,0)[r]{\strut{} 500}}%
      \csname LTb\endcsname%
      \put(1210,3565){\makebox(0,0)[r]{\strut{} 600}}%
      \csname LTb\endcsname%
      \put(1210,3969){\makebox(0,0)[r]{\strut{} 700}}%
      \csname LTb\endcsname%
      \put(1210,4372){\makebox(0,0)[r]{\strut{} 800}}%
      \csname LTb\endcsname%
      \put(1210,4776){\makebox(0,0)[r]{\strut{} 900}}%
      \csname LTb\endcsname%
      \put(1342,924){\makebox(0,0){\strut{}\small$0{\cdot}10^{0}$}}%
      \csname LTb\endcsname%
      \put(2033,924){\makebox(0,0){\strut{}\small$5{\cdot}10^{5}$}}%
      \csname LTb\endcsname%
      \put(2724,924){\makebox(0,0){\strut{}\small$1{\cdot}10^{6}$}}%
      \csname LTb\endcsname%
      \put(3415,924){\makebox(0,0){\strut{}\small$2{\cdot}10^{6}$}}%
      \csname LTb\endcsname%
      \put(4106,924){\makebox(0,0){\strut{}\small$2{\cdot}10^{6}$}}%
      \csname LTb\endcsname%
      \put(4797,924){\makebox(0,0){\strut{}\small$2{\cdot}10^{6}$}}%
      \csname LTb\endcsname%
      \put(5488,924){\makebox(0,0){\strut{}\small$3{\cdot}10^{6}$}}%
      \csname LTb\endcsname%
      \put(6179,924){\makebox(0,0){\strut{}\small$4{\cdot}10^{6}$}}%
      \csname LTb\endcsname%
      \put(6870,924){\makebox(0,0){\strut{}\small$4{\cdot}10^{6}$}}%
      \put(440,2960){\rotatebox{90}{\makebox(0,0){\strut{}Seconds}}}%
      \put(4106,594){\makebox(0,0){\strut{}$n$}}%
    }%
    \gplgaddtomacro\gplfronttext{%
      \csname LTb\endcsname%
      \put(5527,173){\makebox(0,0)[r]{\strut{}CGMLib Euler Tour (mmap) P=4}}%
    }%
    \gplbacktext
    \put(0,0){\includegraphics{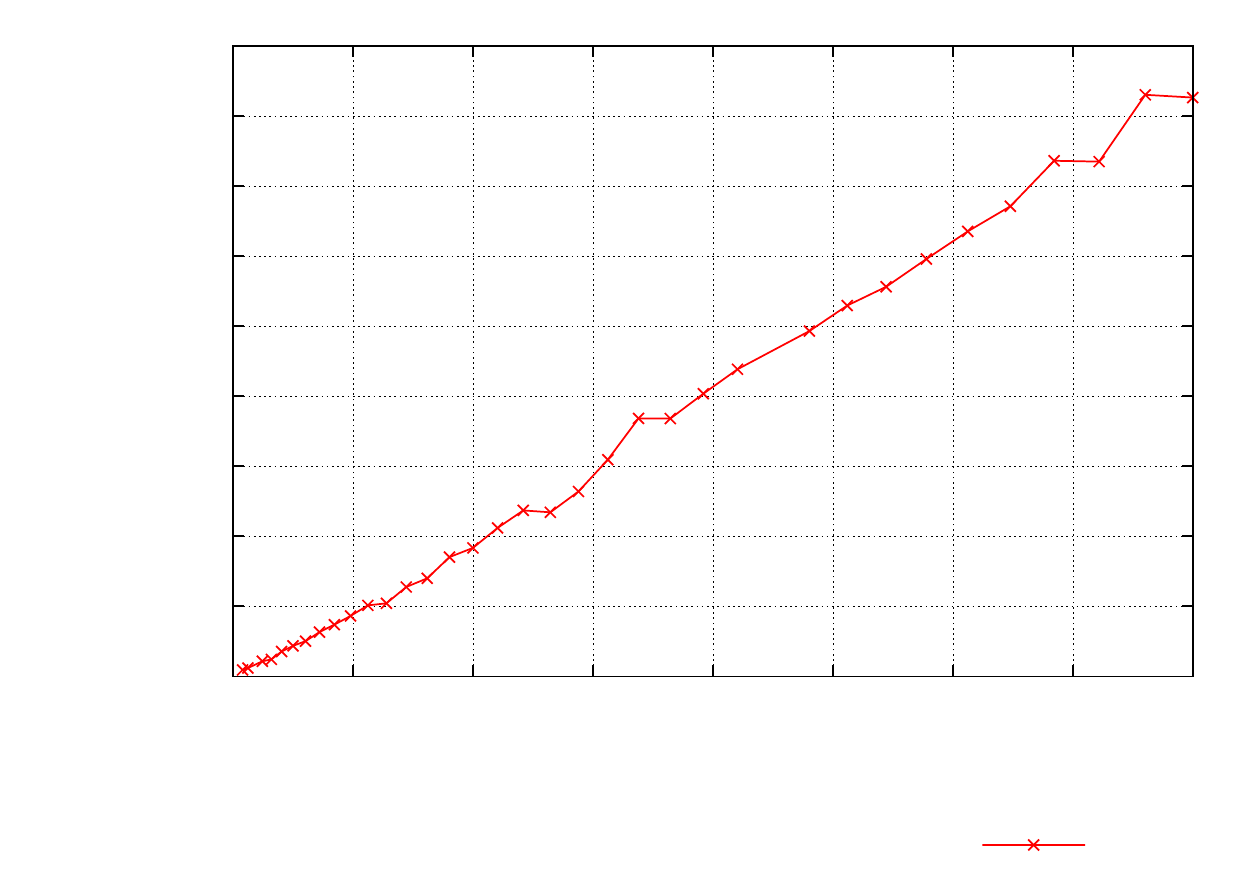}}%
    \gplfronttext
  \end{picture}%
\endgroup

%% file: ext-tex.tex
\begin{figure}[ht]
	\begin{center}
	\resizebox{!}{0.4\textheight}{
		\input{ext3vs4-tex}
	}
	\\
n = 128000000
,  v = 32
,  $P = 4$
,  $\frac{v}{P} = 8$
,  k = 4
,  io = unix

	\caption{ext3 vs ext4}
	\label{ext3vs4}
	\end{center}
\end{figure}

%% file: ext3vs4-tex.tex
\begingroup
  \makeatletter
  \providecommand\color[2][]{%
    \GenericError{(gnuplot) \space\space\space\@spaces}{%
      Package color not loaded in conjunction with
      terminal option `colourtext'%
    }{See the gnuplot documentation for explanation.%
    }{Either use 'blacktext' in gnuplot or load the package
      color.sty in LaTeX.}%
    \renewcommand\color[2][]{}%
  }%
  \providecommand\includegraphics[2][]{%
    \GenericError{(gnuplot) \space\space\space\@spaces}{%
      Package graphicx or graphics not loaded%
    }{See the gnuplot documentation for explanation.%
    }{The gnuplot epslatex terminal needs graphicx.sty or graphics.sty.}%
    \renewcommand\includegraphics[2][]{}%
  }%
  \providecommand\rotatebox[2]{#2}%
  \@ifundefined{ifGPcolor}{%
    \newif\ifGPcolor
    \GPcolortrue
  }{}%
  \@ifundefined{ifGPblacktext}{%
    \newif\ifGPblacktext
    \GPblacktexttrue
  }{}%
  \let\gplgaddtomacro\g@addto@macro
  \gdef\gplbacktext{}%
  \gdef\gplfronttext{}%
  \makeatother
  \ifGPblacktext
    \def\colorrgb#1{}%
    \def\colorgray#1{}%
  \else
    \ifGPcolor
      \def\colorrgb#1{\color[rgb]{#1}}%
      \def\colorgray#1{\color[gray]{#1}}%
      \expandafter\def\csname LTw\endcsname{\color{white}}%
      \expandafter\def\csname LTb\endcsname{\color{black}}%
      \expandafter\def\csname LTa\endcsname{\color{black}}%
      \expandafter\def\csname LT0\endcsname{\color[rgb]{1,0,0}}%
      \expandafter\def\csname LT1\endcsname{\color[rgb]{0,1,0}}%
      \expandafter\def\csname LT2\endcsname{\color[rgb]{0,0,1}}%
      \expandafter\def\csname LT3\endcsname{\color[rgb]{1,0,1}}%
      \expandafter\def\csname LT4\endcsname{\color[rgb]{0,1,1}}%
      \expandafter\def\csname LT5\endcsname{\color[rgb]{1,1,0}}%
      \expandafter\def\csname LT6\endcsname{\color[rgb]{0,0,0}}%
      \expandafter\def\csname LT7\endcsname{\color[rgb]{1,0.3,0}}%
      \expandafter\def\csname LT8\endcsname{\color[rgb]{0.5,0.5,0.5}}%
    \else
      \def\colorrgb#1{\color{black}}%
      \def\colorgray#1{\color[gray]{#1}}%
      \expandafter\def\csname LTw\endcsname{\color{white}}%
      \expandafter\def\csname LTb\endcsname{\color{black}}%
      \expandafter\def\csname LTa\endcsname{\color{black}}%
      \expandafter\def\csname LT0\endcsname{\color{black}}%
      \expandafter\def\csname LT1\endcsname{\color{black}}%
      \expandafter\def\csname LT2\endcsname{\color{black}}%
      \expandafter\def\csname LT3\endcsname{\color{black}}%
      \expandafter\def\csname LT4\endcsname{\color{black}}%
      \expandafter\def\csname LT5\endcsname{\color{black}}%
      \expandafter\def\csname LT6\endcsname{\color{black}}%
      \expandafter\def\csname LT7\endcsname{\color{black}}%
      \expandafter\def\csname LT8\endcsname{\color{black}}%
    \fi
  \fi
  \setlength{\unitlength}{0.0500bp}%
  \begin{picture}(7200.00,5040.00)%
    \gplgaddtomacro\gplbacktext{%
      \csname LTb\endcsname%
      \put(1210,1144){\makebox(0,0)[r]{\strut{} 50}}%
      \csname LTb\endcsname%
      \put(1210,1548){\makebox(0,0)[r]{\strut{} 100}}%
      \csname LTb\endcsname%
      \put(1210,1951){\makebox(0,0)[r]{\strut{} 150}}%
      \csname LTb\endcsname%
      \put(1210,2355){\makebox(0,0)[r]{\strut{} 200}}%
      \csname LTb\endcsname%
      \put(1210,2758){\makebox(0,0)[r]{\strut{} 250}}%
      \csname LTb\endcsname%
      \put(1210,3162){\makebox(0,0)[r]{\strut{} 300}}%
      \csname LTb\endcsname%
      \put(1210,3565){\makebox(0,0)[r]{\strut{} 350}}%
      \csname LTb\endcsname%
      \put(1210,3969){\makebox(0,0)[r]{\strut{} 400}}%
      \csname LTb\endcsname%
      \put(1210,4372){\makebox(0,0)[r]{\strut{} 450}}%
      \csname LTb\endcsname%
      \put(1210,4776){\makebox(0,0)[r]{\strut{} 500}}%
      \csname LTb\endcsname%
      \put(1342,924){\makebox(0,0){\strut{} 0}}%
      \csname LTb\endcsname%
      \put(1956,924){\makebox(0,0){\strut{} 500}}%
      \csname LTb\endcsname%
      \put(2570,924){\makebox(0,0){\strut{} 1000}}%
      \csname LTb\endcsname%
      \put(3185,924){\makebox(0,0){\strut{} 1500}}%
      \csname LTb\endcsname%
      \put(3799,924){\makebox(0,0){\strut{} 2000}}%
      \csname LTb\endcsname%
      \put(4413,924){\makebox(0,0){\strut{} 2500}}%
      \csname LTb\endcsname%
      \put(5027,924){\makebox(0,0){\strut{} 3000}}%
      \csname LTb\endcsname%
      \put(5642,924){\makebox(0,0){\strut{} 3500}}%
      \csname LTb\endcsname%
      \put(6256,924){\makebox(0,0){\strut{} 4000}}%
      \csname LTb\endcsname%
      \put(6870,924){\makebox(0,0){\strut{} 4500}}%
      \put(440,2960){\rotatebox{90}{\makebox(0,0){\strut{}Seconds}}}%
      \put(4106,594){\makebox(0,0){\strut{}$\mu$}}%
    }%
    \gplgaddtomacro\gplfronttext{%
      \csname LTb\endcsname%
      \put(3251,173){\makebox(0,0)[r]{\strut{}PEMS2 (ext3)}}%
      \csname LTb\endcsname%
      \put(5690,173){\makebox(0,0)[r]{\strut{}PEMS2 (ext4)}}%
    }%
    \gplbacktext
    \put(0,0){\includegraphics{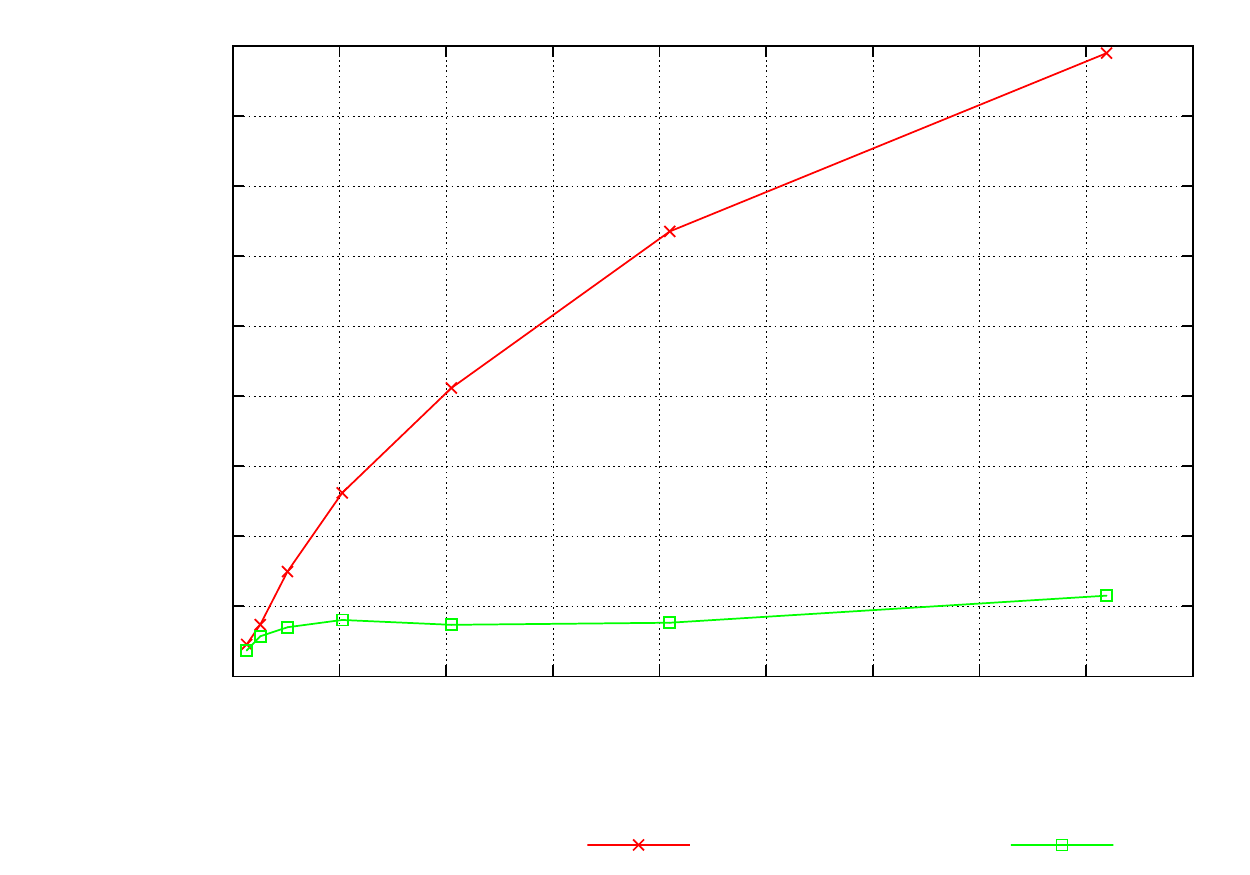}}%
    \gplfronttext
  \end{picture}%
\endgroup